\documentclass{report}
\usepackage{cuthesis}

\usepackage[square,comma,numbers,sort&compress]{natbib}
\usepackage{amssymb}
\usepackage{rotating}
\usepackage{tensor}
\usepackage{aas_macros_pk}

\usepackage{color}
\usepackage{times}
\usepackage{graphicx}
\usepackage{float}
\usepackage{alltt}
\usepackage{listings}
\usepackage{enumerate}
\usepackage[T1]{fontenc}
\usepackage{ae}
\usepackage{subfigure}
\usepackage{amsmath}

\usepackage{longtable}
\setlongtables

\usepackage{hyperref}
\hypersetup{backref=true, colorlinks=true}

\def\be{\begin{equation}}
\def\ee{\end{equation}}
\def\beg{\begin{align}}
\def\eeg{\end{align}}
\def\bi{\begin{itemize}}
\def\ei{\end{itemize}}
\def\ben{\begin{enumerate}[1.]}
\def\een{\end{enumerate}}
\def\i{\item}

\newcommand{\abs}[1]{\left| {#1} \right|}

\newcommand{\del}[2]{\frac{\partial #1}{\partial #2}}

\newcommand{\bo}{\raise-1mm\hbox{\Large$\Box$}}

\newcommand{\cc}[1]{{#1}^{\ast}}

\newcommand{\sci}[2]{#1 \times 10^{#2}}
\newcommand{\rthz}{\mathrm{Hz}^{-\frac{1}{2}}}
\newcommand{\hrss}{h_{\mathrm{rss}}}
\newcommand{\hrssf}{h_{\mathrm{rss}}^{50\%}}
\newcommand{\hrssn}{h_{\mathrm{rss}}^{90\%}}

\newcommand{\egwn}{E_{\mathrm{GW}}^{90\%}}
\newcommand{\egw}{E_{\mathrm{GW}}}
\newcommand{\eem}{E_{\mathrm{EM}}}
\newcommand{\lten}{\mathrm{L}_{10}}

\newcommand{\solarmass}{M_{\odot}c^2}
\newcommand{\msun}{M_{\odot}c^2}
\newcommand{\msunonly}{M_{\odot}}

\newcommand{\hp}{h_{+}}
\newcommand{\hc}{h_{\times }}
\newcommand{\fp}{F_{+}(\theta, \phi, \psi)}
\newcommand{\fc}{F_{\times}(\theta, \phi, \psi) }

\renewcommand{\thesisauthor}{Peter M. Kalmus}
\renewcommand{\thesisyear}{2008}

\renewcommand{\thesistitle}{Search for Gravitational Wave Bursts from\\ Soft Gamma Repeaters}
\renewcommand{\thesisdedication}{To my family}
\renewcommand{\advisorone}{Szabolcs M\'{a}rka}
\renewcommand{\advisoronedep}{Department of Physics}
\renewcommand{\advisortwo}{William Zajc}
\renewcommand{\advisortwodep}{Department of Physics}
\renewcommand{\advisorthree}{Allan Blaer}
\renewcommand{\advisorthreedep}{Department of Physics}
\renewcommand{\advisorfour}{Caleb Scharf}
\renewcommand{\advisorfourdep}{Department of Astronomy}
\renewcommand{\advisorfive}{David Schiminovich}
\renewcommand{\advisorfivedep}{Department of Astronomy}

\renewcommand{\depchair}{Andrew J. Millis}
\renewcommand{\thesisdedication}{ To my parents, who never stop believing in me \vspace{6mm} \\ \indent  And to Sharon, for insight in the dark and celebrating when the light shines down}

\begin{document}

\thesistitlepage

\thesiscopyrightpage

\begin{thesisabstract}

We present the results of a LIGO search for short-duration
gravitational waves associated with soft gamma repeater (SGR)
bursts, and a method for calibrating gravitational wave detectors
via photon actuators.

Photon calibrators provide an independent calibration of LIGO's
three gravitational wave detectors. Their nominal $2\sigma$
confidence error bars are currently estimated to be $\sim$3\%. The
photon calibrators have provided a valuable check on the official
calibration, uncovering problems that may otherwise have gone
unnoticed.

We also present the first gravitational wave search sensitive to
neutron star $f$-modes, usually considered the most efficient
gravitational wave emitting modes. We find no evidence of
gravitational waves associated with any SGR burst in a sample
consisting of the 2004 December 27 giant flare from SGR 1806$-20$
and 190 lesser events from SGR 1806$-20$ and SGR 1900+14 which
occurred during the first year of LIGO's fifth science run.
Gravitational wave strain upper limits and model-dependent
gravitational wave emission energy upper limits are estimated for
individual bursts using a variety of simulated waveforms. The
unprecedented sensitivity of the detectors allows us to set the most
stringent limits on transient gravitational wave amplitudes
published to date. We find upper limit estimates on the
model-dependent isotropic gravitational wave emission energies (at a
nominal distance of 10\,kpc) between $\sci{3}{45}$ and $\sci{9}{52}$
erg depending on waveform type, detector antenna factors and noise
characteristics at the time of the burst.  These upper limits are
within the theoretically predicted range of some SGR models.

Finally, we propose a new method which extends the initial SGR burst search, exploring the possibility that SGR sources emit similarly in gravitational waves from burst to burst by ``stacking" potential gravitational wave signals.  We show that gains in gravitational wave energy sensitivity of $N^{1/2}$ are  possible, where $N$ is the number of stacked SGR bursts.   Estimated sensitivities for a mock search for gravitational waves from the 2006 March 29 storm from SGR 1900+14 are presented for two stacking scenarios: the ``fluence-weighted" scenario and the ``flat" (unweighted) scenario.

\end{thesisabstract}

\tableofcontents

\listoftables

\listoffigures

\begin{thesisacknowledgments}  {

What a privilege it is to do science!  I'm grateful to live in a time and a place where I could have this opportunity.  Science stands not only on
the shoulders of giants, but of all people.

Hundreds of folks in LIGO and at Columbia have
helped me fulfill this dream in one way or another.  I'm especially
grateful to two groups: the people who worked to make the
LIGO detectors sensitive enough to allow this work to be interesting; and the Burst Working Group, whose members forced me to defend my ideas and
ultimately supported my work.   I am also grateful for the support of the United
States National Science Foundation under cooperative agreements
PHY-04-57528/PHY-07-38147/PHY-07-57982, the LIGO Scientific Collaboration, Columbia University, and the California Institute of Technology.

A handful of individuals have been especially helpful. Foremost
among them is my advisor, Szabolcs~M\'arka, who guided me through the perilous phase transitions needed to become a scientist. Szabi rarely lost patience explaining something to me for
the $N^{\mathrm{th}}$ time, and provided a foundation on which to
do good work.  I also owe thanks to Sergey Klimenko at the
University of Florida, who was a kind of second advisor. Sergey generously taught me much of what I know about
gravitational wave burst analysis.   I am grateful to Kipp Cannon
and Ben Owen, who formally reviewed the SGR search presented in Chapter\,\ref{chapter:search} on behalf of the LSC, a huge
job which they performed efficiently.    The work in Chapter\,\ref{chapter:search} relied on S. Barthelmy, N. Gehrels, K. C. Hurley, G. Lichti, S. Mereghetti, D. Palmer, D. Smith, and the Konus-Wind team for electromagnetic SGR burst data.  It was a pleasure working with these satellite folks, and I look forward to further collaboration.   I thank Yoichi
Aso for comments on Chapter\,\ref{chapter:detector}, and Drew Keppel for comments on Chapter\,\ref{chapter:gw}.  I am grateful to Andy Millis and Amber Miller at Columbia for helping me find my way back to physics.  I am also
grateful to Lyman Page at Princeton, for showing me that physics is
fun, for encouraging me to return to graduate school, and for his benediction as I started this segment of my life: ``Do good work.''  Finally, and above all, I thank Sharon for putting up with this project as we
started our family.

 }
\end{thesisacknowledgments}

\clearpage

\thesisdedicationpage
 \clearpage

\chapter{The Universe Through New Eyes}

We are poised to enter the age of gravitational wave astrophysics.
Gravitational waves are perturbations in the spacetime metric propagating at the speed of light, predicted by
Einstein in 1914 by his theory of general
relativity\,\cite{einstein1916}.  In GR, massive objects cause a
curvature distortion in four-dimensional spacetime, and
gravitational force is our perception of something trying to follow
a geodesic in the curved spacetime.  When a massive object moves it
drags the distortion with it, sending out ripples in spacetime.  Due
to the weakness of the gravitational force, gravitational waves have
not yet been directly detected.  An attempt to directly detect
gravitational waves, and to extract science from upper limits on
gravitational wave emission, is the subject of this work.

\begin{figure}[!h]
\begin{center}
\includegraphics[angle=0,width=90mm, clip=false]{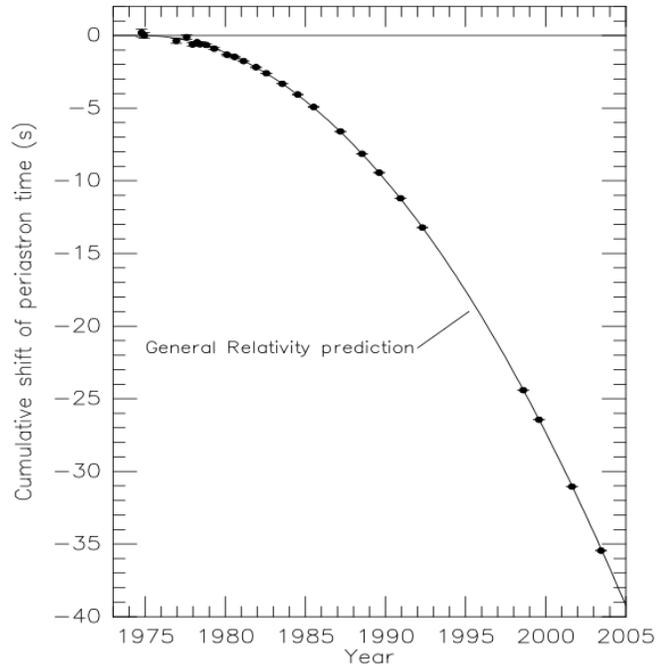}
\caption[Energy loss from gravitational wave emission in PSR
1913+16] { Accumulated shift in the times of periastron in the PSR
1913+16 system, relative to an assumed orbit with constant period.
The parabolic curve represents GR's prediction for energy losses
from gravitational radiation.  (Figure from\,\cite{weisberg04}.) } \label{fig:psr1913}
\end{center}
\end{figure}

We do have indirect evidence for the existence of gravitational
waves.  In a beautiful observational confirmation of Einstein's
prediction, R. Hulse and J. Taylor measured the time derivative of
the orbital period of PSR 1913+16, a pulsar in a binary system with
a second neutron star, by observing the Doppler shifting of the
pulsar signals due to orbital velocity.  They confirmed that the
orbital velocity increase due to loss of gravitational potential
energy matches what would be expected in general relativity if the
system were losing energy due to gravitational wave emission to
within experimental precision, better than 0.5\%
(Figure\,\ref{fig:psr1913})\,\cite{hulse1975, taylor79, taylor82,
taylor94, will06}. More recent observations place the agreement to
within about 0.2\%\,\cite{weisberg04}.  These stars will collide and
merge in about 300 million years due to losses from gravitational
wave emission.

Because of this indirect detection of gravitational waves, and also
because of other experimental verifications of general relativity\,\cite{will06}, we are confident that gravitational waves do
exist.  One of the biggest goals in physics today is to directly
detect gravitational waves in order to see what the information they
carry can teach us about the universe.  The science that will come
from such a discovery is thrilling to contemplate.

For the first time, the LIGO Scientific Collaboration has gathered
more than a year of data with laser interferometric gravitational
wave detectors with sensitivities and frequency bands such that a
detection might not require an extraordinary event\,\cite{S5}. By
2009 planned upgrades on these detectors are expected to improve
amplitude sensitivity by a factor of $\sim 3$, and additional upgrades planned
to be completed by 2013 will give us advanced detectors about 10
times more sensitive to gravitational wave strain than current
detectors, and therefore with about 1000 times the reach in terms of
astrophysical volume\,\cite{aligosite}. Meanwhile, plans for both
new and existing detectors in other countries will strengthen the
emerging global interferometeric gravitational wave detector network
(see Table\,\ref{table:globaldetectors}).  Global networks 
reduce the false detection rate and improve the source sky
localization.

\begin{table}[h]
\begin{center}
\caption[Existing or planned gravitational wave
interferometers]{Interferometric gravitational wave detectors which
exist already or are planned, with expected dates of operation.  }
\begin{tabular}{r|rrr}
 \hline \textbf{detector} & \textbf{location} &  \textbf{date} & \textbf{arm length [m]} \\
   \hline \hline

LIGO \cite{ligoWeb} H1 & Hanford, WA & operational & 4000 \\
 L1 & Livingston, LA & operational & 4000 \\
H2 & Hanford, WA & operational & 2000 \\
 \hline

GEO600 \cite{geo} & Hannover, Germany & operational & 600  \\
 Virgo \cite{virgo} & Cascina, Italy & operational & 3000  \\
TAMA300 \cite{tama} & Mitaka, Japan & operational &  300  \\
LCGT \cite{lcgt} & Kamioka Mine, Japan & not clear &  3000  \\
LISA \cite{lisaNasa, lisaAei} & space & $\sim$2018 & $\sci{5}{9}$  \\  

\end{tabular}
\label{table:globaldetectors}
\end{center}
\end{table}

Lack of a detection with the network of advanced detectors would be
very interesting.  The more likely scenario of routine detection may
revolutionize astrophysics. Each new portion of the electromagnetic
spectrum, when opened to astrophysical observation, has resulted in
unimagined discoveries. Observing the universe through gravitational
radiation ought to be even more radical than looking with a new
color of light.  Table\,\ref{table:emVsgw} compares
gravitational waves to electromagnetic waves as carriers of
astrophysical information.

\begin{table}[h]
\begin{center}
\caption[Astrophysics probed with gravitational wave vs.
electromagnetic waves]{Comparison of gravitational waves and
electromagnetic waves as carriers of astrophysical information. This
comparison motivates our opinion that looking at the universe as
portrayed in gravitational radiation is more radical than looking
with a new color of light.}
\begin{tabular}{p{1.3in}|p{2in}|p{2in}}
 \hline
 \textbf{characteristic} & \textbf{gravitational wave} &  \textbf{electromagnetic waves} \\
   \hline \hline

 medium & spacetime & space \\
\hline
 source & coherent quadrupole motions of black holes, stars, galaxies,
 etc. & incoherent dipole motions of electrons, other charged particles
 \\
\hline
 interaction with intervening matter & insignificant; could probe supernova centers, very early universe, etc. & absorbed and scattered by intervening
 matter; carry information from outer layer of objects only. \\
\hline
 frequency & $< 10^{4}$\,Hz & $>10^7$\,Hz \\
\hline
 detectors & omnidirectional & unidirectional \\
\hline
 quantum mechanics & spin 2 graviton & spin 1 photon \\
\hline
\end{tabular}
\label{table:emVsgw}
\end{center}
\end{table}

We have contributed to this massive effort in both the hardware and
data analysis domains.  We have improved the LIGO gravitational wave
detectors through advancement of a calibration technique that uses
photons to drive the interferometer test masses. In the course of
this work we discovered significant discrepancies in measurements of
the response function magnitude with the official calibration via
coil actuators, which ultimately led to an improved understanding of
the detector calibration.  We also used the
photon calibrators to help uncover a significant
error in the official timing calibration.

We have also developed and used novel data analysis techniques to
search LIGO data for gravitational wave bursts associated with soft
gamma repeaters (SGRs), a bizarre and enigmatic class of astrophysical
sources and a promising source for gravitational waves. Soft gamma
repeaters sporadically emit short energetic bursts of soft
gamma rays.  The bursts may be driven by violent interaction between
the most intense magnetic fields known in the universe and the solid
crust of a neutron star\,\cite{thompson95, schwartz05}. These
catastrophic events could excite the star's nonradial
modes\,\cite{andersson97, pacheco98, ioka01} which are damped via
gravitational wave emission\,\cite{andersson02, horvath05,
pacheco98, ioka01}. We have performed an electromagnetically
triggered search for gravitational waves associated with SGR burst
events in a sample which includes 214 bursts from the first year of
LIGO's fifth science run (S5y1) and the 2004 December 27 giant flare
from SGR 1806--20. This is the first search sensitive to
$f$-modes, usually considered the most efficient gravitational wave
emitters\,\cite{andersson97}. The unprecedented sensitivity of the
detectors and of our analysis pipeline allows us to set the most
stringent limits on transient gravitational wave amplitudes
published to date. These upper limits already begin to constrain
some SGR models.   

There is a possibility that potential gravitational wave emission from SGR bursts is similar from burst to burst.  We have also developed a search method which explores this possibility, by attempting to ``stack" potential gravitational wave  signals from multiple SGR bursts with the aim of digging deeper into the noise and increasing the probability of a detection.   This new method extends the individual burst search described in the last paragraph.

This thesis is organized as follows.

In Chapter\,\ref{chapter:gw} we give a brief theoretical description
of gravitational waves to facilitate understanding of gravitational
wave detectors and searches.  We also discuss some of the sources of
potentially detectable gravitational waves.

In Chapter\,\ref{chapter:detector} we briefly describe laser
interferometric gravitational wave detectors as implemented by LIGO,
the Laser Interferometric Gravitational Observatory.  While other
types of detectors exist, such as bar\cite{ju00} and
spherical\,\cite{coccia98} detectors, we believe that
interferometric detectors are currently most likely to yield
interesting science, and we limit our discussion to them. We focus
in particular on the methods used to measure relative length changes
in an interferometer that are 1000 times smaller than a proton
diameter, the sensitivity required to begin to have some chance of
detecting gravitational waves.

In Chapter\,\ref{chapter:pcal} we describe the LIGO photon
calibrator system, and compare photon calibrator measurements to
measurements made with the traditional coil calibrators.  These
measurements revealed a significant discrepancy with the official
detector response calibration magnitude. We also describe photon
calibrator measurements of the detector timing, which revealed a
significant discrepancy with the official timing calibration. We
discuss the future utility of photon calibrators in LIGO, suggesting
that they could be a candidate for the Advanced LIGO primary
calibration system.

In Chapter\,\ref{chapter:flare} we describe in detail a simple but
powerful general purpose coherent gravitational wave analysis
pipeline for externally triggered searches, the Flare pipeline.  We
present the Flare pipeline as a complete and automated analysis
system which, given inputs including data from one or two
interferometric gravitational wave detectors and information
describing one or more electromagnetic triggers, produces a
statement of detection or non-detection and upper limits for a
variety of simulation waveform types on gravitational wave strain
and isotropic gravitational wave emission energy from the source.

In Chapter\,\ref{chapter:validation} we describe careful
characterization and validation of the Flare pipeline which was
carried out before and during official review by the LIGO Scientific
Collaboration (LSC).

In Chapter\,\ref{chapter:sgrs} we describe soft gamma repeaters.
Knowledge of SGRs was used in designing a compelling search for
gravitational waves associated with their bursts. 

In Chapter\,\ref{chapter:search} we describe the search for
gravitational waves associated with individual SGR bursts using the
Flare pipeline.  We describe the sample of SGR triggers and give the
results of the search.

In Chapter\,\ref{chapter:stack} we describe the search for
gravitational waves associated with \emph{multiple} SGR bursts using
the Stack-a-flare pipeline.  This is forward-looking work with
interesting problems which will benefit from collaboration with the
community of SGR theorists.  We describe and characterize the
pipeline, and we present estimated sensitivities for a mock search for gravitational waves from the 2006 March 29 storm from SGR 1900+14.

Finally, in Chapter\,\ref{chapter:conclusion} we summarize our work,  describe potential
future extensions, and conclude our thesis.
\clearpage

\chapter{Gravitational Waves}
\label{chapter:gw}

In this chapter we give a brief introduction to the theory of
gravitational waves.  The intention here is to provide a foundation
for understanding gravitational wave detectors and data analysis. We
then survey interesting potential astrophysical sources of
gravitational wave emission.

An introduction to general relativity is given in
Schutz\,\cite{schutz};  an intermediate treatment is given in
Carroll\,\cite{carroll}; and a definitive reference is
Misner, Thorne and Wheeler\,\cite{misner}.  We have followed these
texts in our discussion.

\section{Gravitational waves in general relativity} \label{section:gr}

In general relativity Einstein sought to present a self-consistent
theory of gravity in which no frame of reference is favored, based
upon the postulates of special relativity: the relativity of
velocity and the universality of a finite speed of light.  In so
doing he revolutionized intuitive conceptions of space and time.

Newton's theory of gravity predicts instantaneous transmission of
information. Einstein's field equations in general relativity allow
wave solutions, with gravity propagating at the speed of light.
These gravitational wave solutions are analogous to electromagnetic
waves described by Maxwell, and the wave solutions which emerge from
the descriptions of gravity and electromagnetism propagate at the
same speed.

Though there is a deep connection between the forces, there are major
differences on the surface.  For one, gravity is much weaker than
electromagnetism: the ratio of electromagnetic force to the
gravitational force between the proton and electron in a hydrogen
atom is $\sim10^{40}$.  This is why detection of gravitational waves
has been beyond our technology until now.  For another,  there is only one sign of gravitational
charge, which is also the source of inertia. It is from this dual
role of gravitational charge that the unique connection between
gravity and the geometry of spacetime arises. 

\subsection{Perturbations in spacetime}

Far from their source, we can treat gravitational waves as small
perturbations propagating through an otherwise flat four-dimensional
manifold called spacetime.  The interval between two spacetime events is
coordinate-independent, and is given in a specific coordinate system
$\{ x^{\nu} \}$ by

\be
  ds^2 = g_{\mu\nu} dx^{\mu} dx^{\nu}, \label{eq:interval}
\ee where $g_{\mu\nu}$ is the metric tensor describing the geometry
of the spacetime.  We will use Einstein's summation convention, let Greek
spacetime indices run from 0 to 3 with 0 representing the time
coordinate, and let latin indices will run from 1 to 3.  For flat
spacetime in Cartesian coordinates the metric has components
 \be
 g_{\mu\nu} = \eta_{\mu\nu} = \begin{pmatrix}
-1 & 0 & 0 & 0\\
0 & 1 & 0 & 0 \\
0 & 0 & 1 & 0 \\
0 & 0 & 0 & 1
 \end{pmatrix}.
 \ee
Equation\,\ref{eq:interval} for the interval with the Minkowski
metric for flat space expands to \be
 ds^2 = - c^2 dt^2 + dx^2 + dy^2 + dz^2.
\ee Here we have written $c$ explicitly, but in what follows we
sometimes use geometrized units in which the speed of light $c=1$
and the gravitational constant $G=1$.

Einstein's field equations describe the relationship between sources
of gravity and the geometry of the spacetime manifold.  This
relationship is analogous to the relationship between mass and the
gravitational potential in Newtonian gravity,
 \be
 \nabla^2 \phi = 4 \pi \rho, \label{eq:newton}
 \ee
where $\rho$ is the mass density and $\phi$ is the potential.
General relativity must reduce to this form in the Newtonian limit,
where gravity is weak and velocities are small.  We therefore seek a
generalization of both the second-order differential operator on the
left and the source term on the right.

We first consider the source term.  GR prefers no reference frame,
so the generalization we seek must be coordinate-independent.  In
special relativity the mass density generalizes to the energy
density, which is the $00$ component of the symmetric stress-energy
tensor -- $T^{00}$.  $T^{00}$ is not a coordinate-independent
quantity, however, and Einstein's insight was to take as the source
the whole second rank symmetric stress-energy tensor $\mathbf{T}$,
which \emph{is} coordinate-independent.  The $T^{ii}$ components
represent pressure, the $T^{i0}$ components
represent momentum density, and the $T^{ij}$ components
represent shear stress.  These can all be thought of as sources of the field in addition to
energy density.

Having chosen a plausible source term which is a symmetric rank 2
tensor, we need to equate it to a second-order differential operator
which produces a rank 2 symmetric tensor encoding the spacetime
geometry, which we can call $G^{\mu\nu}$. The natural choice for
$\mathbf{G}$ is the Ricci tensor, a second rank tensor which is a
contraction of the Riemann tensor,
 \be
 R_{\mu\nu} = R^{\alpha}_{\mu\alpha\nu}.
 \ee
The Riemann tensor gives the geodesic deviation in a curved
spacetime, and may be written in terms of the metric connection as
 \be
 R^{\alpha}_{\beta\mu\nu} = \Gamma^{\alpha}_{\nu\beta,\mu} -
 \Gamma^{\alpha}_{\mu\beta,\nu} + \Gamma^{\alpha}_{\mu\lambda}
 \Gamma^{\lambda}_{\nu\beta} - \Gamma^{\alpha}_{\nu\lambda}
 \Gamma^{\lambda}_{\mu\beta}.
 \ee
Here, indices following a comma indicate partial differentiation
with respect to the coordinate represented by the index.
The connection, in turn, may be written in terms of the metric as
 \be
 \Gamma^{\lambda}_{\mu\nu} = \frac{1}{2} g^{\lambda\sigma}
 \left(g_{\nu\sigma,\mu} + g_{\sigma\mu,\nu} - g_{\mu\nu,\sigma}
 \right).
 \ee
Thus the Ricci tensor involves second order derivatives of the
metric, as sought.  We can raise the indices of the Ricci tensor
using the metric,
 \be
 R^{\mu\nu} = g^{\mu\alpha} g^{\beta\nu} R_{\alpha\beta}.
 \ee
The conservation laws for energy and momentum can be expressed in
terms of $\mathbf{T}$ as
 \be
 T^{\mu\nu}_{;\nu} = 0,
 \ee
where the semicolon indicates covariant differentiation with the
metric connection, that is
 \be
 A^\nu_{;\mu} = A^\nu_{,\mu} + \Gamma^{\nu}_{\mu\lambda}A^\lambda,
 \ee
for some tensor $\mathbf{A}$.  In a flat space the covariant
derivative reduces to the partial derivative.  The conservation laws
require that $G^{\mu\nu}_{;\nu}$ must also vanish.  This is not true
of the Ricci tensor; but the Bianchi identity
 \be
 \left(R^{\alpha\mu} - \frac{1}{2} g^{\alpha\mu} R \right)_{;\alpha}
 = 0
 \ee
makes it clear that the tensor
 \be
 G^{\mu\nu} = R^{\mu\nu} - \frac{1}{2} g^{\mu\nu} R
 \ee
fulfills our requirements.  We thus have
 \be
 G^{\mu\nu} = 8 \pi T^{\mu\nu},
 \label{eq:einstein}
 \ee
where the Einstein tensor $\mathbf{G}$ encodes the geometry of
spacetime, and the constant $8\pi$ was chosen to give
Equation\,\ref{eq:newton} in the Newtonian limit.

We can represent gravitational waves as weak perturbations to flat
spacetime.  In this case a convenient coordinate system can be found where
 \be
  g_{\mu\nu} = \eta_{\mu\nu} + h_{\mu\nu} \label{eq:perturbedmetric}
 \ee
with $\abs{h_{\mu\nu}} \ll 1$ throughout spacetime.  We now want to see what form the Einstein equations take
in the weak field approximation, where only terms to first order in
$\mathbf{h}$ are kept.

In the vacuum $T^{\mu\nu} = 0$, and we have
 \be
 R^{\mu\nu} - \frac{1}{2} g^{\mu\nu} R = 0.
 \ee
After some manipulation, we have to first order in $h_{\mu\nu}$
 \be
 R_{\alpha\beta\mu\nu} = \frac{1}{2}
 \left(h_{\alpha\nu,\beta\mu}+h_{\beta\mu,\alpha\nu} -
 h_{\alpha\mu,\beta\nu} - h_{\beta\nu,\alpha\mu} \right).
 \ee

A convenient choice of gauge is the transverse traceless (TT) gauge,
in which coordinates are determined by world lines traced in
spacetime by freely falling masses.  In the TT gauge, and in the
weak field limit where higher order terms in $h$ are dropped,
Einstein's equations take the form
 \be
\left( \nabla^2 - \del{}{t^2}  \right)  h_{\mu\nu}  = 0,
 \ee
which has solutions of the form
 \be h(\mathbf{x}, t) = h_0 e^{(i \omega t
- \mathbf{k} \cdot \mathbf{x})}
 \ee
with $\omega = \abs{\mathbf{k}}$, describing a wave propagating in
the direction of the wave vector $\mathbf{k}$ at the speed of light.
It is only natural that gravitational waves must propagate with
speed $c$, as $c$ is the only relevant speed in the theory,
appearing in the spacetime interval itself. The explicit component
form in the wave frame is
 \be
 h_{\mu\nu}(z, t) = \begin{pmatrix}
0 & 0 & 0 & 0\\
0 & -h_+ & h_{\times} & 0 \\
0 & h_{\times}  & h_+ & 0 \\
0 & 0 & 0 & 0
 \end{pmatrix} \cos \omega \left(z/c -t \right),
 \label{eq:transformed}
 \ee
with $h_+$ and $h_{\times}$ representing the two polarization states
(``plus'' and ``cross'') of the wave.

\subsection{Effect of gravitational waves on free test particles}

We now wish to use the metric to describe how a passing
gravitational wave measurably affects free test particles. Imagine
two free massive test particles, separated by a distance $L$. Define
a coordinate system such that the first mass is at the origin and
the second mass is at $x=L$.  We can measure the distance between
the two particles by timing light emitted from the first mass,
reflecting off of the second mass, and returning to the first mass.
If space is flat, the Minkowski metric allows us to relate the
measured time to a distance using the speed of light, that is,
$\Delta x = c \Delta t$.

One way to think about this is as a calculation of the interval
 \be
 \int_0^T \sqrt{\abs{\eta_{tt}}} dt,
 \ee
where $\sqrt{\abs{\eta_{tt}}} = c$ and the spatial terms were not
written since the worldline begins and ends at the same spatial
coordinates. Instead, let's think directly in terms of the proper
distance between the two masses
 \be
 \Delta l = \int_0^L \sqrt{\abs{\eta_{xx}}} dx.
 \ee
If there is a perturbation in spacetime due to a passing
gravitational wave, we need to use the appropriately perturbed
metric $g$ given by Equations\,\ref{eq:perturbedmetric}
and\,\ref{eq:transformed} instead. If the gravitational wave has the
plus polarization, $h_{xy} = h_{yx} = 0$, then
 \begin{align}
 \Delta l_x & = \int_0^L \sqrt{\abs{g_{xx}}} dx \\
 & = \int_0^L \sqrt{1+h_{xx}} dx \\
 & \approx L \sqrt{1+h_{xx}} \\
 & \approx L (1-\frac{1}{2} h_{+}),
 \end{align}
where in the penultimate step we have used the fact that the
perturbation is small. We can express the ratio
 \be
  \frac{\Delta L}{L} = \frac{h_+}{2}.
 \ee
We conventionally refer to $h_{+}$ and $h_{\times}$ as gravitational
wave ``strains.''

We can now repeat this thought experiment for a pair of test masses
oriented along the y-direction, finding
 \be
 \Delta l_y \approx L (1+\frac{1}{2} h_{yy}) = L (1+\frac{1}{2} h_{+}).
 \ee

\begin{figure}[!t]
\begin{center}
\includegraphics[angle=0,width=110mm, clip=false]{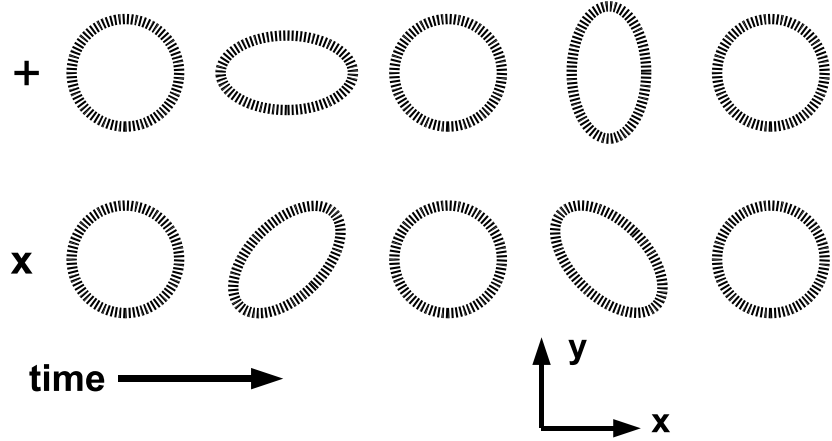}
\caption[Effect of gravitational wave on test particles] { Effect of
a passing gravitational wave on test particles.  The gravitational
wave is traveling in the z-direction (out of the page).  The effect
of plus and cross polarization components on a circle of point-like test
particles is shown in the top and bottom rows of time snapshots,
respectively, as the wave passes.  The effect is hugely
exaggerated.} \label{fig:pluscrossEffect}
\end{center}
\end{figure}

The situation is illustrated in the top portion of
Figure\,\ref{fig:pluscrossEffect}, with the addition of many more test
masses. A similar logic could be applied to the case of a
cross-polarized gravitational wave ($h_{xx} = h_{yy} = 0$),
illustrated in the bottom portion of Figure\,\ref{fig:pluscross}.

We emphasize that the coordinate positions of these particles are
not changed by the gravitational wave.  The particles sit at their
respective positions in spacetime, but spacetime changes and we can
measure the changes in \emph{proper} distances between the
particles.  The masses therefore experience no acceleration in the
conventional sense.

We could also think of these changes in proper distances in terms of
their effect on the phase of the plane wave traveling between the
test masses. This is the fundamental principle behind
interferometric gravitational wave detectors.

\subsection{What gravitational waves can we expect?}
\label{section:emissionEstimate}

Now that we have developed some understanding of the nature of
gravitational waves, a natural question to ask is, how strong can we
reasonably expect them to be?  This is an important question for
direct gravitational wave detection.

The three conservation laws, energy, momentum, and angular momentum,
eliminate the monopole, dipole, and magnetic dipole gravitational
radiation moments, respectively.  If we assume that a typical source
emits gravitational waves only via its leading term (the quadrupole
moment), we can estimate a strain amplitude upper bound using a
dimensional argument\,\cite{sigg98}. The dimensionless strain
amplitude of gravitational waves is defined as the ratio of the
fractional change in a proper distance $L$ (as opposed to coordinate
distance, which gravitational waves preserve):
 \be
 h \equiv \frac{\Delta L}{L}.
 \ee
The strain by our assumption is proportional to the second time
derivative of the quadrupole moment $Q$ and by conservation of
energy must go as $1/r$:
 \be
 h \sim \frac{G\ddot{Q}}{c^4 r}.
 \ee
We need an estimate for the magnitude of $\ddot{Q}$.  We can
identify $\ddot{Q}$ as the spherically asymmetric part of the source
kinetic energy $E_{\mathrm{asym}}$, and then we can write
 \be
 h \sim \frac{G E_{\mathrm{asym}}/c^2}{c^2 r} \sim 10^{-21} \left(
 \frac{M}{M_{\odot}} \right)  \left(  \frac{16 \enspace \mathrm{Mpc}}{r}  \right).
 \ee
$M c^2$ is the non-symmetric kinetic energy, and 16\,Mpc is the
distance to the center of the Virgo cluster. From this we see that
we are much better off searching for gravitational waves of
astrophysical origin rather than gravitational waves we could
produce ourselves in a laboratory. Imagine a quadrupolar source
(e.g. a spinning dumbell) with 1000\,kg weights fixed 1\,m apart
spinning at 100\,Hz, placed $r=2.5$\,m from an interferometric
detector.  The ratio of 16\,Mpc to 2.5\,m gives a gain of about
$10^{22}$, but the ratio of $(I \omega^2 / c^2) / \msunonly$ gives a
loss of about $10^{38}$, making the manmade gravitational wave
source some 16 orders of magnitude less detectable than the perhaps
optimistic $\msun$ event in Virgo. Even at a source distance of
2.5\,m, test mass displacement due to gravitational radiation at
twice the spin frequency would be less than $10^{-33}$\,m.  However,
the situation is actually much worse: in the near field, such a
device would produce easily detectable periodic gravitational
gradients at twice the spin frequency (which could be used to
very precisely calibrate an interferometric gravitational wave
detector\,\cite{rotor}).  These local gradients, which go as
$1/r^4$, would dominate the gravitational radiation until the source
is a few hundred kilometers from the detector.

Since gravitational waves are produced by motions of entire
astrophysical objects such as stars, black holes and galaxies, we do
not expect gravitational wave periods shorter than the light travel
time around the circumference of the smallest scale associated with
a source, given by the source's Schwarzschild radius $2GM/c^2$,
where $M$ is the source mass. This gives an upper bound on the
frequency
 \be
 f = \frac{1}{T} < \frac{c^3}{4 \pi GM} \sim 10^4 \mathrm{Hz}
 \left( \frac{\msunonly}{M} \right).
 \ee

\section{Astrophysical sources of gravitational waves} \label{section:sources}

In this work we are principally interested in specific burst sources
(SGRs) which will be described in more detail later.  However, we
wish to put these sources into a context of gravitational wave
sources.

In this section we briefly describe some of the most promising non-SGR
sources of gravitational waves. (For a detailed surveys of astrophysical gravitational wave sources see\,\cite{300years}.)  We classify the
astrophysical sources in the subsections below according to the
four major LIGO Scientific Collaboration data analysis working
groups:  Burst, Compact Binary Coalescence, Continuous, and
Stochastic.

\subsection{Burst} \label{section:burstSources}

Burst sources, such as GRBs and supernovae, are characterized by
transient gravitational wave signals of duration $\lesssim$1\,s.
Typically, emission from these sources is difficult to predict
with precision.  Often the information from theorists involves a
 frequency range and duration and not much more.  The
notable exception is compact binary coalescence, which can be approached from a burst perspective and is such an
important source class that
it has a dedicated working group within the LSC.

Searches for gravitational waves associated with GRBs are quintessential burst-type searches.  GRBs are the
most electromagnetically luminous events in the universe after the Big Bang.  Given
their typical cosmological distances, the energies involved could be
larger than $10^{50}$\,erg.  However, the mechanism behind
them is still mysterious, and there are no precise models of
gravitational emission.  Long duration ($\gtrsim$2\,s) bursts are
thought to be massive stars collapsing into black holes, and short
duration bursts may be CBC events.  There is the intriguing
possibility that some portion of the short burst population could be
due to SGR giant flares.  A LIGO gravitational wave search (see
Section\,\ref{section:070201}) added compelling evidence to this
hypothesis, as it excludes the possibility that GRB 070201
coincident with the Andromeda Galaxy was a CBC event, to high
confidence\,\cite{S5GRB070201}.

Supernovae are attractive gravitational wave source candidates
because of the possibility of nearly relativistic aspherical
collapse of large amounts of matter ($\msun$ or more) potentially
followed by energetic ejection of stellar matter. If the collapse or
bounce has a quadrupole moment gravitational waves will be emitted.
The events are accompanied by electromagnetic and neutrino emission
so some advantages of triggered searches can be reaped; however, the
events often go unnoticed until some point late in the light curve,
making extrapolation back to the collapse event problematic.
Detection of gravitational waves from a supernova event would shed
light on these events, as it would provide information from the core
impossible to obtain otherwise, even from neutrino observations.

Attempts to predict gravitational wave emission from supernova have
been made\,\cite{ott04, dimmelmeier02, zwerger97}.  However, such
predictions are still far from the precision and certainty necessary
for templated searches. Therefore, burst search methods such as the
one at the core of this work are necessary for supernova
gravitational wave searches.

Burst searches can increase sensitivity by partitioning the
two-dimensional time/frequency search space in such a way that
single units (sometimes ``tiles'' or ``pixels'') of the space can
contain the target signal with a minimum of extra noise. Clustering
algorithms which can join adjacent significant pixels, and
multi-resolution analysis are common techniques used to accomplish
this partitioning despite lack of precise prior knowledge of the
extent of the expected signal in time and frequency.

Burst searches come in two varieties:  all-sky searches and
externally triggered searches.  All-sky searches operate over long
durations of data such as an entire science run, and treat all sky
locations equally.  Externally triggered searches use
non-gravitational wave observations, from gamma ray satellites,
telescopes or neutrino detectors for example, to reduce the scope of
the search in the time dimension and possibly the sky location
dimension as well. The SGR search which forms the core of this work
is an externally triggered burst search.

Either variety of burst search can attempt to be sensitive to a wide
parameter space which includes most model predictions of most source
classes. This can be considered an advantage of searches for excess
power in the gravitational wave data; they trade sensitivity for
generality, as compared to templated searches such are used for CBC
sources. For a quantitative comparison of burst searches to CBC
templated searches in a limited portion of the CBC search space see
Section\,\ref{section:complementingInspirals}.

\subsection{Compact binary coalescence} \label{section:cbc}

Compact binary coalescence (CBC) events, the inspiral and merger of
binary systems of compact objects, are a primary target for
gravitational wave searches\,\cite{S1inspiral, S2inspiral, S2macho,
S2inspiralLigoTama, S2inspiralBBH, S3S4inspiral, S3spinningInspiral,
S5GRB070201}. Searches for gravitational waves from mergers of
compact binary systems have been performed for systems of two
neutron stars (BNS), two black holes (BBH), neutron star and black
hole (NS-BH), and primordial black holes (PBH). CBC sources are
among the most promising for a first direct gravitational wave
detection.  The gravitational wave emission from some parts of the
life cycle of compact binary systems is well-modeled relative to
other burst sources, and the expected frequencies of some systems
are near the sweet spot of ground-based interferometric
gravitational wave detectors\,\cite{cutler93}.

Compact binary systems are expected to emit gravitational waves in
three distinct stages: the inspiral
stage, the merger stage, and the ringdown stage.  During the
inspiral stage, the members of the binary system are well-separated
in space and the system evolves in an orderly fashion as the binary
orbit decays due to loss of energy via gravitational wave emission.
Gravitational wave emission from the inspiral stage is modeled well
enough that searches relying on signal templates are feasible.  For non-spinning systems the
inspiral strain waveform at the Earth can be
written\,\cite{S3S4inspiral}
 \be
  h(t) = \frac{1 \mathrm{Mpc}}{D_{\mathrm{eff}}} A(t) \cos(\phi(t) -
  \phi_0). \label{eq:inspiral}
 \ee
The functions $A(t)$ and $\phi(t)$ depend on the masses and spins of
the binary members and $\phi_0$ is an unknown phase parameter. The
effective distance $D_{\mathrm{eff}}$ is the distance at which a
merger event could be detected if the binary system would be
optimally oriented and located relative to the gravitational wave
detector --- that is, at a sky position directly on zenith or nadir
and orbiting in a plane parallel to the detector's plane.
$D_{\mathrm{eff}}$ is always greater than or equal to the physical
distance,
 \be
  D_{\mathrm{eff}} = D / \sqrt{F_{\times}^2 (1+\cos^2 \iota)^2/4 +
  F_{+}^2\cos^2 \iota} ,
 \ee
where $F_+$ and $F_{\times}$ are the detector's antenna responses to
the plus and cross polarizations\,\cite{300years} and $\iota$ is the
inclination angle between the binary system and the detector.
Matched filter searches typically ignore the effects of
spin\,\cite{S3S4inspiral}, arguing that it not
significant\,\cite{apostolatos95}.  Effects from tidal coupling and
other effects which depend on the component objects' equations of
state are also thought to be insignificant\,\cite{bildsten92}. Thus,
in practice, the parameter phase space for templated merger searches
is defined by the binary system masses.  Extrinsic parameters such as source effective distance, the inclination of the
system $\iota$,  and the unknown orbital
phase $\phi_0$ do not increase the dimensionality of the template space.  In the LIGO S3-S4 inspiral
search\,\cite{S3S4inspiral}, for the PBH and BNS cases, search
templates implementing Equation\,\ref{eq:inspiral} are constructed from
second order restricted post-Newtonian
approximations\,\cite{blanchet96, droz99, blanchet06}. For the BBH
case the template family was phenomenological as described
in\,\cite{buonanno04} due to uncertainties in the templates.

During the merger stage, the two compact objects fall into each
other's event horizons and merge into a single black hole. This
stage is difficult to model, and is the focus of much ongoing
research.  During the ringdown stage, the single black hole is in an
excited state and decays through gravitational wave emission from
damped non-spherically-symmetric ringdown modes.  In this work we
focus on the inspiral stage.

The gravitational wave energies emitted in compact binary
coalescence (CBC) events are large because the second derivative of
the quadrupole moment $\ddot{Q}$ is large. This is due to the
compactness of the systems.  Neutron stars, for example, have radii
of order 10\,km and members of a BNS pair can orbit at close range
and at high frequencies of up to $\sim500$\,Hz.  Before merger, the
frequency of the system and the gravitational wave
amplitude increase with time, resulting in a ``chirp''
signal.

Rates for such events are typically given in terms of $\lten$,
$10^{10}$ times the blue solar luminosity. The Milky Way Galaxy has
a luminosity of about 1.7$\lten$.  Merger rates are assumed to
depend on the rate of star formation in a volume, which is measured
by the blue luminosity in that volume.  BNS merger rates can be
estimated from four binary pulsar systems, and are between
$\sci{10-170}{-6}\,\mathrm{yr}^{-1}\mathrm{L}_{10}^{-1} $ at 95\%
confidence in one plausible model\,\cite{S3S4inspiral, kalogera04}.
Merger rates for BBH systems and hybrid NS-BH systems are based on
theoretical populations studies (for a review see\,\cite{postnov06})
and are between
$\sci{0.1-15}{-6}\,\mathrm{yr}^{-1}\mathrm{L}_{10}^{-1} $ and
$\sci{0.15-10}{-6}\,\mathrm{yr}^{-1}\mathrm{L}_{10}^{-1} $ at 95\%
confidence, respectively\,\cite{S3S4inspiral}.  Converting these
rates into LIGO detection rates is complicated, as the detection
range depends on the choice of SNR threshold, on the detector
sensitivity as a function of frequency and the component masses
which set the frequencies tracked by the inspiral waveform. To give
a sense, the LIGO S3-S4 inspiral paper\,\cite{S3S4inspiral} gives an
S4 horizon distance (the distance at which the detector would detect
an optimally oriented and located binary merger with SNR of 8) for
the 4\,km Hanford LIGO detector of $\sim$15\,Mpc for compact binary
systems with 2.8$\msunonly$ total mass.  This horizon distance
corresponds roughly to an effective cumulative blue luminosity (the
cumulative blue luminosity as a function of effective distance, as
opposed to physical distance) of $\sim100\lten$.   In terms of
volume, the reach of the LIGO detectors improved by an order of
magnitude between S4 and S5\,\cite{S5, S5strain}.  Very roughly,
this implies that the S5 detectors could expect to detect a BNS at
SNR of 8 at a rate of $\sim \sci{5}{-2}\,\mathrm{yr}^{-1}$, assuming
100\% duty cycle. The Advanced LIGO detectors are expected to give
an additional factor of $10^3$ in volume, which would give an SNR 8
detection rate of $\sim 50\,\mathrm{yr}^{-1}$.

When CBC event inspiral waveforms are finally observed by
gravitational wave detectors, we will be able to extract information
about the source system parameters, such as the masses of the
compact objects, their spins, the eccentricity of the orbits, and
event the compact object equation of state.

\subsection{Continuous}

No astrophysical source of gravitational waves is truly
monochromatic, as the emission of gravitational waves removes energy
from the mechanism which is producing them.  The gravitational waves
are coupled to the spinning system and will increase its period in
the case of orbital systems where the energy reservoir is
gravitational potential energy, or decrease its period in the case
of rotating objects where the energy reservoir is rotational kinetic
energy.

One source for continuous nearly monochromatic gravitational waves
is binary star systems.  Systems comprised of ordinary stars will
emit gravitational waves at twice their orbital frequency; as this
is typically less than $10^{-3}$\,Hz these sources are not available
to ground-based detectors, though they will be available to LISA.

Spinning neutron stars comprise a more promising class of sources for LIGO. A
non-axisymmetric spinning neutron star is a spinning quadrupole, and
to the extent that its spin rate $f_0$ is constant, it would emit
gravitational waves continuously at $2f_0$. Deviation from axial
symmetry could be caused by misalignment of the principle axis
with the spin axis, by  strong magnetic fields, or by a
``mountain'' on the neutron star. The gravitational wave strain from
such a source is approximately\,\cite{300years}
 \be
 \sci{2}{-26} \left( \frac{f_0}{1 \mathrm{kHz}} \right)^2 \left(
 \frac{10 \mathrm{kpc}}{r}
 \right) \left( \frac{\epsilon}{10^{-6}}  \right),
 \ee
where $r$ is the distance to the spinning star and $\epsilon =
(I_{xx}-I_{yy}) / I_{zz}$ is the ellipticity.

Searches for gravitational waves from such a source are referred to
as ``pulsar searches,'' and have several advantages over other
gravitational wave searches:
 \ben
 \i $f_0$ is precisely known for gravitational wave searches involving known
 pulsars, which is the typical case.  Therefore, the expected signal
 is well understood;
 \i Integration over entire gravitational wave detector science runs is possible,
 allowing much smaller gravitational wave strain amplitudes to emerge from the noise
 than other types of searches;
 \i Historically, the rate limiting step in gravitational wave analysis has often been
 completion of detector calibration.  It is much easier to produce a
 detector calibration at a single frequency than over the entire
 band of the detector.
 \een

Observations of pulsars show that their spin rates decrease over
time (spindown).  This decrease is thought to be due to a
combination of mechanisms:  magnetic dipole radiation, particle
acceleration in the magnetosphere, and emission of gravitational
waves\,\cite{S5crab}.  An upper limit on gravitational wave emission
can thus be set by measuring the spindown rate via electromagnetic
observation.

LIGO has published several pulsar searches beginning with the S1
pulsar search\,\cite{S1pulsar}.  The most recent result is the Crab
pulsar search from the first nine months of S5\,\cite{S5crab}, which
presented upper limits on gravitational wave emission which beat the
spin-down upper limit.

\subsection{Stochastic}

There are two main classes of stochastic gravitational wave sources,
confusion noise stochastic backgrounds and primordial stochastic
backgrounds.

Since interferometric gravitational wave detectors have such a wide
antenna pattern, a sufficiently sensitive detector would suffer from a
continuous and unpredictable bubbling of a large number of discrete
foreground events, including events of the types we have already
described --- supernovae, GRBs, SGRs, CBC events, etc.  In fact, a
major challenge anticipated by LISA data analysts is sorting through
this confusion noise stochastic background in the detector's
sensitive frequency band in order to extract interesting
information\,\cite{lisaMDC}.  For LISA, gravitational wave confusion
noise is expected to be the dominant noise source at some
frequencies\,\cite{confusionNoise}.

The other major expected source of a stochastic gravitational wave
background is primordial: gravitational waves left over from the
very early universe, just after the Big Bang.  A primordial
stochastic gravitational wave background could be caused by
inflation-amplified zero-point quantum mechanical metric
fluctuations, by cosmic strings, and by phase transitions in the
early universe.

Analysis strategies with  cross-correlation between two or more
detectors can help distinguish both foreground and cosmological
stochastic gravitational wave backgrounds from non-astrophysical
stochastic processes causing noise in the detectors, after
accounting for different antenna factors, different gravitational
wave crossing times for a given sky location due to different
detector locations, and different detector sensitivities.  For the
case of two equivalent detectors with search bandwidth $\Delta F$
and integration time $T$, cross-correlation can increase search
sensitivity by a factor of $\sqrt{T\Delta F}$ over a single
detector.  This strategy was used in the LIGO S3 cosmological
gravitational wave stochastic background analysis setting an upper
limit on $\Omega_0$\,\cite{S3stochastic}.

When (and if) the primordial stochastic background is finally detected, the
cosmological implications will be astounding.  Whereas photons
began free streaming some $10^5$\,years after the Big Bang when
electrons and protons condensed into atomic hydrogen, the last
scattering of gravitational waves occurred about $10^{-22}$\,s after
the Big Bang\,\cite{battye96}.  Gravitational waves are therefore an
excellent way to shed light on the very early universe.
\chapter{Interferometric Gravitational Wave Detectors}
\label{chapter:detector}

In this chapter we introduce interferometric
gravitational wave detectors, and describe  the LIGO
detectors at the time of LIGO's fifth science run. One goal of this
chapter is to communicate the remarkable technical achievement of
modern gravitational wave interferometry: the ability to measure
relative displacement changes several orders of magnitude smaller
than the diameter of a proton occurring over kilometer-scale
lengths.  As we saw in Section\,\ref{section:emissionEstimate}, we
require this sensitivity in order to have some chance of detecting
gravitational waves.

\section{Overview of the LIGO detectors} \label{section:detectors}

The LIGO detectors\,\cite{S5} are sensitive
Michelson interferometers. Since Michelson interferometers are good
at measuring differential length changes between their two arms,
they are ideal for detecting passing gravitational waves, which
cause time-dependent quadrupolar spatial deformations as discussed
in Chapter~\,\ref{chapter:gw}. In principle, turning an
interferometer into a gravitational wave detector only requires the
interferometer optics to double as test masses.  Such detectors
are made on Earth, for example, by suspending the optics from
wires. However, making a detector sufficiently sensitive to have a
reasonable chance of observing astrophysical gravitational wave
signals is difficult.

The LIGO observatory includes three detectors at two sites.  The
Hanford, Washington site is home to a 4\,km arm-length detector (H1)
and a 2\,km arm-length detector (H2), which share the same
ultra-high-vacuum beamtube enclosures.  The Livingston Parish,
Louisiana site is home to a single 4\,km arm-length detector (L1).
The L1 detector is slightly misaligned relative to the H1 and H2
detectors, primarily due to the Earth's curvature.

\begin{figure}[!t]
\begin{center}
\includegraphics[angle=0,width=130mm, clip=false]{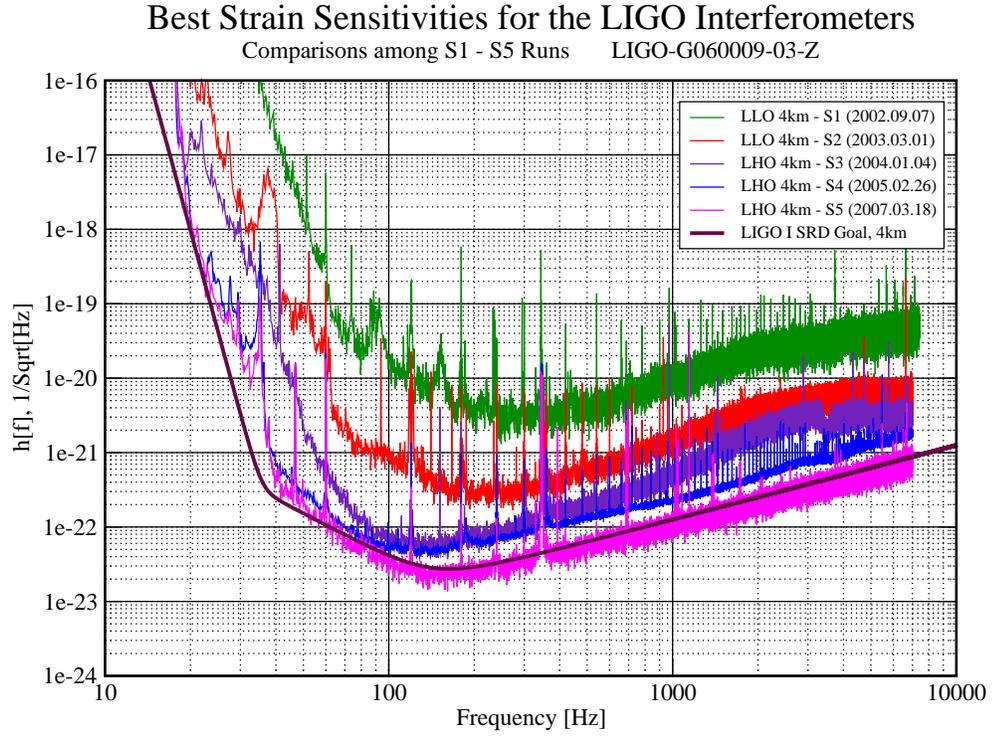}
\caption[Progression of strain noise in LIGO science runs]
{Progression of strain noise in LIGO science
runs\,\cite{ligoNoiseRuns}.  Detector noise is expressed
as amplitude spectral density (the square root of the power
spectrum) since the detectors measure strain amplitudes.  Four
relatively short science runs occurred during the commissioning
period, as the detectors' sensitivity consistently was improved. The
solid line shows the design goal, which was met in S5 across the band, except for the region below $\sim 70$\,Hz.
Seismic noise was about ten times worse than expected near the
seismic wall frequency of $\sim 45$\,Hz\,\cite{S5}. }
\label{fig:runs}
\end{center}
\end{figure}

\begin{figure}[!t]
\begin{center}
\includegraphics[angle=0,width=120mm, clip=false]{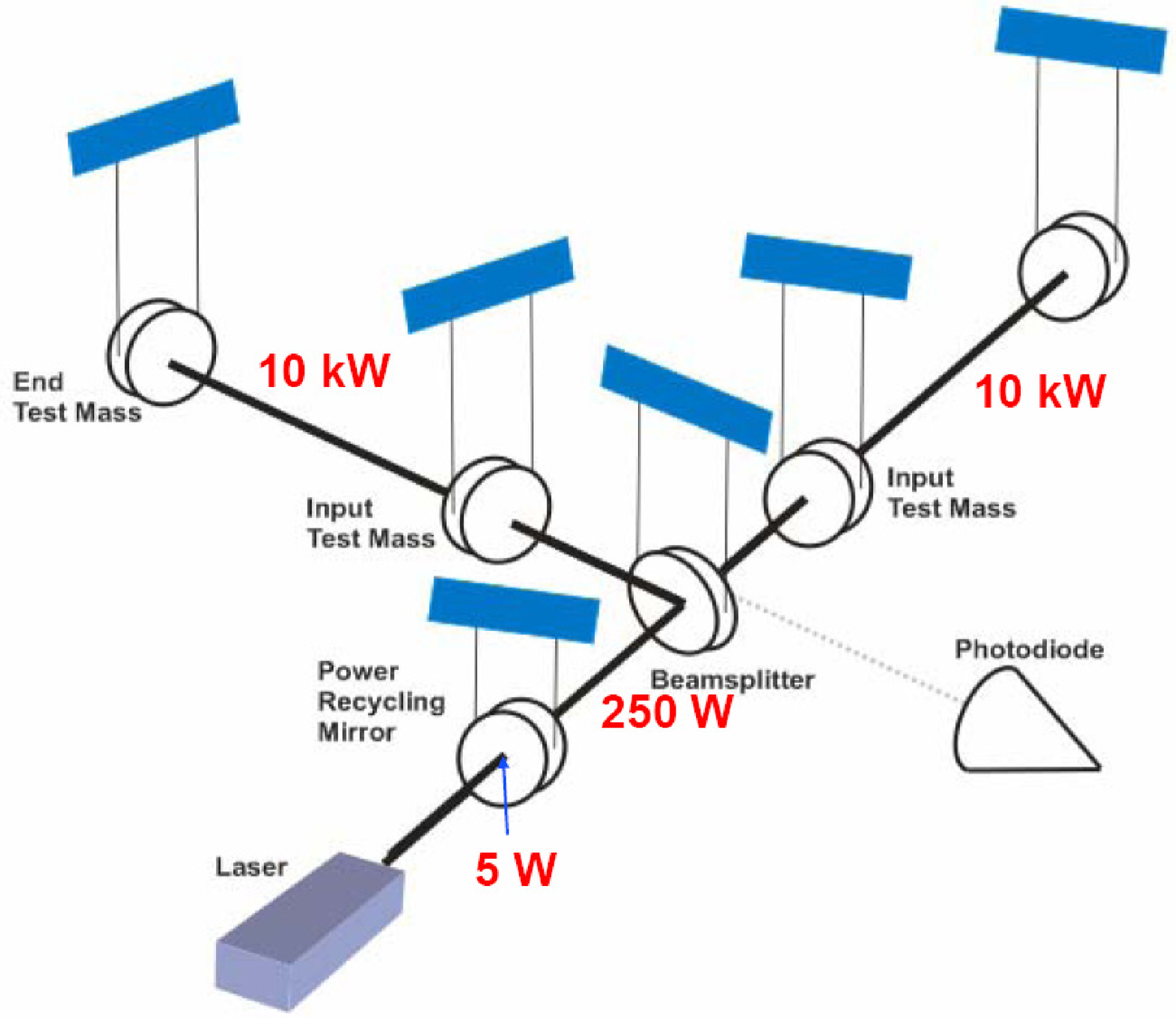}
\caption[Schematic diagram of a LIGO detector] {Simplified schematic
diagram of a LIGO detector. This diagram shows the basic
configuration, including the laser; the Fabry-Perot cavities between
the input test masses (ITMs) and end test masses (ETMs) for
amplifying phase difference between arms; the photodetector on the
dark anti-symmetric (AS) port; and power recycling mirror (PRM)
sending reflected light from the beam splitter (BS) back into the
interferometer, thereby increasing laser power in the arms and
minimizing shot noise. The six optics shown are suspended. }
\label{fig:ifo}
\end{center}
\end{figure}

LIGO finished its fifth science run (S5) on 11 November 2007. S5 met
a major goal of the initial phase of the LIGO project: collecting
one year's worth of triple-coincident data at design sensitivity
across the sensitive band.  The detectors measure gravitational wave
strain amplitude (a unitless quantity), so sensitivity as a function
of frequency is characterized by the amplitude spectral density of
strain equivalent noise, in units of $\rthz$. Figure\,\ref{fig:runs}
shows improvement of the LIGO 4\,km interferometers' noise floors as
a function of frequency over the five science runs.  The initial
sensitivity goals were detection of gravitational wave strain
amplitudes as low as $10^{-21}$\,\cite{S5}, and instrument strain
noise as low as $10^{-21}$ rms integrated in the 100--200\,Hz
range\,\cite{fritschel01, abramovici92}.  This sensitivity level
requires interferometers capable of detecting differential
displacements on the order of $10^{-18}$\,m, approximately one
thousandth the diameter of a proton. Achieving this goal required
several engineering ``tricks'' and perseverance in reducing noises
in the detectors. Noises are discussed in detail in
Section\,\ref{section:noise}; here we describe the basic
interferometer configuration, shown in Figure\,\ref{fig:ifo}.

A key extension to the basic Michelson interferometer
configuration is the addition of multiple coupled optical cavities.
These cavities increase sensitivity, but maintaining resonances poses a significant
control problem. The interferometers consist
of three primary resonant cavities. The first is the power recycling
cavity, which sends light from the symmetric port back into the interferometer.
The power recycling cavity is thus contained
between the power recycling mirror (PRM) an the beamsplitter (BS).  

The other two resonant cavities are Fabry-Perot cavities in the two
interferometer arms, contained between an end test mass (ETM) and in
input test mass (ITM). The ETMs have a multilayer highly reflective
coating, and the ITMs are designed to let a small fraction of
light into and out of the cavities.  When the cavities are on
resonance, the laser light is trapped for order of hundreds of round
trips, depending on the quality of the optical coatings.  Relative
phase differences (for gravitational waves of adequately low
frequencies) are amplified by a factor on the order of the finesse
of the cavity. Power in the arms is also amplified, which reduces
photon shot noise (Section\,\ref{section:noise}).

The light source for the S5 interferometers is nominally a 10\,W
Nd:YAG stabilized laser at 1064\,nm\,\cite{savage98}. The laser is
pre-stabilized in frequency against a reference cavity using the
Pound-Drever-Hall technique\,\cite{drever83} and passed through a
spatial mode cleaner before entering the
interferometer\,\cite{ruldiger81}.

Power is split into two sets of sidebands which are used as error
signals to control the interferometer optics' degrees of freedom.
The carrier controls the differential arm (DARM) degree of freedom,
i.e. the relative difference in length between the two
interferometer arms.  The interferometer is operated so that the
antisymmetric port is maintained, or ``locked,'' on a dark fringe,
and the antisymmetric error signal is used as the gravitational wave
signal readout.

\section{Noise} \label{section:noise}

\begin{figure}[!t]
\begin{center}
\includegraphics[angle=0,width=110mm, clip=false]{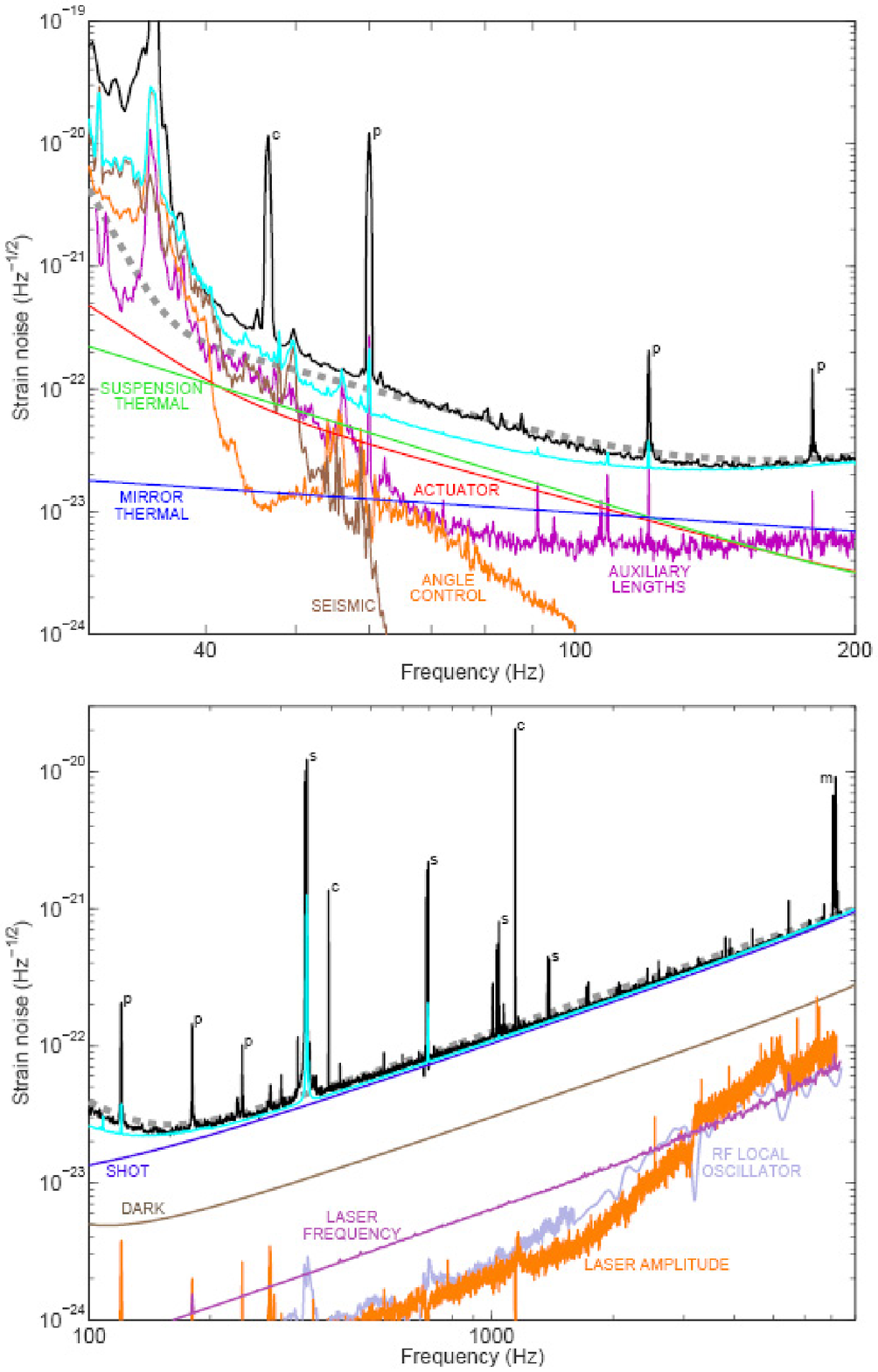}
\caption[Noises in the LIGO detectors] { Noises in the LIGO
detectors.  The plot shows the major contributors to the H1
detector's strain noise in S5.  The top plot shows the force noises
and the bottom plot shows the sensing noises.  The cyan curves show
the root square sum of the known noise components, and the black
curve is the measured noise.  Letters on the spectral peaks denote:
c -- calibration line; p -- power line harmonic; s -- suspension
vibrational mode; m -- mirror vibrational mode.  Figure taken
from\,\cite{S5}.  } \label{fig:noiseBudget}
\end{center}
\end{figure}

Any phenomenon other than a passing gravitational wave that breaks
symmetry between the interferometer's two arms and moves the
antisymmetric port from the dark fringe will cause noise. Noises
fall into one of two types: force noise --- extraneous motions of
the test masses; and sensing noise --- noise inherent in measuring
the test mass displacement. In general force noises dominate at
lower frequencies and sensing noises dominate at higher frequencies,
with the transition at order 100\,Hz.  The major technical effort in
creating gravitational wave interferometers with astrophysical
sensitivity is understanding and overcoming these noises.

There are two noises which primarily limit the sensitivity of the
LIGO detectors: seismic and acoustic noise at low frequencies, and
photon shot noise in the antisymmetric port photodetector at high
frequencies.  Figure\,\ref{fig:noiseBudget} shows the S5 noise
budget for H1\,\cite{S5}.

\subsection{Seismic noise}

Seismic noises are fundamental limiting noises for all ground-based
interferometric detectors at low frequencies.  They arise from
anthropogenic activity, wind coupling to the ground through various
mechanisms, earthquakes, waves lapping against distant shorelines
(producing the ``microseism''), and the like. One of the principle
reasons for putting an interferometric gravitational wave detector
in space (the LISA project) is to escape these noises.

Seismic/acoustic noises couple into the interferometer through the
suspension to the optics, which must be adequately isolated from the
seismically active ground and from the external acoustic
environment. They make the interferometer cavities more difficult to
bring into resonance via the various control systems(``lock''), they
can cause loss of lock, and they can limit the sensitivity to
gravitational waves.  Isolation is provided by stacks
supporting the optic suspensions. The L1 detector has an additional
active pre-isolation stage, due to higher levels of anthropogenic
and microseismic noises at the Louisiana site. This pre-isolation
senses seismic motions and compensates by moving in the opposite
direction.

\subsection{Shot noise}

Photon shot noise arises from the quantum mechanical discreteness of
light and is characterized by poisson statistics.  The shot-noise
amplitude noise density for the detector is\,\cite{meers88, S5}
 \be
 \tilde{h}(f) = \sqrt{\frac{\pi \hbar \lambda}{\eta P_{\mathrm{BS}} c}}
 \frac{\sqrt{1+(4\pi f \tau_s)^2}}{4 \pi \tau_s},
 \ee
where $\tau_s$ is the arm cavity storage time in the Fabry-Perot
cavities, $f$ is the gravitational wave frequency, $P_{\mathrm{BS}}$
is the power incident on the beamsplitter, $\lambda$ is the
frequency of the interferometer carrier light, and $\eta$ is the
photodetector quantum efficiency. Shot noise can thus be reduced
both by increasing $P_{\mathrm{BS}}$, the photon number density at
the beamsplitter, and by increasing the finesse in the Fabry-Perot
cavities, which increases $\tau_s$. This is the reason for the power
recycling resonant cavities described in
Section\,\ref{section:detectors}.

\subsection{Intrinsic thermal noise}

A mechanical system such as a LIGO optic or its suspension is
coupled to the external heat reservoir.  Mechanical vibrations in
the systems can be damped dissipatively through this connection, but
thermal fluctuations in the reservoir can also couple back into
mechanical energy.  This connection is quantified in the fluctuation
dissipation theorem\,\cite{callen51}.

The thermal noise power spectrum is proportional to the temperature,
and this is the reason some advanced detectors propose to use
cryogenically cooled optics and suspensions\cite{lcgt}.  It also
depends on the quality factor of the system.  A higher quality
factor means lower dissipative loss, which conversely means less
coupling to the external reservoir and less thermal fluctuation
forcing to the mechanical system.  Therefore, designs with high
quality factors are sought for the optics and suspensions.

Suspension thermal noise appears most obviously in the spectrum from
the interferometer's gravitational wave channel as forests of peaks
at around 350\,Hz, and their harmonics. Mirror resonances tend to be
at higher frequencies outside of the band relevant to this work.

\subsection{Laser noise}

Noise from the laser originates upstream from the beamsplitter, and
in a perfect interferometer laser frequency and intensity noises
would both be ``common mode'' in the arms and would not appear in
the gravitational wave channel.  However, due to asymmetries for
example in alignment this is not the case.  Therefore the laser is
stabilized both in frequency and in amplitude.

\subsection{Other noises}

There are many other potential sources of noise.  Gas and dust in
the interferometer beams could cause transient symmetry breaking
between the arms by obfuscating the interferometer beam, resulting
in sensing noise. Enclosing the beam in ultra-high vacuum
($<10^{-8}$\,Torr) effectively solves this problem.  Control system
circuits introduce noise.  Gravitational gradients from sources such
as clouds and density changes in the crust under the site produce
low frequency noises in the regime that is already dominated by
seismic noises. Ambient magnetic field gradients could couple to the
magnets affixed to optics as part of the coil driving system.
Ambient electric fields could couple to static charges on the
optics. Sources of noise such as these become prevalent from time to
time in the system, and must be chased down by commissioning
workers.

\section{Detector calibration} \label{section:calibration}

The LIGO detectors are designed to sense passing gravitational waves
by monitoring the induced differential separation of suspended test
masses~\cite{stan03, barish99}.  A detector calibration response
function is necessary to interpret the detector's output signal,
obtained from a photodetector at the interferometer's antisymmetric
port, as differential length changes $\Delta L = L_x - L_y$ measured
in meters and thus the relative strain $h=\Delta L /L$ where $L$ is
the average arm length. Fiducial detector response functions as a
function of frequency are occasionally measured by moving test
masses by a known amount determined by counting fringes.  The
detector response changes by small amounts due to drifting optical
alignment which affects the light power in the arms, laser power
drift, and other causes. These changes are monitored via constant
sinusoidal calibration lines injected via actuation of test masses.

The LIGO detectors are conventionally calibrated via voice coil
actuators coupled to magnets directly affixed to the test
masses~\cite{sigg03, S4, S1}. This procedure is discussed in
Section\,\ref{section:coilActuators}.   An alternative
actuation technique makes use of so-called ``photon calibrators,''
which we discuss in detail in Chapter~\ref{chapter:pcal}.  Before
discussing these particular calibration techniques, we describe in
Section\,\ref{section:DARMservoloop} the differential arm servo loop
which produces the error signal we wish to calibrate.

\subsection{DARM servo loop}\label{section:DARMservoloop}

We now present a brief description of the DARM (differential arm)
loop, the primary control loop that maintains the differential
interferometer arm length~\cite{S4}. The gravitational wave output
signal obtained from the photodetector at the interferometer's
antisymmetric port, is used as the loop error signal. We relate this
signal ($D_{\textrm{ERR}}$) to gravitational wave strain by
\begin{equation}
 h(\omega,t) \equiv X_\emph{ext}(\omega,t) / L =
 R(\omega,t)D_{\textrm{ERR}}(\omega,t) \label{eq:strain}
\end{equation}
where $h(\omega,t)$ is the strain, $X_\emph{ext}$ is the
differential test mass displacement due to an externally induced
motion, $L$ is the interferometer arm length and $R(\omega,t)$ is
the interferometer response function in units of strain per
$D_{\textrm{ERR}}$ count.

\begin{figure}
\begin{center}
\includegraphics[angle=0,width=100mm, clip=false]{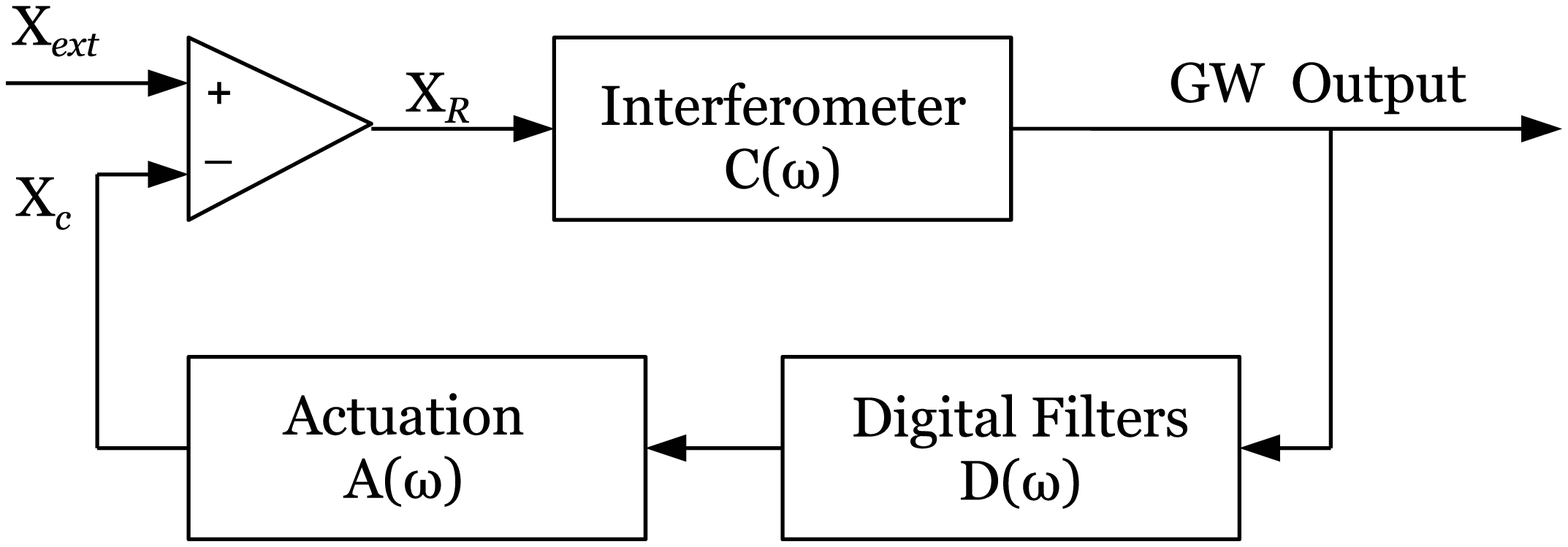}
\caption[Block diagram of the interferometer servo loop]{Block
diagram of the interferometer servo loop. The gravitational wave
output signal is also the loop error signal $D_{\textrm{ERR}}$.
} \label{fig:Loop}
\end{center}
\end{figure}

A detector block diagram is shown in Figure~\ref{fig:Loop}. $X_c$ is
the corrective displacement due to the servo which attempts to drive
the residual displacement $X_R=X_\emph{ext} - X_c$ to zero.  The
components of the loop are the interferometer sensing transfer
function $C(\omega)$ relating differential test mass displacement to
the gravitational wave output; the digital filter transfer function
$D(\omega)$ relating gravitational wave output counts to voice coil
drive counts; and the voice coil actuation transfer function
$A(\omega)$ relating voice coil drive counts to test mass
displacement.

Following the loop (Figure~\ref{fig:Loop}), we find
\begin{equation}
D_{\textrm{ERR}}(\omega,t) =
\frac{C(\omega)}{1+G(\omega)}X_\emph{ext}(\omega,t)/L,
\end{equation}
where the open loop gain, $G(\omega)$, is the (dimensionless)
product of the three loop transfer functions,
\begin{equation}
G(\omega) \equiv D(\omega)A(\omega)C(\omega). \label{eq:olg}
\end{equation}
Using Equation~\ref{eq:strain}, we find the detector's response
function,
\begin{equation}
R(\omega) = \frac{1+G(\omega)}{C(\omega)}. \label{eq:response}
\end{equation}

\subsection{ Calibration via coil actuators}
\label{section:coilActuators}

The goal of calibrating the detector is to produce the response
function $R(f)$ of Equation\,\ref{eq:response}. If a technique is
available for displacing the test mass by a known magnitude and
phase, then $R(\omega)$ can be measured directly. This is the
situation with the photon calibrator, as will be described in
Chapter\,\ref{chapter:pcal}.

However, such direct measurement is not possible with conventional
calibration via coil actuators.  Instead, the response function is
calculated from measurements of the open loop gain $G(\omega)$ and
the actuation function $A(\omega)$ and knowledge of the digital
filter transfer function $D(\omega)$ via equations~\ref{eq:response}
and~\ref{eq:olg}.  This procedure is described in detail
in~\cite{S1,S2,S3,S4,fringe}.

This can be done using the coil actuators affixed to the test
masses, as follows. The coil actuator transfer function $A(\omega)$
(converting coil drive counts to displacement in meters) can be
estimated for the \emph{input test masses} at frequencies below
$\sim$40 Hz, at a reference time. Measurement techniques used (for
example fringe counting and sign toggling~\cite{fringe}) require
test mass displacements on the order of $\lambda = 1064$ nm. Input
test masses are used instead of end test masses because the
interferometer must be placed in a ``simple Michelson''
configuration consisting only of the beamsplitter and the input test
masses.

$A(\omega)$ is assumed to be linear down to displacements relevant
to astrophysical gravitational wave events, $\sim10^{-18}$ m, some
12 orders of magnitude. It is extrapolated from near-DC up to 7\,kHz
to cover the bandwidth of the detector.  This transfer function for
the input test mass is then bootstrapped to the end test masses by
closing a feedback loop on the optical cavity of a single
interferometer arm, driving the input test mass at a known
amplitude, and measuring the response of the end test mass. With an
estimate of $A(\omega)$ in hand, $R(\omega)$ is estimated at all
frequencies at the reference time.

The interferometer sensing function $C(\omega)$ is sensitive to
angular alignment and heating of the test masses, and changes
slightly on a timescale of minutes. Injected calibration lines are
used to track changes in $R(\omega)$ due to these changes in
$C(\omega)$ relative to the fiducial calibration at $t_0$,
$G(\omega,t_0) = C(\omega,t_0)A(\omega)D(\omega)$. Calibration lines
are produced in the gravitational wave channel output by constant
excitation of an end test mass at chosen frequencies.  Changes in
the amplitudes of the lines in the gravitational wave channel output
are monitored. Assuming that fluctuations in the sensing function
can be parameterized by a scalar multiplicative factor
$\alpha(t)$~\cite{S1}, where $\alpha(t_0)$ is taken to be 1, the
calibration line amplitude is given by
\begin{equation}
A_{cal}(t) = s_{cal}\frac{\alpha(t)C(\omega_{cal},t_0)}{1+
\alpha(t)G(\omega_{cal},t_0)}
\end{equation}
where $A_{cal}$ is  the amplitude of the calibration line the
gravitational wave channel output and $s_{cal}$ is the strain due to
the calibration line, which is constant in time and will factor in
the ratio $A_{cal}(t) / A_{cal}(t_0)$.  This ratio determines
$\alpha(t)$ from directly measurable quantities.  The digital filter
transfer function also changes occasionally when digital filters are
changed; these changes can be parameterized in a second scalar
coefficient $\beta(t)$.  Then the response function propagated to a
time $t$ from the fiducial transfer functions is given by
    \be R(\omega, t) = \frac{1+ \alpha(t) \beta(t) G(\omega, t=0)}{\alpha(t) C(\omega,
    t=0)}. \ee

\subsection{S5 strain-calibrated data} \label{section:hOfT}

The raw gravitational wave signal from the interferometer's dark
port must be calibrated before astrophysical interpretation is
possible.  This can be done once the detector response function
$R(t, f)$ is known (Section\,\ref{section:calibration}. To
facilitate analysis of data collected during S5, a strain-calibrated
time series referred to as ``h(t)'' was produced from the raw signal
and stored in the data archives\,\cite{siemens04}. This calibrated
data may then be used as the input to an S5 analysis.  The Flare
pipeline can use either strain-calibrated data or raw data plus the
detector response functions.

\section{Data quality flags} \label{section:dq}

During data collection during a science run, a detector and its
environment are continuously monitored by multiple channels.
Abnormalities observed in these auxiliary channels can be used to
flag data segments that should not be used in astrophysical
analysis, or that should only be used with care.
 
Some data quality (DQ) flags are set in real time by data
monitoring tools, while others are set after research by
the Detector Characterization Group.  There are four categories of DQ flags.  For the purposes of the S5 SGR search presented in
Chapter\,\ref{chapter:search}, category 1 and category 2 flags mean science is impossible
with the flagged data (e.g. H1:OUT\_OF\_LOCK). If active category 1
or 2 flags are found in an on-source region, that search is aborted.
If they are found in a background region those portions are
excluded.  Category 3 and category 4 flags (e.g.
H1:WIND\_OVER\_30MPH) are less serious and can be handled in
post-processing.  We chose to ignore them for upper
limits, as they will tend to make loudest event upper limits more
conservative.  Candidate detection analysis events in on-source
regions would need to consider them in the follow-up.  Lists of DQ
flags used in the analysis may be obtained from\,\cite{v3S5dq}.

\section{ Antenna pattern of interferometric detectors}

\begin{figure}[!t]
\begin{center}
\includegraphics[angle=0,width=120mm, clip=false]{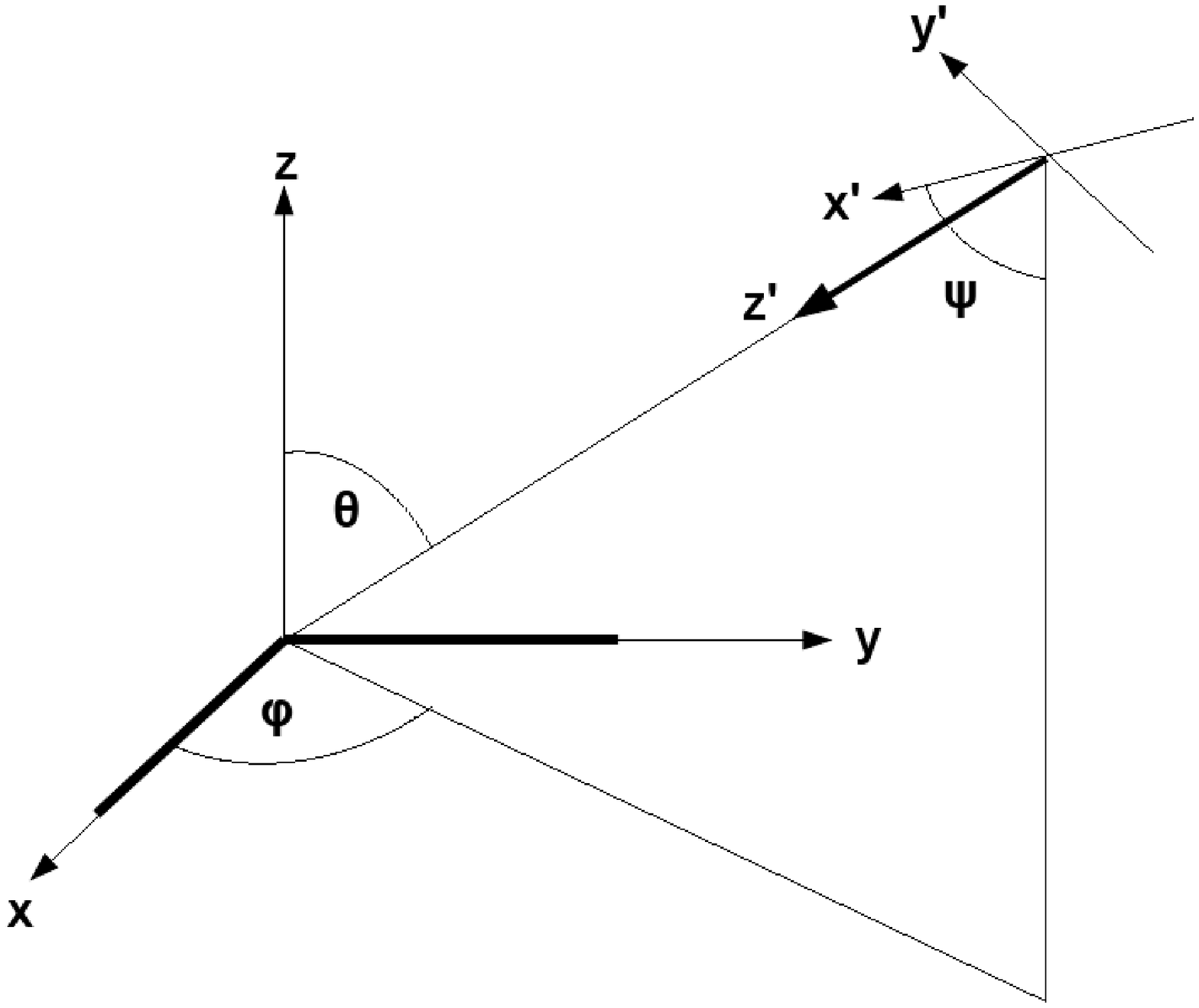}
\caption[Angles used in describing detector antenna pattern] {
Angles used in describing detector antenna pattern.  The detector is
located at the origin of the unprimed coordinates, with arms
pointing in the x and y directions.  The gravitational wave travels
along the $z'$ direction, with polarization ellipse axes aligned
with the $x'$ and $y'$ axes.  $\theta$ and $\phi$ are the standard
altitude and azimuth angles defining the direction of the incoming
wave relative to interferometer.  $\psi$ gives the angle between the
$z$-axis and one of the polarization ellipse axes.  Figure
follows\,\cite{saulson}.} \label{fig:psi}
\end{center}
\end{figure}

Passing gravitational waves causing spatial distortions aligned with
the interferometer's arms will be optimally detected; such is the
case with waves arriving from the detector's zenith or nadir with
polarization aligned to the detector arms.  Passing gravitational
waves arriving from a location along the arm bisector in the plane
of the interferometer, on the other hand, cannot be detected.  Thus,
sensitivity to gravitational waves depends on the gravitational wave
source location relative to the detector.  The relevant angles are
defined in Figure\,\ref{fig:psi}.

\begin{figure}[!t]
\begin{center}
\includegraphics[angle=0,width=140mm, clip=false]{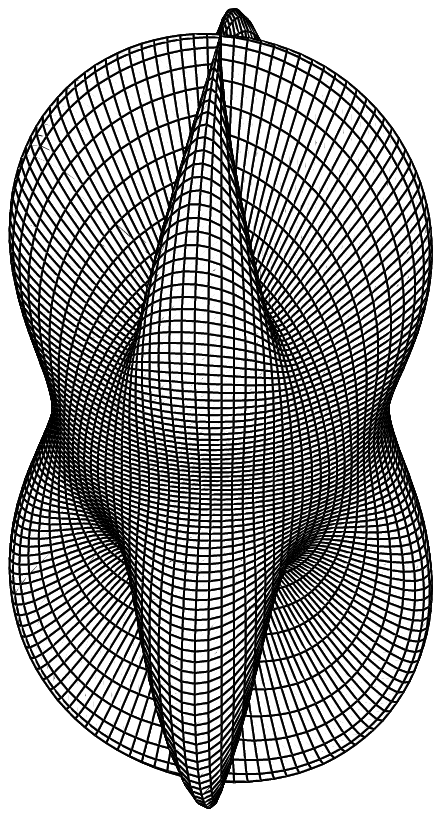}
\caption[Antenna pattern for plus polarization] { Plus polarization
antenna pattern for an interferometric gravitational wave detector.
The plot shows $F_{+}^2$ with $\psi=0$ (changing $\psi$ would scale
the plot by a constant amount). The interferometer would be in the
center of the plot, with the view along the bisector of the arms.
The distance from a point of the plot surface to the interferometer
is a measure for the gravitational wave sensitivity in this
direction. } \label{fig:plus}
\end{center}
\end{figure}

\begin{figure}[!t]
\begin{center}
\includegraphics[angle=0,width=140mm, clip=false]{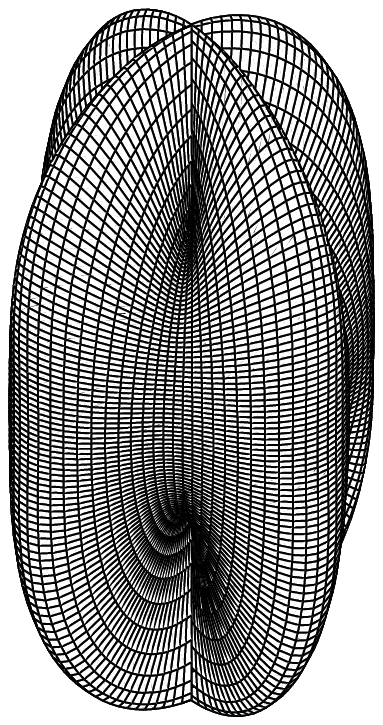}
\caption[Antenna pattern for cross polarization] { Cross
polarization antenna pattern for an interferometric gravitational
wave detector. The plot shows $F_{\times}^2$ with $\psi=0$ (changing
$\psi$ would scale the plot by a constant amount). The
interferometer would be in the center of the plot, with the view
along the bisector of the arms.  The distance from a point of the
plot surface to the interferometer is a measure for the
gravitational wave sensitivity in this direction. }
\label{fig:cross}
\end{center}
\end{figure}

\begin{figure}[!t]
\begin{center}
\includegraphics[angle=0,width=140mm, clip=false]{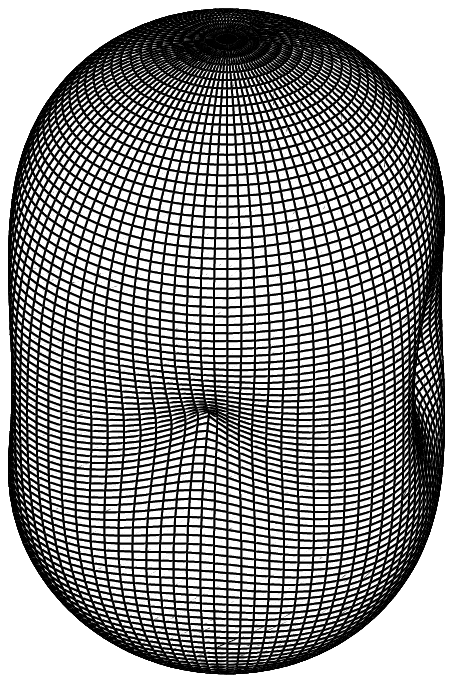}
\caption[Antenna pattern for unpolarized waves] { Antenna pattern
for unpolarized gravitational waves for an interferometric
gravitational wave detector. The plot shows $F_{\times}^2+F_{+}^2$,
which is independent of $\psi$. The interferometer would be in the
center of the plot, with the view along the bisector of the arms.
The distance from a point of the plot surface to the interferometer
is a measure for the gravitational wave sensitivity in this
direction. } \label{fig:pluscross}
\end{center}
\end{figure}

The incoming gravitational wave is subject to antenna functions
describing the detector response and given by~\cite{anderson01}
 \begin{align}
 F^{+}_d(\theta, \phi, \psi) & = \frac{1}{2} \cos2\psi \left(1+\cos^2\theta\right) \cos2\phi - \sin2\psi \cos\theta \sin2\phi \\
 F^{\times}_d(\theta, \phi, \psi) & = -\frac{1}{2} \sin2\psi \left(1+\cos^2\theta\right) \cos2\phi - \cos2\psi \cos\theta
 \sin 2\phi.  \label{eq:antenna1}
 \end{align}
The detector response is then
 \be
 h(t) = F^{+}_d(\theta, \phi, \psi) h_+(t)   +  F^{\times}_d(\theta, \phi, \psi)
 h_{\times}(t).
 \ee
We have plotted the antenna factors in spherical coordinates below.
Figure\,\ref{fig:plus} shows the detector response to a linearly
plus-polarized gravitational wave in coordinates defined in
Figure\,\ref{fig:psi}. Figure\,\ref{fig:cross} shows the detector
response to a linearly cross-polarized gravitational wave.
Figure\,\ref{fig:pluscross} shows the detector response to
unpolarized waves, that is, gravitational waves with equal
amplitudes in plus and cross polarizations.

\section{Future interferometric gravitational wave detectors}
\label{section:futureDetectors}

Because interferometric gravitational wave detectors are sensitive
to gravitational wave strain amplitudes and have an omnidirectional
antenna pattern, increase in astrophysical reach -- the volume of
space in which an astrophysical event of a particular class could be
detected -- goes as the cube of the detector's sensitivity.

Here we briefly describe two major advances in the state-of-the-art
of interferometeric gravitational wave detectors which are currently
being implemented. Both are improvements to the initial LIGO
interferometers used in the SGR search presented here: ``Enhanced
LIGO'' (2009)\,\cite{eligo} and ``Advanced LIGO''
($\sim$2014)\,\cite{aligo, aligosite}. 

\begin{figure}[!t]
\begin{center}
\includegraphics[angle=0,width=140mm, clip=false]{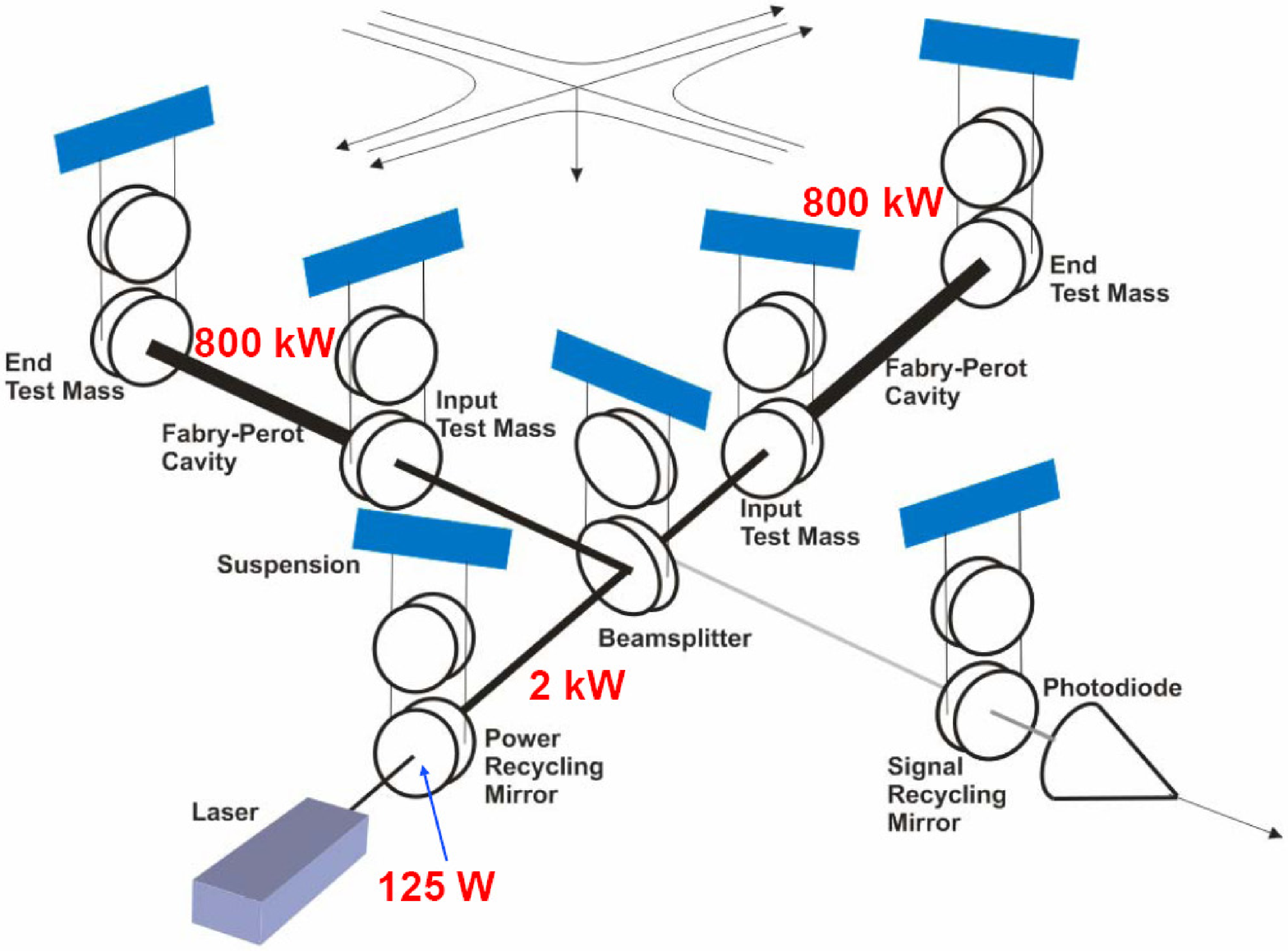}
\caption[Schematic diagram of Advanced LIGO detector] {Simplified
schematic diagram of a possible Advanced LIGO detector
configuration.  The top set of optics is for a suspension point
interferometer which is one advanced strategy for reducing seismic
noise\,\cite{yoichiThesis}.  Compare this figure to
Figure\,\ref{fig:ifo}. } \label{fig:aligo}
\end{center}
\end{figure}

Advanced LIGO's goal is to provide routine detection of
gravitational waves, based in current predictions, by increasing
amplitude sensitivity of the initial LIGO detectors by a factor of
$\sim$10.  This will be accomplished by reducing noise sources, by
lengthening the Hanford 2\,km interferometer to 4\,km, and by
building in flexibility to the optical configuration to allow
optimization of the noise spectrum for different target sources. The
laser power will be increased to almost 200\,W, reducing shot noise.
More massive optics, improved optical coatings, larger beam sizes on
the test masses, and improved suspensions incorporating fused silica
fibers will be used, to decrease thermal noise.  A more stable lock
acquisition system will improve uptime. Active seismic isolation,
possibly including SPI, will decrease seismic noise across the
seismic band and also push the ``seismic wall'' down to about
10\,Hz.  An output mode cleaner and DC readout will reduce noise.
Table\,\ref{table:aligo} compares initial LIGO and advanced LIGO.
Figure\,\ref{fig:aligo} gives a schematic of the Advanced LIGO
configuration. Science operation is currently planned for 2014.

\begin{table}[h]
\begin{center}
\caption[Advanced LIGO compared to initial LIGO]{Comparison of
Advanced LIGO to initial LIGO.}
\begin{tabular}{p{2in}|p{1.2in}|p{1.2in}}
 & \textbf{initial LIGO} &  \textbf{Advanced LIGO} \\
   \hline \hline

 Minimum strain noise & $\sci{3}{-23}$ Hz$^{-1/2}$ & $\sci{2}{-24}$ Hz$^{-1/2}$ \\
\hline
 NS inspiral range & 15 Mpc & 175 Mpc \\
 \hline
 Input laser power & 10 W & 180 W \\
 \hline
 Power in arm cavities & 15 kW & 800 kW \\
 \hline
 Test masses & 10\,kg glass & 40\,kg glass \\
 \hline
 Test mass Q & $\sim10^6$  &  $\sci{2}{8}$ \\
 \hline
 Suspension fiber Q & $\sim10^3$  &  $\sim\sci{3}{7}$ \\

 \hline
\end{tabular}
\label{table:aligo}
\end{center}
\end{table}

Enhanced LIGO is an intermediate station on the way to Advanced
LIGO, with two goals.  First, Enhanced LIGO will double initial
LIGO's amplitude sensitivity in the two 4\,km interferometers. This
sensitivity increase will then be used in an S6 science run, which
will have about an order of magnitude higher probability of a
detection per unit time (roughly determined by the cube of the
sensitivity improvement). Second, the Enhanced LIGO upgrades will
also be used in Advanced LIGO.   These include a 35\,W laser (the
first stage of the Advanced LIGO laser); an advanced Electro-optic
modulator for the laser frequency pre-stabilization system; active
seismic isolation; a DC readout; and an improved thermal compensation
system.  Therefore, Enhanced LIGO provides an opportunity to develop
and test these technologies.

\clearpage
\chapter{Photon Calibrators} \label{chapter:pcal}

In this chapter we describe the LIGO photon calibrator (``pcal'')
system~\cite{justice03,goetz04, goetz07} and discuss advances we
have made in detector calibration.

Radiation pressure provides a relatively straightforward means of
calibrating an interferometric gravitational wave detector.  Photons
from a laser transfer momentum to a test mass whose displacement can
be easily calculated. Radiation pressure calibration has significant
advantages in addition to providing a physically independent check
on the conventional voice coil calibration.  Photon calibrators have
also been implemented at the Glasgow 10\,m gravitational wave
detector~\cite{clubley01} and the GEO600 gravitational wave
detector~\cite{geo06}.

Section~\ref{section:Theory} continues discussion of the calibration
of the LIGO detectors begun in Section\,\ref{section:calibration},
focusing on photon calibrators. Section~\ref{section:Implementation}
describes the LIGO photon calibrator implementation.
Section~\ref{section:recentAdvances} presents early measurements leading to
discovery and characterization of a discrepancy  with the
conventional voice coil calibration.
Section\,\ref{section:pcalTimeDelay} describes a recent use of the
photon calibrators to precisely calibrate the time delay in the
detector response, uncovering a second discrepancy with the
conventional calibration.  Section~\ref{section:pcalFuture} briefly
describes the current status and suggests future directions.

\section{Principles of operation}\label{section:Theory}

The goal of calibration is knowledge of the detector's response
function $R(\omega)$, which converts the detector's gravitational
wave output signal to differential changes in arm length at any
frequency in the detector's band, making physical and astrophysical interpretation possible. Both the voice coil and
pcal calibration methods rely on an absolute calibration of the
actuator used to move the test mass.

\subsection{A photon actuator}\label{section:Theory1}

A beam of $n$ photons of frequency $\nu$ reflecting with angle of
incidence $\theta$ from the surface of a test mass transfers
momentum
\begin{equation}
p_{\gamma} = 2 \frac{h\nu n}{c} \cos\theta
\end{equation}
where $h$ is Planck's constant and $c$ is the speed of light. The
beam induces a force
\begin{equation}
F(t) = \frac{dp_{\gamma}}{dt} = \frac{2 \cos\theta}{c} \frac{d
\left(h\nu n\right)}{dt} = \frac{2 \cos\theta}{c} P(t)
\end{equation}
where $P(t)$ is the total reflected power.  If we drive the test
mass sinusoidally,
\begin{equation}
P(t) = P_{\textrm{dc}} + P \sin \left(\omega t\right),
\label{eq:Power}
\end{equation}
where $w$ is  the angular frequency of the beam power modulation.
$P_{\textrm{dc}}$ pushes the test mass with a constant force which
is compensated for by the detectors's length sensing and control
system, and is therefore unimportant in principle.

The test mass equation of motion, assuming a simple pendulum, is
\begin{equation}
\frac{F(t)}{M} = \frac{2 \cos\theta}{M c} P \sin (\omega t) =
\ddot{x} + \gamma \dot{x} +  \omega_{p}^{2} x \label{eq:Motion}
\end{equation}
where $M$ is the mass of the test mass, $\gamma$ is a damping
coefficient, and $\omega_{p}^2 = g/l$. If $\omega$ is much greater
than the pendulum resonance frequency, the solution is
\begin{equation}
x(\omega) \simeq -\frac{2 P \cos\theta}{M c \omega^2}.
\label{eq:master_norotation}
\end{equation}

\begin{figure}[!htb]
\begin{center}
\subfigure[]{
\includegraphics[angle=0,width=90mm,clip=false]{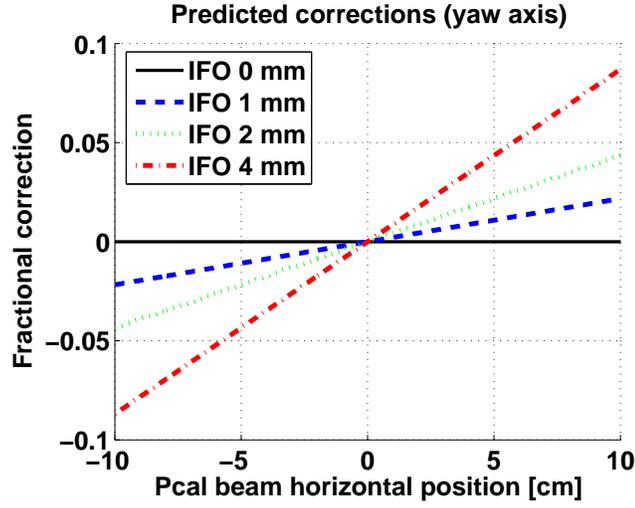}}
\subfigure[]{
\includegraphics[angle=0,width=90mm,clip=false]{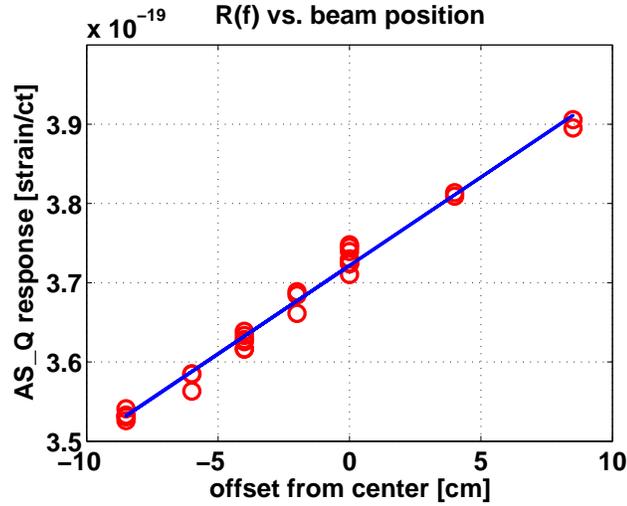}}
\caption[Effect of beam offsets in pcal]{(a) The theoretical
prediction of the yaw correction term $1+a_x b_x M / I_x$ for
various interferometer beam offsets as a function of photon
calibrator beam offset.  (b) Experimental verification of the yaw
correction term in equation~\ref{eq:master} due to off-centered
beams.  The x-axis shows approximate horizontal position of the
photon calibrator beam relative to the center of the test mass, and
the y-axis shows the magnitude of the response function $R(\omega)$.
We swept the beam back and forth across the test mass several times
to provide a sense of precision and check for any systematic
hysteresis.  The fit to experimental data indicates that the main
interferometer beam at the time of the measurement was horizontally offset on the test mass by 2.8
mm to left of center. } \label{fig:offctrdfit}
\end{center}
\end{figure}

Finally, we correct for beam positions on the test mass.  If the
photon calibrator beam is not centered horizontally (vertically) on
the test mass, it will cause an angular motion of the test mass at
frequency $\omega$ in yaw (pitch). If the main interferometer beam
is perfectly centered there will be no net effect in the
gravitational wave output signal; however, if the main beam is
\emph{not} centered, the interferometer will interpret the angular
motion as longitudinal motion.  For photon calibrator beam offsets
$a_x$ and $a_y$, and main interferometer beam offsets $b_x$ and
$b_y$, we have (to lowest order in the offsets)
\begin{equation}
x(\omega) \simeq -\frac{2 P \cos\theta}{M c \omega^2} \left(1 +
\frac{a_x b_x M}{I_x} + \frac{a_y b_y M}{I_y} \right),
\label{eq:master}
\end{equation}
where $I_x$ and $I_y$ are the test mass moments of inertia around
the yaw and pitch axes.  The yaw term is illustrated in
Figure~\ref{fig:offctrdfit} with measurements we took in the summer
of 2005.  In practice, it is difficult to precisely know the
location of the interferometer beam, and these correction terms are
treated as a source of uncertainty.

\clearpage

\subsection{Advantages of the photon calibrator}
\label{section:pcalAdvantages}

With the photon calibrator, we have an estimate for the test mass
displacement $X_{\emph{ext}}(\omega_0)$ for the duration of an excitation at a frequency $\omega_0$ via
equation~\ref{eq:master} and we only need measure the gravitational
wave output $D_{\textrm{ERR}}(\omega_0)$ to obtain $R(\omega_0)$ via
equation~\ref{eq:strain}.  Repeating this procedure at frequencies
across the detector's sensitive band allows interpolation of the
response function $R(\omega)$ at the time of measurement.

In the case of calibration via voice coil actuators~\cite{S1, S4},
which are part of the servo loop unlike the photon actuators,
$R(\omega)$ is estimated from measurements of $G(\omega)$ and
$A(\omega)$ and knowledge of $D(\omega)$ as described in
Section\,\ref{section:calibration} and Equation~\ref{eq:response}.
$A(\omega)$ is estimated for the input test masses at a reference
time. Measurement techniques use the interferometer laser wavelength
as a standard and require test mass displacements on the order of
$\lambda = 1064$ nm~\cite{fringe}. $A(\omega)$ is assumed to be
linear down to displacements relevant to astrophysical gravitational
wave events, $\sim10^{-18}$ m, some 12 orders of magnitude. The
transfer function for the input test mass is then propagated to the
end test masses by closing a feedback loop on the optical cavity of
a single interferometer arm, driving the input test mass at a known
amplitude, and measuring the response of the end test mass~\cite{S1,
S4}.

The photon calibrator excitations, on the other hand, occur on a length scale similar to expected test mass motions
due to passing gravitational waves.  Furthermore, excitations
are applied directly to the end test mass, not the input test mass,
so no bootstrapping is required. Finally, unlike the voice coil
calibration method, the pcal method works with the detector in the
same state as used for collection of science quality data.  This
turns out to be significant, as described in
Section\,\ref{section:pcalFuture}.

Other advantages of photon calibrators include actuation without any
need to attach objects to the interferometer's test masses;
actuation via devices located outside of the test mass vacuum
enclosure; ability to precisely measure and calibrate the
interferometer delay in a relatively simple way (see
Section\,\ref{section:pcalTimeDelay}, and production of calibration
results relatively quickly.

\section{Implementation}\label{section:Implementation}

Two photon calibrator units are mounted on each of the three LIGO
interferometers, one near each end test mass.  The laser of each
photon calibrator is aimed at an end test mass high reflectivity (HR) surface.
Either photon calibrator can be used to measure the response
function of the  given interferometer; one on each end test mass
provides redundancy and a consistency check.

The major components of the system are a $\sim500$ mW 1047 nm Nd:YLF
laser; an acousto-optic modulator (AOM) which modulates the laser
beam power; and a photodetector which monitors a small fraction of
the beam power transmitted by a partially reflecting mirror.
Monitoring of the sample beam allows estimation of the output power
of the system, and thus the amount of test mass displacement, via
Equation{~\ref{eq:master}.

The setup is shown in Figure~\ref{fig:Layout}. First, the
beam is sent through a polarizer to prevent drift in beam
polarization. The beam is then focused onto the AOM which modulates
the power into the deflected beam. Either the undeflected beam or
the first order deflected beam (H1X, H1Y and H2X use the deflected
beam while H2Y, L1X and L1Y use the undeflected beam) encounters a
pickoff mirror that transmits a small fraction of the beam to the
photodetector and reflects the rest out of the enclosure, through
the vacuum viewport, and onto the center of the test mass as shown
in Figure ~\ref{fig:testmass}.  The beam is focused so that the spot
size on the test mass is $\sim1$ cm in diameter. We note here that
there are alternatives to sending the beam onto the center of the
test mass. For example, it has proven useful to split the beam into
two spots around the center of the optic, as illustrated in
Figure~\ref{fig:2beamsetup} and this choice is discussed further in
Section~\ref{section:results}.

These components are mounted on an optical table which is enclosed
in a box attached to the beam tube near a fused silica glass
viewport (see Figure\,\ref{fig:pcalmount}) which provides a view of the test mass
at an angle of incidence of less than 10 degrees.

Two dedicated fast DAQ channels (16384 samples per second) have been
commissioned for each unit. One channel carries an excitation signal which drives
the AOM, thereby controlling the laser power; the other carries the output from the
photodetector readout.  These channels are stored to tape.

\begin{figure}[!htb]
\begin{center}
\includegraphics[angle=0, width=80mm, clip=false]{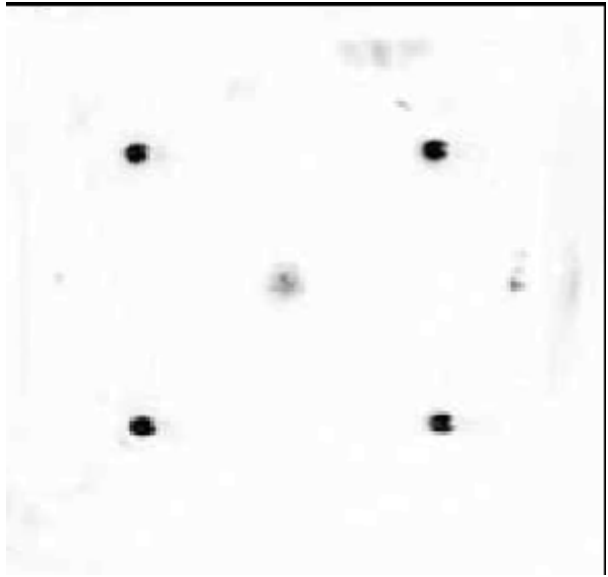}
\caption[Photograph of a LIGO ETM with incident photon calibrator
beam]{Photograph of a LIGO ETM with an incident photon calibrator
beam faintly visible in the center.  The edge of the optic is not
clearly visible; the four bright spots are from the OSEM sensors
behind the optic, near its edge.  The image has been color-inverted
for clarity. } \label{fig:testmass}
\end{center}
\end{figure}

\begin{figure}[!htb]
\begin{center}
\includegraphics[angle=0, width=80mm, clip=false]{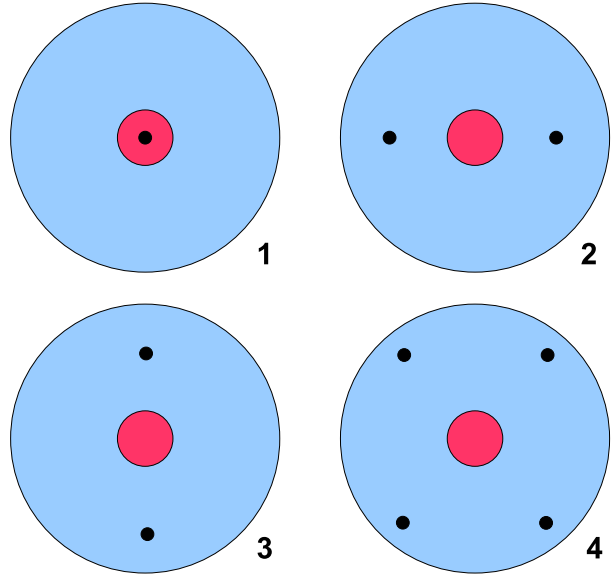}
\caption[Diagrams of single beam and split beam photon calibrator
setups]{Diagrams of single beam (left) and split beam (right) photon
calibrator setups, showing the main interferometer beam (large red
circle centered on optic) and the photon calibrator beam(s) (smaller
black circle(s)) incident on a test mass. The four-beam
configuration has not been implemented. } \label{fig:2beamsetup}
\end{center}
\end{figure}

\begin{figure}[!htb]
\begin{center}
\includegraphics[angle=0, width=110mm, clip=false]{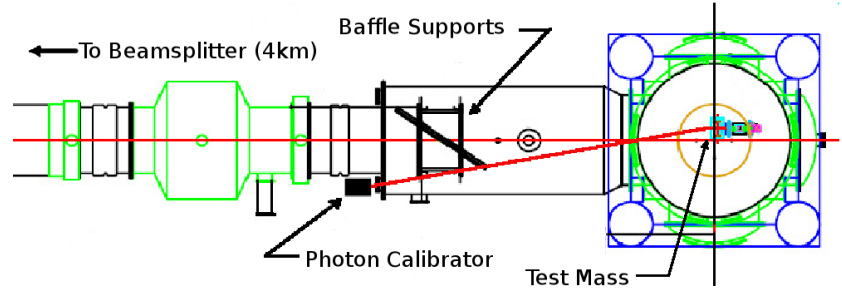}
\caption[Top view of photon calibrator enclosure mounted near an end
test mass]{Top view of photon calibrator enclosure mounted near an
end test mass. A beam enters the vacuum chamber through a glass
viewport and is aimed as close to the center of the test mass as
possible. In the case of the Hanford 4 km detector (shown here) the
beam must pass between two vertical baffle supports.}
\label{fig:pcalmount}
\end{center}
\end{figure}

\begin{figure}[!htb]
\begin{center}
\includegraphics[angle=0, width=80mm, clip=false]{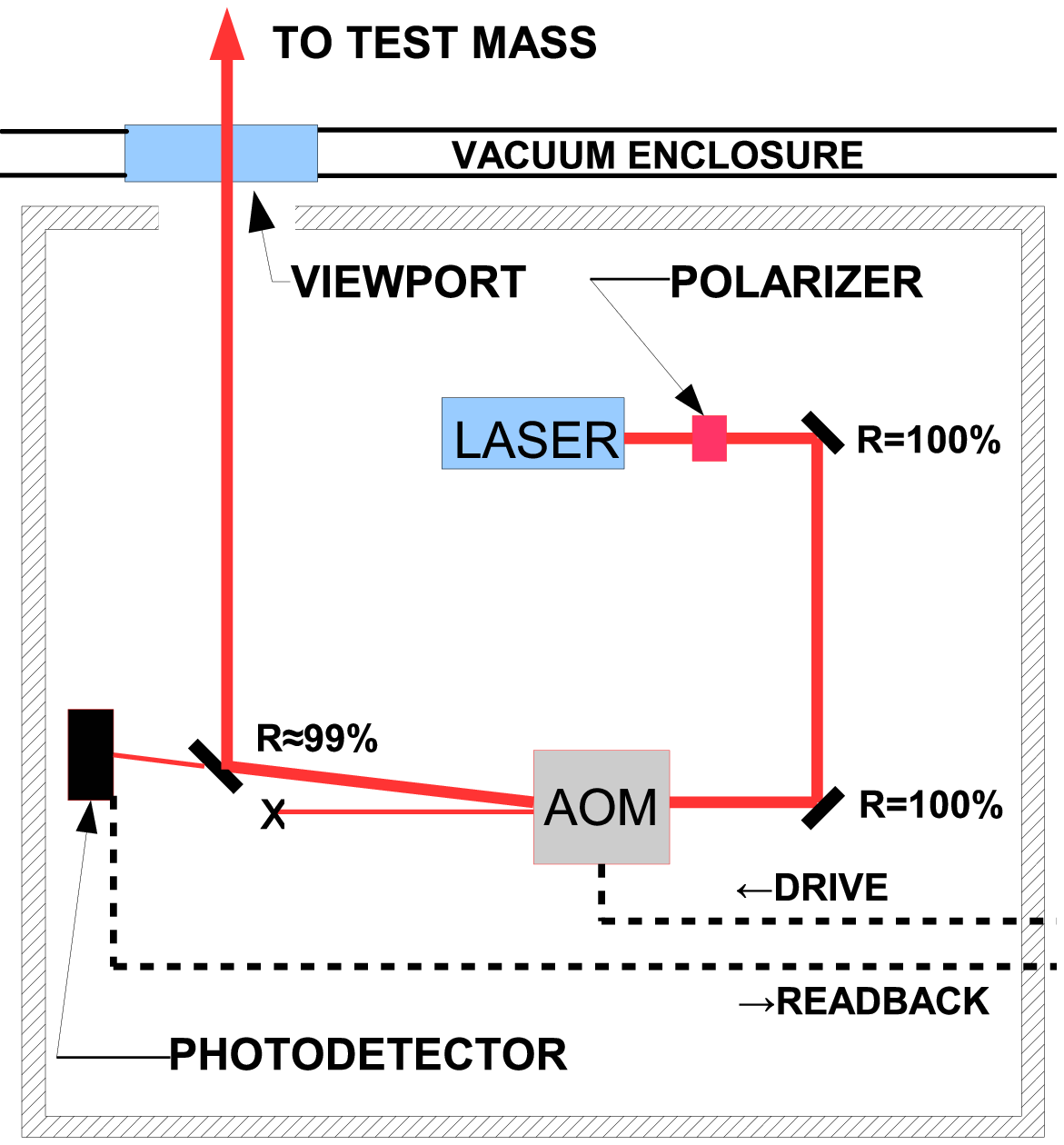}
\caption[Layout of a photon calibrator optical table]{Layout of a
photon calibrator optical table showing major components of the
system, as commissioned in the summer of 2005. The beam passes
through a polarizer and the AOM before being emitted from the
enclosure, through the viewport into the vacuum system and onto the
test mass. A small fraction of the beam is continuously picked off
for readback by the photodetector, so that power incident onto the
test mass can be estimated.} \label{fig:Layout}
\end{center}
\end{figure}

\section{Discovering a discrepancy} \label{section:results}

In this section, we summarize results of early pcal calibration
measurements of the LIGO detectors, estimate precision, and make a
comparison with the conventional calibration with voice coil
actuators.  This is work we carried out between the summer of 2005
and the summer of 2006.

\subsection{Initial photon calibrator commissioning}

We commissioned all four photon calibrator systems at the LIGO
Hanford observatory in the summer of 2005\,\cite{kalmus05}.  These commissioning decisions were then replicated at the two Livingston photon calibrator
systems.  Commissioning involved installation of the systems,
development of robust measurement techniques, and characterization
of the systems and measurements made with them.  The six the photon
calibrators were ready in time for the start of the S5 run, although outstanding
mysteries remained.

After installing the Hanford pcal systems, we calibrated the
immediate laser power out of the pcal units (before the beam enters
the vacuum system, hereafter ``immediate power''), in terms of the
AOM drive input DAQ channel counts, using a hand held laser power
meter. We refer to this as the ``photodetector calibration,'' and it
produces a number -- the calibration factor $\alpha_{c}$.

The calibration factor is directly proportional to the power
incident on the optic, which is related to the immediate power
\begin{equation}
P_{\emph{optic}} =
\left(T_{\mathrm{VP}}\right)\left(R_{\mathrm{TM}}\right)P_{\emph{i}}
\label{eqn:Power}
\end{equation}
where the two multiplicative factors, viewport transmission
$T_{\mathrm{VP}}$ and test mass reflectivity $R_{\mathrm{TM}}$ are
slightly less than 1, and it is assumed that there is no other power
loss between the enclosure and the optic. $P_{\emph{i}}$, the
immediate power emerging from the enclosure, can be written in terms
of the directly measurable photodetector readout channel:
\begin{equation}
P_{\emph{box}} = \alpha_{c} P_{\emph{PD}}
\end{equation}
where $P_{\emph{PD}}$ is the number of analog-to-digital converter
counts returned by the photodetector readout channel, and
$\alpha_{c}$ is a conversion factor, which is also measurable.

Any uncertainty or error in either $P_{\emph{PD}}$ or $\alpha_c$
therefore contributes to the overall uncertainty or error in the
calibration. Uncertainty and error in $P_{\emph{PD}}$ is negligible,
but $\alpha_{c}$ is a principle source of both error
and uncertainty at low frequencies $\lesssim800$ Hz.

To measure $\alpha_{c}$, a handheld power meter was placed in front
of the beam immediately before it leaves the enclosure. ADC counts
from the photodetector and the power (in mW) measured by the power
meter were recorded for several DC AOM driver inputs. These
measurements were made at DC; the transfer function between the
input to the AOM driver and the beam power incident on the power
meter is flat from DC up to the highest frequency we are interested
in, $\sim$2\,kHz. These measurements are plotted, and fit with a
line. An example is given in Figure~\ref{fig:Photodetector}.

\begin{figure}[]
\begin{center}
\includegraphics[angle=0,width=90mm, clip=false]{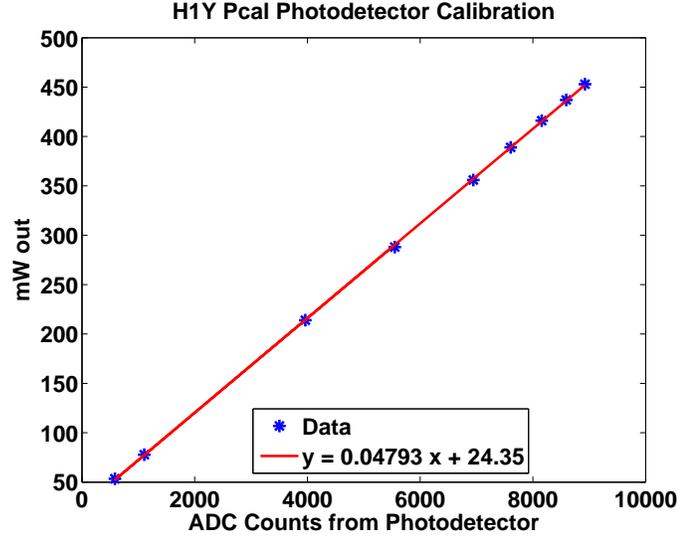}
\caption[Example photodetector conversion factor plot]{Example
photodetector conversion factor ($\alpha_{c}$) plot. Only the slope
of the line is important, as measurements will always be
peak-to-peak.} \label{fig:Photodetector}
\end{center}
\end{figure}

We measured $\alpha_c$ on each of the pcal units several times
during the course of the summer to begin quantifying the uncertainty
in this measurement, which dominates the pcal uncertainty error
budget due to the uncertainty in the laser power measurement using
the hand held power meter.  We initially set the statistical
uncertainty at 3\% at a $1\sigma$
level (Table\,\ref{table:prelimPcalErrors}). In addition to the statistical
uncertainty in $\alpha_{c}$ measurements, there is a $\sim$3\%
systematic error due to the absolute calibration of the power meter
used in the measurement.

We next quantified the transmission of the immediate power to power
reflected off of the test mass, $P$ in Equation\,\ref{eq:master},
for each of the four units.  The immediate power suffers losses
through the vacuum viewport and imperfect reflectivity of the test
mass.  We inferred the transmission through each of the four pcal
viewports by measuring a reflected pcal laser beam from the outer
surface with the hand held power meter.  The results are shown in
Table\,\ref{table:Viewports}; surprisingly, we determined that one of the Hanford viewports was flawed and had much higher reflectivity than
the others, though this had only a small effect on the pcal
uncertainty. Both Livingston viewports were later found to suffer from the same effect.  We inferred the reflectivity of the test mass optic by
direct measurements of test mass witness plates in the laboratory.
The small optical witness plates were given identical coatings to
the actual optics at fabrication time. The results of these
measurements are shown in Figure\,\ref{fig:witness}.  Both of these
effects are accounted for in the pcal displacement calculation;
uncertainties in these measured values increase the uncertainty in
the final pcal displacement, and hence calibration.

\begin{table}[h]
\begin{center}

\caption[Photon calibrator viewport reflectivities]{Viewport
reflectivities for the four Hanford photon calibrator units and the
two Livingston units. Apparently, not all viewports received the
same coatings.  A 10\% uncertainty on a reflectivity measurement of
0.01 corresponds to an uncertainty in the transmitted power of
0.1\%; even with our conservative uncertainty estimate, reflectivity
measurements of the three $\sim1\%$ viewports contribute negligibly
to the overall uncertainty. A 10\% uncertainty on a measurement of
0.07 gives (rounding up) a 1\% contribution to the overall
uncertainty in the calibration factor.}
\begin{tabular}
     {@{\extracolsep{\fill}}ccc}
     \hline
     \hline
Viewport & Reflectivity [\%] &  Uncertainty [\%] \\
\hline
H1X & 7.1 & 1 or less\\
H1Y & 1.1 & negligible\\
L1X & 6.7 & 1 or less \\
L1Y & 7.1 & 1 or less \\
H2X & 1.1 & negligible\\
H2Y & 0.8 & negligible\\
\hline \hline
\end{tabular}

 \label{table:Viewports}
 \end{center}
\end{table}

\begin{figure}[!htb]
\begin{center}
\includegraphics[angle=0,width=110mm, clip=false]{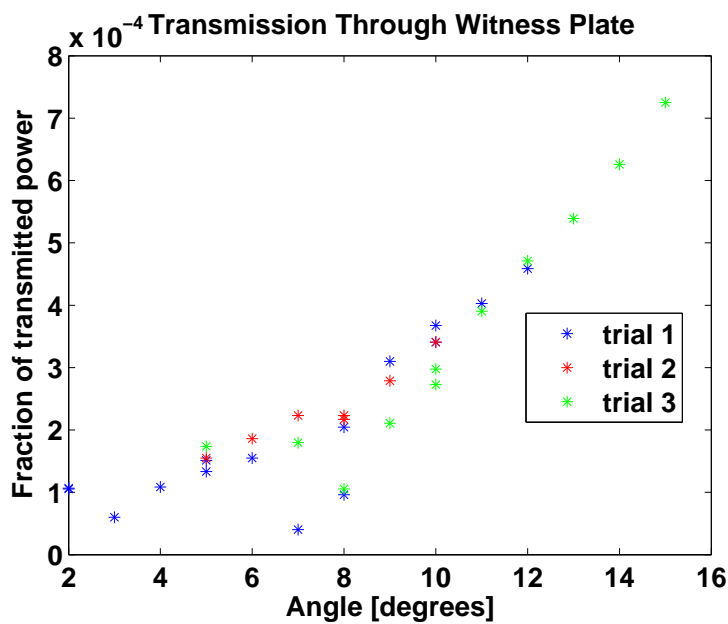}
\caption[Transmission fraction through an end test mass witness
plate]{Transmission fraction through an end test mass witness plate
as a function of beam incident angle.  Transmission was measured
with a spare pcal laser in the laboratory and a hand held power
meter.} \label{fig:witness}
\end{center}
\end{figure}

We then made initial estimates of uncertainties from other
measured quantities in Equation\,\ref{eq:master}. Uncertainty in
measurements of the detector response to the input excitation in the
AS\_Q gravitational wave channel depend on the SNR of the pcal
signal; the test mass displacement amplitudes from the $\sim$300\,mW
beam excitations fall of as $1/f^2$, affecting the SNR
proportionally.   This uncertainty was estimated from populations of
measurements, and ranged from negligible at the lower frequencies to
dominant at higher frequencies.  Estimates of the masses $M$ of the
optics were carefully checked to avoid systematic errors.
Determination of the mass of the optics introduces an estimated random error of
$<0.1$\% which is not significant compared to the other errors. The
same is true of the angle of incidence $\theta$; uncertainty in the length
measurements from the LIGO as-built drawings negligibly affect the
pcal calibration.  However, we were forced to use beams which did
not strike the center of the test mass at both EndX and EndY due to
baffles support frames installed in the beam tubes between the
viewport and the optic.  In this initial measurement, we did not
account for this affect, although we developed the means for doing
so in the future.  Using the measurements in
Figure\,\ref{fig:offctrdfit}, we estimated systematic error from
this source, for these two pcal systems, to be 3\%.  These initial
estimates, summarized in Table\,\ref{table:prelimPcalErrors}, were
conservative; follow-up measurements were planned for
the future.

\begin{table}[h]
\begin{center}

\caption[Initial uncertainties and errors for Hanford pcal
units]{Initial conservative estimates of  significant statistical
uncertainties [\%] and systematic errors [\%] for the Hanford photon
calibrator units at the $1\sigma$ level, made during the summer of 2005. Uncertainty in measurement of the gravitational wave
channel line peak is not included here; for typical integration times and pcal power levels, it
can range between negligible and $\gtrsim$10\% depending on
frequency.   We have added the major individual
sources of uncertainty listed linearly.  }
\begin{tabular}{l|cccc}
\hline \hline
  &  H1X & H1Y & H2X & H2Y \\
\hline
Photodetector Calibration & 3 & 3 & 3 & 3 \\
Viewport Transmission & 1 & - & - & -\\
Power Meter Systematic & 3 & 3 & 3 & 3 \\
Off-centered Beam Systematic & 3 & 3 & 0 & 0\\
\hline
Overall & 10 & 9 & 6 & 6 \\
\hline \hline

\end{tabular}
 \label{table:prelimPcalErrors}
 \end{center}
\end{table}

Next, we developed software capable of making response function
measurements with the photon calibrators in an automated fashion
from the control room.  This software
performed excitations and measured their response using
``tdsresp,''\,\cite{tdsresp} a simple channel excitation and
readback code in C; and reduced those measurements in a Matlab
script with specific knowledge of  the various pcal
units.

Finally, we used these tools and measurements to create plots of
$R(f)$, and compare them to the coil calibration
numbers\,\cite{kalmus05}. Figure\,\ref{fig:prelimDiscrepancy} shows
the ratio of the pcal measurements to the propagated coil actuator
measurements. A significant systematic discrepancy is readily
apparent.  The pcal response measurements put the detector
sensitivities lower than the conventional coil
actuator response measurements. Furthermore, it is clear that the
discrepancy grows with frequency. 

This 2005 result was the first measurement of a significant
discrepancy between calibration via photon calibrators and
calibration via coil actuators of an interferometric gravitational wave detector.   Other documented photon calibrator
measurements did not uncover a significant discrepancy, although the
setups were subject to the same underlying problem.
In\,\cite{clubley01}, measurements made with the Glasgow 10\,m
interferometer are not compared with coil calibration.  Measurements
made with the GEO600 interferometer and published in 2006 found
``reasonably good agreement,'' although the coil calibration
measurements of the detector response did ``tend to be lower'' than
the photon calibrator measurements\,\cite{geo06}. It would take a
few years for this situation to be resolved.

\begin{figure}[!htb]
\begin{center}
\includegraphics[angle=0,width=110mm, clip=false]{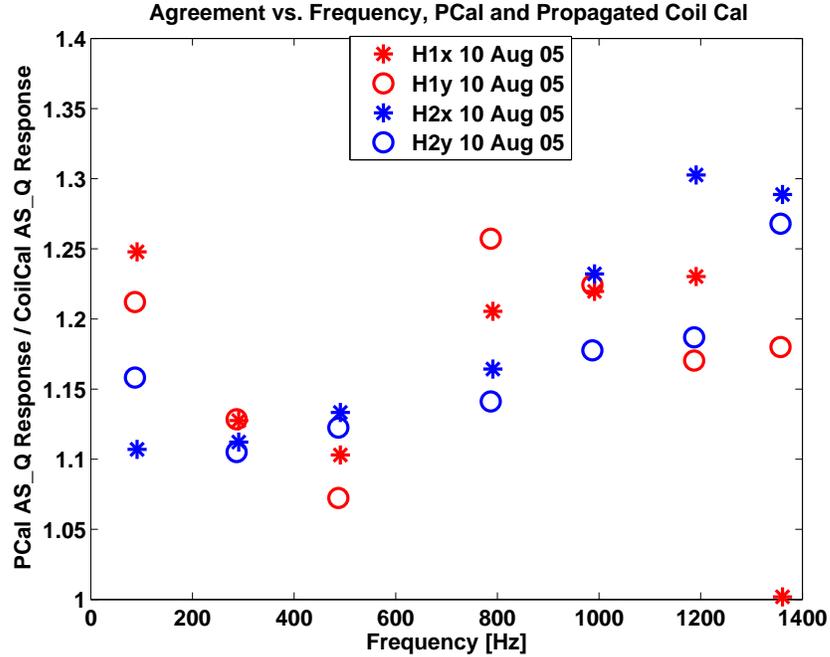}
\caption[Ratio of pcal response to the propagated coil actuator
response] { The ratio of the pcal AS\_Q response measurements to the
propagated coil actuator measurements, presented to the LIGO Scientific Collaboration in the summer
of 2005\,\cite{kalmus05}.     A discrepancy that grows
generally with frequency is readily apparent in each of the four
sets of ratios. The pcal response function puts the detector
sensitivities lower than the conventional coil
actuator response function.  This was the first measurement of a
discrepancy.} \label{fig:prelimDiscrepancy}
\end{center}
\end{figure}

\subsection{Towards resolving the discrepancy}

While still in Hanford during the summer of 2005, we noticed that there
was a $\sim$10\% disagreement in measuring photon calibrator beams
between the power meter used in the $\alpha_c$ pcal calibration
measurements (made by Ophir Optronics), and another hand-held power
meter (made by Scientech Inc.).  After the eventual recalibration of
both units by Scientech it was found that the Scientech unit had
been reading 6.70\% low in laser power measurements and the Ophir
unit had been reading 4.04\% high in power\,\cite{scientechEmail}.
This accounted for 4\% of the H1 and H2 photon calibrator
discrepancies with the conventional coil calibration.  It also made
it clear that a better absolute laser power measuring system for
determination of $\alpha_c$  would be beneficial.

Back in New York in the autumn of 2005, we requested that sinusoidal
excitations be inserted into the interferometers near 1600\,Hz,
using each of the six photon calibrators during the early months of
S5. The purpose of these lines was twofold:   to monitor the
detector responses with an alternative method, and to continue
characterizing the discrepancy. The line locations are given in
Table\,\ref{table:pcal1600}; we chose operating frequencies near
1600\,Hz to be high enough to limit harmonics below 2\,kHz, but to
be low enough for reasonable integration times ($\sim$1000\,s).
Access to the interferometers during S5 for experiments leading to
the resolution of the discrepancy was restricted; a handful of
lines across the spectrum would have been useful, but would have
polluted the S5 noise spectra.

To efficiently utilize continuous excitations, we developed an
automated data monitoring tool (DMT) named PhotonCal with which we
could examine archived data remotely and calculate the
interferometer response function as measured by any photon
calibrator unit.  This tool builds on the Matlab tool we created for
local use at the Hanford site, but it is implemented within the C
DMT framework\,\cite{dmt} which provides standardized services for
data acquisition, graphical display and distribution of
measurements. PhotonCal repeatedly measures $R$ at any pcal
excitation frequency by measuring the DARM\_ERR gravitational wave
channel and one of the four pcal excitation channels, and reducing the measurements with
dictionary files constructed by a Matlab script containing measured
pcal system parameters specific to each of the six units (H1X, H1Y,
H2X, H2Y, L1X, and L1Y) .  In addition, PhotonCal monitors the
detector state vector, so that pcal measurements made while a
detector is out of lock or otherwise not in science mode can be
flagged and discarded.  An example of measurements made with 1024\,s
integrations at 1618.9\,Hz using the L1Y pcal is shown in
Figure\,\ref{fig:dmtexample}.

PhotonCal provides the capability to monitor trends in the long
running $\sim$1600\,Hz pcal lines over periods of months or years.
Using PhotonCal we found the discrepancy to be stable to better than
10\% over a 6.5 month period from the beginning of the pcal lines in
early November 2005 until mid-May
2006\,\cite{kalmus06commissioning}.

\begin{figure}[!htb]
\begin{center}
\includegraphics[angle=0,width=110mm, clip=false]{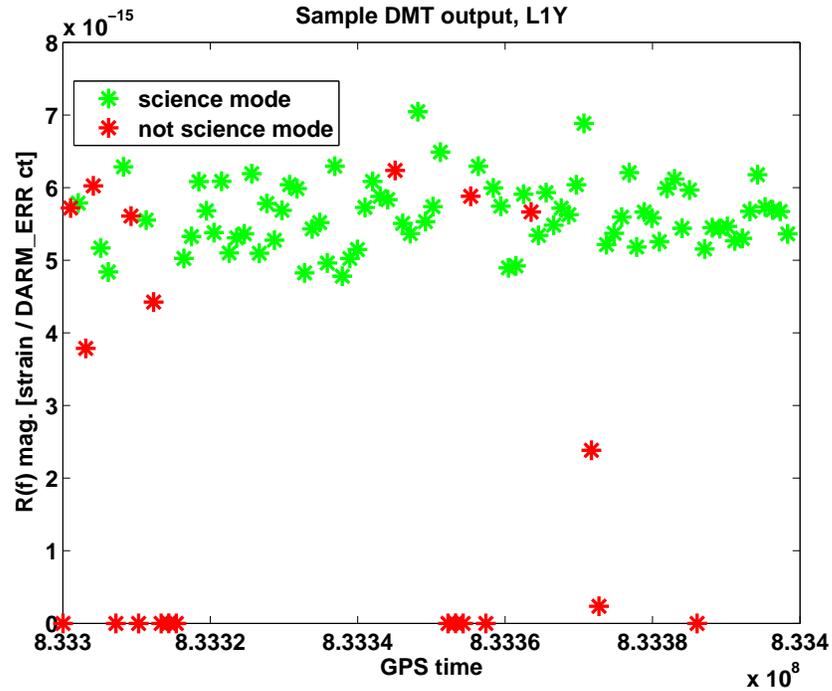}
\caption[Sample PhotonCal DMT measurements] {An example of PhotonCal
DMT measurements made with 1024\,s integrations at 1618.9\,Hz using
the L1Y pcal unit.  The time range on the x-axis spans from 2006
June 02 16:13:38 UTC (GPS 833300032) to 2006 June 03 19:32:02 UTC
(GPS 833398336). } \label{fig:dmtexample}
\end{center}
\end{figure}

With the PhotonCal tool in hand, we made large collections of pcal
measurements near 1600\,Hz for each of the six pcal units, and
compared them to the conventional coil response function at the
corresponding pcal frequency (Table\,\ref{table:pcal1600}).  Large
collections allowed us to measure a time-averaged pcal response
function near 1600\,Hz with precision limited by the pcal systems
themselves, not by the SNR ratio of the pcal excitation in the
gravitational wave channel DARM\_ERR.  Results for the three
interferometers are shown in
Figures\,\ref{fig:h11600},\,\ref{fig:h21600}, and\,\ref{fig:l11600},
and are summarized in Table\,\ref{table:pcal1600}.  We note that the
X and Y pcal units agree to 12\% in H1, 1\% in H2, and 3\% in L1.
This agreement between X- and Y-arm pcals is within the
uncertainty budget given in Table\,\ref{table:prelimPcalErrors}.  The
relatively large disagreement in H1 could be related to the
constraint imposed on the H1 systems by the beam tube baffle
supports, which require a significantly de-centered pcal beam
location on the test masses.  This could cause the pcal beam to overlap the main interferometer beam more or less, which would affect the magnitude of the discrepancy as described below.

\begin{table}[h]
\begin{center}
\caption[Summary of photon calibrator discrepancy near 1600\,Hz and
700\,Hz]{Summary of photon calibrator discrepancy near 1600\,Hz and
700\,Hz. We chose pcal excitation frequencies  $f_{\mathrm{pcal}}$
near 1600\,Hz to be high enough to limit harmonics below 2\,kHz, but
to be low enough for reasonable integration times ($\sim$1000\,s).
Slightly different frequencies were chosen near 1600\,Hz for each
pcal unit so that confusion between units would be impossible.  The
719.1\,Hz measurements were made with special lines which were only
left on for $\sim$hours. $\abs{\bar{R}_{\mathrm{pcal}}}$ is the
average value of the pcal response function magnitude measurements
at $f_{\mathrm{pcal}}$. $\abs{R_{\mathrm{coil}}}$ is the fiducial
response function measured via coil actuators at
$f_{\mathrm{pcal}}$.  We verified that the difference between the
fiducial and propagated coil calibration is insignificant. }
\begin{tabular}{r|cc|cc|cc}
\hline \hline
 &  H1X & H1Y & H2X & H2Y & L1X & L1Y \\
 \hline
 $f_{\mathrm{pcal}}$ [Hz] & 1605.7 & 1609.7 & 1622.9 & 1626.7 & 1613.9 & 1618.9 \\
 $\abs{\bar{R}_{\mathrm{pcal}}}$ & $\sci{1.6}{-14}$ & $\sci{1.9}{-14}$ & $\sci{1.0}{-14}$ & $\sci{9.9}{-15}$ & $\sci{5.8}{-15}$ & $\sci{5.6}{-15}$ \\
 $\abs{\bar{R}_{\mathrm{pcal}}} / \abs{R_{\mathrm{coil}}}$ & 1.36 & 1.56 & 1.62 & 1.61 & 1.38 & 1.33 \\
 $\sigma$ [\%] & 8 & 7 & 10 & 8 & 9 & 8 \\
 N & 235 & 126 & 214 & 148 & 49 & 76 \\
 & & & & & & \\
 \hline
 & & & & & & \\
 $f_{\mathrm{pcal}}$ [Hz] & 719.1 & 719.1 & 719.1 & 719.1 & &  \\
 $\abs{\bar{R}_{\mathrm{pcal}}}$ & $\sci{4.8}{-15}$ & $\sci{4.6}{-14}$ & $\sci{2.3}{-15}$ & $\sci{2.4}{-15}$ & &  \\
 $\abs{\bar{R}_{\mathrm{pcal}}} / \abs{R_{\mathrm{coil}}}$ & 1.34 & 1.29 & 1.05 & 1.09 & & \\
 $\sigma$ [\%] & 2 & 2 & 2 & 1 &  &  \\
 N & 27 & 75 & 77 & 75 &  &  \\
 \hline \hline
\end{tabular}
 \label{table:pcal1600}
 \end{center}
\end{table}

\begin{figure}[!htb]
\begin{center}
\includegraphics[angle=0,width=110mm, clip=false]{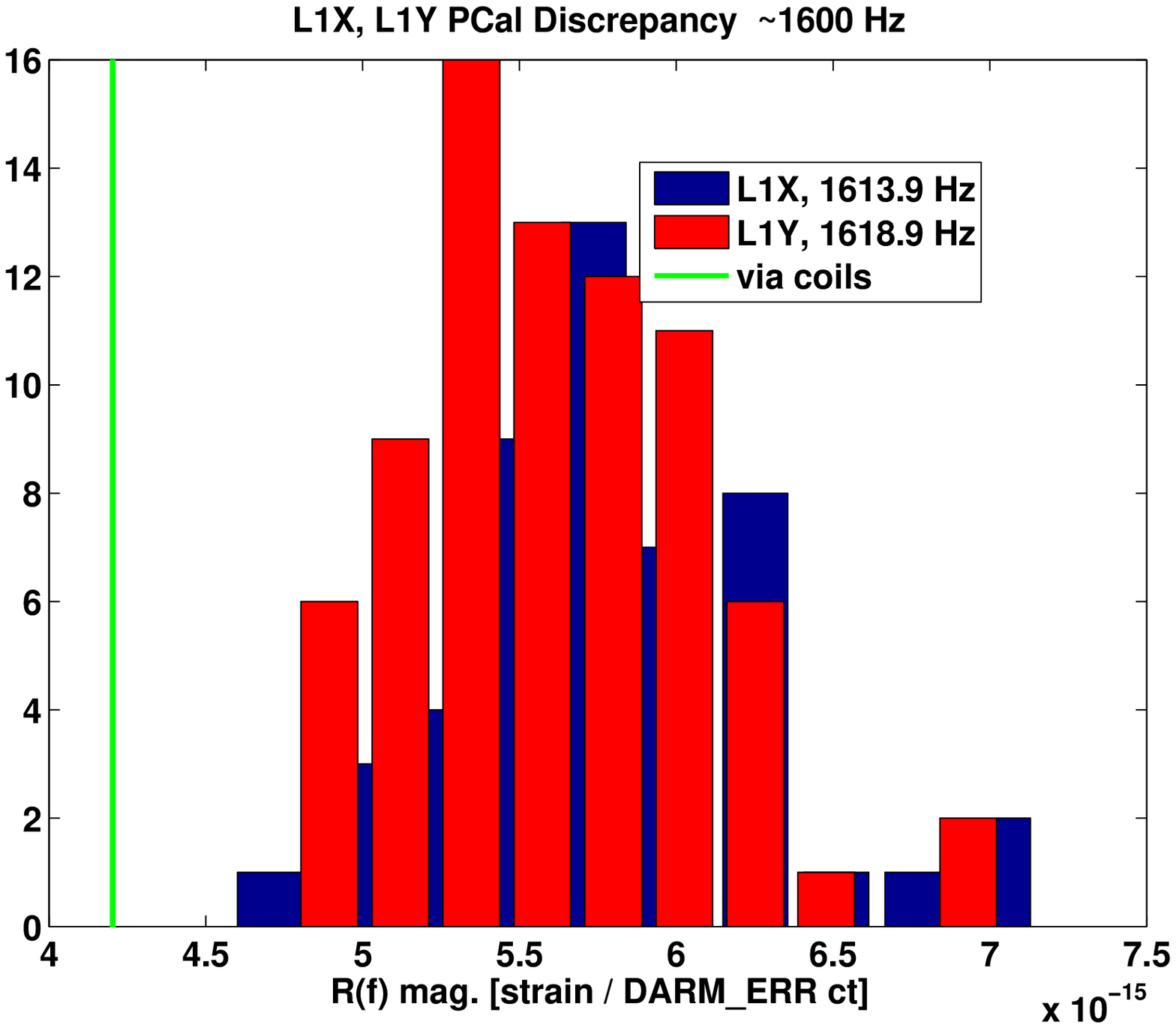}
\caption[L1X and L1Y pcal discrepancy near 1600\,Hz] {L1X and L1Y
pcal discrepancy near 1600\,Hz.  The vertical line is the fiducial
response function magnitude near the pcal excitation frequencies.  }
\label{fig:l11600}
\end{center}
\end{figure}

\begin{figure}[!htb]
\begin{center}
\includegraphics[angle=0,width=110mm, clip=false]{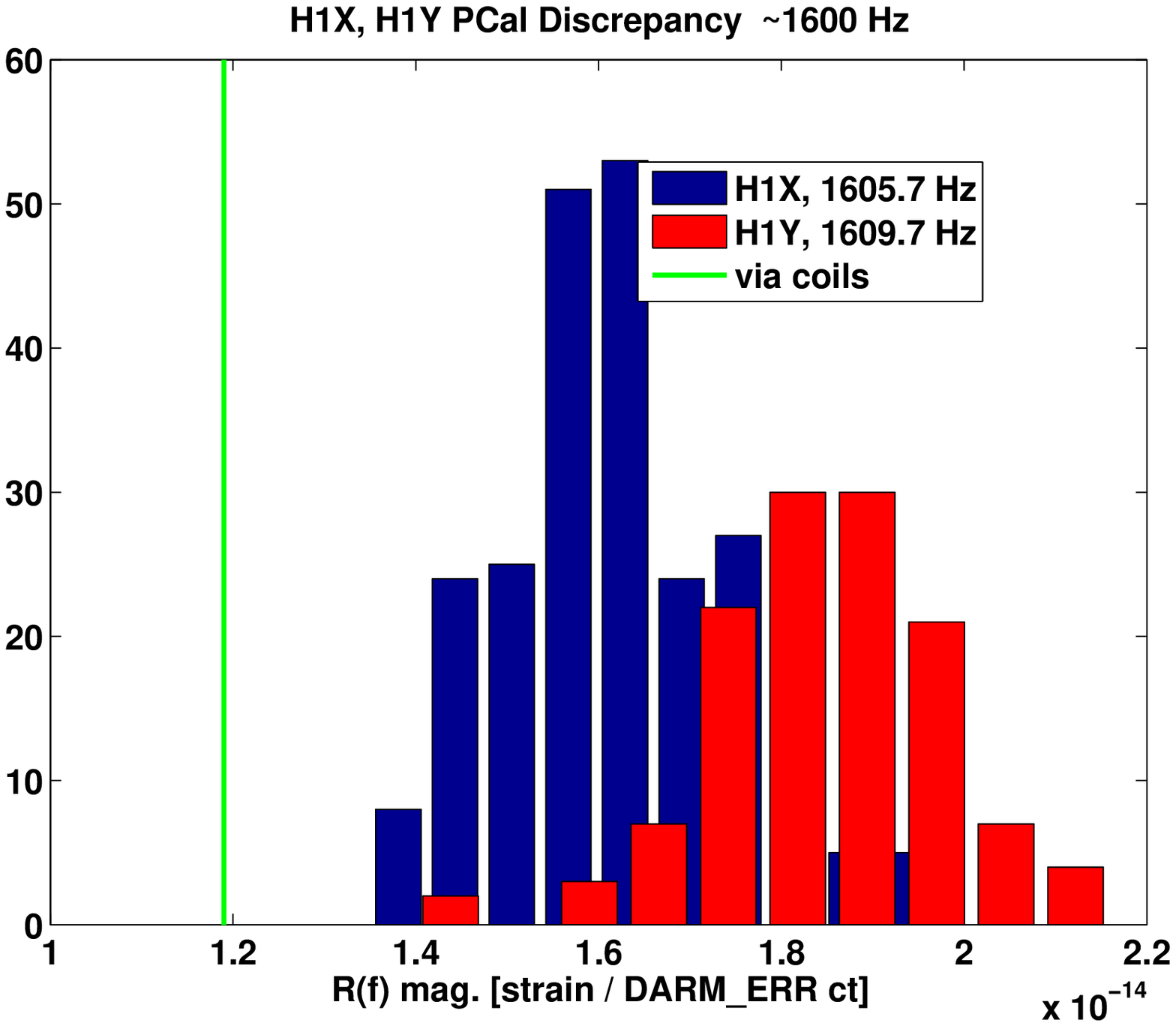}
\caption[H1X and H1Y pcal discrepancy near 1600\,Hz] {H1X and H1Y
pcal discrepancy near 1600\,Hz.  The vertical line is the fiducial
response function magnitude near the pcal excitation frequencies.  }
\label{fig:h11600}
\end{center}
\end{figure}

\begin{figure}[!htb]
\begin{center}
\includegraphics[angle=0,width=110mm, clip=false]{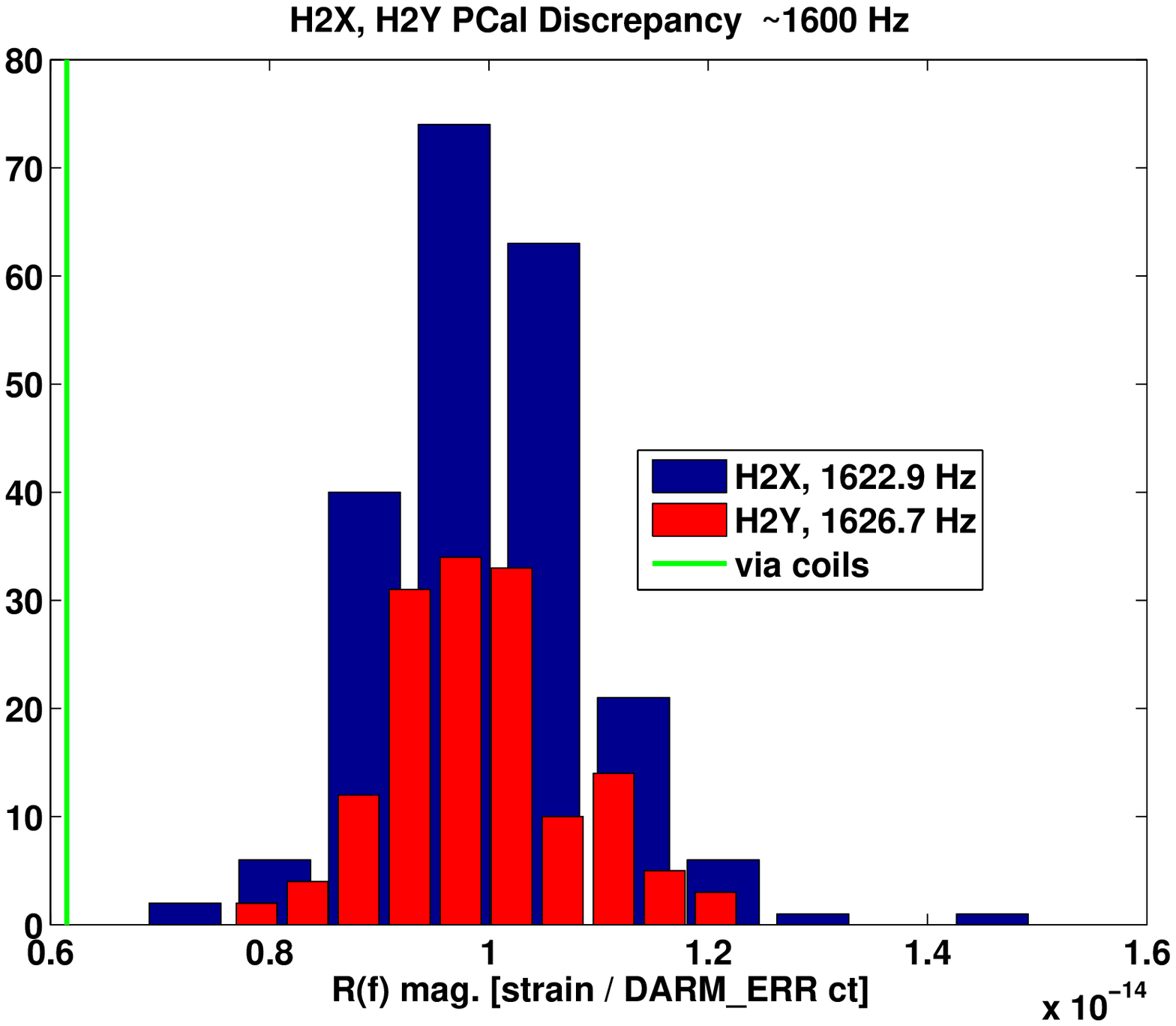}
\caption[H2X and H2Y pcal discrepancy near 1600\,Hz] {H2X and H2Y
pcal discrepancy near 1600\,Hz.  The vertical line is the fiducial
response function magnitude near the pcal excitation frequencies.  }
\label{fig:h21600}
\end{center}
\end{figure}

We then set out to make more detailed measurements of the
discrepancy dependence on frequency, which was difficult due to
restrictions on access to the S5 detectors.  We obtained permission
to place lines at 719.1\,Hz in H1 and H2 for a short time
($\sim$day). The results of these measurements and comparisons to
the measurements near 1600\,Hz are shown in
Figures\,\ref{fig:pcal719h1} and\,\ref{fig:pcal719h2} and summarized
in Table\,\ref{table:pcal1600}.  The H2 detector results show the
same increase in discrepancy with frequency evident in
Figure\,\ref{fig:prelimDiscrepancy}.  The H1 results on the other
hand show less of a dependence on frequency (which was also evident
in Figure\,\ref{fig:prelimDiscrepancy}).  This relative lack of
frequency dependence compared to the H2 detector is also likely related
to the off-centered pcal beams in the H1 detector.

\begin{figure}[!htb]
\begin{center}
\subfigure[]{
\includegraphics[angle=0,width=90mm,clip=false]{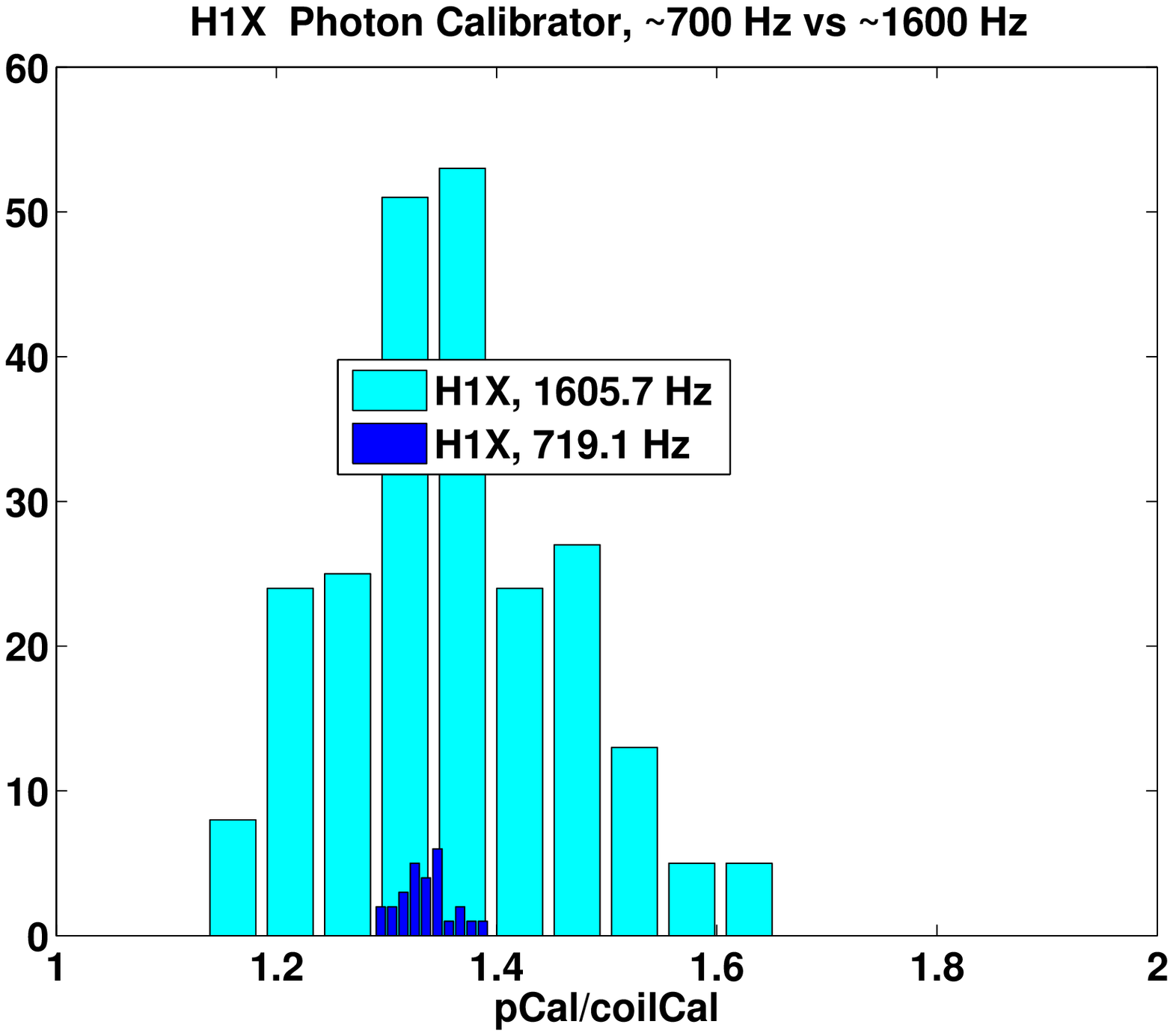}}
\subfigure[]{
\includegraphics[angle=0,width=90mm,clip=false]{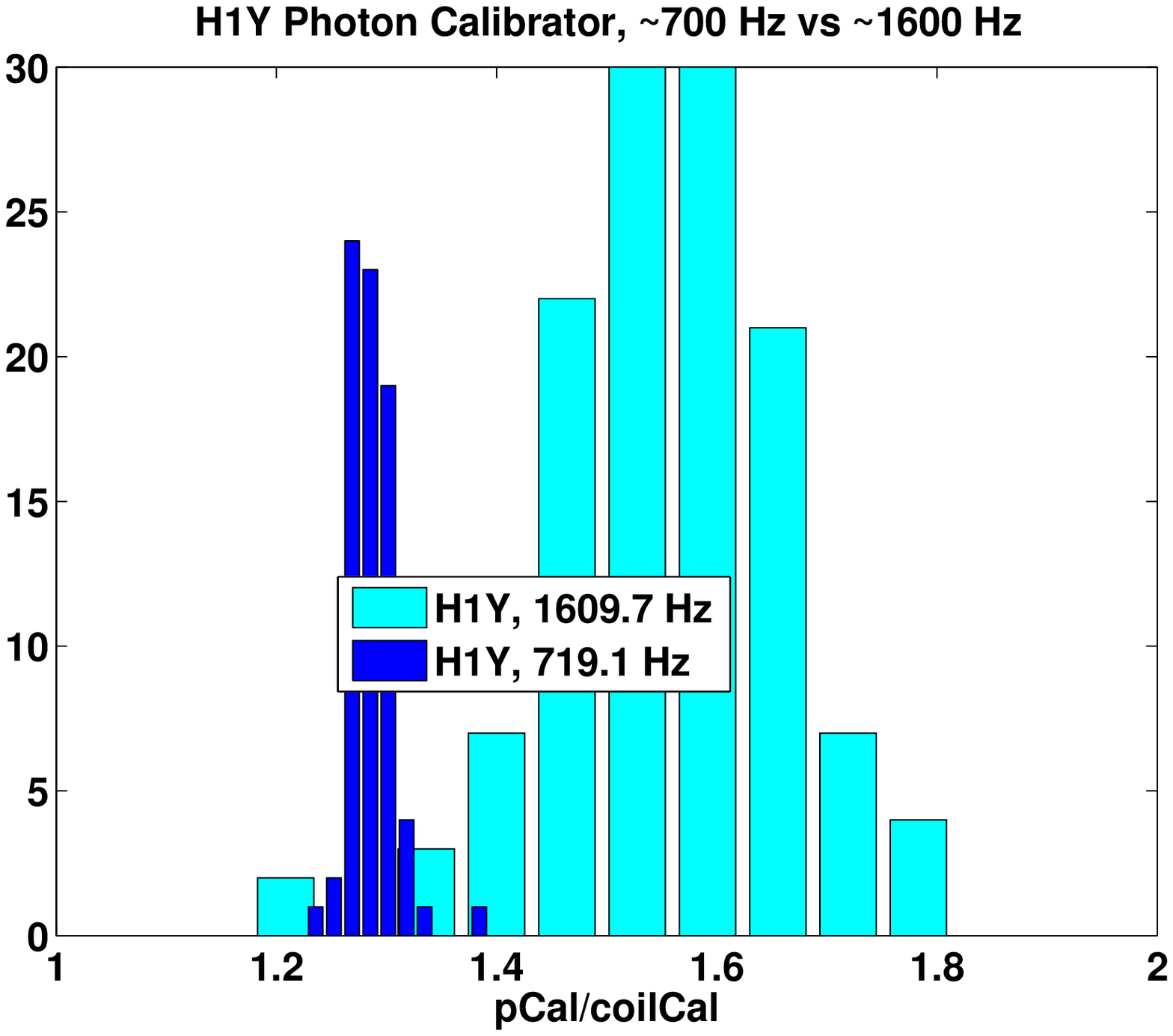}}
\caption[H1X and H1Y pcal discrepancy near 700\,Hz and 1600\,Hz]
{H1X and H1Y pcal discrepancy near 700\,Hz and 1600\,Hz.  Top plot
shows H1X and bottom plot shows H1Y.  The relative lack of frequency
dependence compared to the H2 detector turned out to be related to
the off-centered pcal beams in the H1 detector.}
\label{fig:pcal719h1}
\end{center}
\end{figure}

\begin{figure}[!htb]
\begin{center}
\subfigure[]{
\includegraphics[angle=0,width=90mm,clip=false]{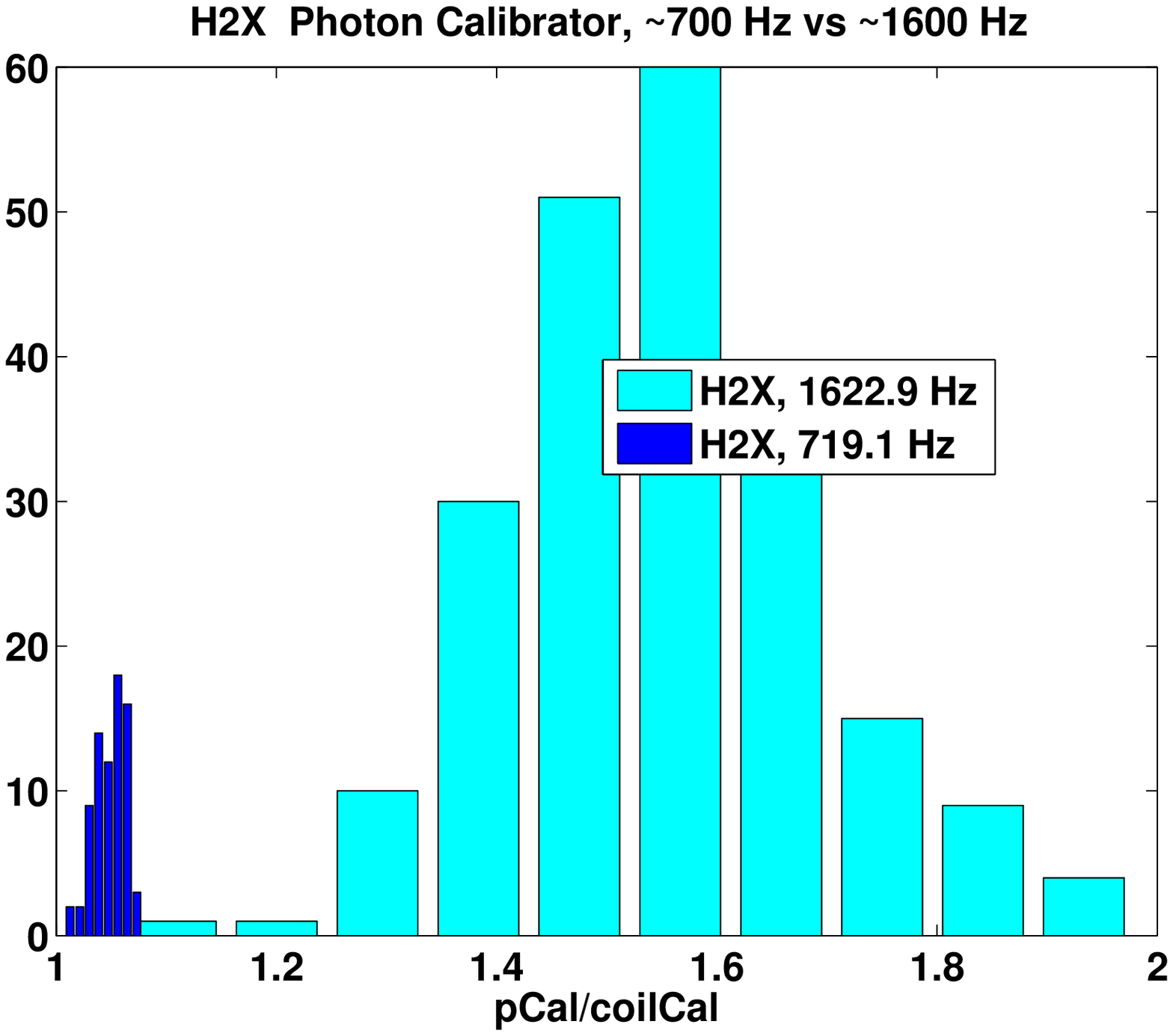}}
\subfigure[]{
\includegraphics[angle=0,width=90mm,clip=false]{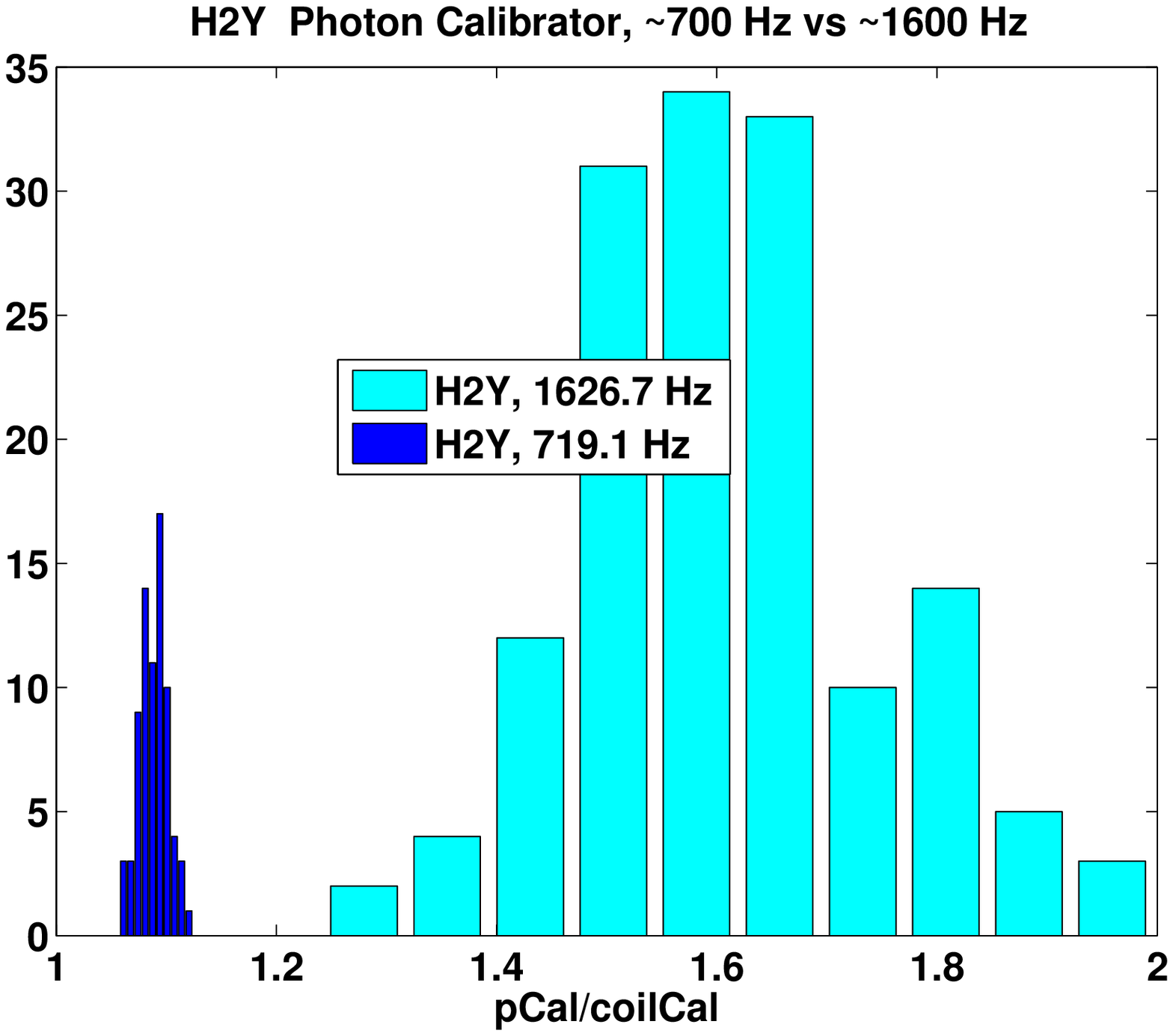}}
\caption[H2X and H2Y pcal discrepancy near 700\,Hz and 1600\,Hz]
{H2X and H2Y pcal discrepancy near 700\,Hz and 1600\,Hz.  Top plot
shows H2X and bottom plot shows H2Y.  } \label{fig:pcal719h2}
\end{center}
\end{figure}

The 719.1\,Hz measurements allowed us to convince the S5 Run
Committee to give us IFO maintenance time to repeat the ``swept
sine'' measurements made in the summer of 2005
(Figure\,\ref{fig:prelimDiscrepancy}) with relatively high frequency
resolution. In the summer of 2006, in collaboration with R. Savage
and E. Goetz at the Hanford site and B. O'Reilly at the Livingston
site, we made swept sine pcal measurements at multiple frequencies
up to about 1500\,Hz.  The results are shown in
Figure\,\ref{fig:secondDiscrepancy}.  These measurements confirmed
the growth of the discrepancy with frequency.  They also confirmed that the discrepancy has a minimum value frequency near 400\,Hz, which may depend on the detector.  We began to think in
terms of a bipartite discrepancy, composed of a frequency dependent
part and a constant bias, possibly with different underlying
mechanisms.

\begin{figure}[!htb]
\begin{center}
\includegraphics[angle=0,width=110mm, clip=false]{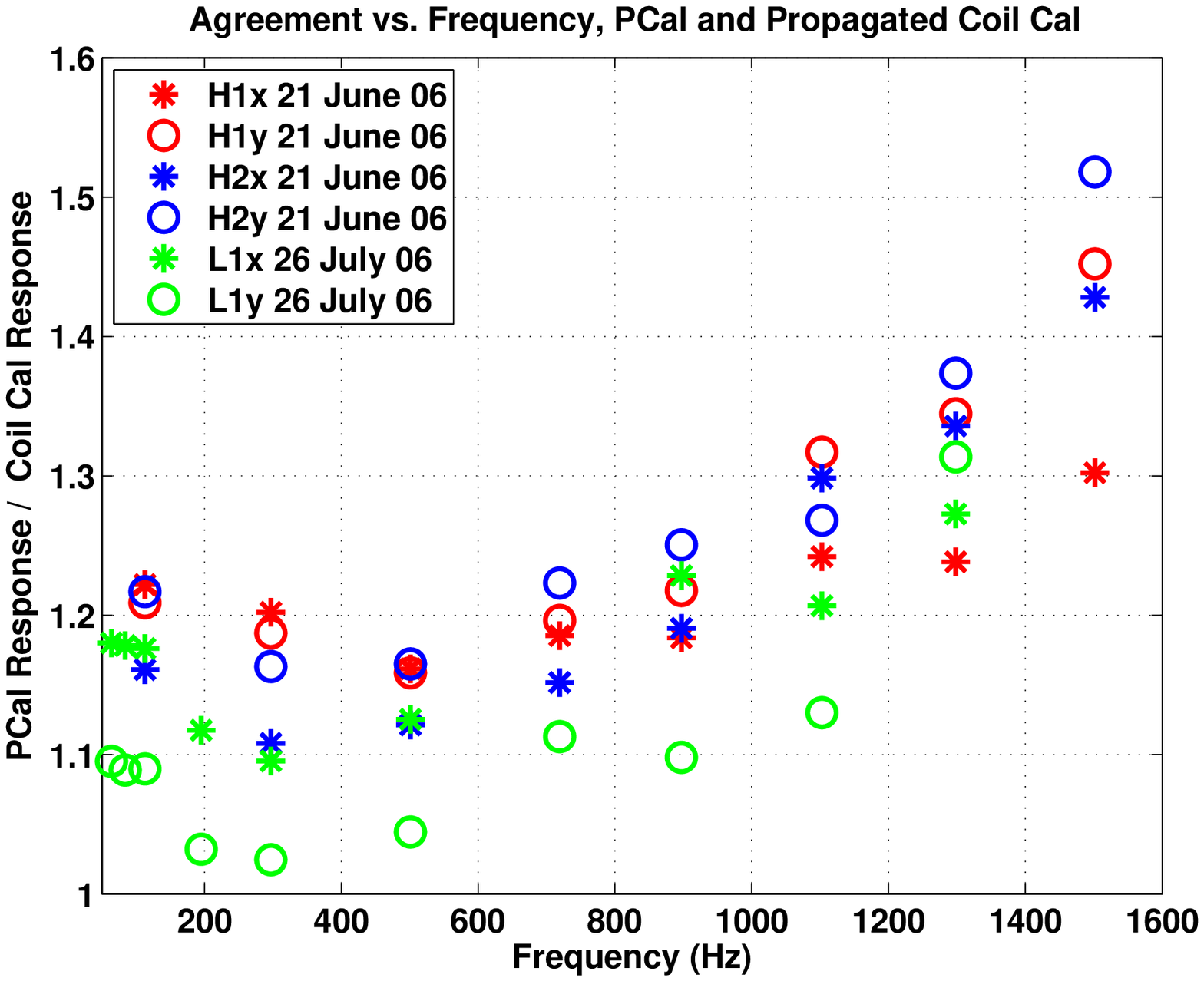}
\caption[Improved measurement of the calibration ratio versus
frequency] { The ratio of pcal DARM\_ERR response measurements to
the propagated coil actuator measurements, made in the summer of
2006. } \label{fig:secondDiscrepancy}
\end{center}
\end{figure}

As work progressed and our confidence in the pcal measurements
grew, we continued to search for explanations for the discrepancy.
One possibility for the frequency dependent part was thermal
expansion in the test mass substrate and HR coating.  Since the pcal
beam was nominally coincident with the main interferometer beam on
the optic, any bulges it caused could be sensed by the
interferometer and interpreted as a relative length change.  We used
the formalism in\,\cite{winkler91} to make a simple model of the
substrate bulging, which we found to be negligible.  We then used
the formalism in\,\cite{raoThesis} to model a bulge in the HR
coating, assuming a ``worst case'' 180 degree phase relative to the
test mass displacement.  We found that this effect could not be
ruled out by our simple model.  We therefore decided to see whether
aiming the pcal spot away from the IFO beam would affect the discrepancy\,\cite{kalmus06}.  

A ``split beam'' pcal configuration was implemented by R. Savage and
E. Goetz at Hanford\,\cite{goetz06} (see
Figure\,\ref{fig:2beamsetup}).  Measurements made in this
configuration showed that it eliminates the frequency
dependent component of the discrepancy with the coil calibration.
The accepted explanation, however, turned out not to be thermal
expansion effects, but mechanical deformations in the test mass
caused by radiation pressure\,\cite{hild07}.  The frequency
dependent mechanism in the long-standing discrepancy had been found.

The low-frequency part of the discrepancy as of 2007 April is
summarized in Table\,\ref{table:actuationCoefficients}.  These
measurements were made by R. Savage and E. Goetz using a procedure
that involves performing two swept-sine transfer function
measurements between an actuator drive signal and DARM\_ERR, one
with the photon actuator and the other with the voice coil actuator.
The ratio of photon actuator transfer function and voice coil
actuator transfer function measurements yields the coil actuator
calibration in meters per drive count, which can be compared to measurements made via voice coil actuators.

\begin{table}[h]
\begin{center}
\caption[Calibration of coil actuators, pcal vs. coils]{Summary of
photon calibrator agreement with standard calibrations of the voice
coil actuators, for the six pcal units, as of April
2007\,\cite{goetz07}. These results are averages over multiple
measurements between 50 and 400\,Hz.  Measurements for H1X had not
yet been performed.  }
\begin{tabular}{c|ccc}
 \hline \hline
 \textbf{Test Mass} & \textbf{Pcal ETM Cal} & \textbf{V2 Standard ETM Cal} & \textbf{Pcal / Standard} \\
  &  [nm/ct]  & [nm/ct]  & \\

   \hline
H1 ETMX & N/A   & 0.470 & N/A\\
H1 ETMY & 0.567 & 0.489 & 1.16\\
H2 ETMX & 0.559 & 0.482 & 1.16\\
H2 ETMY & 0.612 & 0.523 & 1.17\\
L1 ETMX & 0.291 & 0.255 & 1.14\\
L1 ETMY & 0.258 & 0.239 & 1.08\\
 \hline \hline
\end{tabular}
\label{table:actuationCoefficients}
\end{center}
\end{table}

\subsection{Recent advances} \label{section:recentAdvances}

After the S5 run ended on 2008 October 1, the detectors were
available for a few weeks before the interferometer rebuild for
Enhanced LIGO and S6.  During this period we worked with R. Savage
and E. Goetz to test and improve the precision of the photon
calibrators. At the Hanford site, we helped to characterize two new
Labsphere (http://www.labsphere.com/) integrating spheres.  A ``gold standard'' integrating sphere was calibrated
at the National Institute of Standards and Technology (NIST) with an
accuracy of 0.88\% at the $2\sigma$ level\,\cite{goldStandardNIST}.
We used the gold standard to calibrate a ``working standard,'' which
we carried to the end stations to calibrate the pcal systems.  These
integrating sphere measurements of $\alpha_c$ are good to about 1\%
at the $2\sigma$ level.  This advance gives an overall estimated
photon calibrator precision of $\sim$3\% at the $2\sigma$
level\,\cite{goetz08}.  This precision is much higher than what can be acheived with the voice coil calibrators (6\%--8\% at 1$\sigma$).

Work done during this period by R. Savage and E. Goetz provides
evidence that the residual frequency independent part of the
discrepancy might be due to an error in the official calibration
caused by assuming that measurements made with electronics in
``acquire'' mode are valid in ``science'' mode\,\cite{goetz08}.

\section{Time delay measurement} \label{section:pcalTimeDelay}

Precise absolute timing is crucial for coherent analysis of
gravitational wave data, for example the analysis we present in
Chapter\,\ref{chapter:search}.  In
Section\,\ref{section:s5calibration} we show that a residual
30\,$\mu$s timing error between pairs of detectors degrades the
performance of our search by more than 10\% at the highest
frequencies.  This error was uncovered with the photon calibrators.

The photon calibrators were used to make a precise time delay
measurement in the interferometer\,\cite{aso08}.  We performed this
measurement on the Hanford 4\,km detector during a trip to the site
immediately following the S5 science run, in October of 2007, and Y.
Aso subsequently performed the analysis.  The idea was to check the
time delay considerations used in the conventional calibration used
to generate strain-calibrated data (so-called h(t), mentioned in
Section\,\ref{section:calibration}), before the initial LIGO
detectors entered their violent commissioning transformation enroute
to becoming Enhanced LIGO a few weeks after the end of S5. What we
found with the photon calibrator measurements was both surprising
and important.

The measurement was straightforward.  In principle, a sine wave
injected into the photon calibrator AOM input causes a
sine-modulated laser to excite the ETM, thereby inducing a sine wave
response in the detector's gravitational wave output channel, which
is recorded and then converted into h(t) using the conventional
detector calibration at some later time. The photon calibrator
excitation is also recorded digitally.  The relative phases between
the recorded photon calibrator excitation and strain-calibrated data
h(t) can be measured to determine the time delay at that frequency.
In practice, we injected two superposed sine waves, at 110 and
111\,Hz, so that we could determine time delays as large as $\pm
1/(2 \Delta f) = \pm 1/2$\,s by using the beats between the
sinusoids. A single sine wave at 110\,Hz would only allow a time
delay measurement modulo a half cycle, or about $\pm 10$\,ms, at
best.  We make no assumption about the sign of the relative delay.
There could be errors in the calibration model used to generate h(t)
from the raw gravitational wave data. Furthermore, delays in the
pcal pickoff photodetector readback path, or data acquisition
system, could be relatively large.  Details of the measurement and
analysis are given in\,\cite{aso08}.

The h(t) stream attempts to record the strain measured by the
detector via differential test mass motions as a function of time.
Therefore it accounts for sensing delays in the interferometer.
These delays include a data acquisition (DAQ) delay, and the light
travel delay of 13.3\,$\mu$s.   The h(t) should be advanced by this
amount, so that it is synchronized with the test mass excitations.
On the other hand, there are also delays in the pcal excitation
record, relative to the actual test mass excitations:  a separate
DAQ delay and 4\,$\mu$s in the pcal pickoff photodetector.  The pcal
DAQ had been determined to be 25.5\,$\mu$s slower than the
gravitational wave channel DAQ\,\cite{smith08}.  Therefore, we
expect the pcal excitation record to be delayed by
$25.5+4=29.5$\,$\mu$s relative to h(t). Instead we measured a time
difference between the recorded pcal excitation and h(t) of
211.4\,$\mu$s, with an uncertainty in the relative timing of about
1\,$\mu$s, with h(t) advanced relative to the pcal drive. This
implies an error of 182\,$\mu$s in the h(t) timing.

There turned out to be two errors in the calibration model.  First,
the model had incorrectly assigned two additional delays to the
interferometer sensing:  a computer processing delay of 122\,$\mu$s,
and a sample and hold delay of the digital-to-analog (DAC)
converters driving the coils of 30.5\,$\mu$s.  Second, the DAC delay
had been assigned twice due to a sign error.  The total mis-assigned
delay from these causes is thus 183\,$\mu$s.  The actual delay used
in the model was 187\,$\mu$s and the correct delay should be
13.3\,$\mu$s, for a systematic error of 173.7\,$\mu$s.  This means
that there are still about $183-173.7=9.3$\,$\mu$s of error that is
not completely understood.  This is just within LIGO's original
requirement of 10\,$\mu$s.

This discovery resulted in a revised version of h(t). The photon
calibrator once again provided an invaluable independent calibration
check.

\section{Future photon calibrators}\label{section:pcalFuture}

Photon calibrators provide an independent calibration of LIGO's
three gravitational wave detectors. Their nominal $2\sigma$
confidence error bars are currently estimated to be $\sim$3\%, which
is significantly more precise than the conventional voice coil calibration,
which has error bars of between 6\% and 8\% at the $1\sigma$ level
(see Section\,\ref{section:s5calibration}).  Agreement with the
official calibration via voice coil actuators has been achieved to
within error bars.

The photon calibrators have provided a valuable check on the
voice coil calibration.  They have uncovered a significant timing
error. They may also have uncovered an error due to assuming that
measurements made with electronics in ``acquire'' mode are valid in
``science'' mode.  In general they have helped us to better understand the detector calibration.

Advanced LIGO plans call for photon calibrators as a key element in
the detector calibration chain\,\cite{aligoPcal}.  One of the
limitations with the photon calibrators as implemented in initial
LIGO was pcal SNR at high frequencies.  Advanced LIGO's order of
magnitude amplitude sensitivity improvement will give an SNR improvement to the photon calibrators, although we expect the improvement to be closer to a factor of 3, since advanced LIGO optics are more massive than current optics.  SNR could
be further improved by upgrading the pcal lasers.  This simple
calibration system, now well-understood, which operates while the
detectors are in their science-taking mode, which uncovered problems
with the coil actuation calibration, and which offers significantly
higher precision should be considered as the primary Advanced LIGO
detector calibration system.

\clearpage
\chapter{Flare Analysis Pipeline} \label{chapter:flare}

In this chapter we describe the Flare analysis pipeline\,\cite{kalmus07}, an excess
power type search method~\cite{flanagan98a, flanagan98b, anderson01}
designed to search for gravitational wave signals and using loudest
event statistics to estimate upper limits~\cite{brady04} on
gravitational wave emission associated with
astrophysical triggers.  The Flare pipeline
(Figure\,\ref{fig:infoflowchart})  searches in data from
either one or two detectors, and
was the tool used to complete the individual SGR burst search for
the SGR 1806--20 giant flare and bursts occurring during the first
year of S5 (S5y1), described in Chapter\,\ref{chapter:search}.  It
is also the foundation of the multiple SGR burst search pipeline,
Stack-a-flare, described in Chapter\,\ref{chapter:stack}.

Inputs to the Flare pipeline include gravitational wave detector
data, information describing the set of astrophysical trigger
events, and pipeline parameters.  The astrophysical trigger events
and the pipeline parameters are stored in text files which are read
by the pipeline, so it is not necessary to change code or recompile
before running a new externally triggered search.  This allows future externally triggered searches to be published
with minimal LSC review.

Processing includes generation of \emph{analysis events} (the
fundamental objects used in comparisons of signals to noise and
signals to other signals), determination of the significance of the
loudest on-source analysis event relative to the background, and
estimation of upper limits.

Outputs include lists of characteristic properties of on-source
analysis events, including their significance relative to the
background in terms of false alarm rates (FARs); upper limit
estimates for gravitational wave strain amplitude at the detector
and (if source distance is known) upper limit
estimates for isotropic gravitational wave emission energy from the
source, set using various simulation waveform types; and a variety
of plots and lists useful in establishing confidence in the
on-source analysis event significances and upper limits.

\begin{figure}[!t]
\begin{center}
\includegraphics[angle=0,width=100mm, clip=false]{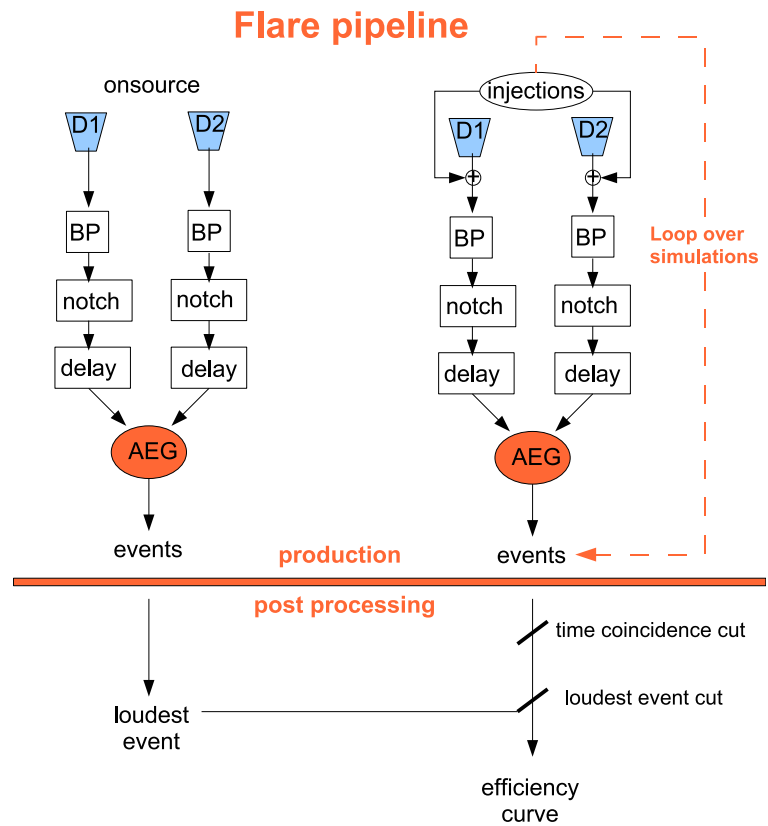}
\caption[Information flow chart of Flare pipeline]{ Information flow
chart of Flare pipeline.  This diagram shows the procedure used to
analyze on-source regions to determine loudest on-source analysis
events, and also the procedure used to perform the Monte Carlo using
simulations injected into the background region to estimate loudest
event upper limits.  Except for the location of the analysis
(on-source region or portion of background region)
and the lack or presence of an injection, the two procedures are
identical. In the post processing stage the
loudest on-source analysis event is used, along with analysis events
associated with injected simulations, to generate an efficiency
curve which yields 90\% detection efficiency loudest event upper
limits.  D1 and D2 represent LIGO detectors (with D2 optional); BP
is bandpass filtering; AEG is the analysis event generator, which is
described in the text.  } \label{fig:infoflowchart}
\end{center}
\end{figure}

\clearpage

\section{A pipeline for triggered searches} \label{section:overview}

It is often the case that astrophysical sources of potential
transient gravitational waves emit electromagnetic bursts
coincidentally or nearly so.  Since electromagnetic and
gravitational waves both travel at the speed of light, knowledge of
the Earth crossing time of the electromagnetic event is thus
knowledge of the Earth crossing time of the gravitational wave.
(Because electromagnetic waves interact with intervening matter,
over cosmological distances the electromagnetic burst could arrive
significantly after the gravitational wave, but this is not a
concern for events occurring in our cosmic neighborhood.) In
addition, modern electromagnetic observations typically reveal the
sky position of the event, as is the case in the search which is the
subject of this work.

Knowledge of time and sky position can be a great advantage to
gravitational wave searches.  It allows us to calculate the detector
response functions, allowing us to estimate upper limits using
simulated signals from the source. Furthermore, in the case of a
global network of gravitational wave detectors, a consistency cut
can be applied to gravitational wave candidate analysis events based
on sky position. This knowledge also significantly decreases the
computation resources necessary for the analysis. Also, strain upper
limits are typically lower than for untriggered ``all-sky''
searches, largely because they are more robust to loud glitches.
Finally, searches can be more scientifically interesting as they
target specific events.  In fact it is possible with triggered
searches to know the distance to the source, which means that the
results can be given in terms of isotropic gravitational wave energy
emitted from the source instead of strain amplitude at the detector.
This ties the search to the astrophysical source instead of the
detector on Earth, which is bound to be more scientifically
interesting.  All of these advantages apply to our search for
gravitational waves from SGR bursts (Chapter\,\ref{chapter:search}).

\section{Overview}

The Flare pipeline can be used to find gravitational wave candidates
and to estimate upper limits using data from gravitational wave
detectors. Available detector data are first divided into an
on-source region in which we might expect to find the gravitational
wave signal associated with the astrophysical trigger, and an
off-source (or background) region which provides a background for
the search.  The on-source region, which is a parameter of the
search, should be large enough to include most model predictions of
the source. The off-source region should be close enough to the
on-source region to ensure similar detector behavior, and large
enough to provide adequate statistics.

On-source and background segments are analyzed identically resulting
in lists of analysis events. The background is used to estimate the
significance of on-source analysis events; significant events, if
any, are subject to additional environmental vetoes and consistency
checks.   Significance is given in terms of FARs estimated from the
background. Calculating FARs in a two-detector search may be
facilitated if necessary by time shifting data streams to increase
the size of the background ensemble, if we assume that the noises
are ergodic. This assumption is commonly made in current LIGO
searches, which use time shifting techniques\,\cite{S4burstAllSky}.

It is useful to compare the loudest on-source event to
pre-determined FAR detection and non-detection thresholds.  These
thresholds are essentially subjective.  A reasonable non-detection
threshold might be one false detection in 10 years; an analysis
event less significant than this is unlikely to persuade the larger
community of a detection.  A reasonable detection threshold might be
one false detection in 100 years of sky observation; an analysis
event more significant than this which passes the detection
checklist tests might persuade the larger community of a detection.
This corresponds to much less than 100 years of data in an
externally triggered search such as ours in which small regions of
data around relatively rare astrophysical triggers are analyzed.

We can also use the Flare pipeline to estimate upper limits on
gravitational wave strain at the Earth via simulated signals
injected into raw data (Section~\ref{section:simulations}.)  If
distance to the astrophysical source is known, upper limits on
isotropic gravitational wave emission energy can also be estimated.
The on-source loudest event is used as a threshold for creating
efficiency curves from the simulations in both cases. (This
efficiency curve threshold is unrelated to the non-detection and
detection thresholds mentioned above.)  We note that upper limit
estimates are in general sensitive to the size of the on-source
region, since longer stretches of noise are likely to produce larger
loudest events. Upper limits also depend on search pipeline
parameter tuning choices, detector sensitivity and antenna factors
at the time of the burst, the loudest on-source analysis event, and
the simulation waveform class used.

These procedures will be explained in detail below.

\section{Input: Astrophysical trigger events list}
\label{section:eventslist}

Any Flare pipeline externally triggered search is controlled by an
ASCII file referred to as an ``events list,'' which gives
information about the astrophysical electromagnetic trigger events
used in the search. The events list for Flare pipeline is
implemented with one line per upper limit. Thus there are M lines
per astrophysical trigger, where M is the number of simulation
waveform classes with which to set upper limits. Each line contains
the following columns:
 \ben
  \i astrophysical trigger name
  \i astrophysical trigger GPS time
  \i detector network that observed the event, after data quality cuts through category 2
  \i source right ascension (degrees)
  \i source declination (degrees)
  \i distance to source (kpc)
  \i on-source region seconds before and after trigger GPS time (two
  numbers separated by a comma)
  \i background region after data quality cuts through category 2
  (Section\,\ref{section:dq})
  (comprised of multiple segments separated by semicolons, each segment defined by start
  and end GPS times separated by a dash)
  \i simulation waveform code, which determines what simulation config
file the pipeline chooses.
 \een
Below is a line excerpted from the events list controlling the S5
first year SGR search presented in Chapter\,\ref{chapter:search}
(the line has been wrapped to fit in the pagewidth):
 \begin{verbatim}
 827345255 827345255.000 L1H1H2 286.80970 9.32225 1.0000e+04 2,2
    827344252-827344724;827344725-827346257 RDL_200ms1090Hz
 \end{verbatim}
The S5y1 SGR events list has 2280 such lines, controlling 190
distinct search on-source regions each with 12 upper limits for 12
distinct waveform classes.

The detector network and background region are determined by
consulting lists of data quality segments for the individual
detectors.

\section{Processing: Generation of analysis events} \label{section:dataProcessing}

Analysis events can be produced from either a single stream of raw
detector data or two synchronized (or time-shifted) streams from two
detectors.

First, data are conditioned via digital filters. In two-detector
searches a time delay is applied as appropriate for the relative
locations of astrophysical source and gravitational wave detectors.
The power spectral density (PSD) or cross PSD is then calculated.
The mean PSD value from off-source data at each frequency bin is
subtracted from elements of that frequency bin to estimate excess
power.

The pipeline can be run with the analysis event generator (AEG) in a
time series mode or in a clustering mode. In the time series mode
the AEG produces as output an excess power type time series (one
detector) or a cross excess power time series (two detectors). This
time series is a projection of a time-frequency matrix onto the time
axis, and this projection can be accomplished by selecting pixels in
different frequency bins at a given time in a variety of ways, as is
optimal for the expected signal.  In the clustering mode the AEG
produces a set of disconnected clusters, which are the analysis
events. The clustering mode typically gives higher sensitivity for
signals that are extended in both the time and frequency dimensions,
as less noise is integrated along with the signal.  The time series
mode is faster and has much tamer memory requirements.

\subsection{Data conditioning}

Data conditioning consists of zero-phase digital filtering in the
time domain~\cite{hamming98}, first with a bandpass filter and then
with a composite notch filter. The raw calibrated LIGO power
spectrum is colored, and is characterized by a sensitive region
between $\sim$60~Hz to $\sim$2~kHz which includes a forest of narrow
lines, with increasingly loud noise on either side of the sensitive
band (see Figure\,\ref{fig:runs}). Search sensitivity is increased
by removing these insensitive regions from the data, which would
otherwise dominate weak signals and destroy bandwidth after
transformation to frequency domain.  We therefore bandpass and notch
the raw data with a 12th-order IIR bandpass filter with 64--2048~Hz
passband and a notch filter ``trained'' on off-source data. Training
consists of creating a high resolution power spectrum and
iteratively finding and removing lines above a specified
significance threshold.

Long duration narrow band signals are not targeted by our search and
their removal in the time domain maximizes the useful bandwidth of
the search. We remove narrow lines associated with the power line
harmonics at multiples of 60~Hz, the violin modes of the mirror
suspension wires, calibration lines, and persistent narrow band
noise sources of unknown origin. In a two-detector search, lines are
found separately for each detector's data stream, and the union of
both sets of lines are used to create a single notch filter.

A buffer interval of 10 seconds on each side of filtered data is
discarded.  This buffer is significantly longer than the
characteristic impulse and step response of the filters, as
discussed in Section\,\ref{section:conditioningValidation}.

After the data conditioning procedure the loudest lines have been
attenuated (Figure~\ref{fig:dataConditioning}), and power on either
side of the sensitive region has been removed.

\begin{figure}[!t]
\begin{center}
\includegraphics[angle=0,width=120mm, clip=false]{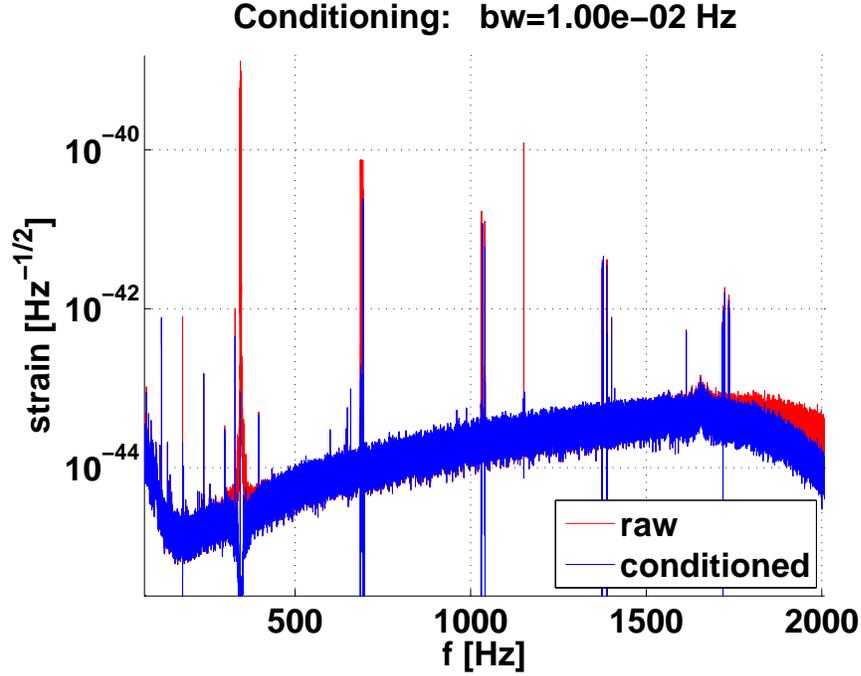}
\caption[Effect of data conditioning on simulated noise spectrum] {
Spectra of simulated L1 noise before data conditioning (background)
and after data conditioning (foreground). Spectral binwidth is
0.01~Hz. The spectra were made with 20 averages. Data conditioning
included application of a 64-2048 Hz bandpass filter and an
automatically generated notch filter which attenuates lines larger
than a specified threshold above the background. The simulated data
used to produce this plot matches the LIGO power spectrum bandpassed
between 64 and 2048~Hz and the data conditioning procedure is
identical as for a real search.} \label{fig:dataConditioning}
\end{center}
\end{figure}

\subsection{Measuring excess power}
\label{section:dataReductionDescription}

Time-frequency spectrograms are then created from conditioned data
for individual detectors from a series of Blackman-windowed discrete
Fourier transforms, of time length $\delta t$ set by the target
signal duration. A \emph{tile} is an estimate of the short-time
Fourier transform of the data at a specific time and frequency. Each
column in the tiling corresponds to a time bin of width $\delta t$
and each row corresponds to a frequency bin of width $\delta f$,
both linearly spaced, with $\delta f \delta t = 1$.  Adjacent time
bins overlap by $0.9 \delta t$ to guard against mismatch between
prospective signals and tiling time bins. Larger overlaps require
more computation and do not noticeably improve sensitivity (see
Section\,\ref{section:overlap}).

In a one-detector search, we then have a complex-valued
time-frequency tiling from which we calculate the real-valued
one-sided PSD for every time bin.  To do this we multiply each tile
value by its complex conjugate and normalize the result to account
for sampling frequency and windowing function. We discard frequency
bins outside of the chosen search band.

\begin{figure}[!t]
\begin{center}
\subfigure[]{
\includegraphics[angle=0,width=100mm,clip=false]{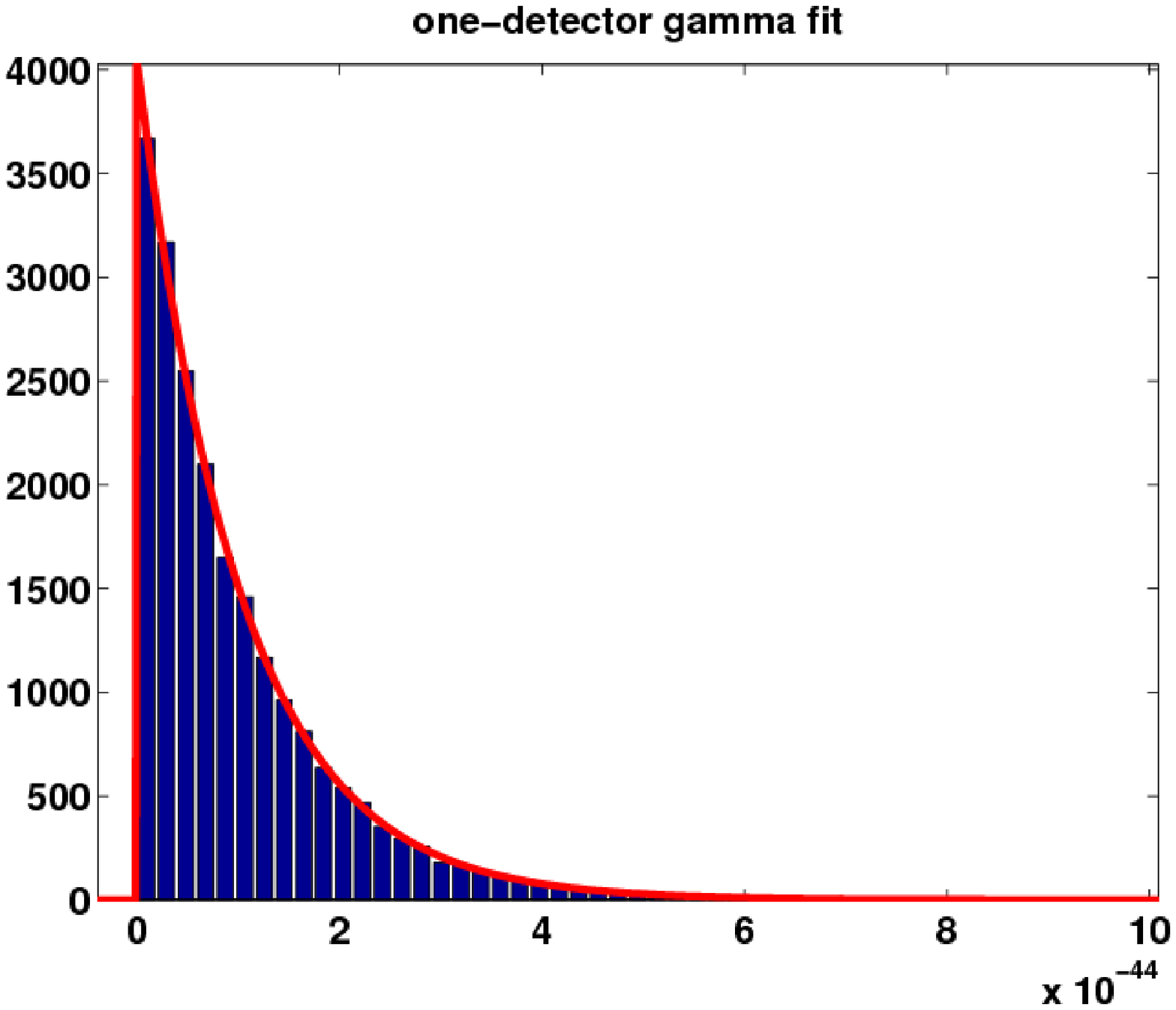}}
\subfigure[]{
\includegraphics[angle=0,width=100mm,clip=false]{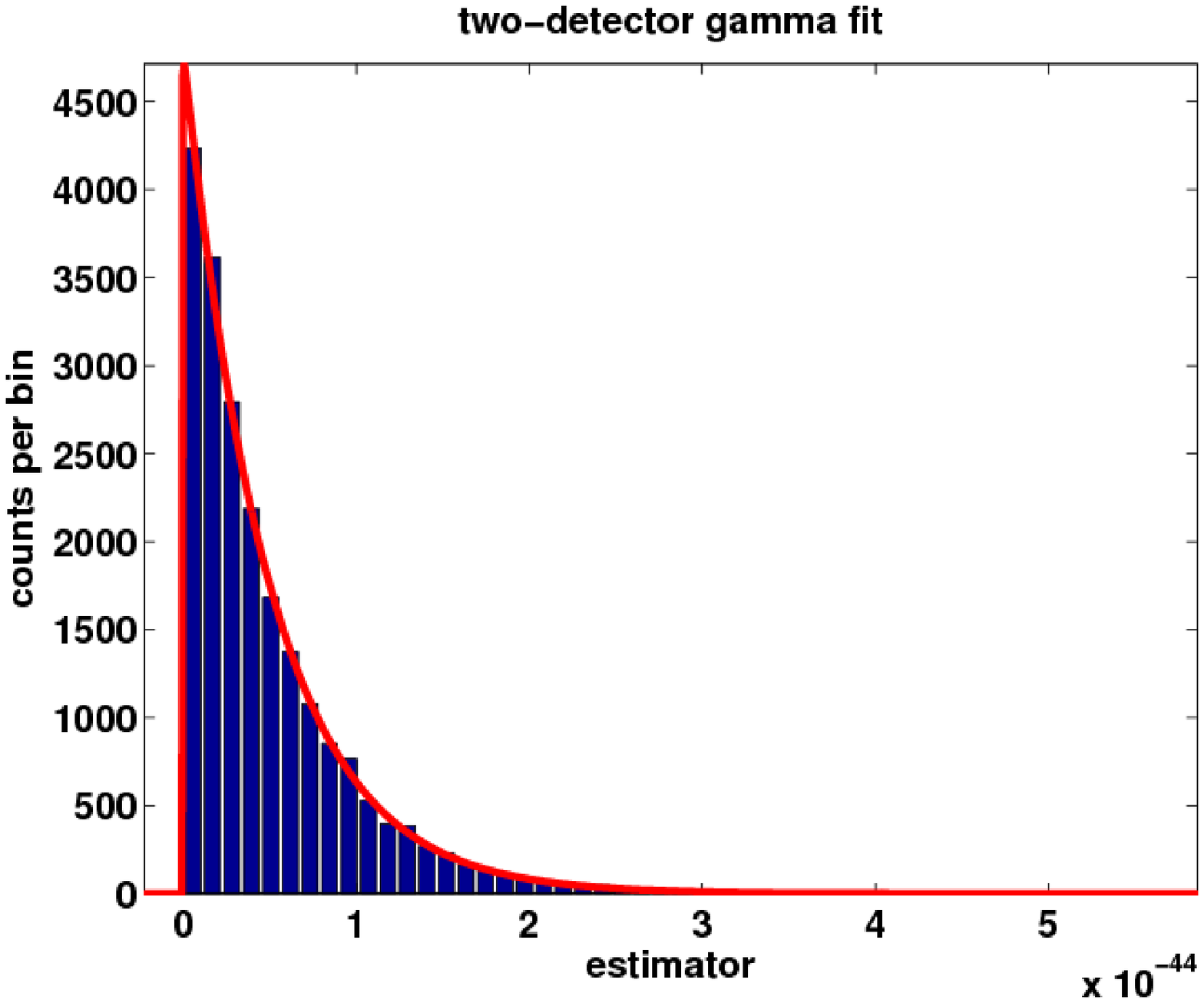}}
\caption[Distribution of power in tiles]{(a) Distribution of power
in tiles in a randomly chosen frequency bin in the one-detector power
tiling. The data was fit to a gamma PDF.  The 90\% confidence
interval values on the maximum likelihood estimates for the
parameters of the gamma distribution fit agree with the parameters
to $\sim$1\%.   (b) Distribution of power in tiles in a randomly
chosen frequency bin in the two-detector tiling. The data was also
fit to a gamma PDF.  The 90\% confidence interval values on the
maximum likelihood estimates for the parameters of the gamma
distribution fit agree with the parameters to $\sim$1\%.   }
\label{fig:dcval}
\end{center}
\end{figure}

In a two-detector search, we have two complex time-frequency tilings
(one for each detector) from which we calculate
 \be
   P^{(12)}_{tf} = \mathrm{Re}\left[T^{(1)}_{tf} \cc{T^{(2)}_{tf}} e^{-i2\pi f \Delta t}\right]
 \ee
where $T$ represents a tiling matrix and $t$ and $f$ are time and
frequency bin indices, and $(1)$ and $(2)$ denote the detector. Here
$\Delta t$ is the gravitational wave crossing time difference
between detectors;  this term takes care of applying the appropriate
time difference between detector data streams in the Fourier space,
with the advantage of permitting sub-sample time delays, which
significantly increases the sensitivity at higher frequencies. The
real part is kept, and normalization is applied as in the
one-detector case. To obtain a positive-definite statistic we take
the absolute value of each tile; this allows sensitivity to both
strongly correlated and strongly anti-correlated signals in the two
(potentially misaligned) detectors.

Next, we use off-source data to remove the background noise power
from each element of the PSD time-frequency tiling.  The elements
are fit to a gamma distribution, and outliers above a threshold
(typically four standard deviations) are discarded.  This process
repeats until no outliers remain.

In the one-detector case the data model could be a chi-square
distribution, which is a special case of the gamma distribution; in
the two-detector case the data model could be a folded normal
product distribution. The gamma distribution fits the data well in
both cases, with 90\% confidence interval values on the maximum
likelihood estimates for the fit parameters constraining those
parameters at the percent level at every frequency bin in typical tilings.

The resulting estimate on the mean is subtracted from each element
of the corresponding frequency bin in the PSD matrix, giving a
matrix of excess power (or ``cross excess power'' in the
two-detector case).  We can also normalize each frequency bin
element in the excess power matrix by the resulting estimate on the
standard deviation, giving a significance matrix.

To create an analysis event with a statistic of event loudness out
of the excess power time-frequency tiling, we project the tiling
onto the time axis. For monochromatic target signals (such as
neutron star ringdowns) we take the two loudest adjacent frequency
tiles in every time bin. Taking two tiles instead of one guards
against mismatch between tiling frequency bin boundaries and signal
location.  For wide-band target signals (such as WNBs) we include
all frequency bins within the search band in the projection.

We can also use a density-based clustering algorithm~\cite{khan07}
which allows retention of signal energy which might otherwise be
fragmented in the case of extended signals in the time-frequency
plane.  The analysis events correspond to discrete clusters found by
the algorithm, and include information on cluster central frequency,
central time, bandwidth, duration, and so forth. The
statistic in this case is the sum over the cluster of tile
significance.

\section{Processing: Significance of on-source analysis events}

We then have an algorithm capable of converting gravitational wave
detector data streams into analysis events with a loudness
statistic.  We can run the implementation of this algorithm on a
search on-source region and on a search background, producing
on-source analysis events and background analysis events.  We can
use the background analysis events to determine the significance of
the loudest on-source analysis event.

Our assumption when choosing the background region was that analysis
events there are not due to gravitational waves associated with the
astrophysical trigger we are examining, and therefore represent
false alarms.  We thus determine from the background region the FAR in Hz as a function of analysis event loudness.  We can use
this knowledge to assign a false rate to the loudest on-source
analysis event, resulting in a statement such as ``We would expect
an event as loud or louder than the loudest on-source event once per
$S^{-1}$ seconds of on-source data analyzed,'' where $S$ is the FAR
corresponding to the loudness statistic of the loudest on-source
event.  We note here that the statement ``once per 10 years of
on-source data analyzed'' is a different statement, for an
externally triggered search, than the statement ``once per 10
years.''  The first statement does not take into account the rate of
astrophysical triggers, whereas the second does. For example, if the
astrophysical trigger rate is 200 per year and the on-source size
is 4\,s, the first statement says we would need to keep our
detectors running at 100\% duty cycle for about $\sci{4}{5}$ years
before we would expect another false on-source analysis event of
that loudness, making such an event well above the detection threshold
of once per 100 years.

On the other hand, if we analyze 100 4\,s on-source regions (400
on-source seconds total) and the loudest analysis event from all
these on-source regions is approximately ``once per 400 seconds of
on-source data analyzed,'' then we cannot claim a detection.  This illustrates how proximity to an external astrophysical trigger increases the significance of an analysis event.

\begin{figure}[!t]
\begin{center}
\includegraphics[angle=0,width=110mm, clip=false]{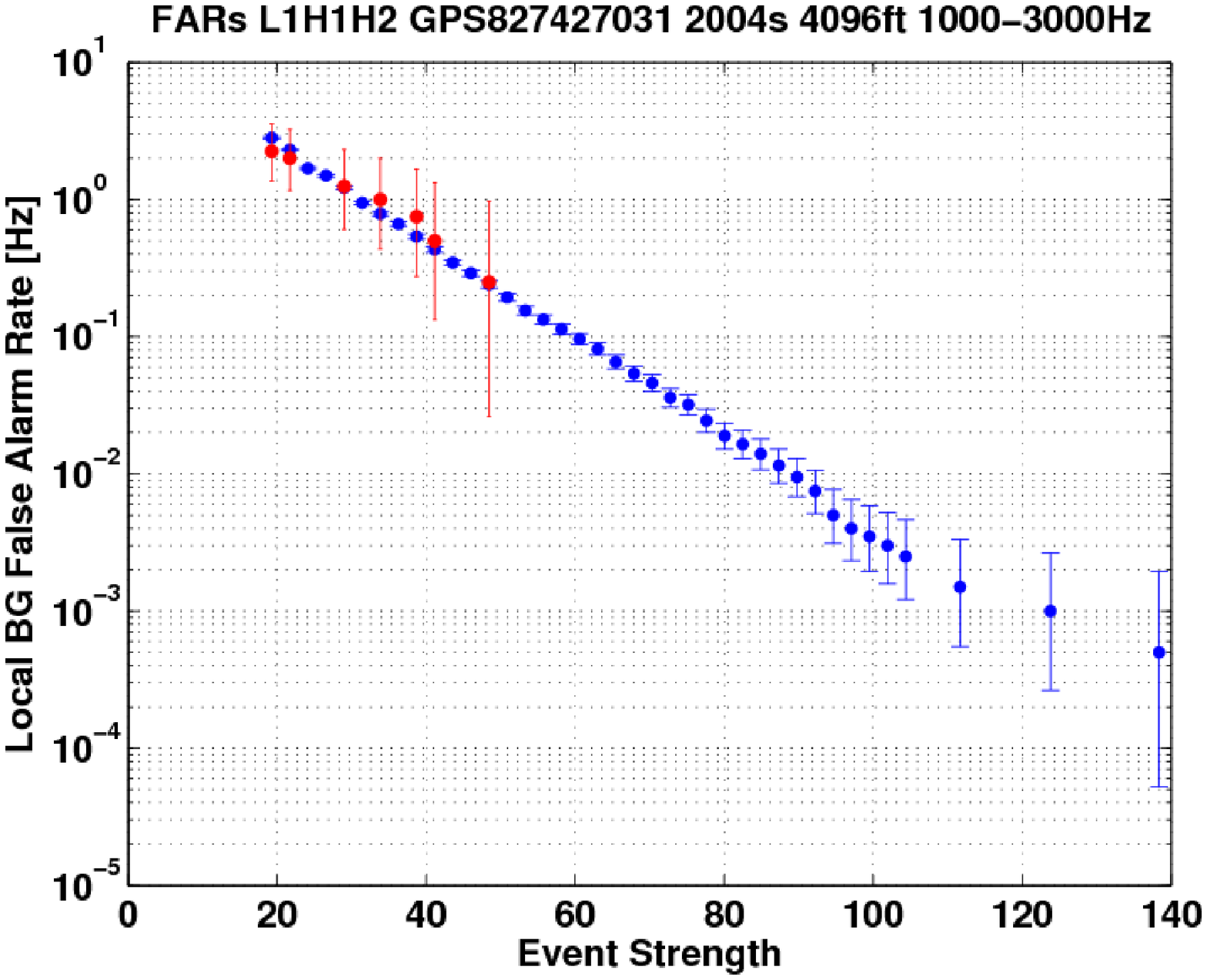}
\caption[Rate versus threshold plot] { Example of a rate versus
threshold plot.  Blue points are the cumulative histogram of the
background region analysis events and red points are the cumulative
histogram of the on-source region analysis events. }
\label{fig:clusterFAR}
\end{center}
\end{figure}

\section{Processing: Estimating upper limits} \label{section:upperLimits}

Upper limits on gravitational wave strain and gravitational wave
energy can be estimated via simulated signals injected into the
detector noise.

\subsection{Simulations} \label{section:simulations}

The magnitude of the response excited in an interferometric detector
by a passing wave depends on the direction from which the wave
arrives relative to the detector, and its polarization state, and is
customarily described by the antenna functions $F^{+}(\theta, \phi,
\psi)$ and $F^{\times}(\theta, \phi, \psi)$. Here $\theta$ is the
altitude of the source relative to the detector's horizon, $\phi$ is
the azimuth of the source relative to the detector's x-arm, and
$\psi$ is the polarization angle (see Figure\,\ref{fig:psi}). In a triggered search the source
location is well-known.

Our goal is to simulate incoming gravitational waves chosen from the
``signal space.'' The pipeline measures the detector output $h_d(t)$
consisting of the detector signal response $\xi_d(t)$ in the
presence of detector noise $n_d(t)$ (assuming a perfectly calibrated
detector):
 \be
 h_d(t) =  n_d(t) +  \xi_d(t),
 \ee
where the detector response $\xi_d(t)$ is given by
 \be
 \xi_d(t) = F^{+}_d(\theta, \phi, \psi) h_+(t)   +  F^{\times}_d(\theta, \phi, \psi)
 h_{\times}(t)  \label{eq:detectorResponse}
 \ee
with two independent polarization states $h_+(t)$ and
$h_{\times}(t)$, and with the antenna functions given
by~\cite{anderson01}
 \begin{align}
 F^{+}_d(\theta, \phi, \psi) & = \frac{1}{2} \cos2\psi \left(1+\cos^2\theta\right) \cos2\phi - \sin2\psi \cos\theta \sin2\phi \\
 F^{\times}_d(\theta, \phi, \psi) & = -\frac{1}{2} \sin2\psi \left(1+\cos^2\theta\right) \cos2\phi - \cos2\psi \cos\theta
 \sin 2\phi.  \label{eq:antenna}
 \end{align}
Note that\,\cite{300years} uses a different coordinate definition
gives the antenna functions in a slightly different form.

We simulate a detector response $\xi^{\mathrm{sim}}_d(t)$ by first
generating waveforms  $h^{\mathrm{sim}}_+(t)$ and
$h^{\mathrm{sim}}_{\times}(t)$.  Generation of waveforms is
discussed for the case of white noise bursts in
Section\,\ref{section:makingWNBs} and for the case of ringdowns in
Section\,\ref{section:makingRDs}. In this work the energy in
$h^{\mathrm{sim}}_+$ and $h^{\mathrm{sim}}_{\times}$ is chosen to be
the same, where the energy in any localized discrete signal $h(t)$
is defined as the square of the root sum square (rss) strain
 \be
    \hrss^2 = \frac{1}{f_s} \sum_i h_i^2, \label{eq:hrss}
 \ee
where $f_s$ is the sampling frequency and $i$ is the discrete time
index.  The total simulation $\hrss$ is then defined as
  \be
    \hrss^{\mathrm{sim}} = \sqrt{ \frac{1}{f_s} \sum_i \left( h^{\mathrm{sim}}_{+i}
    + h^{\mathrm{sim}}_{\times i} \right) }. \label{eq:hrssSim}
  \ee
The $\hrssn$ sensitivities discussed throughout this work are
estimates of the $\hrss$ of an incident wave.

For some polarization angle $\psi$ we next calculate the antenna
factors $F^{+}_d(\theta, \phi, \psi)$ and $F^{\times}_d(\theta,
\phi, \psi)$, and explicitly construct the simulated detector
response

 \be
    \xi^{\mathrm{sim}}_d(t) = F^{+}_d(\theta, \phi, \psi) h^{\mathrm{sim}}_+(t)   +  F^{\times}_d(\theta, \phi, \psi)
 h^{\mathrm{sim}}_{\times}(t).
 \ee

The simulated response $\xi^{\mathrm{sim}}_d(t)$ is then injected at a
random time location into 4~s noise segments, which themselves are
located randomly in time in the off-source region. In a two-detector
search this process is performed for each detector, with identical
simulated waveforms $h^{\mathrm{sim}}_+$ and
$h^{\mathrm{sim}}_{\times}$ and polarization angle $\psi$.

\subsection{Frequency domain gravitational wave crossing time delays}

Because gravitational wave crossing times for gravitational waves
from a particular source in the sky will be different at
non-co-located detectors, it is necessary for simulations to have
the correct gravitational wave crossing time delays applied. We
apply delays relative to the crossing time at the geocenter.

It is perhaps more intuitive to apply delays in the time domain, by
simply shifting the simulation start time relative to a given
detector data stream.  However, this necessarily limits the delay
resolution to $1/f_s = 61 \mu$s.  At 3\,kHz, this is equivalent to
66 degrees of phase, which is not acceptable for a coherent analysis
method.

To circumvent this limitation we apply the gravitational wave
crossing time delay $\Delta t$ in the frequency domain.  This
requires Fourier transformation of the simulation time series,
application of the frequency-dependent term $e^{2 \pi i f \Delta t}$
to the Fourier series, and inverse Fourier transformation back to
the time domain.

The Flare frequency domain time-of-flight delay routine has been
incorporated into the BurstMDC simulation production
package\,\cite{burstmdc}, which is the standard simulations engine used in LIGOÊ all-sky burst searches as
well as triggered searches.

\subsection{Gravitational wave emission energy of a simulation}

The $\hrssn$ upper limit estimates correspond to gravitational wave
emission energy upper limits.  The characteristic isotropic
gravitational wave emission energy $\egwn$ associated with a burst
depends on the simulation waveform and can be estimated via
    \be E_{\mathrm{GW}} = 4\pi R^2 \frac{c^3}{16 \pi G} \int_{-\infty}^{\infty}\left( (\dot{h}_{+})^2 + (\dot{h}_{\times})^2\right) dt. \ee
This follows from the equation for the gravitational wave energy
flux in the direction of propagation~\cite{shapiro83}.  Here $R$ is
the distance to the source.

After the simulation has been constructed, $ \egw / R^2 $ is
calculated from the simulation time series and stored.  In post
processing, efficiency curves can be constructed from these values;
if $R$ is known, efficiency curves can be constructed for $
E_{\mathrm{GW}} $. This is beneficial, as gravitational wave strain $\hrss$
is less familiar to the general astrophysical community than energy.
Energy efficiency curves are a major innovation of the Flare
pipeline.

\subsection{Generating ringdowns} \label{section:makingRDs}

Raw Flare pipeline ringdowns are generated as follows.  First, the
basic ringdown is created from a sine wave with the requested
frequency convoluted with an exponentially decaying envelope with
the requested $\tau$.  Then a second envelope (half of a Hanning
window) is applied to the beginning of the basic ringdown in order
to cause a gradual ramp-up over the course of one cycle.  The
simulation is then normalized to have an $\hrss$ equal to 1.

\subsection{Generating white noise bursts} \label{section:makingWNBs}

Raw Flare pipeline white noise bursts are generated as follows.
First, an adequately long white noise time series is randomly
generated. This time series is iteratively convoluted with a
Blackman window and a bandpass filter.  The Blackman window limits
the white noise in time, and the bandpass filter limits the white
noise in frequency.  The Blackman window is designed to give the
requested WNB duration after iteration with the bandpass filter,
such that 90\% of the final signal energy is contained within the
requested duration.  The bandpass filter has six second order
sections and a passband equal to the requested WNB frequency band.
The time- and band-limited WNB is then normalized to have an $\hrss$
equal to 1.

\subsection{Estimating detection efficiencies}
\label{section:makingEfficiencyCurves}

Post processing consists of constructing efficiency curves by
repeatedly analyzing 4\,s segments, each containing a single
simulation created with a range of $\hrss^{\mathrm{sim}}$ values,
and comparing the loudest simulation analysis event within 100\,ms
(for RDs) or 50\,ms (for WNBs) of the known injection time to the
loudest on-source analysis event (see Section\,\ref{section:injectionCoincidence}). The range of $\hrss$ values must
be chosen so that the smallest value produces simulations that are
always lost in the noise, and the largest value produces simulations
that are typically detected with very large SNRs.  The
$\egw^{\mathrm{sim}}$ or $h_{\mathrm{rss}}^{\mathrm{sim}}$value at
90\% detection efficiency ($\egwn$ or $\hrssn$) occurs where 90\% of
the loudest simulation analysis events are larger than the loudest
on-source event.

For any given on-source region this results in four arrays of
numbers, each of which has length equal to the number of injected
simulations used to estimate the upper limit.  The first contains
the $\hrss$ values of injected simulations.  The second contains the
calculated $\egw$ values of injected simulations, or $\egw/R^2$ if
the distance to the source $R$ is not known.  The third contains the
loudness of the analysis event associated with the injected
simulation.  The fourth contains boolean values indicating whether
the associated analysis event was larger then the loudest on-source
analysis event or not.

The $\hrss$ and loudness arrays can be used to make a plot of
injected $\hrss$ versus detected loudness.  We refer to this as a
``conversion curve,'' since it allows an empirical conversion from
analysis event loudness to waveform $\hrss$.  At very high energies
the loudest analysis event associated with an injection is likely
due to the simulation. At very low energies the simulations are lost
in the noise and the loudest associated analysis event is likely due
to a local noise. The conversion curve is a useful diagnostic tool;
technical problems preventing efficient detection of simulations are
readily revealed in the conversion curve. Examples of conversion
curves are given in Figure\,\ref{fig:hrssEfficiencySampleRDC} and
Figure\,\ref{fig:hrssEfficiencySampleRDL}.  The dotted lines
indicate the threshold used in constructing the efficiency curve,
set by the loudest analysis event in a 4~s on-source region. The
dark solid lines are a curve fit to the data, shown for reference.

The $\hrss$ and boolean (or the $\egw$ and boolean) arrays can be
used to construct the efficiency curve, with the $\hrss$ (or $\egw$)
values on the x-axis.  The y-axis indicates the fraction of analysis
events associated with an injected simulation of $\hrss$ as given by
the x-axis which are louder than the loudest on-source event.  In
the case of simulation $\hrss$ values which range over a discrete
set of scale factors, the y-axis value \emph{is} simply this
fraction. Binomial error bars may be added to these data points
using
 \be
  \sigma = \sqrt{r(1-r)/N}
 \ee
where $N$ is the total number of simulations at a given
$\hrss^{\mathrm{sim}}$ value.

However, we are typically interested in the $\hrssn$ or $\hrssf$
value, that is the $\hrss$ of simulations whose associated analysis
event is louder than the loudest on-source analysis event 90\% or
50\% of the time.  Since we don't know this value ahead of time, it
is necessary to interpolate between the $\hrss$ values associated
with the discrete scale factors.  This is best done by fitting with
a sigmoid function.  The Flare efficiency curve fitting routine uses
two functions to perform these fits:  a four-parameter fit based on
the logistics function, and a five-parameter fit based on the
complementary error function.  The models were chosen
on empirical grounds.

The logistics function fit is given by
 \be
    f(x) = \kappa + \left(e^{-\alpha(x-\beta)} + 1/(1-\kappa) \right)^{-1} \label{eq:logistics}
 \ee
where $\kappa$, $\alpha$ and $\beta$ are fit parameters.  The
variable $x$ is first scaled by a fourth parameter before being
given to Matlab's \emph{nlinfit} routine with the above
three-parameter fit model. This model works well for efficiency
curves with steep transitions between the no-detection $\hrss$
regime and the easily-detected regime, such as circularly-polarized
ringdowns. An example efficiency curve made with the logistics
function fit is given in Figure\,\ref{fig:hrssEfficiencySampleRDC}.

The complementary error function fit uses the variable
 \be
    a \equiv log_{10}(x)-\alpha,
 \ee
with the model given by
\[
f(a) = \left\{ \begin{array}{ll}
  \kappa + \mathrm{erfc}(|a / \beta e^{a \gamma}|) (1/2 - \kappa) & \mbox{if $a<0$} \\
  1 - \frac{1}{2} \mathrm{erfc}(|a / \beta e^{a \delta}|) & \mbox{otherwise}
\end{array} \label{eq:erfc}
\right.
\]
where $\alpha, \beta, \gamma, \delta, \kappa$ fit parameters.  It works well for efficiency curves with shallow transitions between
the no-detection $\hrss$ regime and the easily-detected regime, such
as linearly-polarized ringdowns. An example efficiency curve made
with the logistics function fit is given in
Figure\,\ref{fig:hrssEfficiencySampleRDL}.

In the case of simulation $\hrss$ values which range over continuous
and randomly chosen values, a sigmoid fit is required to interpret
the array of boolean values.  The sigmoid fit models work well with
continuous or discrete simulation $\hrss$ values.  However, use of
randomly ranging simulation $\hrss$ values typically ensures a
robust sigmoid fit with fewer simulations than use of discrete scale
factors, especially for efficiency curves with steep transitions.
For a good fit it is necessary to have at least one measurement on
the transition; if the transition is steep, this requires a fine
spacing of scale factors in a discrete fit.  On the other hand, with
randomly chosen continuous values of $\hrss$ a plot with both the
sigmoid fit and pleasingly congruent measurement values cannot be
produced; an approximation can be made using a running average as in
Figure\,\ref{fig:randomHrssEfficiency}.  Because a referee once
asked for efficiency curves made with randomly chosen values of
$\hrss$ to be replaced with the discrete versions, we now prefer the
discrete versions.  The Flare pipeline can produce either.

The Flare efficiency curve fitting routine has been incorporated
into the X-Pipeline burst search pipeline\,\cite{xphome}, a
testament to its robustness, generality, and ease-of-use.

\begin{figure}[!t]
\begin{center}
\subfigure{
\includegraphics[angle=0,width=80mm, clip=false]{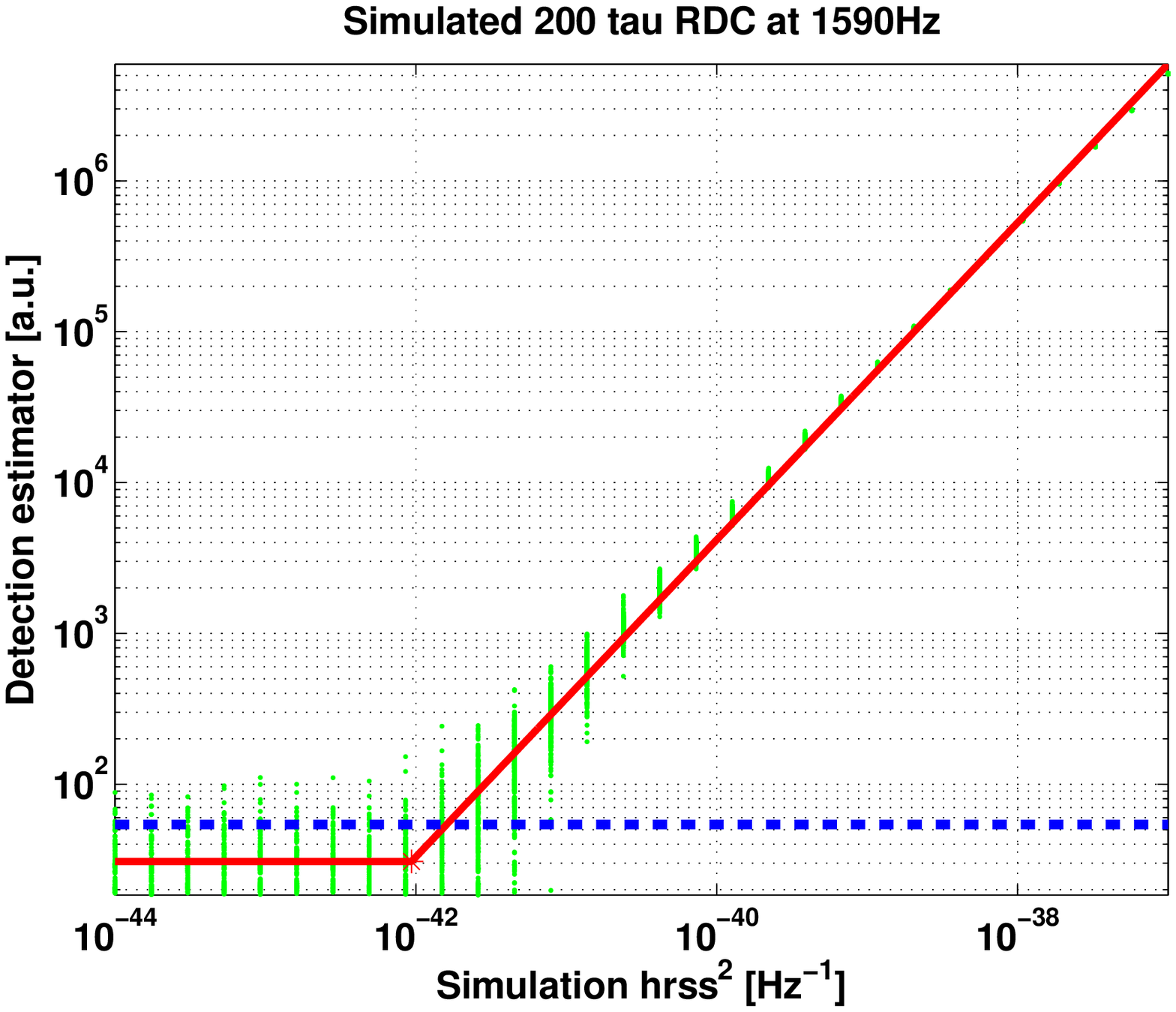}}
\subfigure{
\includegraphics[angle=0,width=80mm, clip=false]{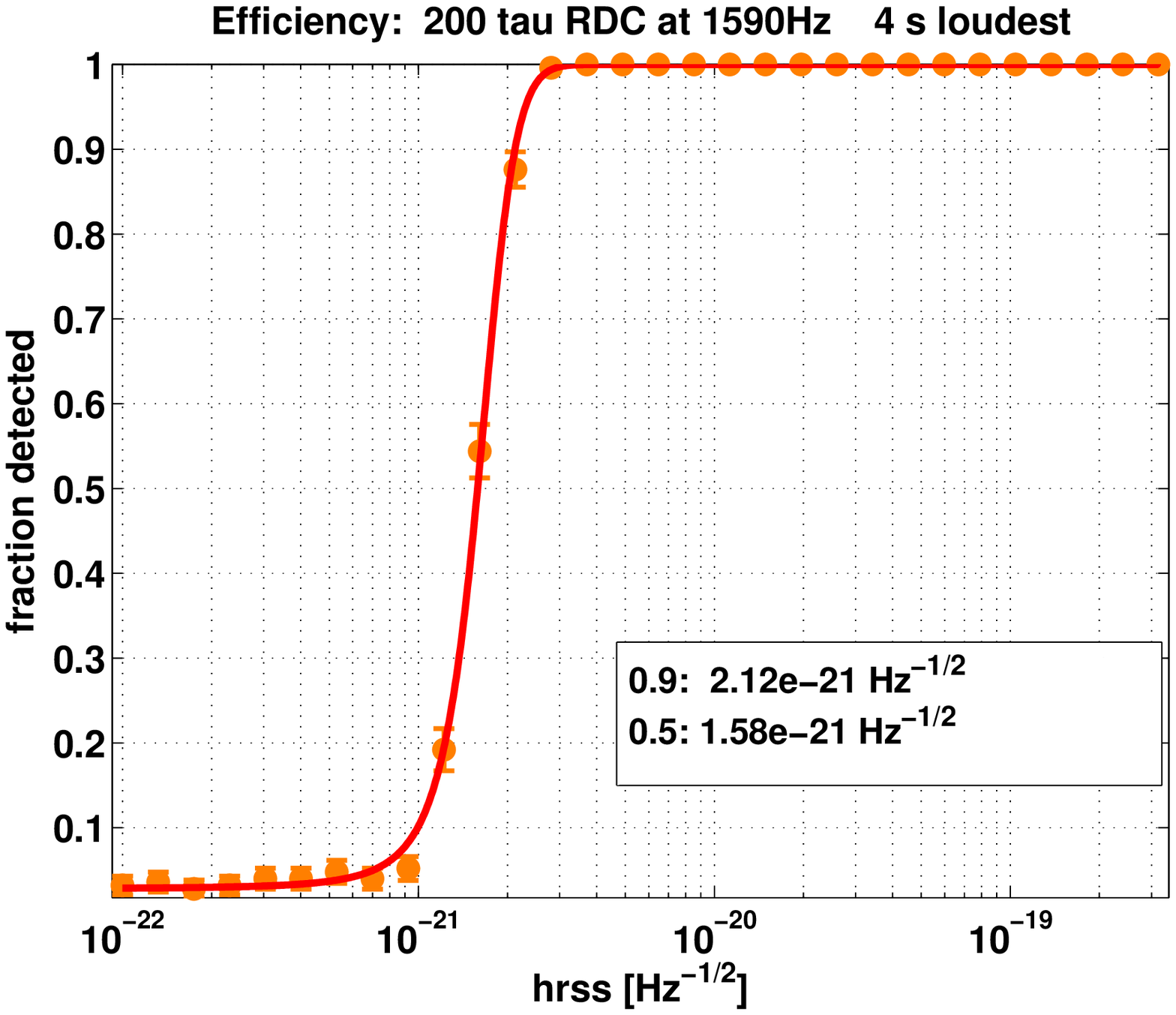}}
\caption[Efficiency curve and conversion curve for circularly
polarized RDs]{ \textbf{(top)} Conversion curve from two-detector
simulated data search for neutron star ringdown simulations at
1590~Hz with circular polarization. Each point represents the
loudest event recovered within 200~ms of an injected simulation. The
dotted line indicates the threshold used in constructing the
efficiency curve, set by the loudest event in a 4~s on-source
region.  Polarization angle $\psi$ was chosen randomly for each
simulation. \textbf{(bottom)} Efficiency curve corresponding to (a).
$\hrssn$ may be obtained by finding the simulation $\hrss$ value at
which the curve crosses the 0.9 fraction detected level.
Detectability of circularly polarized ringdowns does not depends on
$\psi$; this accounts for the relatively steep transition in the
efficiency curves which favors the logistics function sigmoid fit
model. } \label{fig:hrssEfficiencySampleRDC}
\end{center}
\end{figure}

\begin{figure}[!t]
\begin{center}
\subfigure{
\includegraphics[angle=0,width=80mm, clip=false]{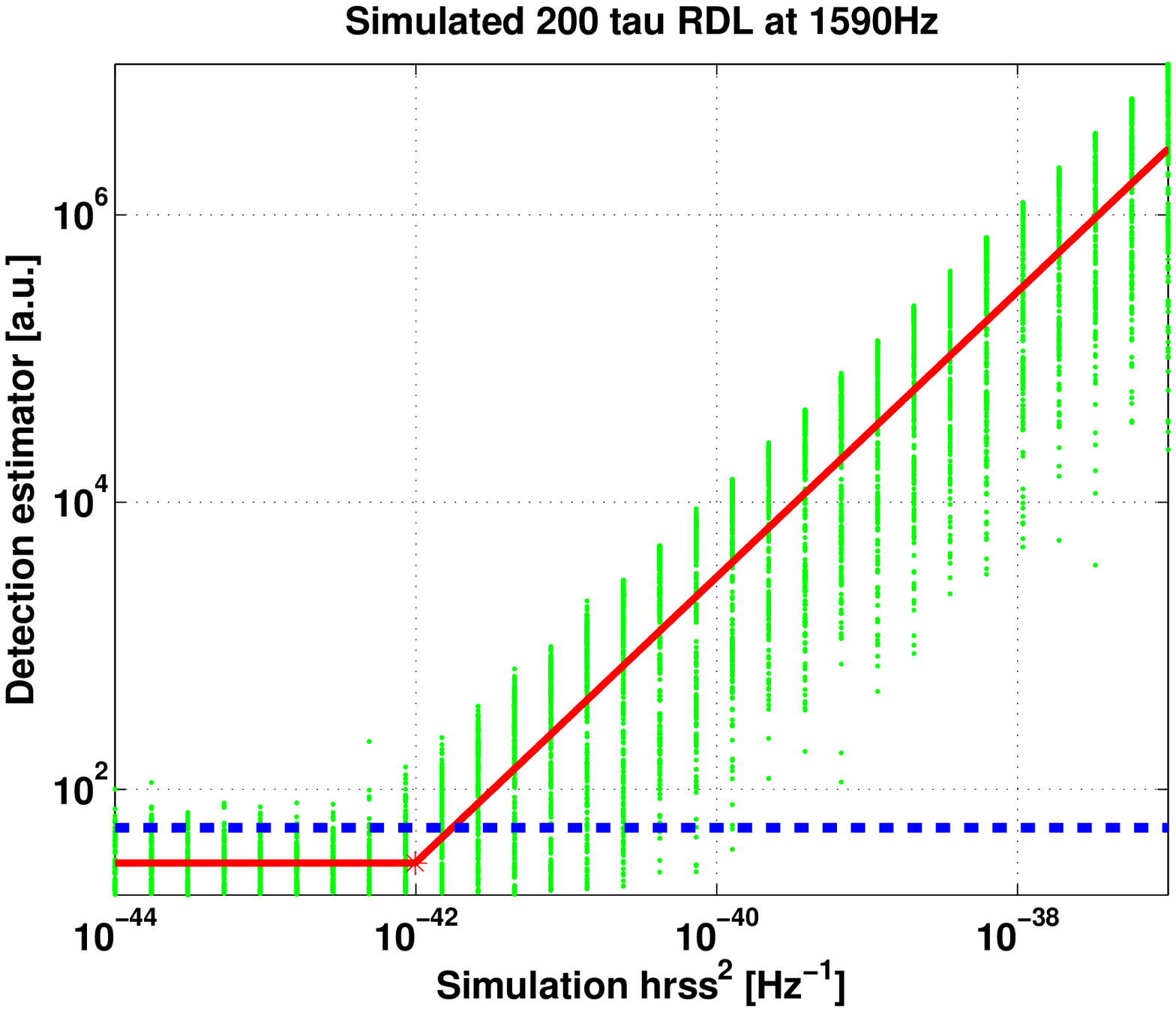}}
\subfigure{
\includegraphics[angle=0,width=80mm, clip=false]{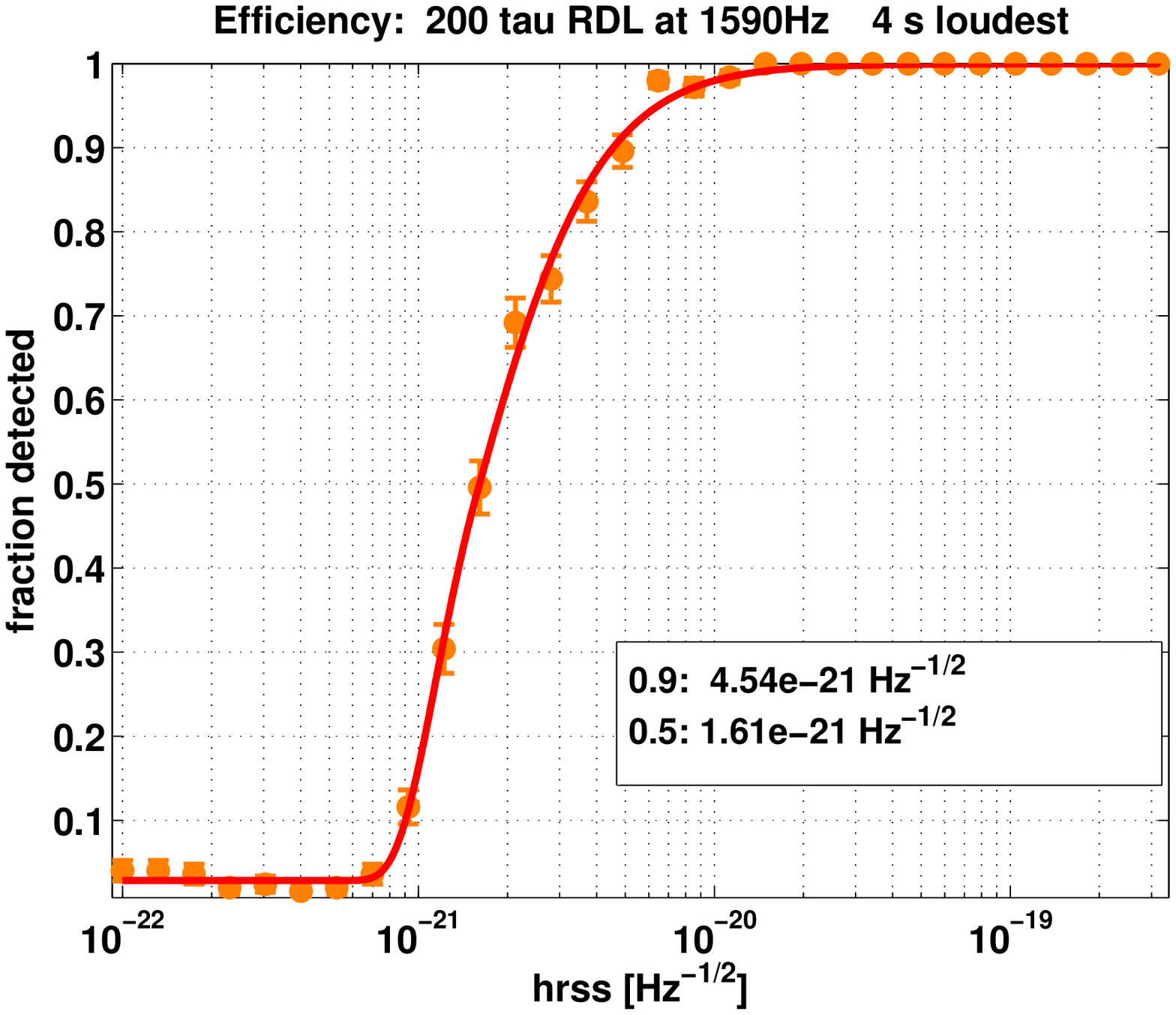}}
\caption[Efficiency curve and conversion curve for linearly
polarized RDs]{Conversion curve \textbf{(top)} and efficiency curve
\textbf{(bottom)} from two-detector simulated data search for
neutron star ringdown simulations at 1590~Hz with linear
polarization.  A 4~s loudest event segment length was used.
Polarization angle $\psi$ was chosen randomly for each simulation.
Detectability of linearly polarized ringdowns depends on $\psi$;
this accounts for the relatively shallow transition in the
efficiency curves, favoring the complementary error function sigmoid
fit model.} \label{fig:hrssEfficiencySampleRDL}
\end{center}
\end{figure}

\begin{figure}[!t]
\begin{center}
\subfigure{
\includegraphics[angle=0,width=80mm, clip=false]{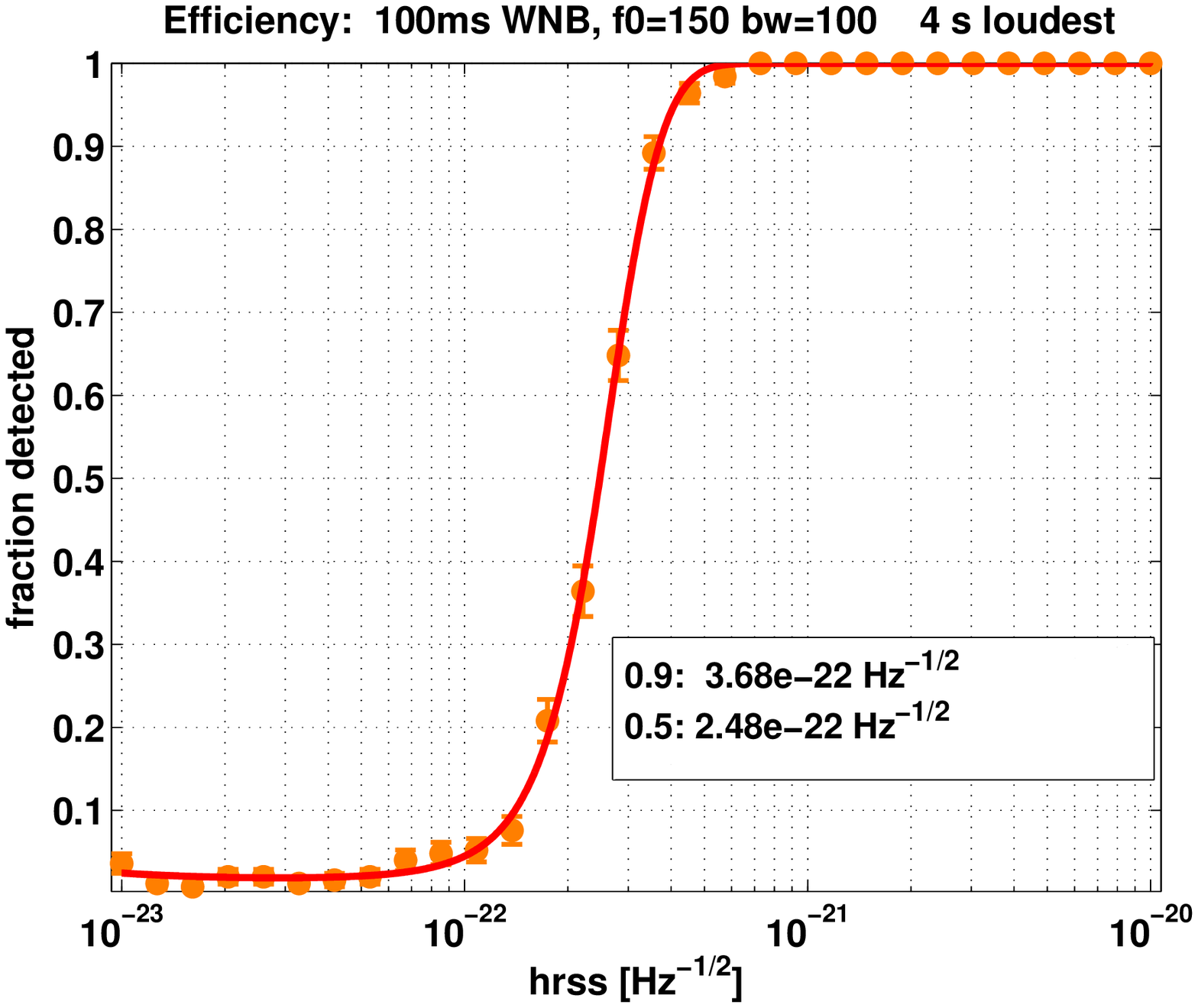}}
\subfigure{
\includegraphics[angle=0,width=80mm, clip=false]{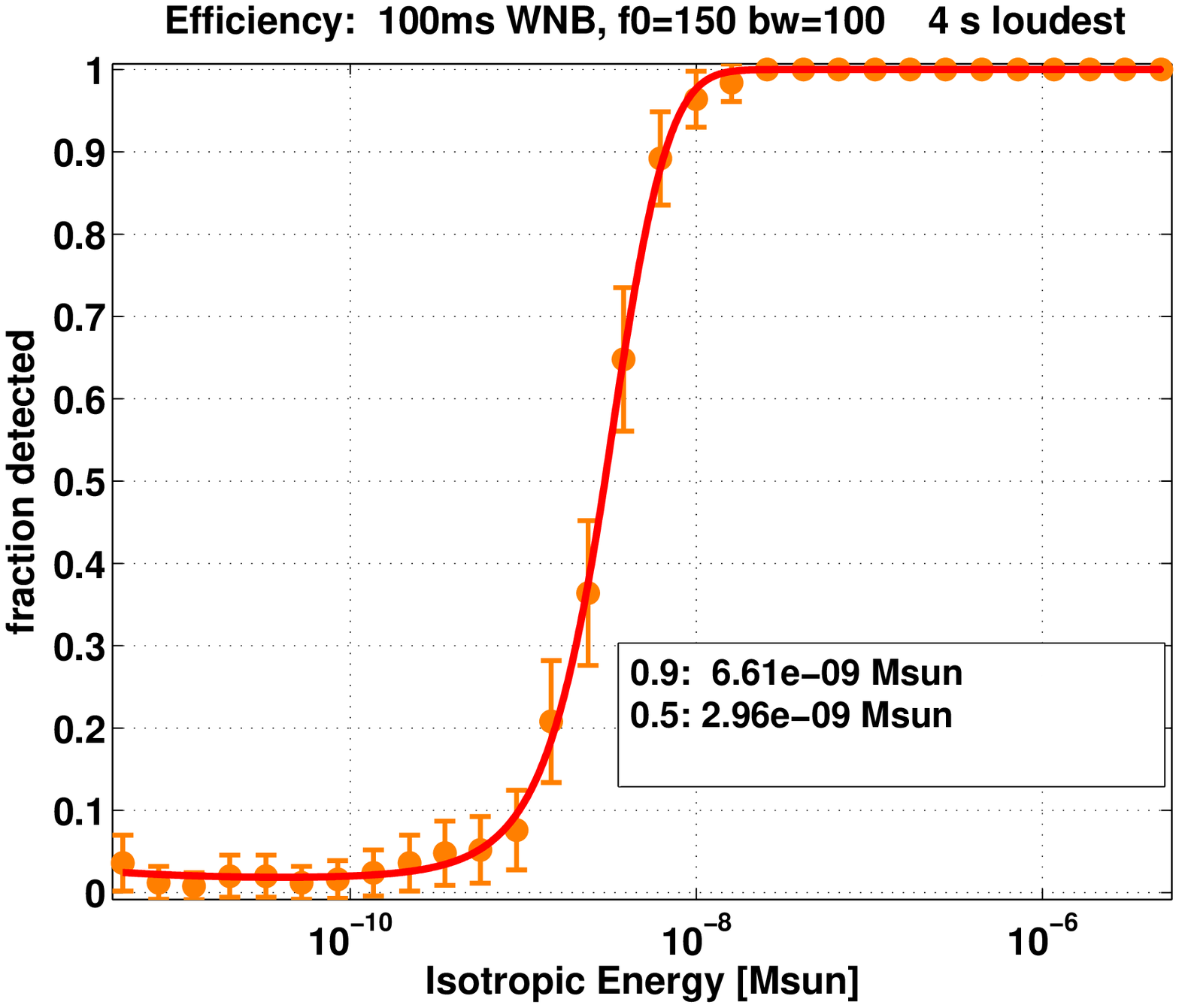}}
\caption[Example efficiency curves 100-200\,Hz 100\,ms WNBs]{WNB
efficiency curves for 100\,ms duration 100-200\,Hz WNBs.
\textbf{(top)} $\hrss$ efficiency curve.  \textbf{(bottom)}
Corresponding $\egw$ efficiency curve. }
\label{fig:efficiencySampleWNBSmall}
\end{center}
\end{figure}

\begin{figure}[!t]
\begin{center}
\subfigure{
\includegraphics[angle=0,width=80mm, clip=false]{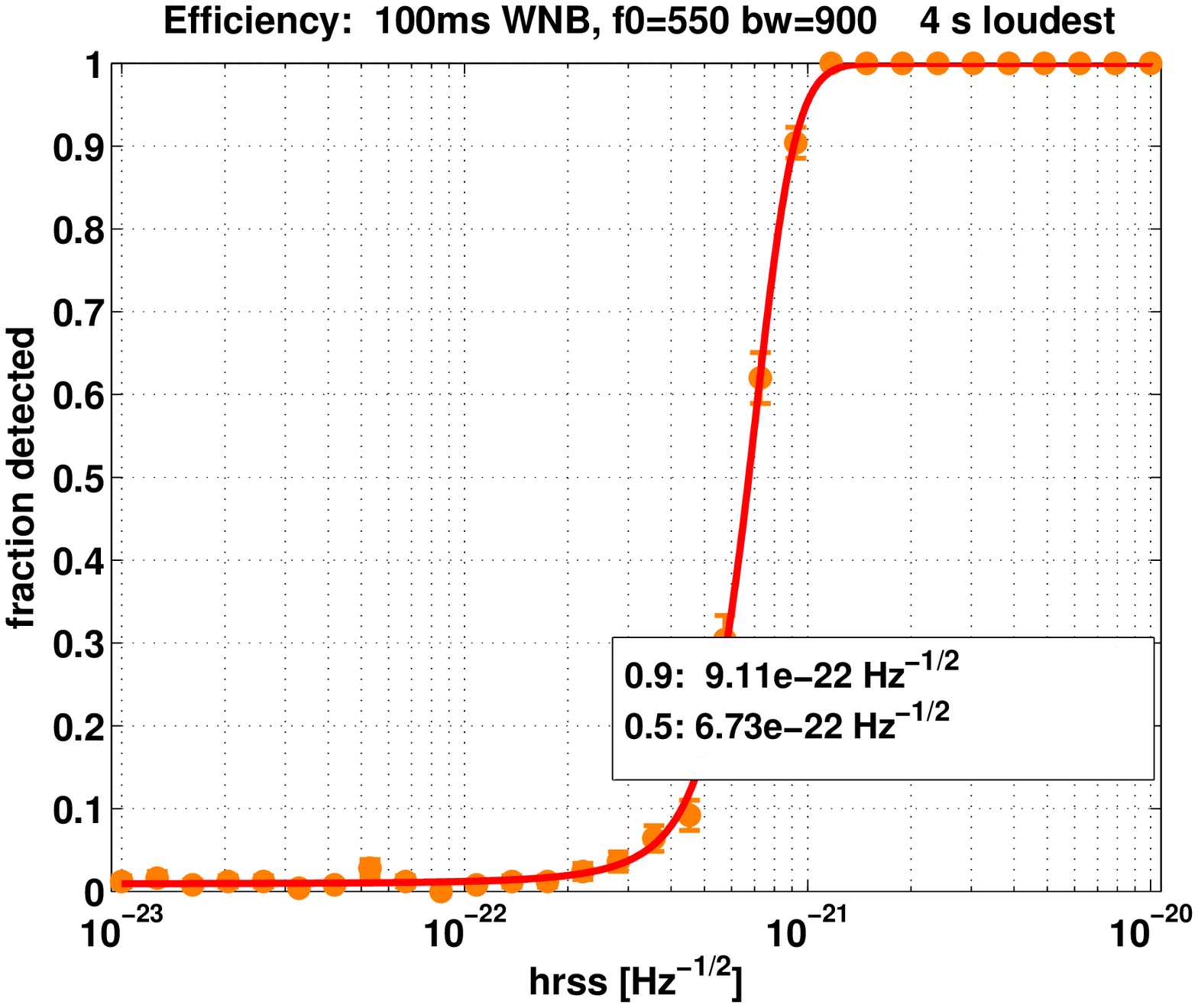}}
\subfigure{
\includegraphics[angle=0,width=80mm, clip=false]{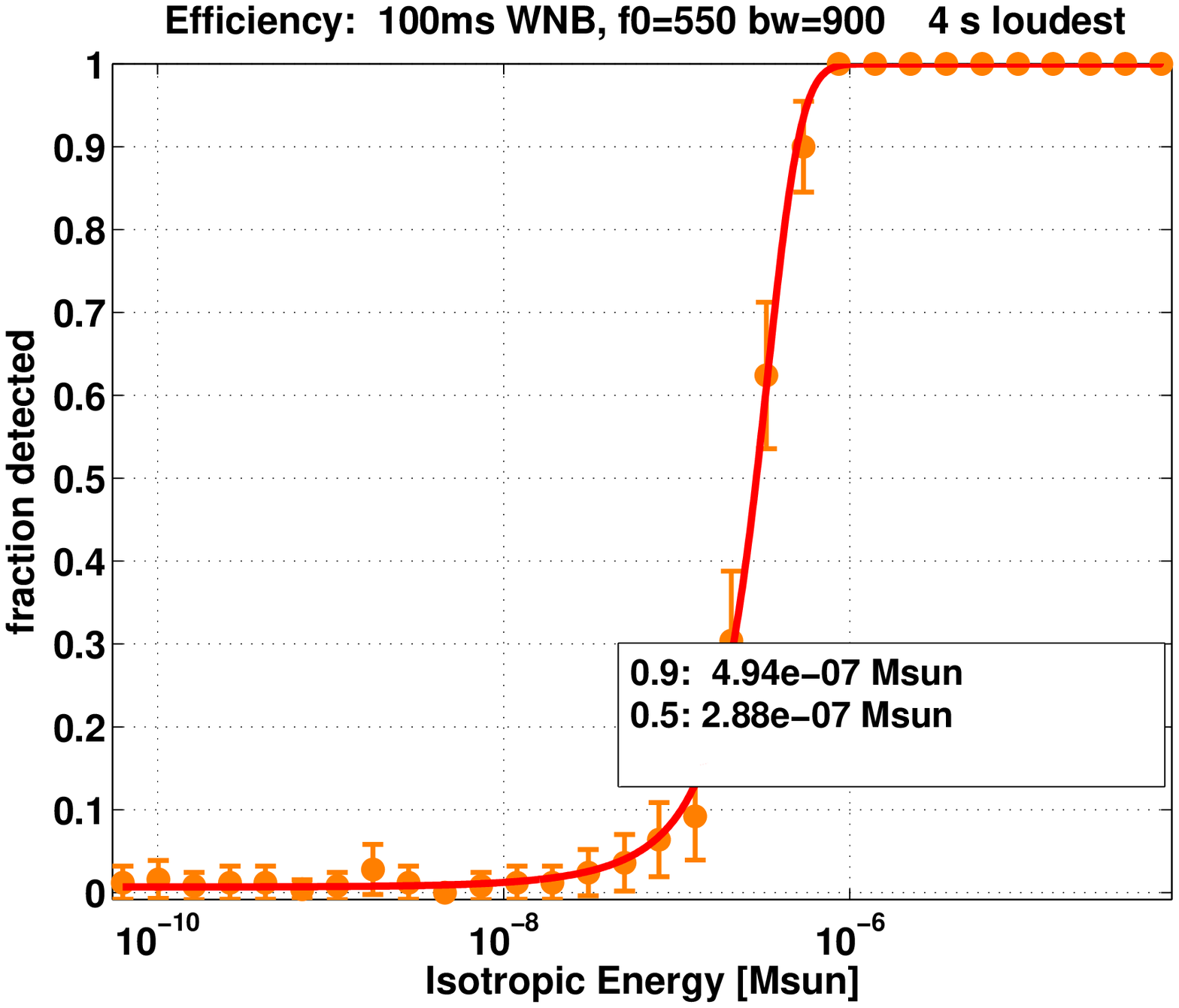}}
\caption[Efficiency curves 100-1000\,Hz 100\,ms WNBs]{WNB efficiency
curves for 100\,ms duration 100-1000\,Hz WNBs. \textbf{(top)}
$\hrss$ efficiency curve.  \textbf{(bottom)} Corresponding $\egw$
efficiency curve. } \label{fig:efficiencySampleWNBLarge}
\end{center}
\end{figure}

\begin{figure}[!t]
\begin{center}
\includegraphics[angle=0,width=100mm, clip=false]{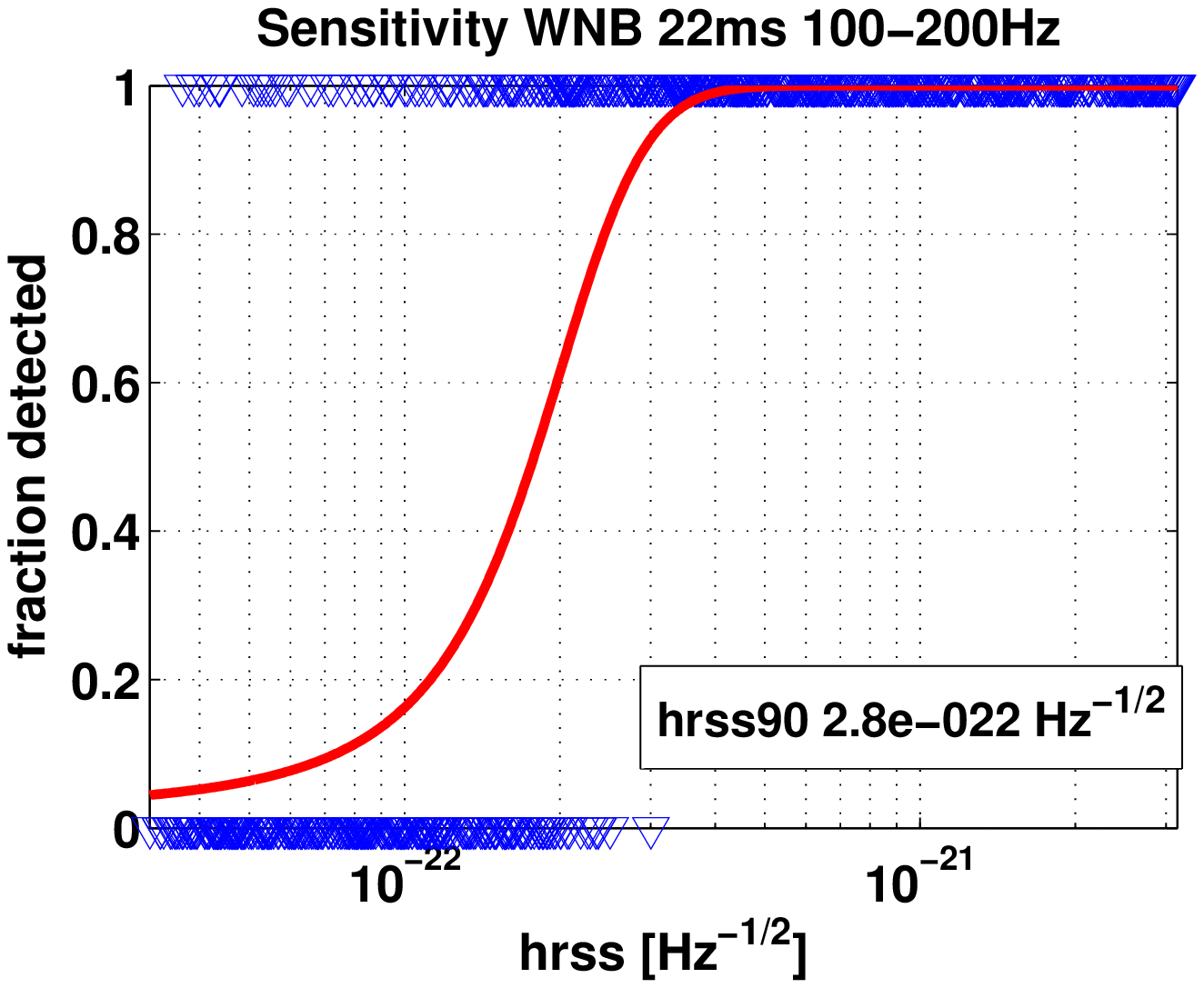}
\caption[Efficiency curve from simulations with randomly chosen
continuous $\hrss$ values.] { Efficiency curve from simulations with
randomly chosen continuous $\hrss$ values. The boolean values from
comparisons between the loudest on-source analysis event and the
analysis event associated with an injected simulation are shown as
triangles.  The fit model is the logistics function in
Equation\,\ref{eq:logistics}.  } \label{fig:randomHrssEfficiency}
\end{center}
\end{figure}

\section{Estimating upper limit uncertainties}
\label{section:uncertaintiesMethod}

Uncertainties in an upper limit are folded in, increasing the upper
limit.  Uncertainty comes primarily from two sources:  the Monte Carlo estimation procedure, and the detector calibrations.

\subsection{Statistical uncertainty for a finite simulation}
\label{section:bootstrap}

Statistical uncertainty arising from using a finite number of
injected simulations may be estimated with the bootstrap method
using $M$ ensembles\,\cite{efron79}.  This is done by running the
efficiency curve fitting routine $M$ times, sampling with
replacement from the original ensemble.

We show two plots from the distribution of $\egw$ at 90\% detection
efficiency from the bootstrap routine used to estimate statistical
uncertainty in post-processing (Figure\,\ref{fig:egwBoot972} and
Figure\,\ref{fig:egwBoot983}). These demonstrate that the sigmoid
curve fitting and bootstrap procedures are well-behaved.

\begin{figure}[!t]
\begin{center}
\includegraphics[angle=0,width=110mm, clip=false]{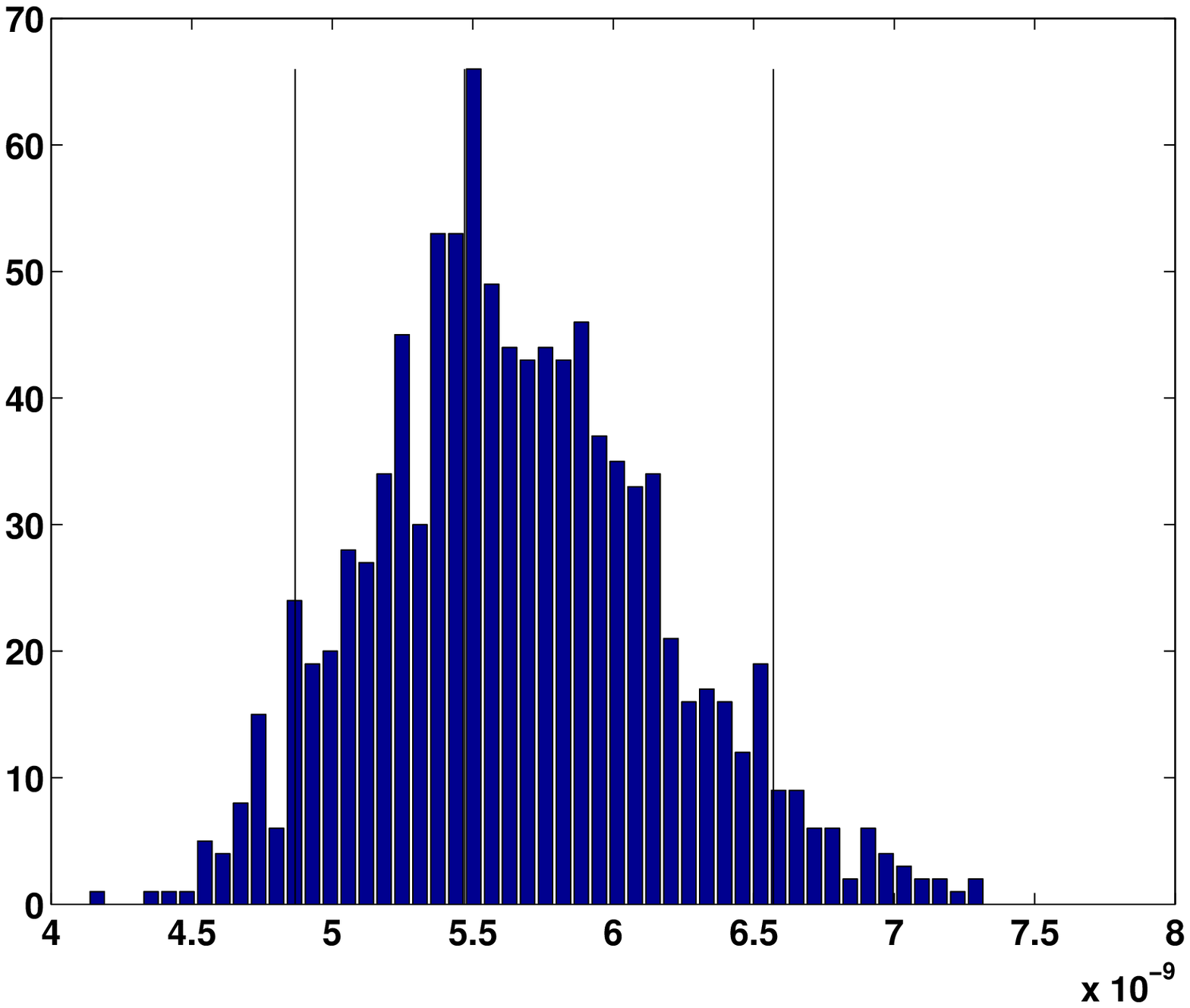}
\caption[Bootstrap histogram for 1000 ensembles for 11ms100-200Hz
WNB] { Histogram of $\egw$ at 90\% detection efficiency for 1000
bootstrap ensembles, for 11ms100-200Hz WNB, for the SGR 1806--20
060806 event. The lines are the 90\% two-sided confidence interval.}
\label{fig:egwBoot972}
\end{center}
\end{figure}

\begin{figure}[!t]
\begin{center}
\includegraphics[angle=0,width=110mm, clip=false]{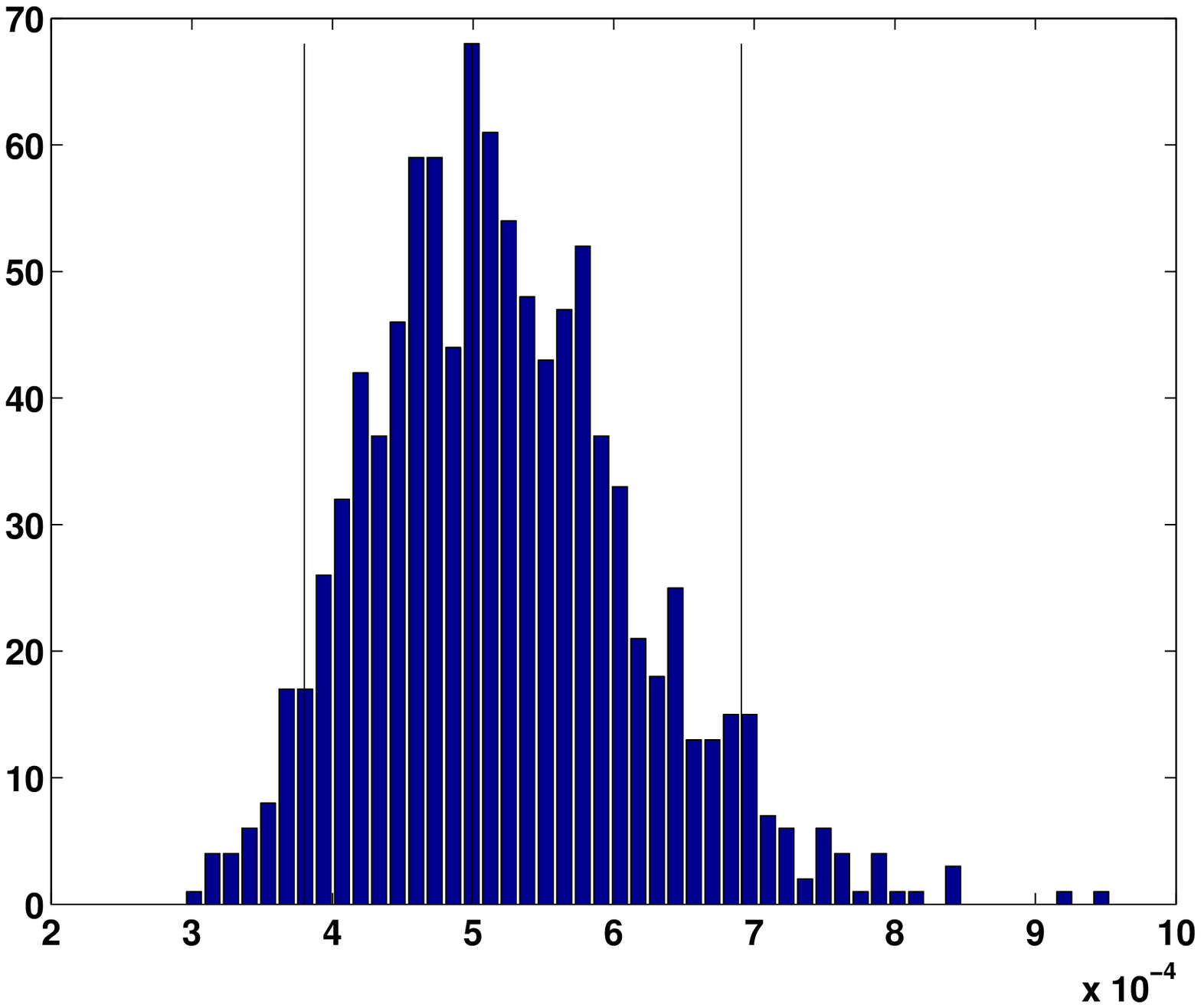}
\caption[Bootstrap histogram for 1000 ensembles for $\tau=$200\,ms
RDL at 2590\,Hz]{ Histogram of $\egw$ at 90\% detection efficiency
for 1000 bootstrap ensembles, for $\tau=$200\,ms linear RD at
2590\,Hz, for the 060806 event from SGR 1806--20. The  lines are the 90\%
two-sided confidence interval. } \label{fig:egwBoot983}
\end{center}
\end{figure}

\subsection{Calibration uncertainty}
\label{section:calibrationUncertainty}

The detector calibration statistical uncertainty is typically
characterized by two numbers: 1-$\sigma$ statistical amplitude
uncertainty and 1-$\sigma$ statistical phase uncertainty in degrees.

Amplitude 1-$\sigma$ uncertainties are multiplied by a factor of
1.28 to get amplitude uncertainty at 90\% confidence.  For upper
limits on two-detector networks, the larger amplitude calibration
uncertainty is chosen.

Phase calibration uncertainty is incorporated into strain upper
limits by estimating its effect on the recoverability of
simulations.  Once known, this value is added in quadrature to the
calibration magnitude uncertainty.

For single-detector searches the calibration phase uncertainty is
assumed to have a negligible effect.  For each simulation type in a
two-detector search, a Monte Carlo simulation is performed.  For
each of many trials a phase error is chosen randomly for simulations
generated for each detector, from a Gaussian distribution with
standard deviation set by the appropriate statistical phase
uncertainty for that detector, multiplied by a factor of 1.28 to
approximate the 90\% confidence level. These phase errors are
converted into timing errors at a characteristic simulation
frequency.  The timing errors increase with frequency. For
monochromatic signals such as RDs, the simulation frequency is
chosen.  For wide band simulations such as band-limited WNBs, the
highest frequency in the WNB band is conservatively chosen.

The simulations are injected into white noise (with simulation
$\hrss$ chosen to give a decent signal-to-noise ratio) each shifted
in time with the chosen timing error, which could be of either sign.
The two noise-plus-simulation streams are then fed through the
pipeline and the resulting loudness statistic $Z$  of the recovered simulation
is recorded. After many trials, an ``error distribution'' in $Z$ can
be examined.

This error distribution is then compared to a ``control
distribution'' created in an identical manner except without any
shifts due to timing errors.  The means of the error distribution
and the control distribution $\mu_e$ and $\mu_c$ are estimated. In
the presence of a significant degradation due to an introduced phase
error we expect $\mu_e < \mu_c$.  The percent difference between the
means of the distributions gives an estimate of the 90\% confidence
uncertainty in simulation amplitude recovery due to calibration
phase uncertainty.  The case $\mu_e > \mu_c$ implies that there is
no significant effect.

The resulting overall effect of calibration statistical uncertainty
on simulation amplitudes is then used to scale strain upper limits.

\subsection{Energy upper limit uncertainty}

For monochromatic simulations such as neutron star ringdowns, $\egw $
is proportional to the square of the simulation strain amplitude.

This is not true for individual large band simulations such as white
noise bursts. However, it is true for the ensemble averages of
independently-generated white noise bursts.

Therefore, so long as adequately many simulations are used, an
uncertainty in a strain upper limit expressed as a multiplicative
factor may be squared and applied to the corresponding energy upper
limit.  Since calibration uncertainties and errors are understood in
terms of their effect on strain upper limits, we use this method to
estimate their effect on energy upper limits.

Because statistical uncertainty due to finite simulations is
estimated directly via bootstrap method, estimates of uncertainty
from this source are independently obtained for the strain and
energy upper limit cases.

\clearpage
\chapter{Flare Pipeline Characterization and Validation} \label{chapter:validation}

In this chapter we describe tests and validations of the Flare
pipeline.  We begin by describing a basic check, recovery of
``hardware injections,'' simulated signals injected directly into
the interferometer via actuators on the optics (e.g. the photon
calibrator or coils), with the Flare pipeline.  We next describe technical checks of the
analysis event generator stage of the pipeline and the internal
simulations engine.   We then test the sensitivity of the pipeline to simulations injected into
simulated data.  Next, we describe early comparisons made between Flare
pipeline and a principle coherent LSC burst pipeline (coherent WaveBurst) using real
data to analyze GRB 070201.  Finally, we describe comparisons between Flare pipeline, the other principle coherent LSC burst pipeline (X-Pipeline), and the LSC matched filter CBC pipeline made with CBC simulations into simulated data.

\section{Hardware injections}

Hardware injections are permanently added to the detector's data
stream using a test mass actuator such as the photon calibrator.
They can take any morphological form, including astrophysically
motivated forms simulating supernovae, CBC events, etc.  Here we
describe Flare pipeline recovery of various sine-Gaussian hardware
injections which were created with coil actuators.

Recovery of hardware injections comprises the most comprehensive
single test of an externally triggered analysis pipeline, as the
pipeline may be run as in a real search.  Knowledge of the hardware
injection time is equivalent to an external trigger time.

The first method of hardware injection recovery, performed with an early version of the Flare pipeline, examined injection sets from the S4 run,
235\,Hz and 914\,Hz sine-Gaussians with known $\hrss$ values of [8,
4, 2, 1, and 0.5]$ \times 10^{-21}$ spaced 10\,s apart.  There were 15 such
sets at each of the two frequencies in S4, 14 of which were usable\,\cite{S4HardwareInjections}.
These 14 sets were recovered with the Flare pipeline running on
uncalibrated S4 gravitational wave data; Flare performs calibration
internally using the response function and cavity gain factors
prepared by the calibration team.  The method was to inject
inverse-calibrated software simulations swept in $\hrss$.  Recovered
loudness values for the software simulations were linear with
injected $\hrss$; this linear relationship was quantified with a
fit, which was then applied to the recovered hardware injections.
The results are plotted against the known $\hrss$ values of the
injections in Figures~\ref{fig:hw235}, ~\ref{fig:hw235prop}
and~\ref{fig:hw914}.  In the figures, the ``cal curve'' referred to on the y-axis is a conversion curve such as described in Section\,\ref{section:makingEfficiencyCurves}.

\begin{figure}[!h]
\includegraphics[angle=0,width=110mm,  clip=false]{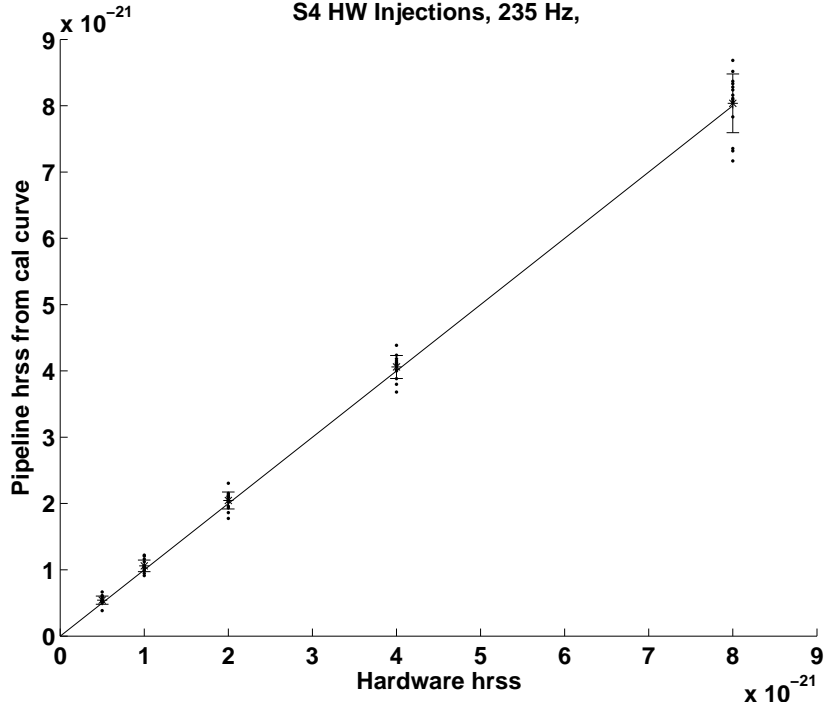}
\caption[Recovery of 235 Hz S4 hardware injections, unpropagated
$R(f)$]{Recovery of 235 Hz S4 hardware injections, using
unpropagated detector response function for inverse calibrating
injections. Agreement between known and measured values for the
largest hardware injection is better than 0.5\%. Agreement between
known and measured values for the smallest injection is 8\%, which
is less than one standard deviation (12\%). The line in the plot has
a slope of unity and represents perfect recovery of hardware
injections.  Parameters used were nfft=2048 samples (1/8\,s), and overlap = 97\%.} \label{fig:hw235}
\end{figure}

\begin{figure}[!h]
\includegraphics[angle=0,width=110mm, clip=false]{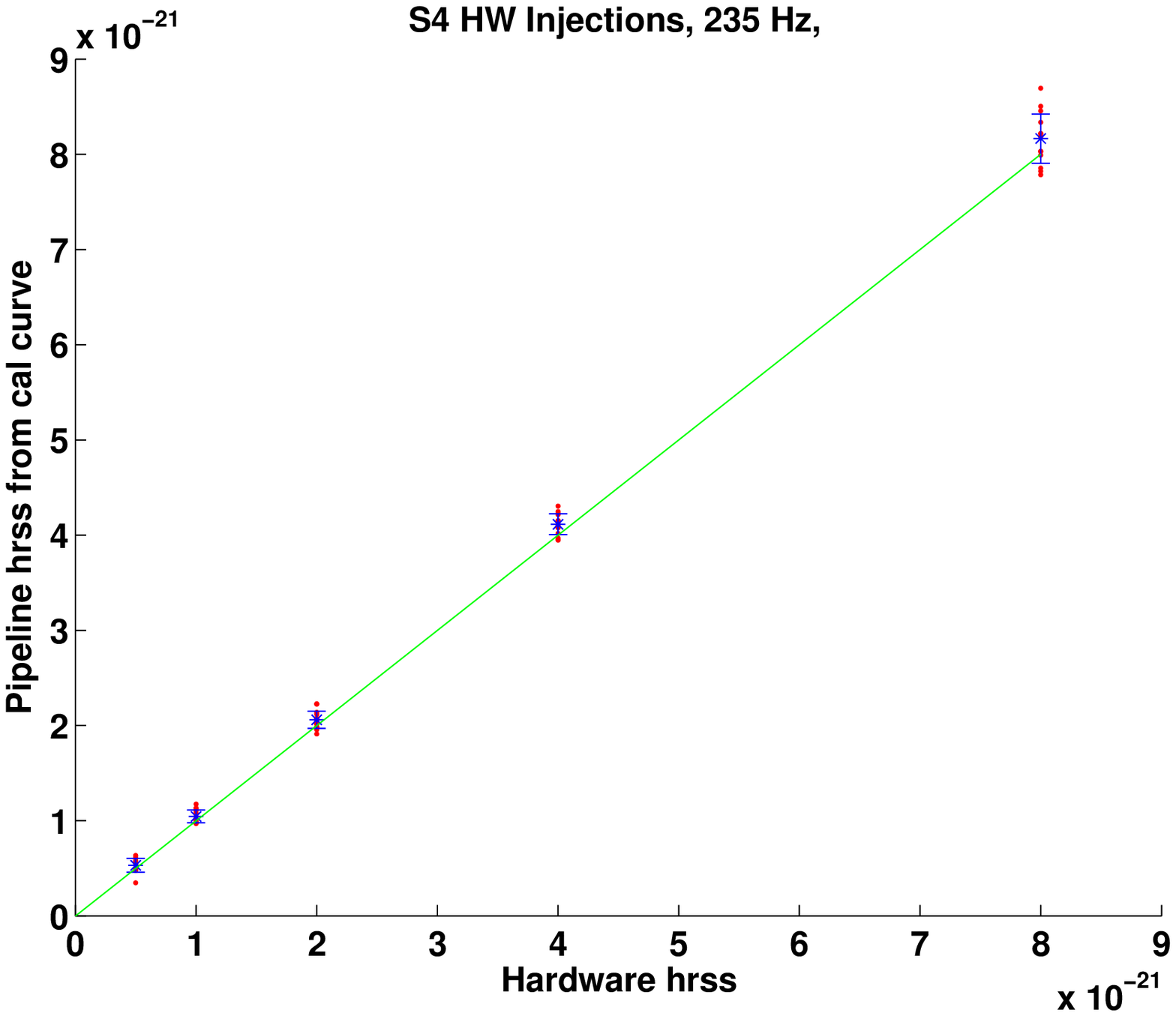}
\caption[Recovery of 235 Hz S4 hardware injections, propagated
$R(f)$]{235 Hz S4 hardware injections,  using propagated detector
response function for inverse calibrating injections. Agreement
between known and measured values for the largest hardware injection is better than 
2\%.  Agreement between known and measured values for the smallest
injection is 6\%, which is less than one standard deviation (14\%).
The line in the plot has a slope of unity and represents perfect
recovery of hardware injections.  Parameters used were nfft=2048
samples, and overlap = 94\%.} \label{fig:hw235prop}
\end{figure}

\begin{figure}[!h]
\includegraphics[angle=0,width=110mm, clip=false]{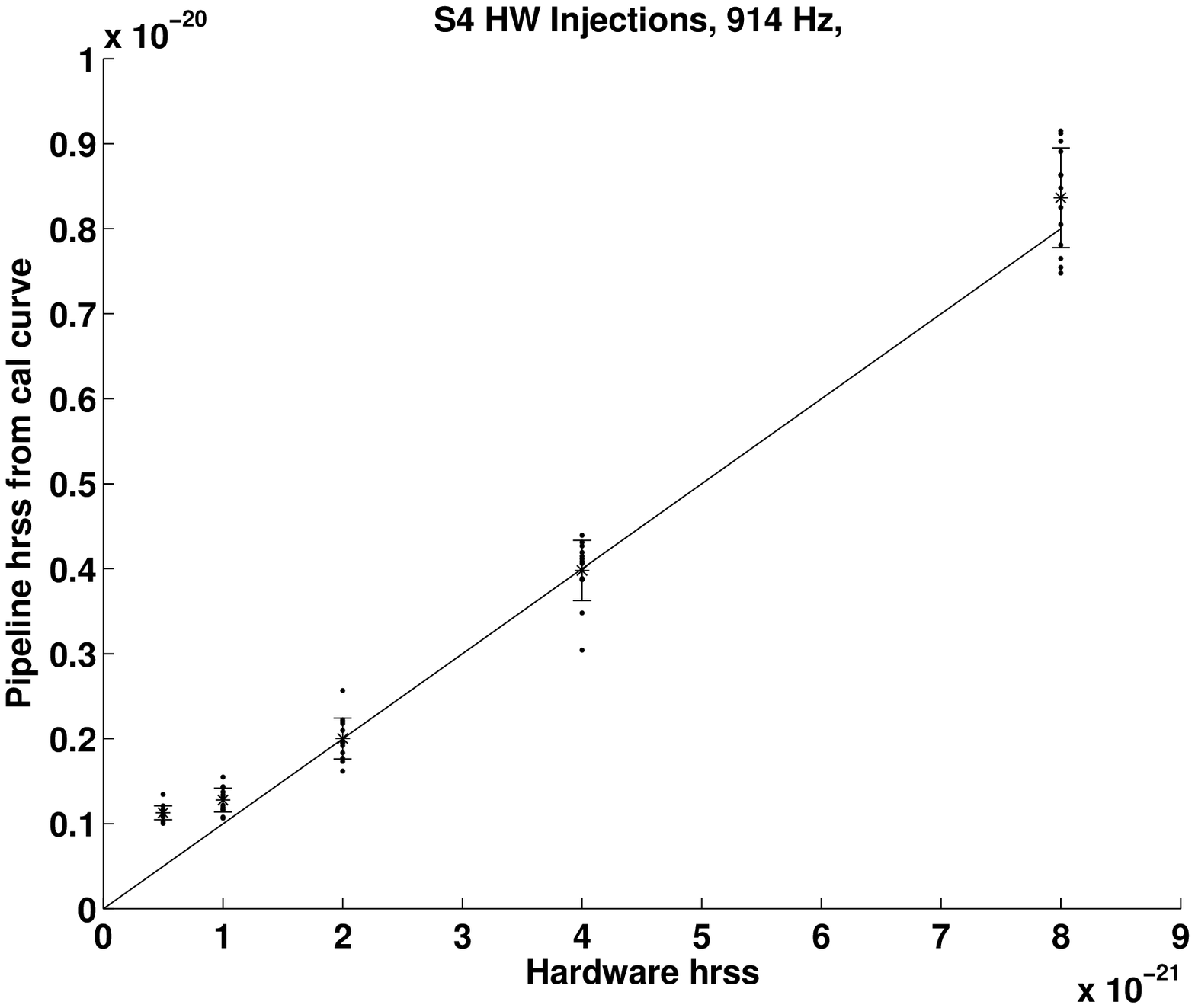}
\caption[Recovery of 914 Hz S4 hardware injections, unpropagated
$R(f)$]{914 Hz S4 hardware injections, using unpropagated detector
response function for inverse calibrating injections.  Agreement
between known and measured values for the largest hardware injection
is better than 5\%.  Known and measured values for the smallest
injection do not agree, because at this frequency this injection is
lost in the noise and undetectable. The line in the plot has a slope
of unity and represents perfect recovery of hardware injections.
Parameters used were nfft=2048 samples (1/8\,s), and overlap = 97\%.} \label{fig:hw914}
\end{figure}

The second method was performed with the mature pipeline, and again
was based on comparison between recovered hardware and software
simulations.  The hardware injection set examined was in S5,
comprised of 914\,Hz Q9 sine-Gaussians of various amplitudes and
relative time offsets between detectors (part of ``S5 burst set 6'').  We examined 114 such
hardware injections placed into S5 data in the three
interferometers between Jan. 19 2006 16:24:36 UTC (GPS 821723090)
and Sep. 25 2007 01:53:59 UTC (GPS 874720453)\,\cite{S5HardwareInjections}.  We
 performed $N$ software injections at the known hardware injection
$\hrss$, and then recovered the hardware injection $N$ times using
time-shifted Flare pipeline TF pixel bin edges.  Each hardware injection
was recovered, when possible, for multiple detector combinations:  the H1 detector only;  the L1
detector only, and the coherent H1--L1 pair.  The result was two
histograms per hardware injection per detector combination; the mean
values were then compared.  An example is shown in
Figure\,\ref{fig:hwSecondMethod}.  A few injections occurred during
stretches of poor quality data, or were near  glitches, and
were discarded.  These  problematic injections caused
bimodal distributions such as shown in Figure\,\ref{fig:hwBimodal}.
After discarding problematic injections average agreement was
within 3\% over all trials..

Hardware injection recovery tests run automatically each night as part of the
test suite described in Section\,\ref{section:testSuite}.

\begin{figure}[!h]
\includegraphics[angle=0,width=110mm, clip=false]{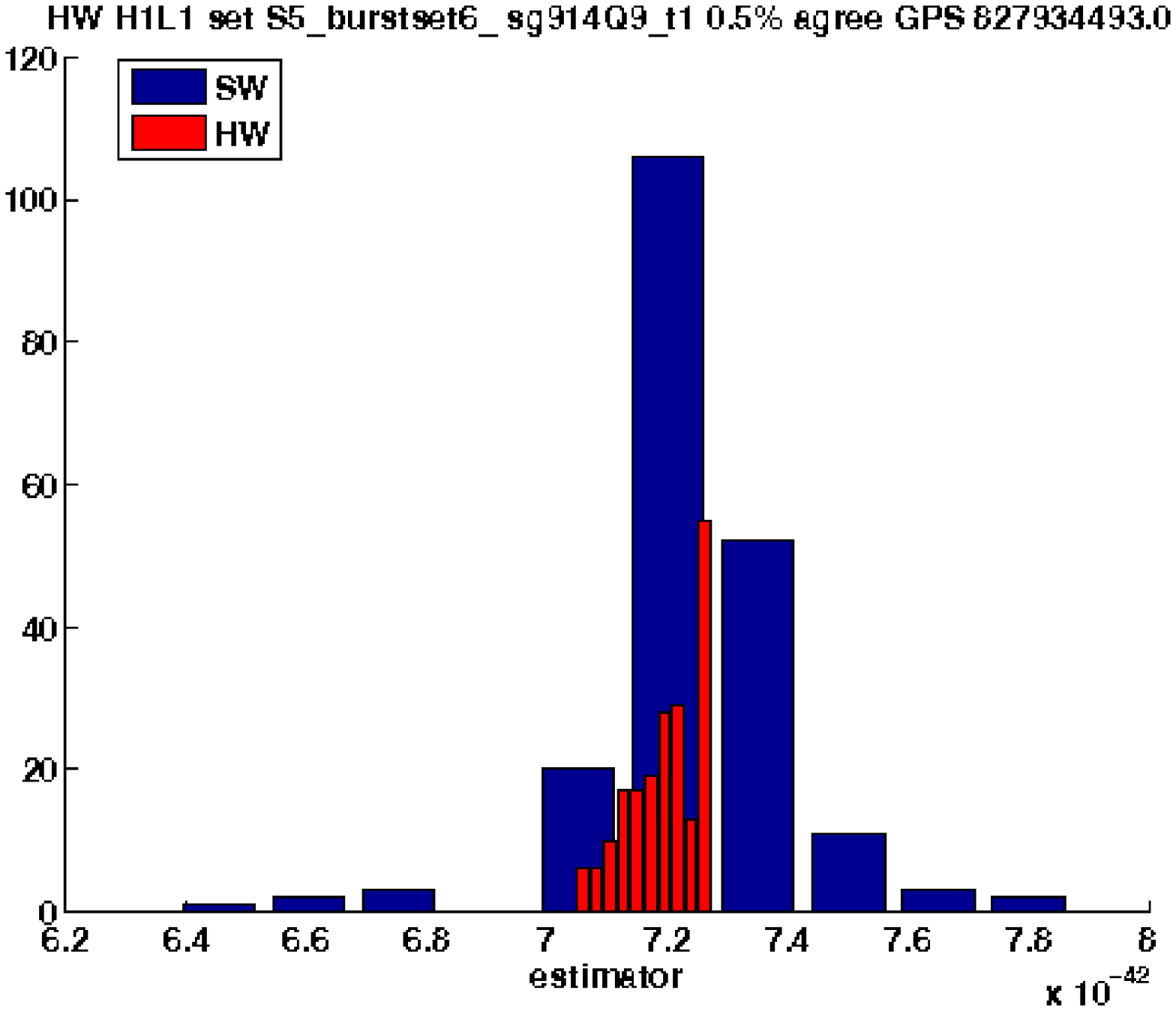}
\caption[Recovery of 914 Hz S5 hardware injection, second
method]{Example of a 914 Hz S5 hardware injection recovered with the
second method described in the text.  The recovered hardware injection distribution is narrower than the recovered software injection distribution because each software injection is added to a different span of noise.  The width of the hardware injection distribution is a measure of the effect of time binning boundaries in the analysis only, as hardware injections are of course fixed relative to the noise background.  Average agreement over all
hardware injections (including problematic injections) was better
than 10\%. After discarding problematic injections average agreement was
within 3\% over all trials.} \label{fig:hwSecondMethod}
\end{figure}

\begin{figure}[!h]
\includegraphics[angle=0,width=110mm, clip=false]{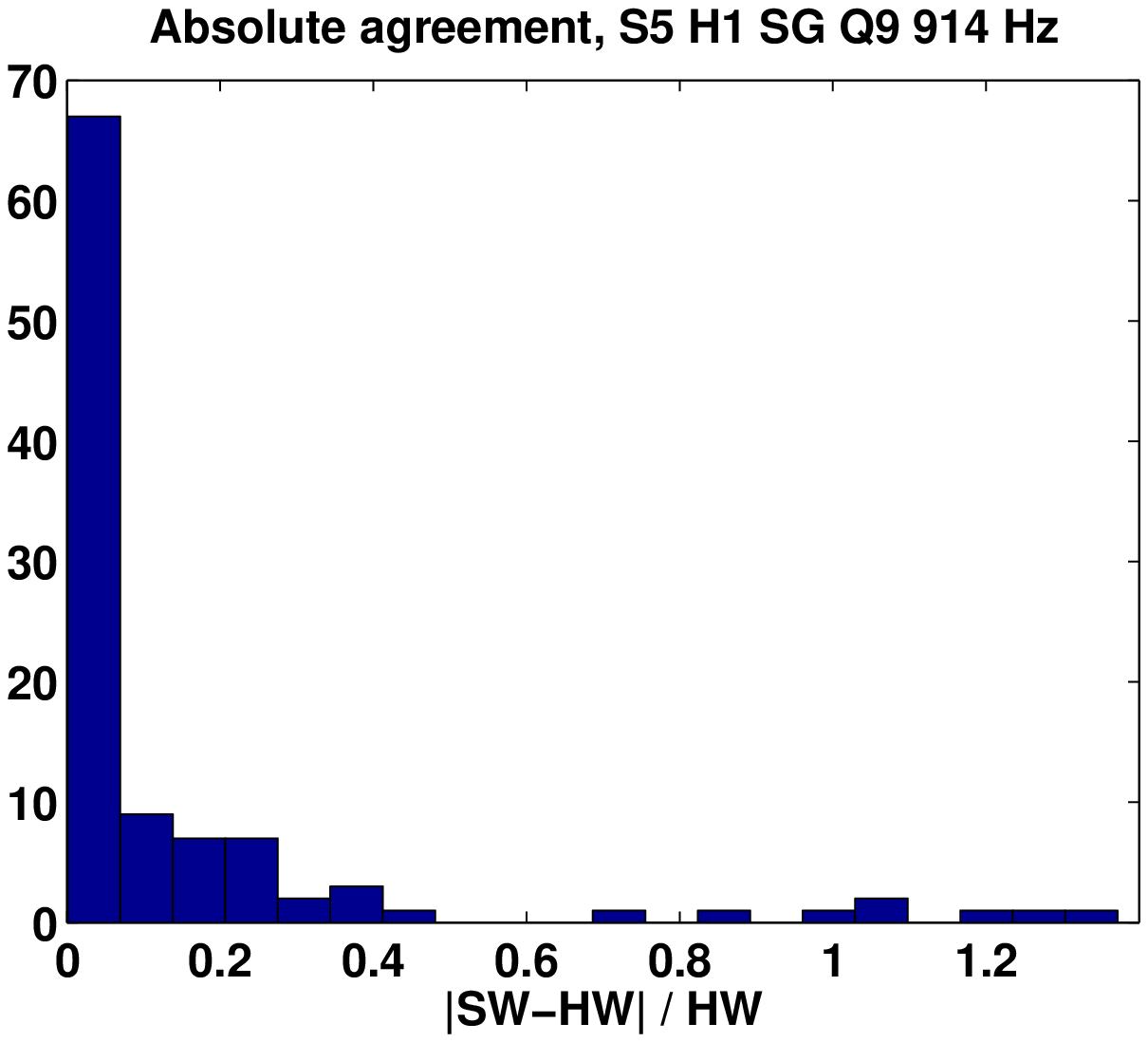}
\caption[Bimodal distribution of HW injection recovery]{Histogram of
914 Hz S5 hardware injections recovered with the second method, H1
detector only.  The bimodal distribution indicates problematic
instances of hardware injections above 0.5, possibly due to noisy
data near the time of the injection.} \label{fig:hwBimodal}
\end{figure}

\section{Technical validations and formal review}

Hardware injections already provide a rigorous validation of  the AEG stage of the
Flare pipeline.  In Appendix\,\ref{appendix:technicalValidations} we present additional technical validations.  The pipeline underwent a formal code review performed by an LSC review committee, and some of these additional checks were done as part of the formal review process.

\section{Choosing pipeline parameters}
\subsection{Fourier transform overlap} \label{section:overlap}

We use an overlap of 90\% when making spectrograms for all searches.
We have performed experiments which show that, in general, larger
overlap values improve sensitivity.  However, overlap values over
90\% do not significantly improve results, but do cause searches to
run significantly more slowly.

As an example, we present simulation results on 64--1024~Hz band
WNBs, with Fourier transform length of 1024 samples (1/16\,s). In this typical
case, we find no significant relationship between increasing overlap
and $\hrssn$ sensitivity beyond overlap of 90\%
(Table\,\ref{table:overlap}).

\begin{table}[h]
\begin{center}
\caption[Overlap effect on sensitivity]{Overlap effect on
sensitivity, for 64--1024~Hz band WNBs with Fourier transform length
of 1024 samples (1/16\,s). We find no significant relationship between
increasing overlap and $\hrssn$ sensitivity beyond overlap of 90\%}
\begin{tabular}{lr}
 \hline \textbf{overlap} & \textbf{sensitivity [strain/rtHz]}  \\
   \hline
 91\% &       $\sci{1.01}{-21}$ \\
 92\% &       $\sci{1.10}{-21}$ \\
 93\% &       $\sci{1.02}{-21}$ \\
 94\% &       $\sci{1.09}{-21}$ \\
\end{tabular}
\label{table:overlap}
\end{center}
\end{table}

Tests such as this have been performed on many different kinds of
waveforms, including other flavors of WNBs, and SGs, and for many
different values of Fourier transform length.

\subsection{Injection coincidence time window}
\label{section:injectionCoincidence}

In order to create an efficiency curve, we need to detect simulated
injections into the noise.  Constructing the curve from a set of
injections has two free parameters.  One is the
threshold for claiming detection of the simulation, which we
determine from the loudest on-source event (or some similar
observable in the case of a simulated sensitivity study).  The
second is the amount of time around the injection in which to look
for events above the threshold.  We refer to this region as the
injection tme coincidence window.

The injection time coincidence window is set by the amount of time
before and after a given injected simulation in which to search for
an event above the efficiency curve threshold.  If the injection
time coincidence window is set to be equal to the on-source length,
then in a loudest event search it is likely to find a false alarm
within the window.  In this case the efficiency curve will go to
100\% at large injection energies, but it will approach some
large fraction $P_{\mathrm{false}}$ as the $\hrss$ of the injections
goes to zero.  The precise value of $P_{\mathrm{false}}$ depends on
the particular loudest event of the search and the FAR rates of the
noise.

Such an efficiency curve may give a misleading estimate of the
actual detection efficiency of injected simulations, since it
counts many false events as detections.  This will give the impression
that the search is more sensitive than it really is.  A more
accurate (and conservative) estimate may be obtained by using a
smaller injection tme coincidence window. The optimum injection
time coincidence window choice is set to be as small as possible
such that the efficiency curve obtains 100\% efficiency at
high injection $\hrss$ values.

In Figure~\ref{fig:testInjection} we show efficiency curves for
injection time coincidence windows of $\pm0.2$, $\pm0.25$, $\pm0.5$,
$\pm1$, $\pm2$, and $\pm4$~s, for loudest event segment sizes of 4
and 180~s, using simulated H1L1 data and 22~ms duration 100-200~Hz
WNBs. Each of the twelve efficiency curves was constructed from the
same injection set.  In each plot the efficiency curve threshold was
set at the 90th percentile of a collection of 10 loudest events
obtained by time shifting H1 and L1 simulated data streams.  The
dramatic difference between the two plots is due to the different
efficiency curve thresholds.  In the 180~s plot, there is agreement
between $\hrssn$ values obtained from the various injection time
coincidence windows because even the largest window is much smaller
than the loudest event segment size. In the 4~s plot, however, the
smallest $\hrssn$ value, obtained from the $\pm4$~s coincidence
window, is 14\% lower than the largest value, obtained from the
$\pm0.2$~s coincidence window.  The systematic error introduced by
using the large coincidence windows would likely be worse in an
actual search, in which the efficiency
threshold, set by a single loudest event instead of the 90th
percentile of a collection of loudest events, would likely be lower.

\begin{figure}[!h]
\begin{center}
\subfigure[]{
\includegraphics[angle=0,width=90mm,clip=false]{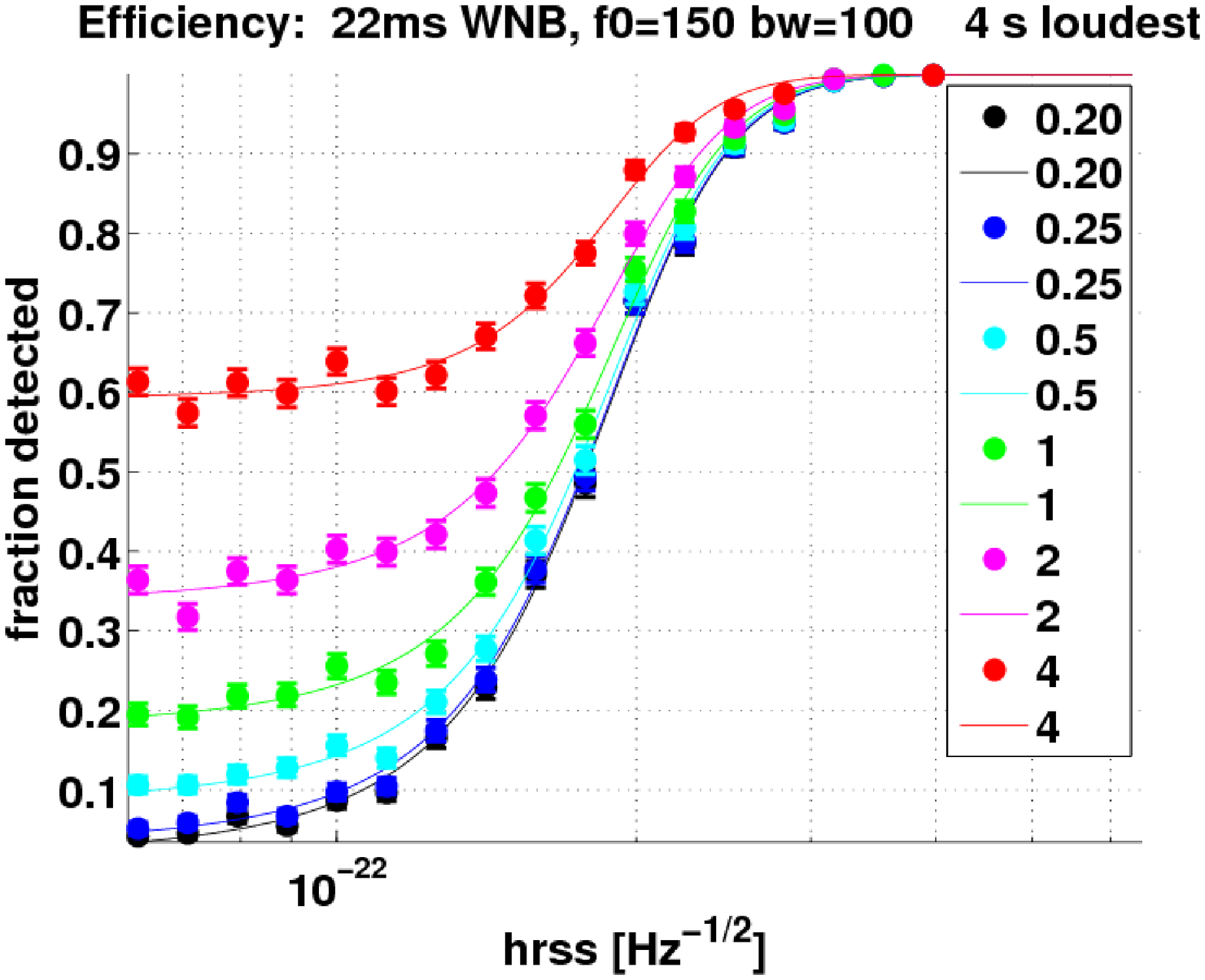}}
\subfigure[]{
\includegraphics[angle=0,width=90mm,clip=false]{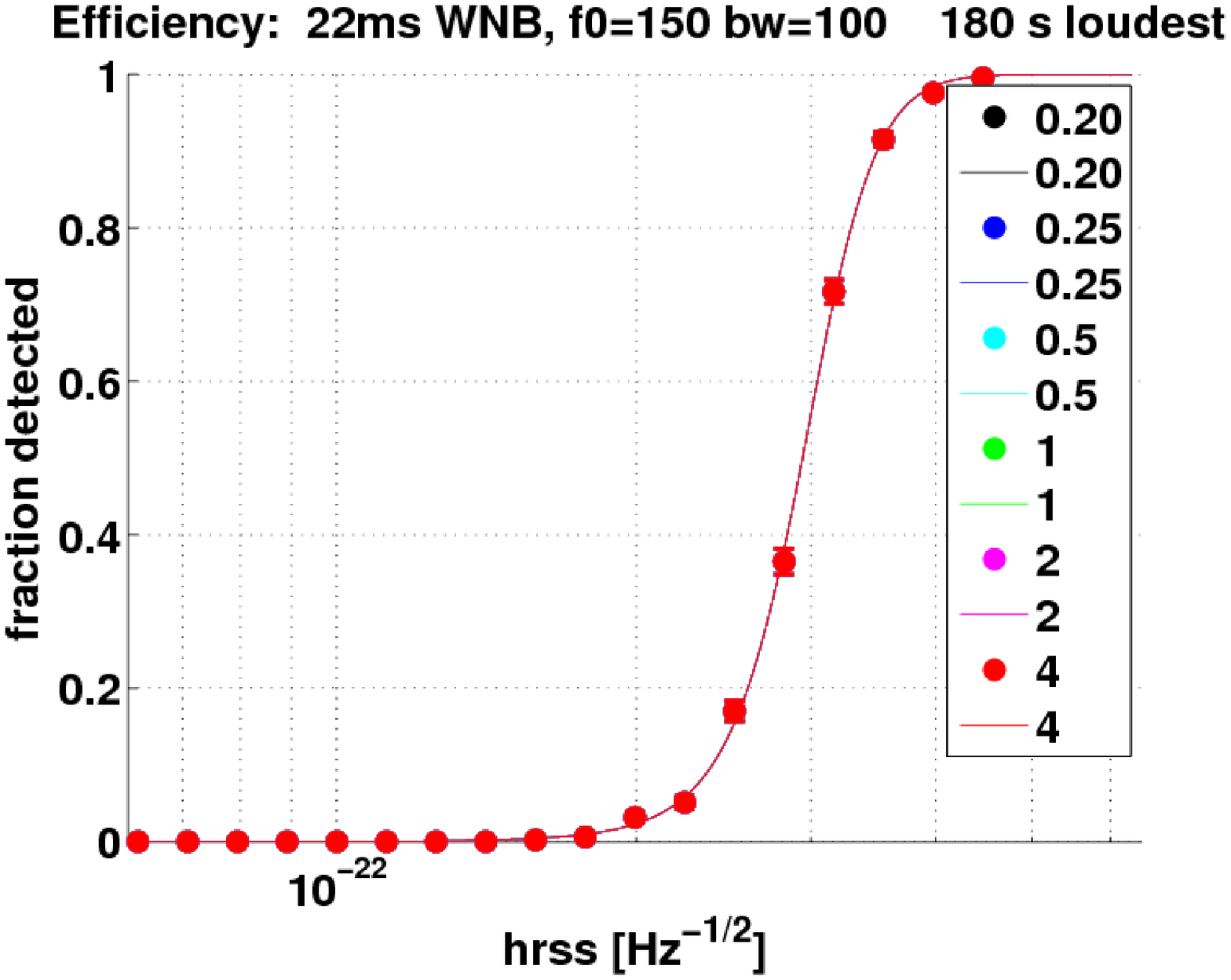}}
\caption[Efficiency curves vs. injection time coincidence]{
Efficiency curves for injection time coincidence windows of
$\pm0.2$, $\pm0.25$, $\pm0.5$, $\pm1$, $\pm2$, and $\pm4$~s, for
loudest event segment sizes of (a) 4 and (b) 180~s, using simulated
H1L1 data and 22~ms duration 100-200~Hz WNBs. Each of the twelve
efficiency curves was constructed from the same set of injected
simulated waveforms. In each plot the efficiency curve threshold was
set at the 90th percentile of a collection of 10 loudest events
obtained by time shifting H1 and L1 simulated data streams. }
\label{fig:testInjection}
\end{center}
\end{figure}

\subsection{Upper limit dependence on duration of on-source region}
\label{section:onsourceDependence}

During the validation and characterization stage of the project, we
performed many closed box SGR searches with different on-source
regions, keeping other aspects of the searches identical. We found
that upper limits estimated from 180\,s on-source durations were
only 20 percent higher on average than those estimated from 4\,s
on-source durations.

\subsection{Off-source segment size} \label{sec:offsourceSize}

If data are stationary, the event-based off-source region is any
usable data region excluding the on-source region; in practice we
choose off-source regions which are contiguous with the on-source
region. The off-source region serves several purposes.  It is used
to determine statistics in each individual frequency bin in the
time-frequency tiling.  The mean and standard deviation from the
background may be used to transform a PSD matrix into an excess
power matrix and a significance matrix.  The background can be
studied to determine the stability of these statistics.  It is also
used to determine a false alarm detection threshold.

These uses effectively set the minimum off-source region length. For
the sake of determining statistics to be applied in a search, we
choose a maximum acceptable tolerance in the gamma-distribution fit
parameters and then require a large enough off-source region to
obtain this tolerance.   The fit must be determined to be acceptable
at all frequencies bins in a given time-frequency tiling; for all
searches, since different searches use different tiling resolutions
and search bands; and for all detector networks, since different
networks in general have different noise characteristics.

For the WNB searches we have examined convergence of the
gamma-distribution fit.  The fit has two parameters, the shape
parameter $\alpha$ and the inverse scale parameter $\beta$.  We
found that convergence of these parameters only depends on the
number of points used in the fit.  That is, a search with a Fourier
transform length of 2048 samples (1/8\,s) takes twice as much background to
converge to the same tolerance as the same search with a Fourier
transform length of 1028 samples (1/16/,s).   

\begin{figure}[!h]
\begin{center}
\subfigure[small band]{
\includegraphics[angle=0,width=90mm,clip=false]{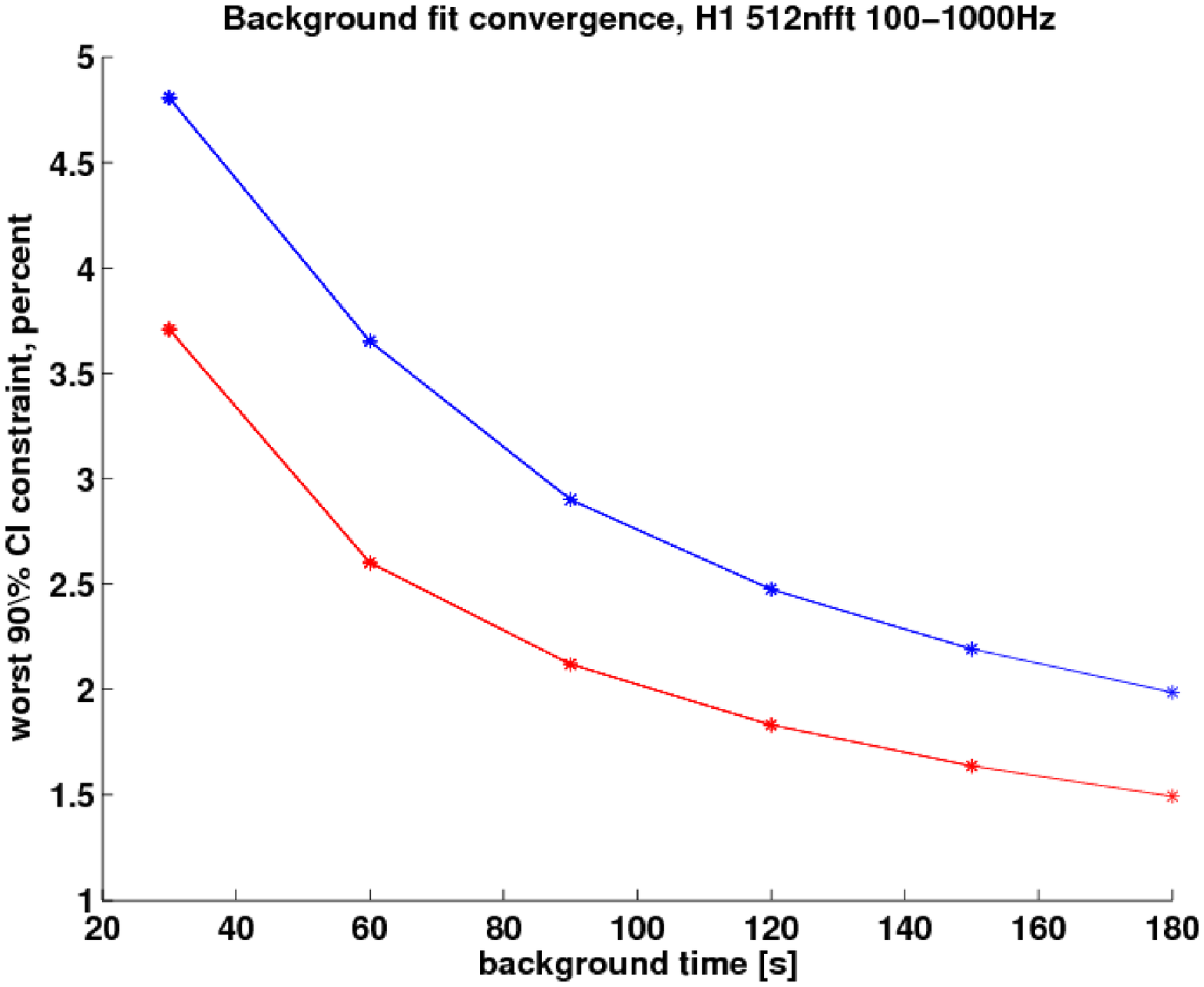}}
\subfigure[large band]{
\includegraphics[angle=0,width=90mm,clip=false]{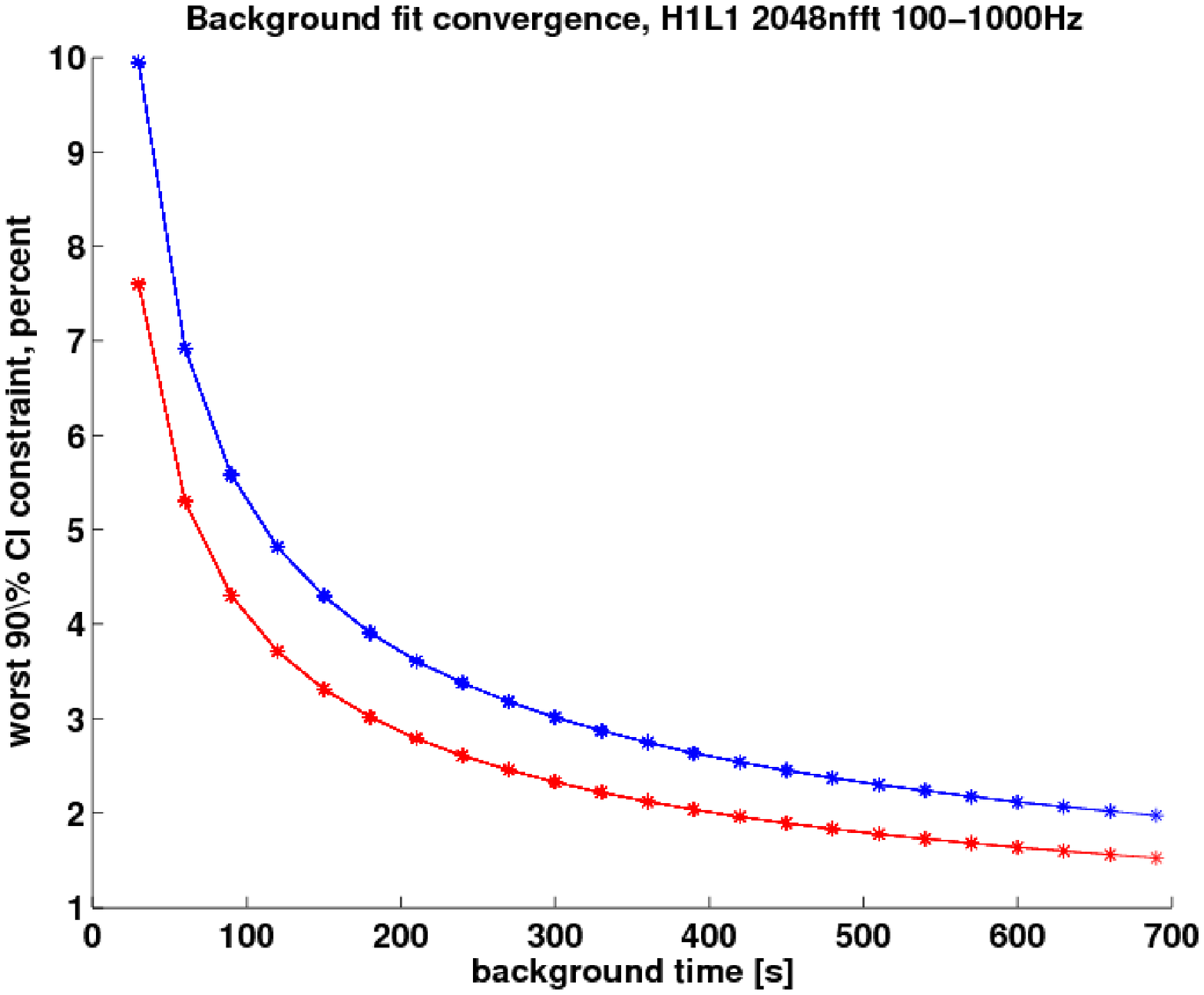}}
\caption[Convergence of gamma-fit parameters]{\textbf{(a)}
Worst-case convergence of gamma-fit parameters at a Fourier
transform length of 512 samples in the 100-1000~Hz band. H1 data
only were used. \textbf{(b)} Worst-case convergence of gamma-fit
parameters at a Fourier transform length of 2048 samples (1/8\,s) in the
100-1000~Hz band. LIGO H1L1 data were used.}
\label{fig:alphabetaconvergeFullBand}
\end{center}
\end{figure}

\begin{figure}[!h]
\begin{center}
\subfigure[alpha]{
\includegraphics[angle=0,width=90mm,clip=false]{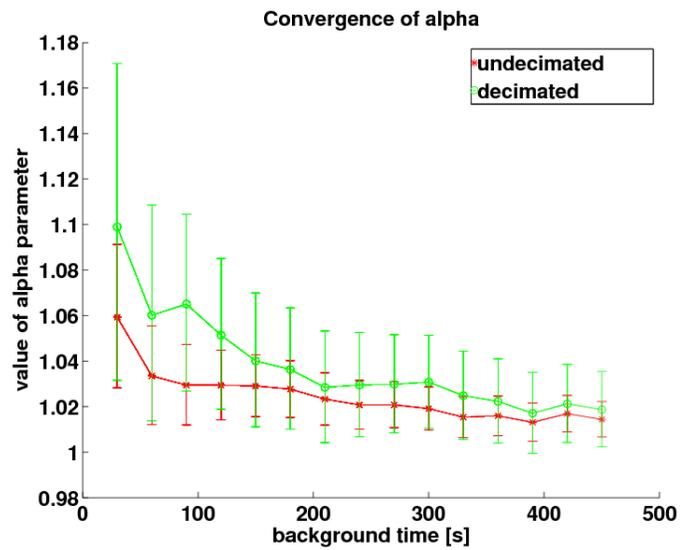}}
\subfigure[beta]{
\includegraphics[angle=0,width=90mm,clip=false]{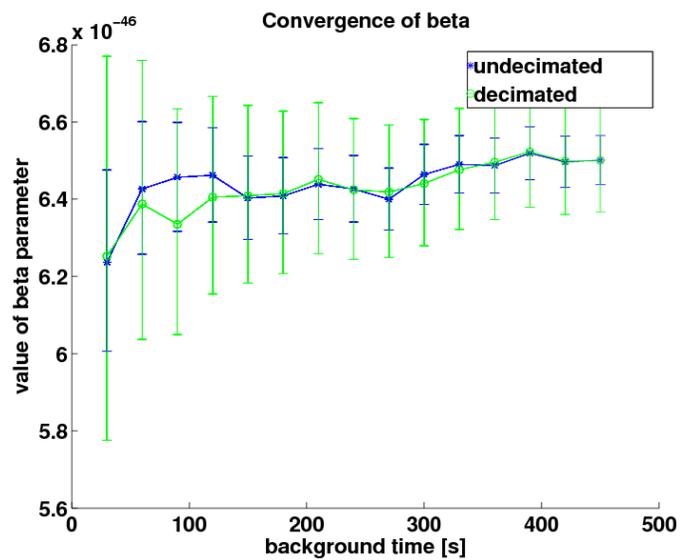}}
\caption[Effect of decimation on gamma-fit convergence]{\textbf{(a)}
Convergence of gamma-fit parameter alpha, with and without
decimation. \textbf{(b)} Convergence of gamma-fit parameter beta,
with and without decimation.  LIGO H1L1 data were used to make both
plots. Error bars in both plots are at the 90\% CL.}
\label{fig:alphabetavalues}
\end{center}
\end{figure}

\begin{figure}[!h]
\includegraphics[angle=0,width=110mm, clip=false]{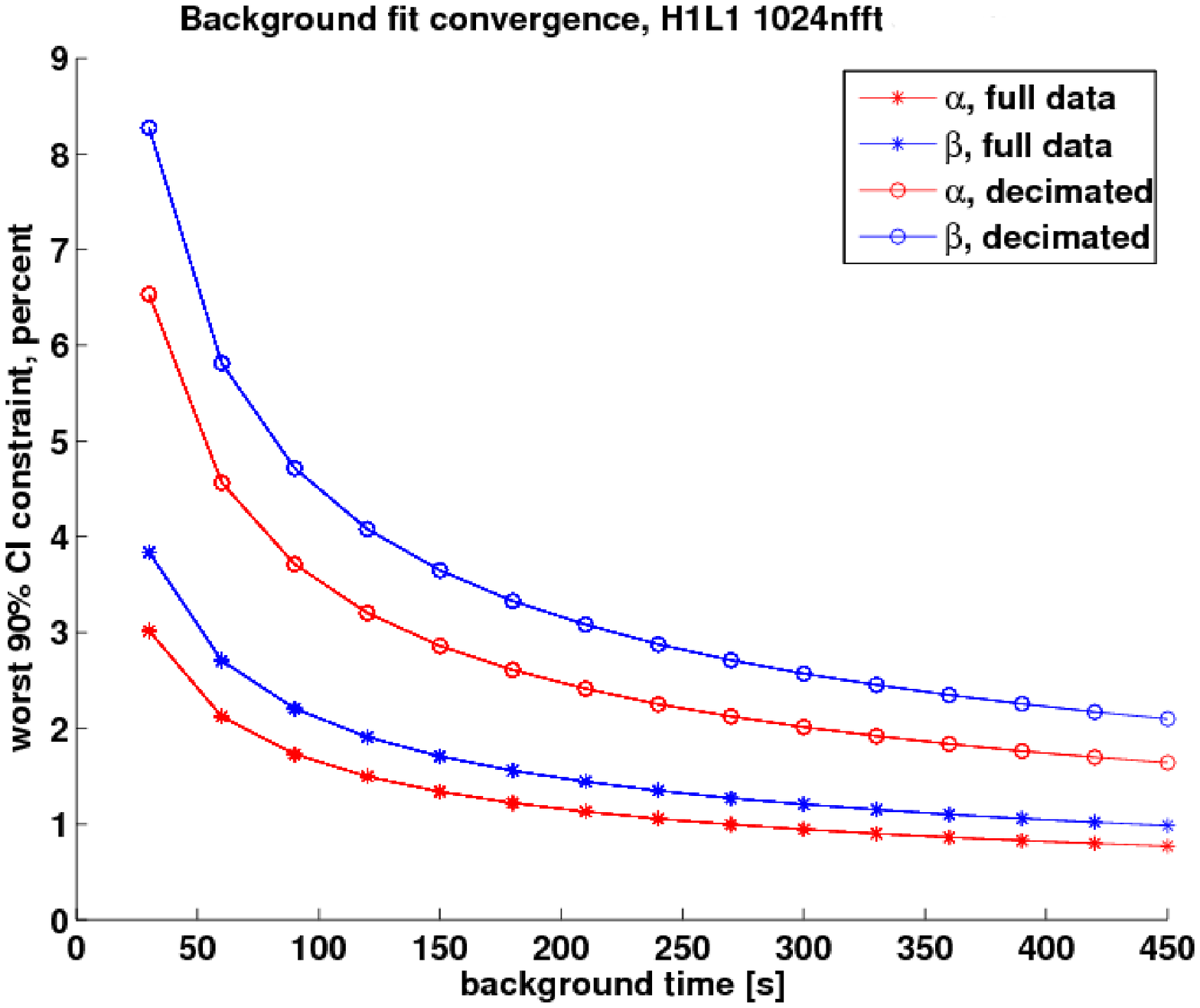}
\caption[Largest error bar of gamma-fit parameters]{ Largest error
bar, in percent, for either alpha or beta for a single frequency bin
(centered at 144~Hz) with and without decimation. Convergence is
relatively fast since only one frequency bin is considered.  }
\label{fig:alphabetaconverge}
\end{figure}

To determine convergence times, we considered the largest 90\% CL
error bar on either $\alpha$ or $\beta$ at any frequency bin in the
search range for a given Fourier transform length
(Figure~\ref{fig:alphabetaconvergeFullBand}). We considered two
search bands, 100-200~Hz and 100-1000~Hz, Fourier transform lengths
of 512, 1024, and 2048 samples (1/8\,s), and networks of H1L1, H1, and
dual-detector white noise. We found no dependence on search band or
detector combination.  Lack of dependence on search band is
explained by the fact that the noisier low frequency bins determine
the worst-case convergence time, and both the large and small band
contain the same low frequency bins. At 512 samples, it took 180~s
for both parameters to be constrained at the 90\% CL to within 2\%
(H1L1 and H1, large band). At 1024 samples (1/16\,s) it took 360~s (H1 and
H1L1, small band). At 2048 samples (1/8\,s) it took 690~s (H1L1, large band).
The same tests were also performed on white noise.

We have considered decimating the background data in an attempt to
account for data dependencies due to Fourier transform overlap.
However, while performing this study it became apparent that this
has no real effect on the statistics of the background, while
requiring more background by a factor equal to the decimation
factor.  At a decimation factor of 3 (set by the auto-correlation
peak) both fit parameters converge to the same values as shown in
Figure~\ref{fig:alphabetavalues}, but the convergence takes 3 times
as long as shown in Figure~\ref{fig:alphabetaconverge}.

We determined that 2000\,s of background is sufficient for determining FAR-equivalent significance of on-source analysis events, except in the case of very large on-source events.  If any such are observed, they can be handled by combining data from multiple background regions or extrapolation to take the background FAR cumulative histogram to lower values.

For clarity, we note that antenna factors do not play a role in
background studies.  The detection procedure
involves studying candidate signals in relation to the noise. Thus
it is not necessary to consider changing antenna factors over the
course of processing large stretches of background.  The upper
limit procedure, which involves simulated gravitational wave
injections, does depend on antenna factors.

\section{Characterizing simulation parameter spaces}

Upper limits on gravitational waves can be estimated using simulated
signals. Any type of simulation has parameters
associated with it. For example, monochromatic ringdown simulations
are parameterized by their frequency $f$ and decay time constant
$\tau$.

\subsection{Ringdown duration} \label{section:ringdownDuration}

We performed an experiment to determine how search sensitivity using
the Flare pipeline depends on ringdown duration $\tau$, for a given
Fourier transform length.  The results are shown in
Table\,\ref{table:rdtau}.  Strain upper limits from values of $\tau$
other than $\tau=200$ in the range 100-300\,ms are within $15$\% of
the 200\,ms value.

\begin{table}[h]
\begin{center}
\caption[Upper limit dependence on ringdown $\tau$]{Flare pipeline
$\hrssn$ upper limit dependence on ringdown $\tau$, at Fourier
transform length of 250\,ms for 1590\,Hz circularly polarized
ringdowns in real H1L1 noise at GPS 827345255.}
\begin{tabular}{ll}
 \hline
 $\tau$ & $\hrssn$ \\
   \hline
100 & $\sci{1.86}{-21}$ \\
150 & $\sci{1.85}{-21}$ \\
200 & $\sci{1.93}{-21}$ \\
300 & $\sci{2.10}{-21}$ \\
350 & $\sci{2.36}{-21}$ \\
\end{tabular}
\label{table:rdtau}
\end{center}
\end{table}

\subsection{WNB simulation duration and search integration length}
\label{section:flareDurations}

We performed an experiment to determine how search sensitivity using
the Flare pipeline depends on WNB duration, for a given Fourier
transform length. The results are shown in
Figure~\ref{fig:WNBdurations}.  The plots were made with real LIGO
L1 data near S5 GPS time 817546378, and using the SGR 1806--20 sky
location.  Efficiency curve thresholds were set from the 90th
percentile of loudest events obtained from the same collection of
100 10~s segments.  This should provide a fair threshold for
comparing different Fourier transform lengths. Therefore, this
experiment is also effectively an optimization experiment for
Fourier transform length.

\begin{figure}[!h]
\begin{center}
\subfigure[]{
\includegraphics[angle=0,width=100mm,clip=false]{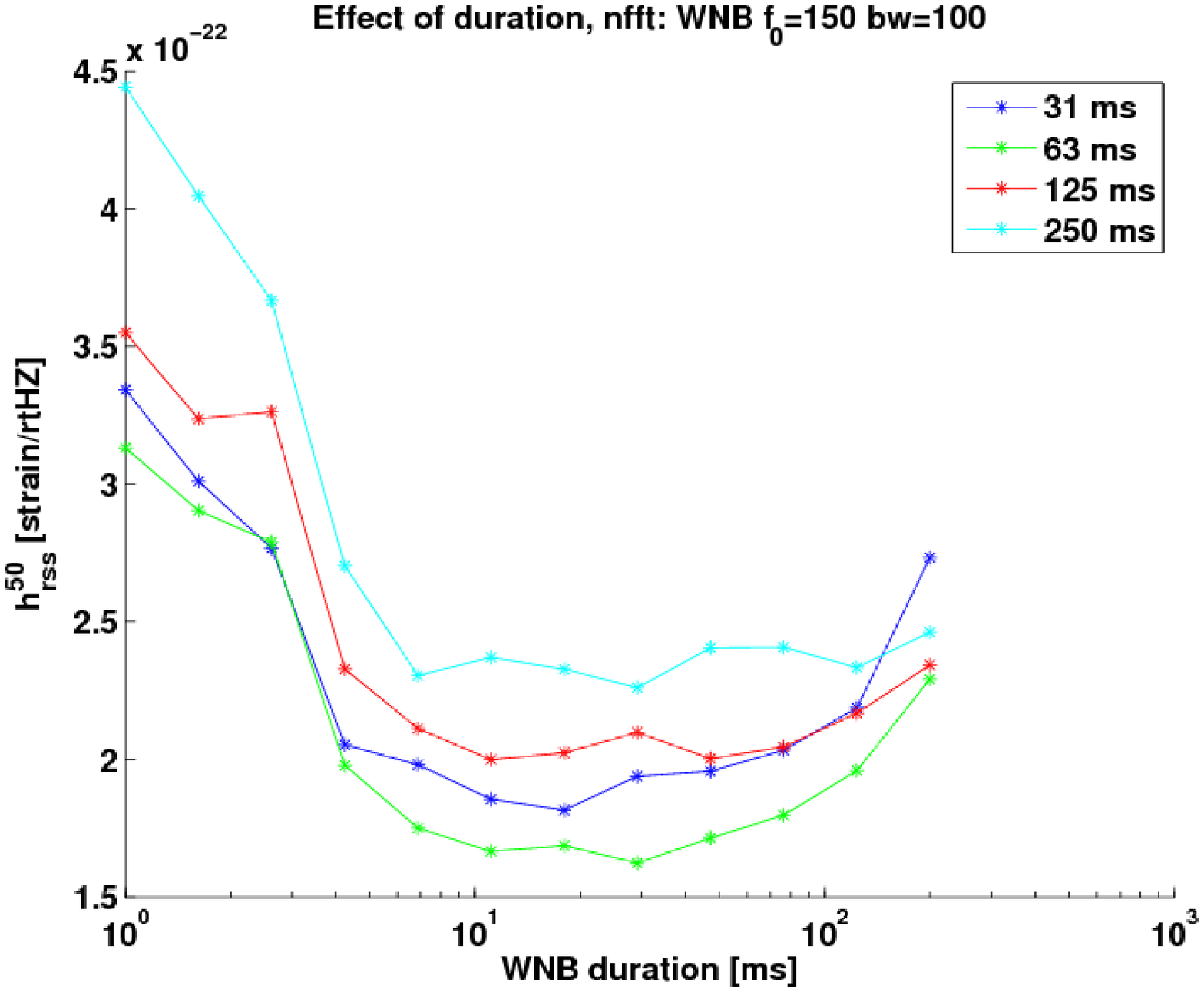}}
\subfigure[]{
\includegraphics[angle=0,width=100mm,clip=false]{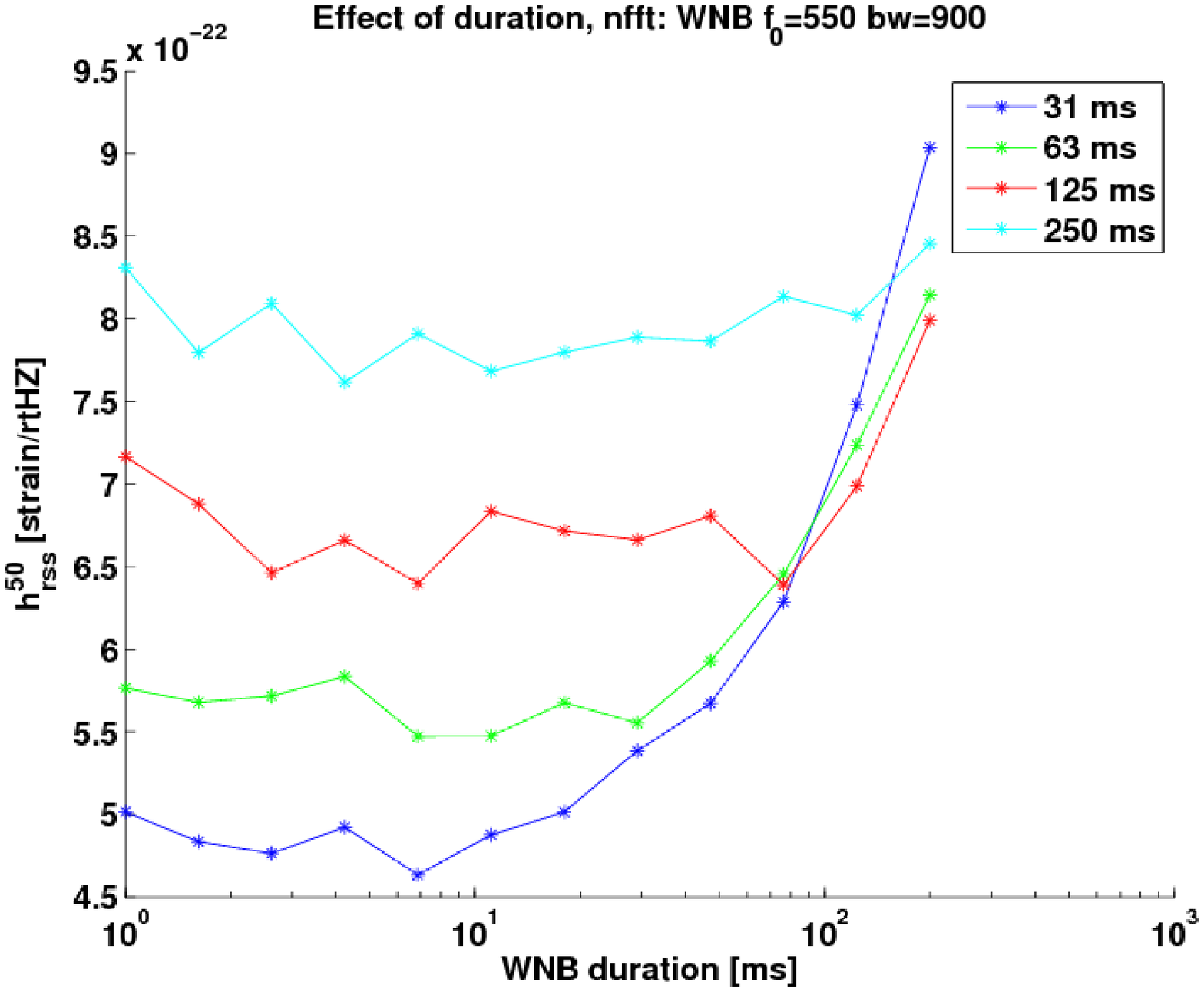}}
\caption[$\hrssf$ sensitivity vs. WNB durations]{\textbf{(a)}
$\hrssf$ sensitivity vs. WNB durations, for a variety of Fourier
transform lengths, for 100-200~Hz WNBs. \textbf{(b)} The same
experiment repeated for 100-1000~Hz WNBs.} \label{fig:WNBdurations}
\end{center}
\end{figure}

In both plots it is apparent that longer Fourier transform lengths
maintain sensitivity to higher WNB durations, and show more
degradation at lower WNB durations, as expected.

We are interested in burst durations spanning from $\sim$5~ms to
$\sim$200~ms, and we would like to adequately cover the duration
space with as few simulated waveforms as possible.  We choose to use
WNB durations of 11~ms and 100~ms. For efficient 
DFTs, we limit Fourier transform lengths, in samples, to powers of
two.

For 100-200~Hz WNBs, the optimal Fourier transform length for both
durations is apparently 1024 samples (63~ms).  The sensitivity
degradation for 5~ms length WNB signals relative to 11~ms signals is
about 13\%. The sensitivity degradation for 200~ms length WNB
signals relative to 100~ms signals is about 18\%.  The sensitivity
degradation for 50~ms length WNB signals relative to 11~ms signals
is about 3\%.

For 100-1000~Hz WNBs, the optimal Fourier transform length for 11~ms
is 512 samples (31~ms) and the optimal length for 100~ms is
apparently 2048 samples (125\,ms).  The sensitivity degradation for
5~ms length WNB signals relative to 11~ms signals is less than 1\%.
The sensitivity degradation for 200~ms length WNB signals relative
to 100~ms signals is about 16\%.  The sensitivity degradation for
50~ms length WNB signals relative to 100~ms signals is about 1\%.

Using these choices of Fourier transform length and WNB simulation
durations, we can effectively cover the duration space while
limiting sensitivity degradations to no more than 20\%.

\section{Automated test suites} \label{section:testSuite}

Many of the above tests are implemented in code in automated test
suites.  We have produced three suites of automated tests for the
pipeline code. These suites are run every night by a cron job,
and the results are captured in daily e-mails. The first suite
consists of unit tests, which examine discrete aspects of individual
code modules, or small groups of modules.  The second suite consists
of ``end-to-end'' tests, which test the pipeline in its complete
state using controlled inputs and checking outputs. The third suite
of tests checks the pipeline's performance on white noise against
theoretical predictions.

Within these test suites there are two types of  automated tests:
those performing general validations of the code and those testing
specific bugs in the code which have been found and fixed along the
way. The ideal way to deal with any bug is to first write an
automated test which will fail when the bug is present and pass when
the bug is eliminated; run the test and watch it fail; fix the bug;
and then run the test and watch it pass. This creates a living
record of the bug's elimination which is exercised every night, and
which is documented in code.

There are currently over 130 daily automated tests.  Some of the
tests are described in the sections below as implementations of
specific validations of the pipeline.

\section{Sensitivity estimates with simulated LIGO noise}

In this section we present preliminary tests of Flare pipeline's
sensitivity on simulated noise modeled after LIGO noise.  In this
case there is no astrophysical trigger and no on-source region, but
we can estimate search sensitivity using a hypothetical burst event
sky location and trigger time. The steps are identical to those used
to estimate an upper limit except the efficiency curve is
constructed from a threshold determined from loudest events in a
collection of data segments instead of a single loudest on-source
event (Section~\ref{section:measuringSensitivity}).  This is done to give a sensitivity estimate less prone to loudest event fluctuations. 

\subsection{Generating simulated LIGO data}
\label{section:SimulatedData}

It is useful to produce simulated LIGO data for estimating pipeline
sensitivity. First, a model segment of LIGO data is obtained, bandpassed and with
transients removed. This segment must be large enough to provide a
histogram and high resolution frequency series from which the
simulated data will be modeled.  The segment is scrambled randomly
to a new vector with similar histogram, but greater entropy.

We then Fourier transform both the model and simulated data vectors,
matching the simulated frequency series in amplitude to the model
frequency series point by point.  Finally, we transform back to the
time domain.

In Figure~\ref{fig:simDataTimeseries}, we show time series of 60 s
of model LIGO data and 60 s of simulated data.  In
Figure~\ref{fig:simDataHistograms} we show histograms for these time
series, and in Figure~\ref{fig:simDataNoNotchStrain} we show
calibrated amplitude spectral density plots for these time series,
without notching the usual 60 Hz and harmonic power, violin mode,
calibration, and unknown lines (which the simulated data preserves).

\begin{figure}[!h]
\includegraphics[angle=0,width=120mm, clip=false]{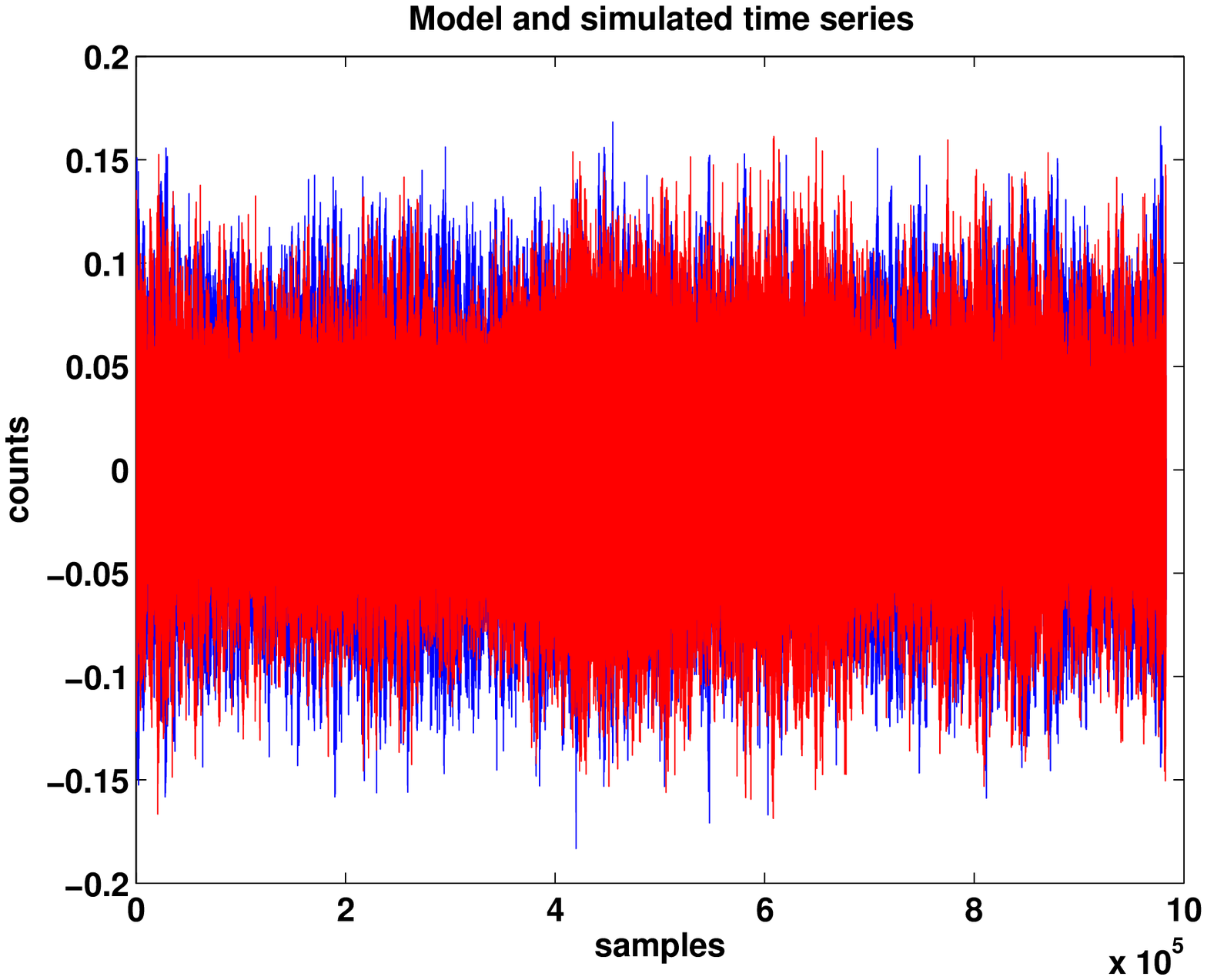}
\caption[Time series of simulated and real LIGO data]{ LIGO model
(blue) and simulated data (red) time series.}
\label{fig:simDataTimeseries}
\end{figure}

\begin{figure}[!h]
\includegraphics[angle=0,width=120mm, clip=false]{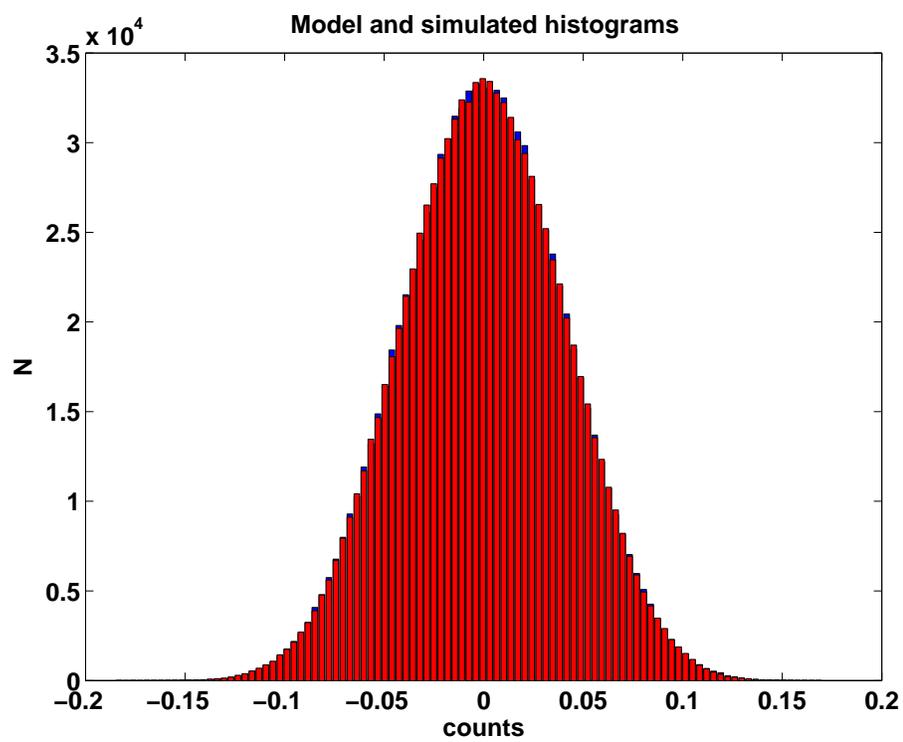}
\caption[Histograms of simulated and real LIGO data]{ LIGO model
(blue) and simulated data (red) histograms. The model data histogram
is almost identical to the simulated histogram.}
\label{fig:simDataHistograms}
\end{figure}

\begin{figure}[!h]
\includegraphics[angle=0,width=120mm, clip=false]{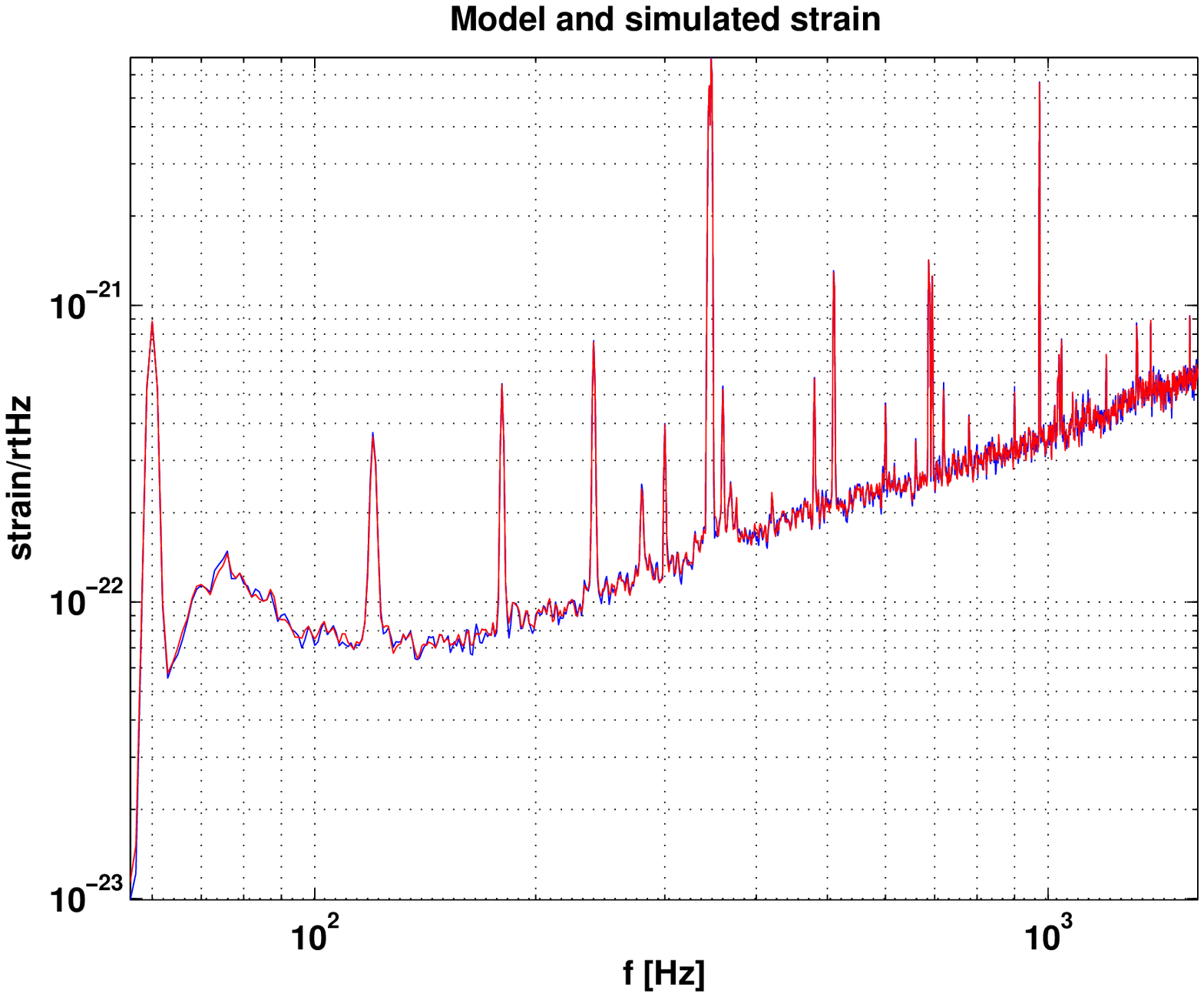}
\caption[Spectra of simulated and real LIGO data]{ LIGO model (blue)
and simulated data (red) and calibrated strain series, without
notching. Agreement is so good that the model data are not always
visible underneath the simulated data.}
\label{fig:simDataNoNotchStrain}
\end{figure}

We remark that this simple algorithm for simulated data does not replicate glitches in the model LIGO data. We
hope in the future to extend this algorithm to generate simulated
data which can model glitches, and would have
false alarm rates matching the model data.  We remark, though, that
the  simple algorithm should be sufficient for the purpose at hand:
estimating single detector pipeline sensitivity for short bursts in
data not overly contaminated with large glitches.

\subsection{Measuring sensitivity} \label{section:measuringSensitivity}

When performing an astrophysical search on \emph{real data} we
choose an efficiency curve threshold equal to the loudest on-source
event (Section~\ref{section:overview}). When using \emph{simulated
data} we instead determine an efficiency curve threshold set
relative to the noise background
(Figure~\ref{fig:thresholdConstruction}). We create a collection of
processed data segments, with lengths equal to some hypothetical
on-source region. (For the two-detector case, this collection can be
created by using relative time shifts between the two streams.) We
choose the loudest event from each segment. We then take the 90th
percentile of this collection as the efficiency curve threshold.

\begin{figure}[!h]
\begin{center}
\includegraphics[angle=0,width=110mm,clip=false]{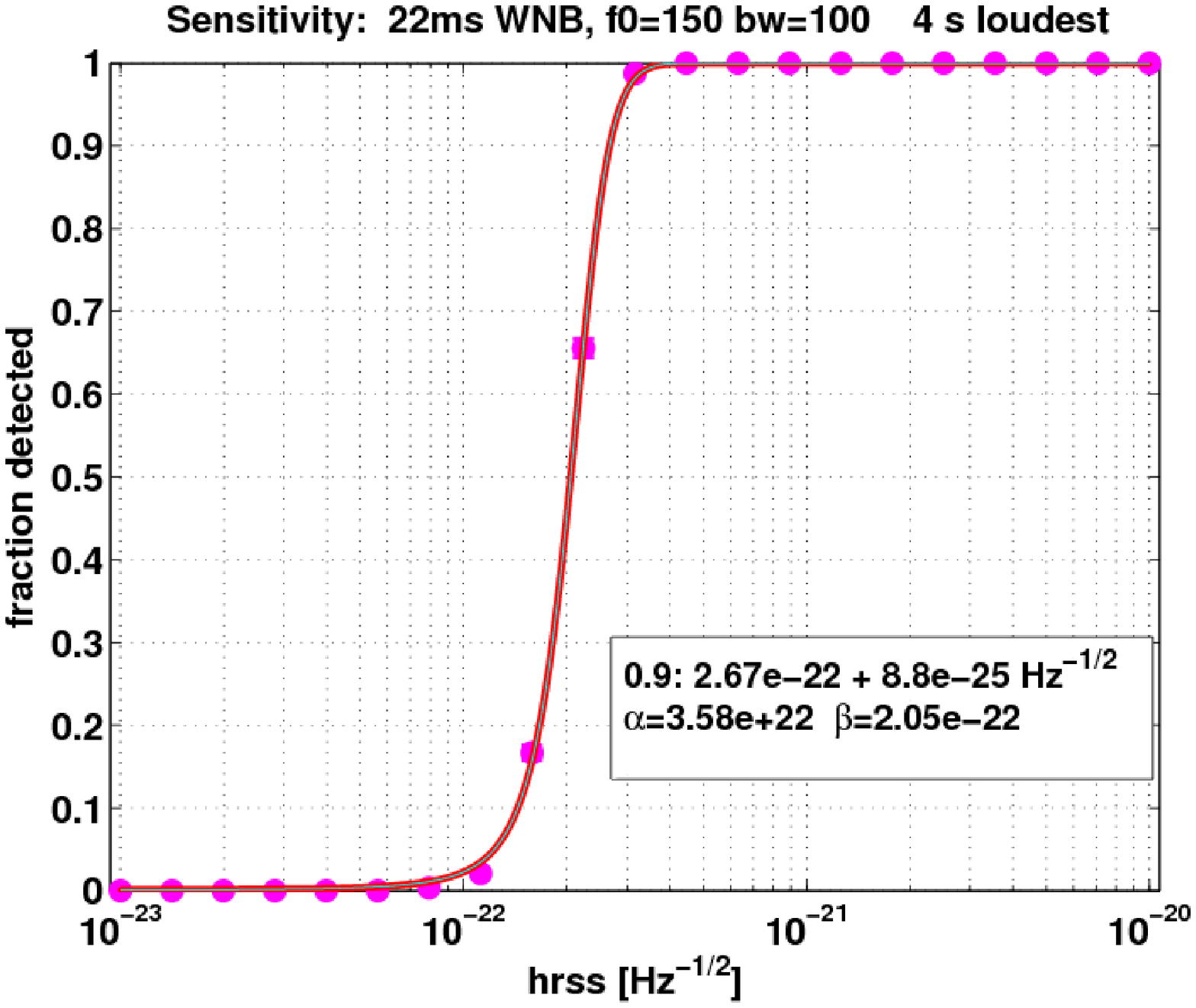}
\caption[Example efficiency curve]{\textbf{(a)} Example efficiency
curve for 22~ms duration WNBs.  Unlike the efficiency curves shown
in Chapter\,\ref{chapter:flare}, this result was made with simulated
noise, and with vertical clustering instead of two-dimensional
clustering.  A two-detector search was simulated.  The threshold
used in constructing the efficiency curve was obtained from a
collection of 4 s loudest event segments as described in the text.
Data points include 1-sigma binomial error bars. $\alpha$ and
$\beta$ parameters of the curve fit are given. The one-sided 95\%
confidence interval for the curve fit at $y=0.9$ is given.  Other
sources of uncertainties are discussed in
Section~\ref{section:exampleResults}. }
\label{fig:thresholdConstruction}
\end{center}
\end{figure}
.

\subsection{Simulated two-detector searches}
\label{section:exampleResults}

We present sensitivity estimates for simulated LIGO-like data, for
4~s and 180~s loudest event segment lengths. The simulated data are
created from white noise with time-series amplitude
distribution and PSD matched to real LIGO sample data (H1 and L1
detectors) taken from early in LIGO's fifth science
run~\cite{noiseCurveEarlyS5}. We present as examples three target
signal classes: 22~ms duration WNBs between 100 and 200~Hz; 100~ms
duration WNBs between 64 and 1024~Hz; and neutron star ringdown (RD)
waveforms with time constant $\tau=150$~ms at 1900~Hz.  These example waveforms were chosen before settling on the final choices used in the S5y1 SGR search.  We expect
some performance degradation when running on real data, since real
data contain non-stationarities and phase relationships not present
in the simulated data used here.

These simulated data searches  were executed as though
for a real event, using the celestial coordinates of SGR 1806--20
and a simulated trigger time chosen to give optimal source location
relative to the LIGO detectors.

The results in Table~\ref{table:simresults} give one-sided 95\%
confidence uncertainties in superscripts.  The first superscript
gives the systematic uncertainty arising from the detector
calibrations, placed at 10\%, the value used in LIGO's fourth
science run all-sky burst search~\cite{s4allsky}.  Our simulated
data will preserve systematic magnitude errors present in the model
data.

The second superscript gives a statistical uncertainty arising from
uncertainty in the estimation of the 90th percentile of the
collection of loudest events used as the efficiency curve threshold.
The collection is fit with a gamma distribution and the 95\% upper
confidence interval on the 90th percentile estimate is determined.
This source of uncertainty would not be present in searches on real
data, since the efficiency curve threshold in that case is set from
the single loudest on-source event.

The third superscript is a statistical uncertainty arising from the
fact that the Flare pipeline's built-in simulations engine is a
Monte Carlo method with a finite number of simulations. It is
obtained by performing each search with at least 2000 injections at
each $\hrss^{\mathrm{sim}}$ value in the efficiency curve, and
grouping these simulations into 10 subgroups. Each subgroup provides
a $\hrssn$ sensitivity estimate, from which the error on the mean is
obtained.  (This preliminary estimation method was later replaced by the bootstrap method.)

The fourth superscript is a statistical uncertainty arising from the
sigmoid fit, as shown in Figure~\ref{fig:thresholdConstruction}.

We note that SGR sky locations have error boxes that are small enough to be insignificant for our purposes.
For example, the position of SGR 1806--20 is known to a few tens of
arcseconds to high confidence~\cite{hurley99WhereIs}.  This
corresponds to an uncertainty in antenna values of a few hundredths
of a percent, which we neglect.

Statistical uncertainties are added in quadrature.  The result is
added to the sensitivity estimate along with the systematic
calibration uncertainty.

\begin{table*}[h]
\caption[Two-detector 90\% detection efficiency Flare pipeline
sensitivity estimates]{ Two-detector 90\% detection efficiency Flare
pipeline sensitivity estimates (strain/$\sqrt{Hz}$) for target
signal classes mentioned in the text.  Results were obtained using
LIGO-like simulated data.  We present results based on two different
loudest event segment lengths, 4 and 180~s (the loudest event
segment length in a simulated search is the equivalent to the
on-source duration used in a real search). Clustered results use a
significance statistic, whereas unclustered results use an excess
power statistic.  Superscripts give one-sided 95\% confidence
uncertainties and are described in
Section~\ref{section:exampleResults}. Polarization angle was chosen
randomly for each injected simulation. Ringdowns (RDC) were
circularly polarized.  }
\begin{center}
\begin{tabular}
    {@{\extracolsep{\fill}} c|c|c|rcr}
 Type & Clustering &  Segment [s]  & \multicolumn{3}{c}{$ \hrss^{90} \qquad [ 10^{-22} ~ ~ \mathrm{strain} \cdot \rthz $] } \\
    \hline \hline
WNB 22ms 100-200Hz & No & 4 & $2.67$ & $^{ +0.27 ~ +0.02 ~ +0.02 ~ +0.01}$ &  $=2.97$ \\
 & No & 180 & $2.91$ & $^{ +0.29 ~ +0.14 ~ +0.02 ~ +0.00}$ &  $=3.34$ \\

  & Yes & 180 & $2.92$ & $^{ +0.29 ~ +0.08 ~ +0.01 ~ +0.00}$ &  $=3.30$ \\
\hline

WNB 100ms 64-1024Hz & No & 4 & $7.84$ & $^{ +0.78 ~ +1.14 ~ +0.14 ~ +0.14}$ &  $=9.79$ \\
 & No & 180 & $11.90$ & $^{ +1.19 ~ +0.01 ~ +0.04 ~ +0.32}$ &  $=13.41$ \\
& Yes & 180 & $6.97$ & $^{ +0.70 ~ +0.11 ~ +0.04 ~ +0.01}$ &  $=7.79$ \\

\hline

RDC 150ms 1900Hz & No & 4 & $17.04$ & $^{ +1.70 ~ +0.05 ~ +0.10 ~ +0.16}$ &  $=18.94$ \\
 & No & 180 & $19.30$ & $^{ +1.93 ~ +0.01 ~ +0.16 ~ +0.03}$ &  $=21.39$ \\

\hline \hline \hline
\end{tabular}
\end{center}
\label{table:simresults}
\end{table*}

\subsubsection{22~ms WNBs between 100~Hz and 200~Hz}
\label{section:22msWNB}

Results for 100--200\,Hz 22\,ms WNBs are given in
Table~\ref{table:simresults}. The absolute cross-correlation between $\hp$ and $\hc$ polarization
components of this example signal class was constrained to be less
than 0.1 (where maximum correlation is +1 and maximum
anti-correlation is -1).  If no cross-correlation constraint is
imposed, search sensitivity to relatively narrow-band white noise
bursts shows a $\psi$-dependence. This can be understood with the
observation that band-limited WNB simulations approach sine-Gaussian
waveforms as the allowed bandwidth approaches zero. The distribution
in cross-correlation between $\hp$ and $\hc$ becomes wider as
simulations approach pairs of sine-Gaussian-like signals with an
undetermined relative phase.  This cross-correlation constraint was not applied to simulations in the S5y1 SGR search.

\subsubsection{100~ms WNBs between 64~Hz and 1024~Hz}

Results for 64--1024\,Hz 100\,ms WNBs are given in
Table~\ref{table:simresults}.

In this case a constraint on the cross-correlation between $\hp$ and
$\hc$ was not applied, since the distribution in cross-correlation
is already sharply peaked around zero.

\subsubsection{Neutron star ringdown waveforms with $\tau=$150~ms}

We present results for circularly polarized ringdowns at 1900~Hz in
Table~\ref{table:simresults}. Figure~\ref{fig:linearRD} shows
several efficiency curves over one period ($\pi/2$) in polarization
angle $\psi$, for both linearly and circularly polarized ringdowns.
We point out that these simulations are close to the edge of the
passband (see Figure~\ref{fig:dataConditioning}).  Using a
64-3000~Hz passband (applied both when creating simulated data and
when performing data conditioning) improves the ringdown results in
Table~\ref{table:simresults} by $\sim$15\%.

\begin{figure}[!h]
\begin{center}
\subfigure[linear polarization]{
\includegraphics[angle=0,width=90mm, clip=false]{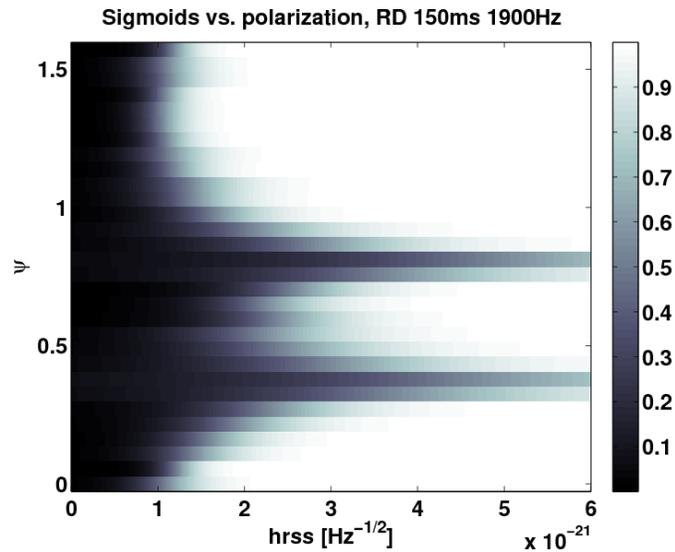}}
\subfigure[circular polarization]{
\includegraphics[angle=0,width=90mm, clip=false]{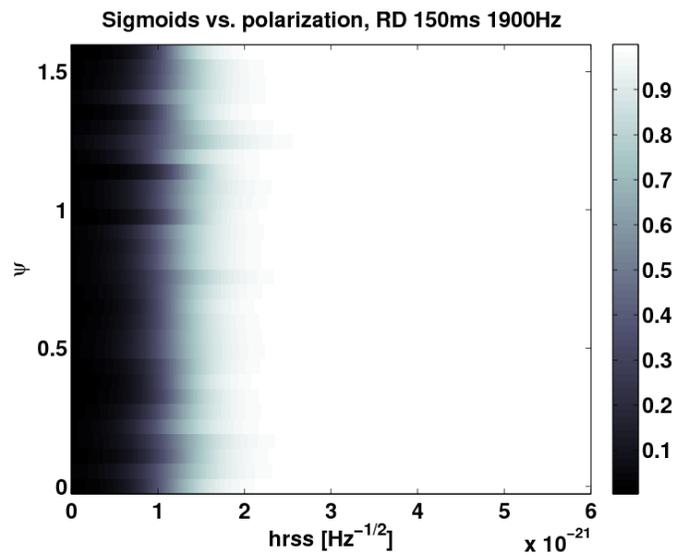}}
\caption[Linear and circular RD efficiency curves vs.
$\psi$]{Efficiency curves from two-detector simulated data search
for neutron star ringdown (RD) waveforms at 1900~Hz with linear
polarization \textbf{(a)} and circular polarization \textbf{(b)} as
a function of polarization angle $\psi$.  A 2~s loudest event
segment length was used; clustering was not used. Each horizontal
stripe in the figure can be thought of as the top view of an
efficiency curve (e.g. Figure~\ref{fig:thresholdConstruction}). The
grayscale depth represents fraction detected.  The sensitivity
minimum in (a) at $\psi \simeq 0.3$ corresponds to the LIGO Hanford
antenna pattern minimum, and the minimum at $\psi \simeq 0.8$
corresponds to the LIGO Livingston antenna pattern minimum.}
\label{fig:linearRD}
\end{center}
\end{figure}

\clearpage

\section{GRB 070201 analysis}
\label{section:070201}

The short, hard GRB 070201, occurring on 2007 February 1, was
interesting as it was coincident with M31, the Andromeda Galaxy.
This makes it very likely that the event occurred remarkably close,
only $\sim$770\,kpc from Earth.  At this distance the isotropic
electromagnetic energy emitted by the event was estimated to be
$10^{45}$\,erg\,\cite{gcn6088}, consistent with an SGR giant flare.
However, this energy is several orders of magnitude less than
typical short hard GRBs with energies in the range
$10^{48}--10^{52}$\,erg.  The most popular progenitor class for
short hard GRBs is compact binary coalescence. LIGO during S5 was
able to exclude a CBC event with $1 \msunonly < m_1 < 3 \msunonly$
and $1 \msunonly < m_2 < 40 \msunonly$ at > 90\%
confidence\,\cite{S5GRB070201}.

We present the analysis of GRB 070201 with the Flare pipeline and
the coherent WaveBurst pipeline. We utilized these pipelines to
validate the published GRB 070201 search.  As usual, data from the
LIGO H1 and H2 detectors were divided into on-source and off-source
regions. The on-source data were chosen to be the interval
$[-120,80]$ seconds around the GRB trigger time. The loudest
on-source event was identified and used to estimate the sensitivity
of the search with simulated injections of gravitational wave bursts
of different types and amplitudes.

The results in this section were not formally reviewed
and do not reflect the scientific opinion 
of the LSC.  The coherent WaveBurst work was carried out in
collaboration with S. Klimenko.

\subsection{Coherent Waveburst (cWB) pipeline} \label{section:cwb}

The Coherent Waveburst (cWB) pipeline was originally designed for
all-sky burst searches \cite{cWBnote, klimenko08}. We implemented
the automated triggered search version of cWB for analysis of
triggered burst events, when the time and sky position of the burst
event is known. The pipeline consists of two stages: a coherent
analysis event production stage, in which analysis events are
generated for a network of gravitational wave detectors; and a post
processing stage, when additional selection cuts are applied to help
distinguish the gravitational wave candidates from background
analysis events. At both stages the pipeline executes coherent
algorithms, based both on the power of individual detectors and the
cross-correlation between the detectors. By using the constraint
likelihood approach \cite{klimenko05}, it coherently combines the
energy of individual detector responses into a single quantity
called the network likelihood statistic, which may be interpreted as
the total SNR of the gravitational wave signal detected in the
network. Coherent analysis events are generated when the network
likelihood exceeds some threshold which is a parameter of the
search.

The off-source region included two time intervals $[-1320,-120]$ and
$[80,1280]$ seconds around the GRB 070201 time and it was used for
background estimation. To increase statistics we performed analysis
at 101 time shifts between the H1 and H2 detectors.  As mentioned
above, the on-source data included the interval $[-120,80]$ seconds
around the GRB 070201 time. It was used to identify the
gravitational wave candidates and estimate the sensitivity of the
search with simulated injections of gravitational wave bursts of
different types and amplitudes.

The cWB pipeline is implemented as a ROOT script which executes
various data analysis algorithms implemented in the Wavelet Analysis
Tool (WAT), which is a part of the LIGO Data Monitoring Tool
~\cite{cvswat}.  It uses calibrated strain ``h(t)'' data which are
resampled from 16384 Hz to 4096 Hz. The data conditioning is
performed in the wavelet (time-frequency) domain. First, predictable
components such as power lines are removed with linear-predictor
filters constructed individually for each wavelet layer. Then the
data are normalized by the variance of the noise estimated for each
wavelet layer. The final product of the cWB data conditioning stage
is whitened time-frequency series of the detector outputs. More
details on data conditioning can be found in the LIGO note
\cite{WBnoteS5}.

The wavelet transformation is used to produce data in the
time-frequency domain. The Meyers wavelet with a  filter consisting
of 1024 coefficients is used. To cover a possible range of the
gravitational wave signal durations the analysis was performed at
the time resolutions of 1/8, 1/16, 1/32, 1/64, 1/128 and 1/256
seconds.

The whitened data from all detectors are combined into a likelihood
time-frequency (LTF) map $L(t,f)$ via the constraint likelihood
approach\,\cite{klimenko05}. In this case the likelihood analysis is
applied to individual TF pixels in different detectors at the same
time-frequency location $(t,f)$. For two aligned detectors, such as
H1 and H2, the $L(t,f)$ statistic is calculated as a projection of
the data vector $\bf{w}$ on the line defined by the antenna pattern
vector  $\bf{F}$
\begin{eqnarray}\label{eq:ltf}
   L(t,f) = \frac{(\bf{w},\bf{F})^2}{||F||^2}, \\
   {\bf{w}} = (w_{H1}(t,f),w_{H2}(t,f)), \\
   {\bf{F}} = (\frac{F_+(H1)}{\sigma_{H1}(t,f)},\frac{F_+(H2)}{\sigma_{H2}(t,f)}),
\end{eqnarray}
where $F_+(H1)$ and $F_+(H2)$ are the antenna patterns for the plus
gravitational wave component calculated in the dominant polarization
frame \cite{klimenko05}, the $\sigma_{H1}$ and $\sigma_{H2}$ are the
rms of the detector noise and $||F||$ is the norm of the vector
$\bf{F}$, which is called the network sensitivity factor. In case of
two aligned detectors  $F_+(H1)=F_+(H2)$.

The cWB clustering procedure is then applied to the 1\% of brightest
pixels. Clustered pixels are called analysis events and they are
reconstructed individually for each time-frequency resolution.

In the post processing we attempt to select an optimal set of
parameters and cuts for rejection of detector glitches.  These
selection cuts are ad hoc and depend on the network configuration
and parameters of the search.

\subsection{GRB 070201 cWB results} \label{section:cwbResults}

The pipeline output  rate estimated in the off-source region is
$0.35$Hz (total live time is 67 hours). The on-source segment yields
the rate of $0.33 \pm 0.05$Hz, consistent with no detection.

The SNR of the loudest on-source analysis event (6.5) is used as a
threshold for estimation of the detection efficiency using simulated
injections.  The search sensitivity was  estimated by using the MDC
waveforms for the threshold of 6.5 on the total SNR.
Figure~\ref{sg250q9} shows the efficiency curve for sine-Gaussian
injections sg250q9 with linear polarization.
\begin{figure}[!h]
\begin{center}
\includegraphics[angle=0,width=100mm, clip=false]{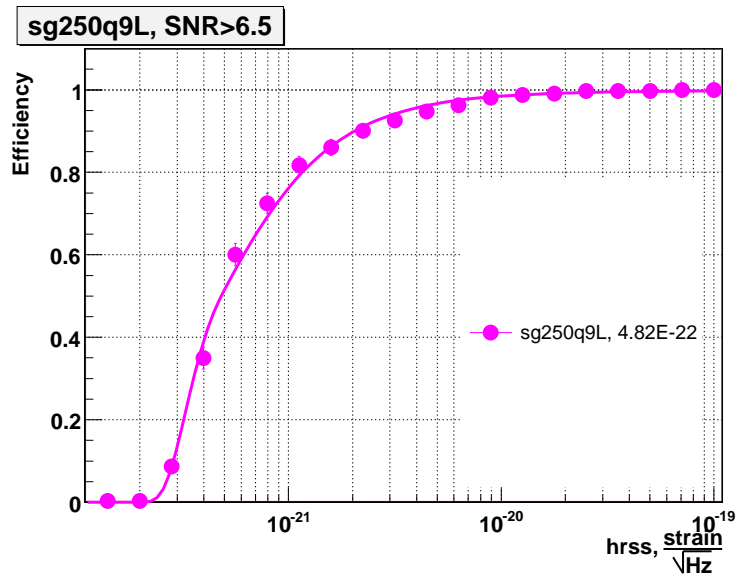}
   \caption[cWB efficiency curve for sg250q9 injections]{cWB efficiency curve for sg250q9 injections.}
   \label{sg250q9}
   \end{center}
\end{figure}

Table~\ref{tab1} shows the sensitivity of the search at 50\% and
90\% of the detection efficiency.

\begin{table}[h]
\begin{center}
\caption[GRB 070201 upper limit results for the cWB pipeline]{  GRB
070201 upper limit results for the cWB pipeline. Polarizations are U
(unpolarized); L (linear); C (circular); and E (elliptical). }
\begin{tabular}{|c|c|c|c|}
 \hline \hline
 MDC waveform & polarization & 50\% efficiency & 90\% efficiency\\ \hline
 SG250Q9  & L & $4.8\cdot 10^{-22} \: \rthz$ & $20.\cdot 10^{-22} \: \rthz$ \\
 SG250Q9  & C & $4.3\cdot 10^{-22} \: \rthz$ & $5.6\cdot 10^{-22} \: \rthz$ \\
 WNB1  & U & $7.2\cdot 10^{-22} \: \rthz$ & $9.8\cdot 10^{-22} \: \rthz$ \\
 WNB2  & U & $29\cdot 10^{-22} \: \rthz$ & $42\cdot 10^{-22} \: \rthz$ \\ \hline
 NSBH 1410 & E  & $14.4$ Mpc & $6.8$ Mpc \\
  NSNS 1414 & E  & $5.0$ Mpc & $2.5$ Mpc \\ \hline \hline
\end{tabular}
  \label{tab1}
  \end{center}
\end{table}

\subsection{GRB 070201 Flare pipeline results}

We present efficiency curves for 1.4-10 solar mass inspiral
waveforms (Figure~\ref{flare1410}), 1.4-1.4 solar mass inspiral
waveforms (Figure~\ref{flare1414}), 250 Hz Q9 sine-Gaussians
(linearly and circularly polarized), and 100 ms duration white noise
bursts centered at 250 Hz with 150 Hz bandwidth
(Figures~\ref{flarewnb1} and~\ref{flarewnb1H1}). Efficiency curves
were made using MDC waveforms prepared by P. Sutton. All MDC
waveforms were prepared with random polarization angles. Inspiral
waveforms were prepared with random inclination angles.

The Flare inspiral search used density-based clustering described
above, with ad-hoc post-processing cuts applied identically to
simulation regions and the on-source region.  The Flare WNB search
used the time series-based pipeline (vertical clustering). A
preliminary search on the WNB MDC waveforms using density-based
clustering did not seem to produce better results. We also plan to
produce results for other MDC waveforms.

Results are summarized in Table~\ref{flareresults}.  We note here
that these results, as with the cWB results presented in Section\,\ref{section:cwbResults}, were prepared with V2 h(t) H1 data, which has a $\sim$20\%
systematic error in magnitude which has not been accounted for.

\begin{table}[h]
\begin{center}
\caption[GRB 070201 upper limit results for the Flare pipeline]{
MDC waveform results for the Flare pipeline, for the H1H2 GRB 070201
search. Polarizations are U (unpolarized); L (linear); C (circular);
and E (elliptical). }
\begin{tabular}{|l|c|c|c|c|c|}
 \hline \hline
 Waveform & Network & Polarization & 50\% efficiency          & 90\% efficiency \\ \hline
 SG250Q9   & H1H2 & L   & $4.7\cdot 10^{-22} \: \rthz$ & $21.\cdot 10^{-22} \: \rthz$ \\
 SG250Q9   & H1H2 & C   & $4.1\cdot 10^{-22} \: \rthz$ & $5.2\cdot 10^{-22} \: \rthz$ \\
 WNB1  &     H1   & U   & $7.8\cdot 10^{-22} \: \rthz$ & $10\cdot 10^{-22} \: \rthz$ \\ \hline
 NSBH 1410 & H1H2 & E   & $14.0$ Mpc                   & $7.5$ Mpc \\
 NSBH 1414 & H1H2 & E   & $6.0$ Mpc                    & $3.1$ Mpc \\ \hline \hline
\end{tabular}
\label{flareresults}
\end{center}
\end{table}

\begin{figure}[ht]
\begin{center}
\includegraphics[angle=0,width=100mm, clip=false]{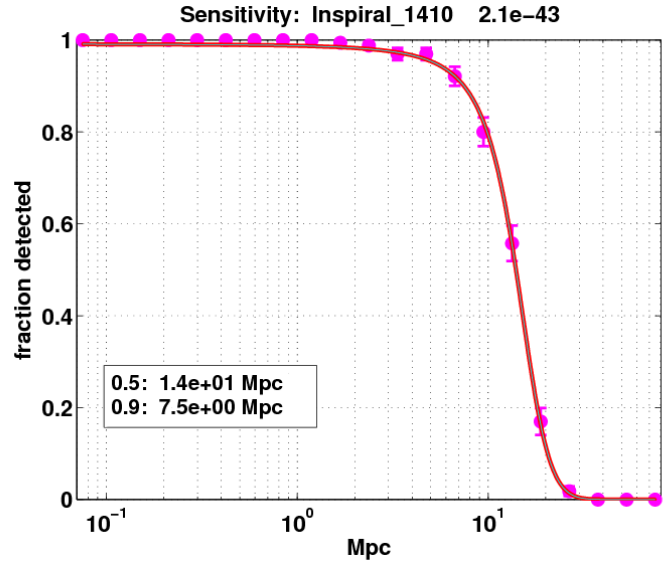}
   \caption[Flare pipeline efficiency curve for 1.4-10 solar mass
inspiral MDCs]{Flare pipeline efficiency curve for 1.4-10 solar mass
inspiral MDCs. The search band is 64-512 Hz.  The x-axis is the
hypothetical scale distance in Mpc at which an inspiral event would
produce the injected simulated waveforms. }
   \label{flare1410}
   \end{center}
\end{figure}

\begin{figure}[ht]
\begin{center}
\includegraphics[angle=0,width=100mm, clip=false]{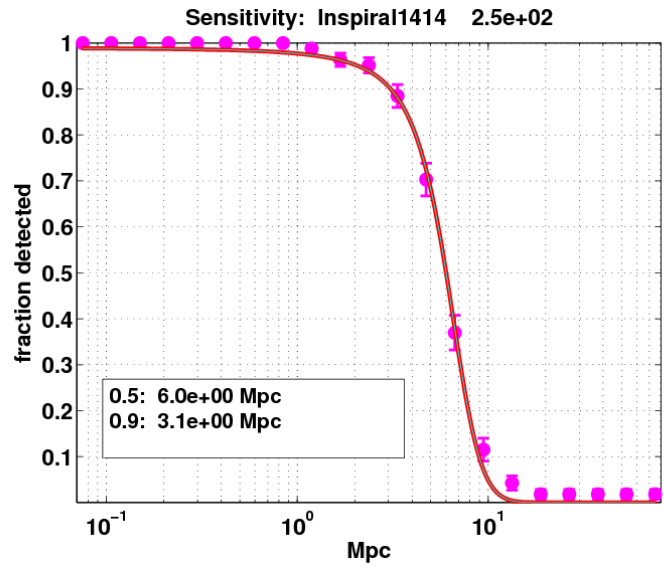}
   \caption[Flare pipeline efficiency curve for 1.4-1.4 solar mass
inspiral MDCs]{Flare pipeline efficiency curve for 1.4-1.4 solar
mass inspiral MDCs. The search band is 64-512 Hz.  The x-axis is the
hypothetical scale distance in Mpc at which an inspiral event would
produce the injected simulated waveforms.  This search uses a
different statistic (tile significance) than the 1.4-10 solar mass
search shown here (which used the excess power statistic).  The
numbers shown in the title of each plot represent the on-source
loudest event statistic threshold used to construct the efficiency
curve. }
   \label{flare1414}
      \end{center}
\end{figure}

\begin{figure}[ht]
\begin{center}
\includegraphics[angle=0,width=100mm, clip=false]{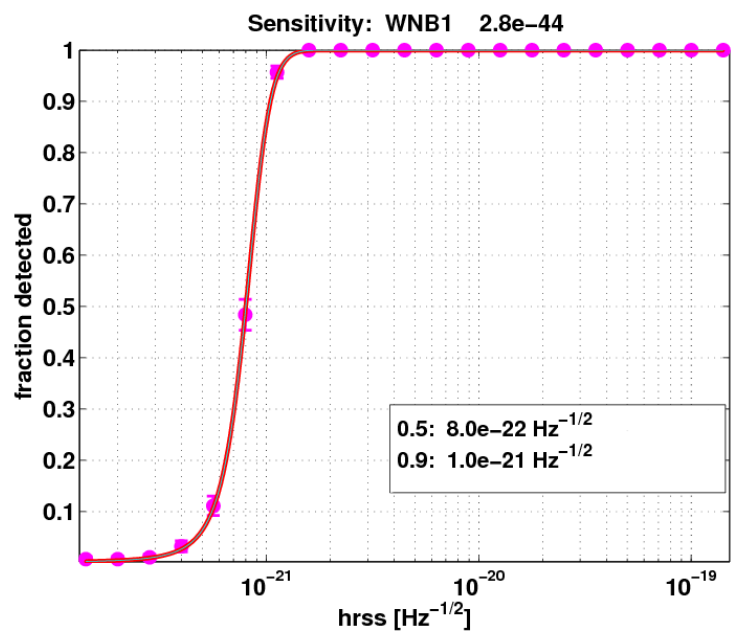}
    \caption[Flare pipeline efficiency curve for WNB1]{Flare pipeline efficiency curve for 100 ms duration white
noise bursts centered at 250 Hz with 150 Hz bandwidth (WNB1). The
search band is 175-325 Hz. Apparently (see Figure~\ref{flarewnb1H1})
the search is being limited slightly by H2. }
    \label{flarewnb1}
       \end{center}
\end{figure}

\begin{figure}[ht]
\begin{center}
\includegraphics[angle=0,width=100mm, clip=false]{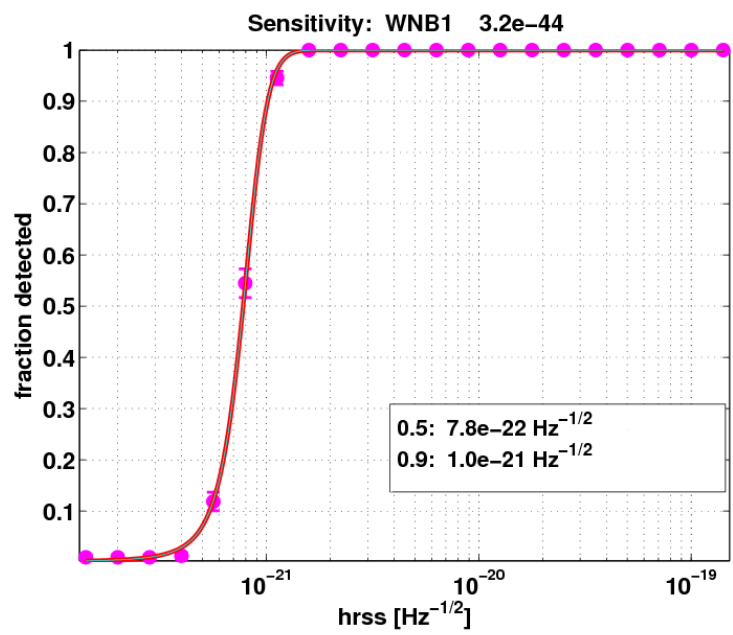}
    \caption[Flare pipeline efficiency curve for WNB1 for the H1 detector]{Flare pipeline efficiency curve for 100 ms duration white
noise bursts centered at 250 Hz with 150 Hz bandwidth (WNB1) for a
single detector (H1).  The search band is 175-325 Hz.  }
    \label{flarewnb1H1}
       \end{center}
\end{figure}

\clearpage

\section{Complementing inspiral searches with burst pipelines}
\label{section:complementingInspirals}

In Section\,\ref{section:cbc}, we introduced the compact binary
coalescence (CBC) class of gravitational wave sources, often
referred to as ``inspiral sources.''

The conventional approach to searches for gravitational waves
associated with inspiral events takes advantage of the well-modeled
inspiral phase of the event by performing a matched filter search.
These searches are ``specialists.'' In a matched filter search, data
from a gravitational wave detector are correlated against templates
matching theoretical CBC waveforms within the relatively narrow
parameter space\,\cite{blanchet06}. Templated searches provide
optimal sensitivity to the target waveforms. However, their
sensitivity to gravitational wave signals outside of the template
bank drops as the correlation decreases.

Coherent burst searches are designed to detect any signal in the
detector's band lasting a few seconds or less.  These searches are
``generalists.'' They use fully coherent addition of gravitational
wave detector data streams to sensitively search small patches of
the sky for gravitational wave bursts without the need for source
modeling. There are many expected gravitational wave burst sources
besides the inspiraling compact binaries (Section\,\ref{section:burstSources}). These burst-type gravitational wave
events are often unmodeled or poorly-modeled. Coherent burst
searches typically don't make assumptions about waveforms beyond
duration and bandwidth ranges.

Externally triggered gravitational wave searches for inspiral events
performed with matched filter pipelines can be complemented with
coherent burst pipelines, which are designed to find short-duration
gravitational wave bursts with little or no further knowledge of the
expected waveform. Though burst pipelines are not as sensitive as
matched filter pipelines to precisely specified waveforms such as
theoretical predictions of inspirals preceding CBC events, their generality may allow them to detect
unpredicted signals or parts of signals (such as the CBC merger phase) that templated searches may miss.

In this section we estimate the sensitivity of the Flare pipeline
and X-Pipeline to the inspiral phase of CBC events using simulated
noise and the GRB 070201 sky position and trigger time.  We make a
rough quantitative comparison to results obtained from the matched
filter search for gravitational waves associated with GRB
070201\,\cite{S5GRB070201}.

\subsection{X-Pipeline}

X-Pipeline is a software package designed to detect unmodelled
gravitational wave bursts in noisy detector data while vetoing
noise-induced glitches\,\cite{chatterji05}. By time-shifting the
data from each detector, X-Pipeline coherently sums the
gravitational wave contributions from a particular sky position O
for each polarization ($\hp$ and $\hc$) and also produces a gravitational
wave-free null stream for consistency testing. Time-frequency maps
are made of the energy in the reconstructed $\hp$, $\hc$, and null
streams. X-Pipeline then identifies clusters of pixels with large
E+, the energy in the $\hp$ stream\,\cite{klimenko05}.

\subsection{Results}\label{section:results}

The error box on the location of GRB 070201 was consistent with the 
position of M31 -- the Andromeda Galaxy, located only 770 kpc from 
Earth.  At the time of the event LIGO's H1 and H2 detectors were 
taking science mode data.  If GRB 070201 did indeed originate from 
M31, and if the progenitor was a binary inspiral event, then LIGO 
would almost certainly detect associated gravitational waves.  A templated inspiral 
search detected no gravitational waves, and a CBC progenitor in M31 was ruled out 
at high confidence\,\cite{S5GRB070201}.  For binary pairs with $m_1$ 
 and $m_2$ in the ranges [1,3] and [10,13] $\msunonly$, the inspiral search yielded physical distance lower limits 
of $\sim$8 and $\sim$15.5\,Mpc at 90\% and 50\% detection efficiencies.
 
 We have performed mock GRB 070201 loudest event searches using
coherent burst pipelines and simulated LIGO noise (produced for
LIGO-VIRGO project Ib\,\cite{beauville07}).  The on-source region was
[-120,60] seconds around trigger time and the search range was 64-1024
Hz. Signals simulating 1.4-10 solar mass inspiral events originating
in M31 were injected into the simulated noise.  Results are shown in
Table\,\ref{table:results}.   An example efficiency curve is shown
in Figure\,\ref{fig:sigmoid1410}. 
 
The Flare pipeline is a simple but effective coherent 
burst pipeline, capable of performing either one- 
detector or two-detector triggered searches [2]. 
The Flare pipeline conditions LIGO data with a 
bandpass filter and a notch filter (generated at runtime

\begin{table}[h]
\begin{center}
\caption[Physical distance lower limits for 1.4-10 solar mass M31
simulated inspirals]{Physical distance lower limits for 1.4-10 solar
mass M31 simulated inspirals.  Binary system orientation and
polarization angle were chosen randomly for each injection.  15\%
has already been subtracted from the results from the two burst pipelines; this is approximately
the same as the overall uncertainty subtracted from matched filter results
in\,\cite{S5GRB070201}. We estimate up to additional ~20\% error in the burst pipeline results
from use of simulated data.  }
\begin{tabular}{l|rr}
 \hline \textbf{Method} & \textbf{90\% efficiency} &  \textbf{50\% efficiency} \\
   \hline \hline
  Matched filter & 8.0 Mpc & 15.5 Mpc \\
 Flare pipeline & 5.4 Mpc & 9.9 Mpc \\
 X-Pipeline & 5.1 Mpc & 10.1 Mpc \\
\end{tabular}
\label{table:results}
\end{center}
\end{table}

\begin{figure}[!h]
\begin{center}
\includegraphics[angle=0,width=100mm, clip=false]{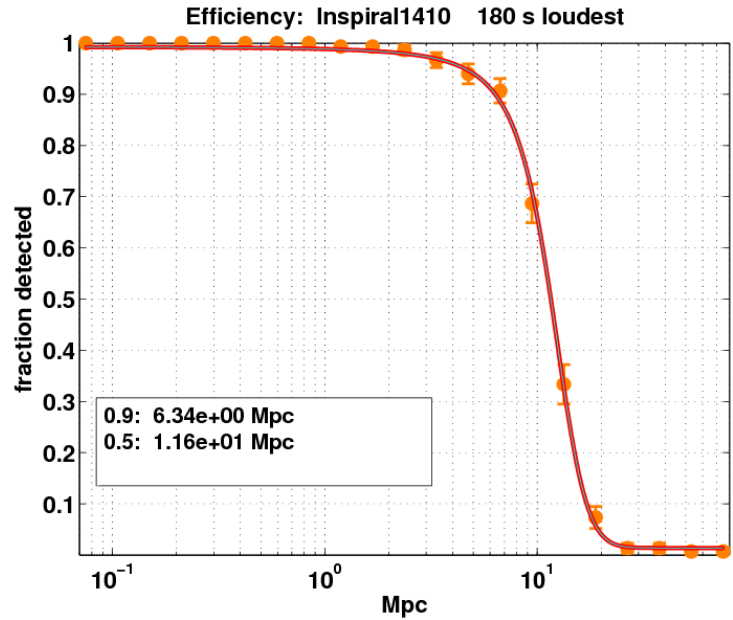}
\caption[Flare efficiency curve for 1.4-10 solar mass inspiral
injections]{ Flare efficiency curve for 1.4-10 solar mass inspiral
injections.  Note that the x-axis is in Mpc of physical distance. } \label{fig:sigmoid1410}
\end{center}
\end{figure}

Coherent burst pipelines are less sensitive to the well-predicted
inspiral phase than templated searches.  However, their larger
search phase space makes them more robust e.g. to non-CBC progenitors or events falling out of the matched filter template bank, and could allow them to exceed matched filter searches in regions of the phase space where the merger stage contributes significantly within the LIGO band. In triggered
searches where the progenitor event may not be CBC For events
outside of the matched filter search template bank (e.g. outside of
the mass or spin range).
\clearpage
\chapter{Soft Gamma Repeaters} \label{chapter:sgrs}

On 2004 December 27, the brightest transient burst ever observed
swept through the solar system~\cite{hurley05}, saturating almost
every satellite instrument viewing the event~\cite{palmer05,
hurley05, terasawa05, schwartz05, RXTEGCN, KonusWindGCN,
IntegralGCN, RHESSIGCN, SwiftGCN}. This gamma ray flare was so
bright that it 
depressed the altitude of Earth's daytime ionosphere\,\cite{inan07}.
Its reflection off the moon was not only readily
observable~\cite{MoonGCN}, but useful in reconstructing the peak
fluence of the event, unobtainable from saturated
detectors~\cite{mereghetti05}. Triangulation~\cite{IPNGCN}, and a
tell-tale 7.56 s modulation in a $\sim$6 minute tail following the
flare~\cite{palmer05, hurley05} (Figure\,\ref{fig:giantFlareTail}),
unambiguously identified the source as SGR 1806--20, one of a class
of rare objects called soft gamma repeaters (SGRs).

SGRs are one of the most bizarre and enigmatic classes of
astrophysical sources.  They are characterized by
sporadic emission of brief ($\approx0.1$\,s) intense bursts of soft
gamma rays with peak luminosities commonly up to $10^{42}$ erg/s
\,\cite{mereghetti08, woods04a}, and are thought to be ``magnetars,'' neutron stars with extraordinarily strong magnetic fields $\sim10^{16}$ G (Section\,\ref{section:magnetarModel}).  Less common intermediate bursts
with greater peak luminosities can last for seconds.  Rare ``giant
flare'' events, some 1000 times brighter than common
bursts\,\cite{palmer05}, have initial bright, short
($\approx0.2$\,s) pulses followed by tails lasting minutes and are
among the most electromagnetically luminous events in the
universe\,\cite{woods04a}. The giant flare tails are modulated at
the rotation period of the star, typically 5--7\,s.  Only five Galactic SGRs have been identified
with confidence~\cite{gcn8112, kouveliotou98}.  SGR 1806--20 and two
others (SGR 1900+14 and SGR 1627--41) are located in our galaxy near
the galactic plane, between 6 and 15\,kpc distant (see
Table\,\ref{table:sgrsummary} and
Section\,\ref{section:sgrDistances}). A fourth, SGR 0526--66, is
located in the Large Magellanic Cloud, about 50\,kpc
away\,\cite{woods04a}.  A fifth, SGR 0501+4516, was discovered on
2008 August 22, only a few days before this writing\,\cite{gcn8112,
gcn8113, gcn8115}, and may be located at a distance of only
1.5\,kpc in the direction of the galactic anti-center\,\cite{gcn8149, leahy95}.  Of the 
confirmed SGRs, SGR 0526$-66$, SGR 1900+14 and SGR 1806$-20$ have
each produced a giant flare since the discovery of SGRs in
1979\,\cite{mazets79,hurley99,hurley05}, making the giant flare rate
on the order of once per 10 years. SGRs have also been associated
with persistent but variable X-ray sources emitting at
luminosities of $10^{34}$ to $10^{36}$\,erg/s in the 1--10\,keV
band\,\cite{woods04a}.

These objects are the astrophysical targets of our gravitational
wave search.  In this chapter we describe them and discuss their
potential for gravitational wave emission, which unfortunately
remains largely unknown. For recent reviews of SGRs and the
magnetar model see\,\cite{woods04a} and\,\cite{mereghetti08}.  In
what follows, luminosities and energies assume isotropic emission
and are for photons above 20\,keV unless stated otherwise.
Table\,\ref{table:sgrsummary} summarizes some electromagnetically
observed properties of the five confirmed SGRs.

\begin{table}
\begin{minipage}{\textwidth}
\begin{center}
\caption[Summary of SGR properties]{Summary of electromagnetically
observed SGR properties for the five confirmed SGRs.  Position and
distance references are given under the uncertainty. }
\begin{tabular}{llrcc}
 \hline
 \textbf{Source} & \textbf{Position} & \textbf{1$\sigma$ uncertainty} & \textbf{Distance [kpc]} & \textbf{Period [s]}  \\
 \hline
SGR 0501+4516 & $05^{\mathrm{h}} 01^{\mathrm{m}} 06.80^{\mathrm{s}}$ & $1.4''$  & 1.5 & 5.8  \\ 
             & $+45^\circ 16' 35.4''$ & \cite{gcn8148}  &  &  \cite{gcn8166} \\
             \hline
SGR 0526--66 & $05^{\mathrm{h}} 26^{\mathrm{m}} 00.89^{\mathrm{s}}$ & $0.6''$  & 50 & 8.0   \\
             & $-66^\circ 04' 36.3''$ & \cite{kaplan01, kulkarni03, klose04}  &  &  \\
             \hline
SGR 1627--41 & $16^{\mathrm{h}} 35^{\mathrm{m}} 51.84^{\mathrm{s}}$ & $0.2''$  & 11 & 6.4   \\
             & $-47^\circ 35' 23.3''$ & \cite{wachter04}  &  &  \\
             \hline
SGR 1806--20 & $18^{\mathrm{h}} 08^{\mathrm{m}} 39.32^{\mathrm{s}}$ & $0.3''$  & 15 & 7.5   \\
             & $-20^\circ 24' 39.5''$ & \cite{kaplan02a, corbel04, fuchs99}  &  &  \\
             \hline
SGR 1900+14  & $19^{\mathrm{h}} 07^{\mathrm{m}} 14.33^{\mathrm{s}}$
& $0.15''$ \footnote{Localization uncertainty for transient radio source associated with the 1998 giant flare\,\cite{frail99}.}  & 15 & 5.2   \\
             & $+09^\circ 19' 20.1''$ & \cite{kaplan02b, frail99, vrba00}  &  &  \\
             \hline

\end{tabular}
\label{table:sgrsummary}
\end{center}
\end{minipage}
\end{table}

\begin{figure}[!t]
\begin{center}
\includegraphics[angle=0,width=160mm, clip=false]{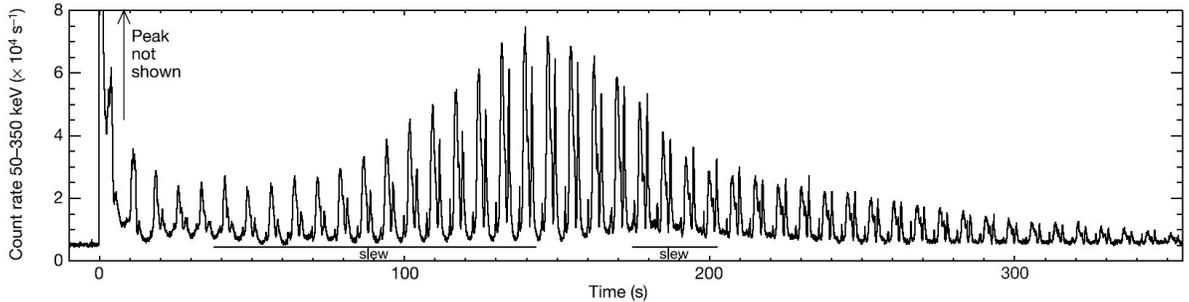}
\caption[SGR 1806--20 giant flare light curve from BAT] { SGR
1806--20 giant flare light curve from Swift/BAT, taken
from\,\cite{palmer05}.  Bin size is 64\,ms and photons with energy
greater than 50\,keV are recorded.  The peak of the prompt burst is
not shown as the detector was saturated. The apparent increase in
the light curve to a peak at 140\,s is due to a preprogrammed
slewing of the spacecraft to another source. The 7.56 modulation due
to the SGR source spin rate is clearly visible.}
\label{fig:giantFlareTail}
\end{center}
\end{figure}

\section{Burst emission} \label{section:sgrbursts}

The defining behavior of SGRs is sporadic burst
emission.  SGR bursts are typically classified as common
bursts, intermediate bursts, or giant flares.  Occasionally SGRs emit many bursts in an unusually short period of time; such events are referred to as ``multi-episodic events'' or ``storms.''  In this section we describe SGR burst emission.

\subsection{Common bursts}

Common bursts from SGRs have peak luminosities up to about $10^{42}$
erg/s and typically last about 100\,ms.  They exhibit nonthermal
spectra peaking in the soft gamma rays.   SGRs have been known to
have active bursting periods lasting for weeks, months, or years and
then to fall into quiet periods which can last for years.  For
example, SGR 0526--66 has been inactive since 1983\,\cite{woods04a}.
Active periods vary widely in terms of total energy released.
Activity levels of different SGRs vary as well
(Section\,\ref{section:burstHistories}).

Light curves of common bursts are similar from one to the next, even
from different SGR sources~\cite{aptekar01, gogus01, woods04a,
kouveliotou87}. Histograms of the durations of common bursts are
sharply peaked near 100\,ms, a characteristic SGR timescale,  and
show mild positive correlation to fluence\,\cite{woods04a, gogus01}.
This correlation means the typical duration may depend on the
sensitivity of the detector used to construct the SGR flare sample.
In the sample presented in\,\cite{gogus01}, the mean durations for
SGR~1806--20 and SGR~1900+14 were 162 and 94\,ms respectively. Rise
times are of the order of a few milliseconds, and decay times are
somewhat longer.

Energies in common bursts follow a power law distribution $dN/dE
\propto E^{-5/3}$ with $10^{35} < E < 10^{42}$\,erg\,\cite{cheng96,
gogus99}.  A similar distribution is observed in earthquake
energies\,\cite{woods04b} and other self-organized critical systems
in which the energy reservoir is much greater than the energy
emitted in individual bursts.  There is no observed correlation
between the energy of a burst and the waiting time to the next
burst.

Some flares apparently have multiple peaks, that is peaks that are
so close to each other in time that the flux does not return to
quiescent levels.  It is likely that these complex bursts are
superpositions of two or more single-peaked flares~\cite{gogus01}.
In the individual burst search described in
Chapter\,\ref{chapter:search}, a single on-source region is used for
such events.

The spectral properties of bursts are also similar from burst to
burst and SGR to SGR.  Above about 25\,keV the spectra are
well-modeled by optically thin thermal bremsstrahlung (OTTB) with
temperatures in the range 20--40\,keV\,\cite{gogus01, aptekar01}.
However this model overestimates the power at lower photon energies
significantly\,\cite{olive03}. An alternate model consisting of the
sum of two blackbodies can fit burst spectra over a wider range of
photon energies\,\cite{olive03, woods04a}.

\subsection{Giant flares} \label{section:giantFlares}

SGRs 0526$-66$, 1900+14 and 1806$-20$ have each produced a giant
flare since 1979. GRBs and blazars are more energetic  sources\,\cite{woods04a}; however, these events occur at cosmological distances, whereas the giant flares occurred in our
Galactic neighborhood. With five
known SGRs, a rough upper limit on the giant flare rate is about one
per 50 years per nearby SGR. The three giant flares were remarkably
similar. They each had similar prompt burst durations of
$\sim$100\,ms, though the flares from SGR 0526$-66$ and SGR 1900+14
were $\sim$100 times less energetic than the SGR 1806--20 giant
flare, which is the subject of our analysis in Chapter\,\ref{chapter:search}.  All three had tails radiating energies of $\sim\sci{5}{43}$\,erg lasting $\sim$6 minutes, modulated with periods of $\sim$5-8\,s.

The giant flare of 1979 March 5 from SGR~0526-66~\cite{mazets79} was
called ``the most singular high-energy astrophysical phenomenon of
the space age''\,\cite{cline82b}.  The energy emitted was
$\sci{5}{44}$\,erg, and the spectral peak temperature
($kT\sim$250--500\,keV) was about ten times higher than typical
common bursts\,\cite{woods04a}. The first evidence that SGR
progenitors might be neutron stars came from observations of the
1979 giant flare from SGR 0526--66\,\cite{mazets79}. Triangulation
of the event associated it with a SNR in the large Magellanic
cloud\,\cite{cline82, golenetskii84}.  An LMC distance put the peak
luminosity at about $\sci{4}{44}$\,erg/s\cite{fenimore81}, more than
a million times the Eddington limit, implying a compact object
source. However, the 8.1\,s modulation in the giant flare tail and
significant structure on a $\sim$2~ms timescale~\cite{woods04a}
ruled out the black hole possibility, and the associated SNR added
further evidence for a neutron star\,\cite{cline82b}.

On 1998 August 27, SGR 1900+14 gave a giant flare, similar to the
1979 giant flare, lasting for about 400\,s with a peak luminosity of
$\sim\sci{4}{44}$\,erg/s and an energy of at least
$10^{44}$\,erg\,\cite{hurley99}. Ionization in the Earth's nighttime
atmosphere was enhanced at altitudes 30--90\,km to daytime levels,
causing disturbance of propagating low frequency signals with a
period of 5.16\,s, equal to the spin period of SGR
1900+14\,\cite{inan99}.  The spin down rate of the SGR more than
doubled over the 80 day period containing the event\,\cite{woods99}.

The 2004 December 27 giant flare from SGR 1806--20 was about 100
times as energetic as the previous two.  The LIGO H1 detector was
operating at the time, and this event is included in the S5y1
individual burst search sample described in
Chapter\,\ref{chapter:search}. It consisted of a short hard burst
lasting $\sim$200 ms with and a peak luminosity of $\sim\sci{2}{46}$\,erg/s\,\cite{rea06} and isotropic energy of $\sim\sci{3}{46}$ erg
($\sim\sci{1}{-8}\solarmass$)  assuming a distance of 15\,kpc to the
source (see Section\,\ref{section:distanceTo1806}), as much energy
as the sun radiates in a quarter million years.  The short burst was
followed by the $\sim$6 minute tail which radiated a total energy of
$\sci{5}{43}$ erg. The tail showed a complicated pulse profile which
evolved in time (Figure\,\ref{fig:gfProfile}.  Quasiperiodic
oscillations were observed at times in the tail emission, the
strongest with frequencies $\sim$92.5 Hz, $\sim$18 Hz and $\sim$30
Hz~\cite{israel05}. In addition there was a $\sim$1 s precursor 142
s before the main flare (Figure\,\ref{fig:precursorLightcurve},
evidence of a $\sim$1 hour X-ray afterglow~\cite{mereghetti05}, and
a radio nebula expanding with a velocity of 0.3 c~\cite{gaensler05}.
Following~\cite{boggs06}, we identify separate stages of the SGR
1806--20 event:
 \bi
 \i the $\sim$1 s duration precursor flare, 142 s prior to the main burst
 (Figure~\ref{fig:precursorLightcurve});
 \i quiet period between the precursor and main burst;
 \i a brief 2.5 ms duration ``fast peak'' immediately preceding the
 main burst~\cite{boggs06};
 \i the main burst (Figure~\ref{fig:gftimescales});
 \i a $\sim$60 s decay period, characterized by nonthermal
 emission~\cite{boggs06};
 \i $\sim$6 minute pulsed tail;
 \i $\sim$1 hour of X-ray afterglow~\cite{mereghetti05};
 \i expanding radio nebula.
 \ei

\begin{figure}[!t]
\begin{center}
\includegraphics[angle=0,width=100mm, clip=false]{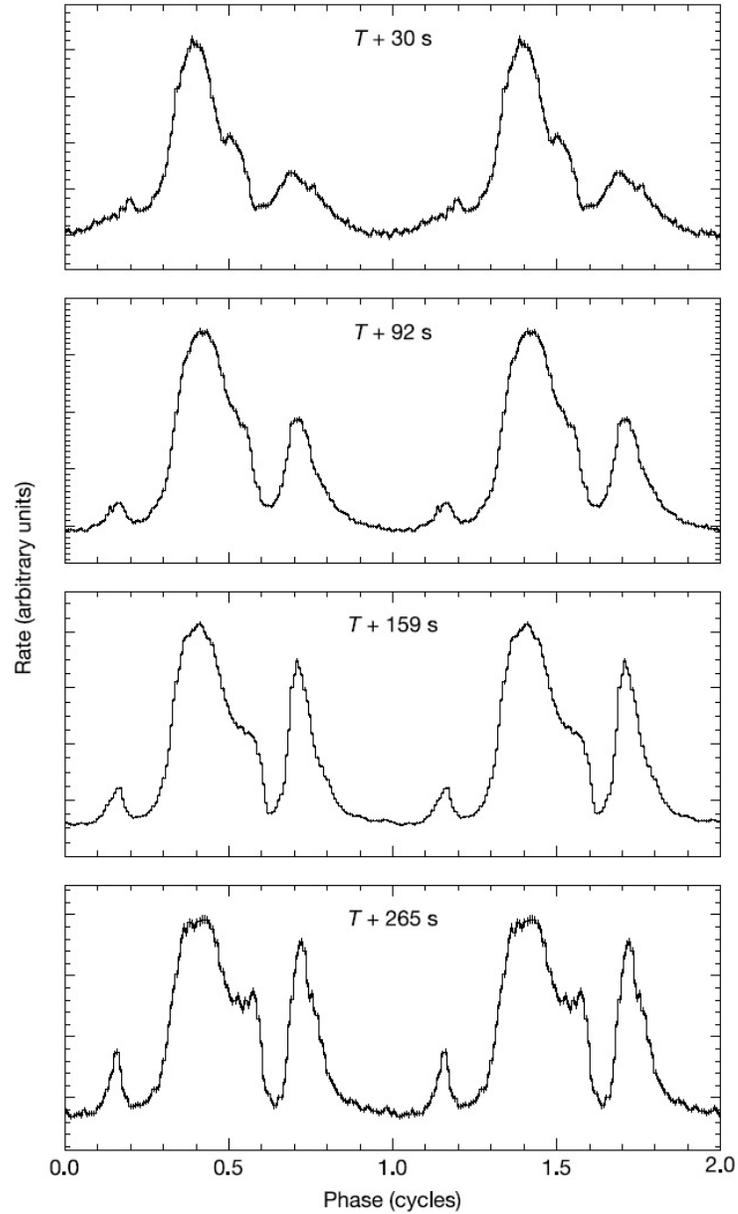}
\caption[Pulse profile evolution of SGR 1806--20 giant flare tail]
{Pulse profile evolution of SGR 1806--20 giant flare tail. Each
panel displays the pulse profile over two pulse cycles at the given
time intervals during the flare. The times are for the midpoint of
each interval relative to the start of the main spike.  The pulse
profile becomes less sinusoidal during the course of the flare. The
phases of the peaks remain fixed, which suggests a finalized
magnetic geometry and emission from a trapped
fireball\,\cite{palmer05}. Figure taken from\,\cite{palmer05}.  }
\label{fig:gfProfile}
\end{center}
\end{figure}

\clearpage

Quasiperiodic oscillations (QPOs) seem to be a regular feature of
giant flare tails. There had been tentative evidence for
quasiperiodic oscillations in the 1979 giant flare tail from
instruments available at the time~\cite{watts06website}; and
oscillations were found in a recent re-examination of data from the
tail of the 1998 giant flare~\cite{strohmayer05}.

\begin{figure}[!t]
\begin{center}
\includegraphics[angle=0,width=110mm, clip=false]{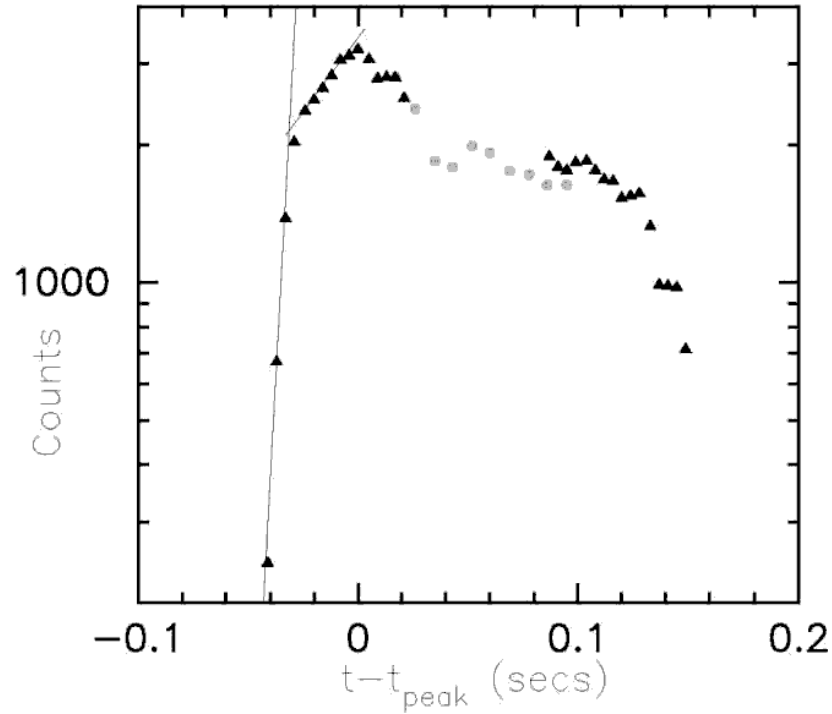}
\caption[SGR 1806--20 giant flare main burst light curve]{ Light
curve of the main burst of the SGR 1806--20 giant flare, taken
from~\cite{schwartz05}. This data was taken by the Cluster C4
(triangles) and Double Star TC-2 (circles) satellites, which were
designed for the study of Earth's magnetosphere. Solid lines in the
figure express timescales in the light curve.  The steep initial
rise is fit by an e-folding time of 4.9\,ms and the second rise is
fit by an e-folding time of 67\,ms\,\cite{schwartz05}.  }
\label{fig:gftimescales}
\end{center}
\end{figure}

\begin{figure}[!t]
\begin{center}
\includegraphics[angle=0,width=120mm, clip=false]{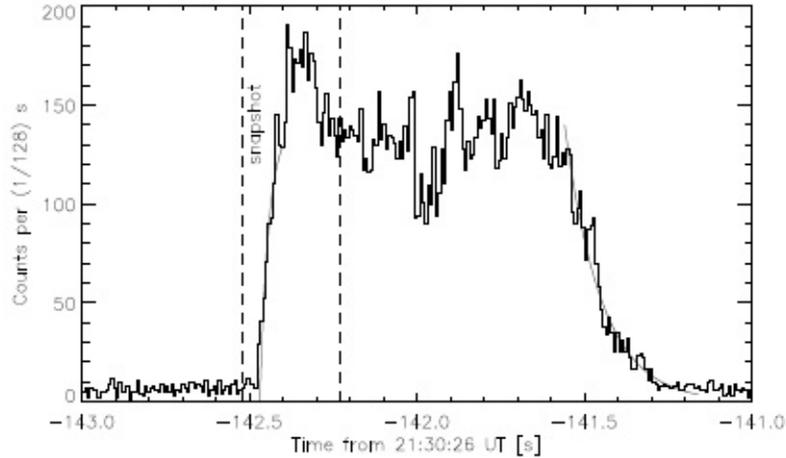}
\caption[SGR 1806--20 giant flare precursor light curve]{ Light
curve of the SGR 1806--20 giant flare precursor, taken from
~\cite{boggs06}. } \label{fig:precursorLightcurve}
\end{center}
\end{figure}

\subsection{Intermediate bursts}

So-called intermediate bursts are characterized by longer durations,
larger peak luminosities, and larger energies than common bursts.
Bursts lasting more than 500\,ms are generally considered
intermediate bursts.  The classification was created after the
observation of an uncommmon 2001 April 18 SGR 1900+14 burst lasting
$\sim$40\,s and with energy greater than $10^{42}$\,erg, which
occurred after a quiet period of almost two years\,\cite{guidorzi04,
mereghetti08}. The rising edges of these bursts look like the abrupt
rising edges of common bursts. They can last longer than the
rotation period of the progenitor, but if they don't they have
abrupt endings as well\,\cite{woods04a}. They often, but not always,
occur in the months following giant flares. \cite{woods04a} suggests
that at least some intermediate flares could be ``aftershocks'' of
giant flares.

The addition of intermediate bursts to the SGR burst menagerie
suggests that there may be a continuum of burst energies from these
objects, from the smallest common bursts to the largest giant
flares\,\cite{olive04}.  This supports using the same gravitational
wave search methodology for all SGR bursts, from the common to the
giant.

\subsection{Burst storms}

The most dramatic SGR activity besides giant flares may be emission
of series of many common and intermediate bursts in short periods of
time lasting on the order of a minute.  These are referred to as
multi-episodic events, storms, or forests.  This section describes a
few of the most spectacular storms (most of which have been given by
SGR 1900+14) but does not attempt to give an exhaustive list.
Reference\,\cite{woods04a} states that most burst storms are seen at
lower peak flux than these.

On 1998 May 30, about three months before giving a giant flare, SGR
1900+14 gave the first observed intense storm
(Figure\,\ref{fig:first1900Storm})\,\cite{hurley99c}. On 1998
September 1, a few days after the giant flare, it gave another
storm (Figure\,\ref{fig:sept98Storm})\,\cite{woods04a}.

\begin{figure}[!t]
\begin{center}
\includegraphics[angle=0,width=120mm,clip=false]{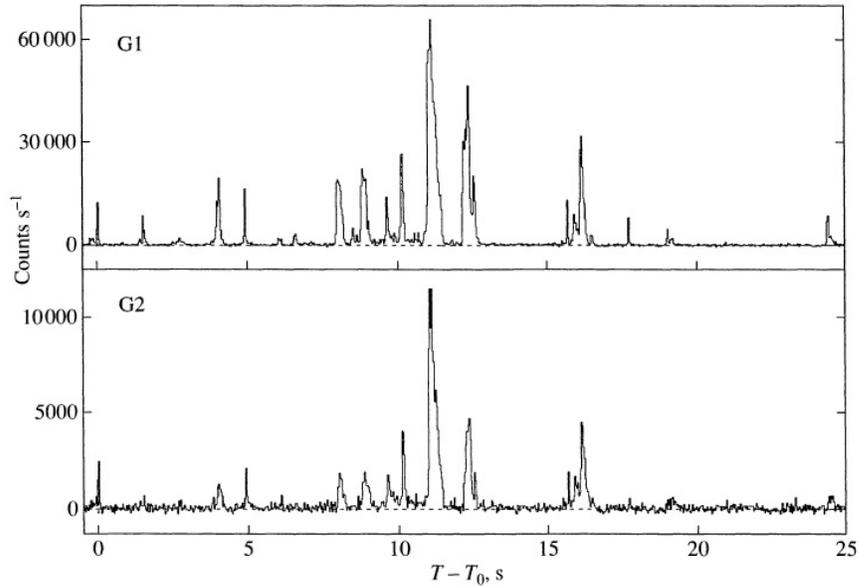}
\caption[Light curve from the 1998 May 30 SGR 1900+14 storm]{ Light
curve from the 1998 May 30 SGR 1900+14 storm, from Konus-Wind, showing the
region of greatest activity in the storm.  G1 and G2 show the 15-50\,keV
and the 50-250\,keV bands, respectively.  Figure
from\,\cite{mazets99}. } \label{fig:first1900Storm}
\end{center}
\end{figure}

\begin{figure}[!t]
\begin{center}
\includegraphics[angle=0,width=120mm,clip=false]{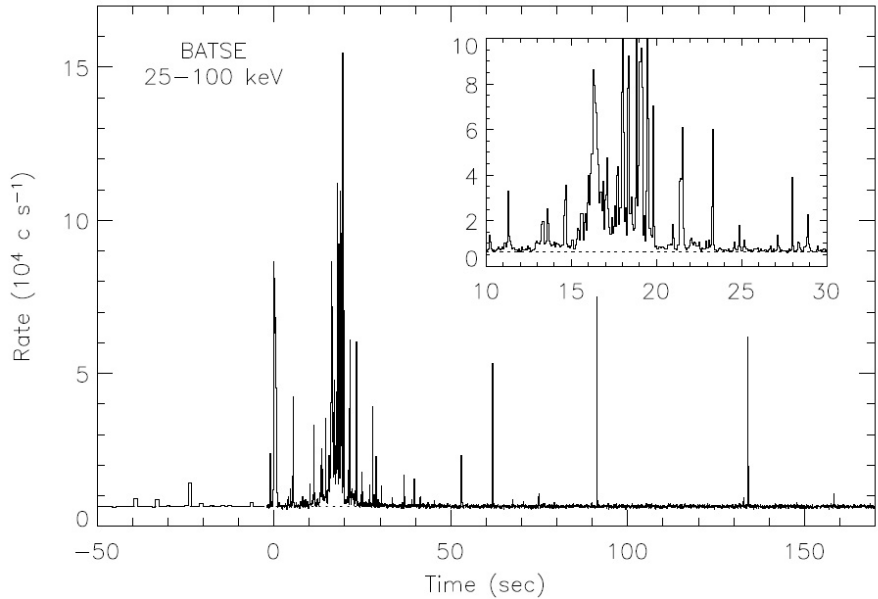}
\caption[Light curve from the 1998 September 1 SGR 1900+14 storm]{
Light curve from the 1998 September 1 SGR 1900+14 storm (BATSE
25--100 keV). The inset shows the most intense part of the storm.
Figure from\,\cite{woods04a}.
 } \label{fig:sept98Storm}
\end{center}
\end{figure}

On 2006 March 29 SGR 1900+14 gave a storm\,\cite{israel08}.  The
lightcurve for the event is shown in
Figure\,\ref{fig:stormLightcurve}.  More than 40 bursts were
detected over the course of $\sim$30\,s, including seven
intermediate flares. The total event released
$\sci{2-3}{42}$\,erg\,\cite{israel08}. Since intermediate flares
are rare, this event was an excellent opportunity to probe model
predictions such as the trapped fireball and twisted magnetosphere,
and to refine spectroscopic measurements, including time-resolved
spectroscopy. These observations provide additional support for the
magnetar model\,\cite{israel08}. Furthermore, they support a
continuum between common bursts and intermediate flares.  The SGR
1900+14 storm light curve is discussed further in
Chapter\,\ref{chapter:stack}.

The renewed activity of SGR 1627--44
(Section\,\ref{section:burstHistories}) also included a burst
``storm'' or ``forest'' of dozens of events, starting on 5 May 2008
10:25:54 UT\,\cite{gcn7777, atel1549}. The BAT light curve of this
storm is shown in Figure\,\ref{fig:1627storm}.

\begin{sidewaysfigure}[!t]
\begin{center}
\includegraphics[angle=0,width=220mm, clip=false]{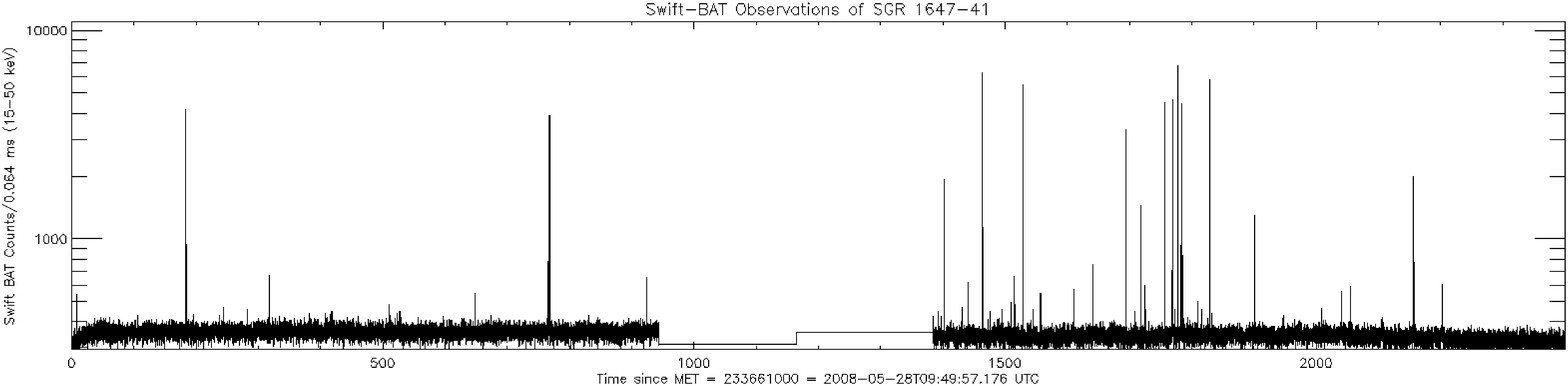}
\caption[Light curve of the May 2008 storm from SGR 1627--44] { BAT
light curve of the May 2008 storm from SGR 1627--44\,\cite{gcn7777}
Two preliminary bursts are visible in the section at the left, and
the storm itself is visible in the section at the right. }
\label{fig:1627storm}
\end{center}
\end{sidewaysfigure}

\subsection{SGR activity histories} \label{section:burstHistories}

While only SGR 1806--20 and SGR 1900+14 were active during S5y1
(Chapter\,\ref{chapter:search}), it is useful to summarize the burst
histories of all four of the SGRs. This gives a rough qualitative
sense of the likelihood of SGR burst activity during future LIGO
science runs. Figure\,\ref{fig:burstHistory} shows histograms of
bursts from four of the five SGRs, from the beginning of observation
to 2005.

SGR 1806--20 was the first SGR to be discovered.  It emitted a burst
of soft gamma rays on 1979 January 7, remaining sporadically active
until a period of intense activity consisting of over 100 detected
bursts in the mid 1980s.  As of November 2006 more than 450 soft
gamma ray bursts had been detected from SGR 1806--20 since its
discovery in 1979~\cite{boggs06}.  SGR 1806--20 gave a giant flare
on 2004 December 27, and was active throughout the LIGO S5
science run from November 2005 to November 2007, during which
$\sim$300 bursts from SGR 1806--20 were detected by the IPN. At the
time of this writing, the last GCN report of burst activity from SGR
1806--20 was published on 2007 September 24, though this does not
necessarily mean activity has ceased.

SGR 0526--66, in dramatic fashion, emitted a giant flare a few
months after the discovery of SGR 1806--20.  It continued to emit
bursts until 1983, and has since been silent\,\cite{golenetskii87,
hurley00}.

SGR 1900+14 gave 3 detected bursts in two days in 1979, a mere nine
days after the SGR 0526--66 giant flare.  In the first three months
of 1979, three of the five confirmed SGRs were discovered.  After
these three bursts SGR 1900+14 was quiet for nearly two decades,
with the exception of 3 bursts detected in 1992. Then in the summer
of 1998 it became very active, giving over 1000 bursts over nine
months, including burst storms and the 1998 August 27 giant
flare\,\cite{israel08}. After a period of inactivity, it gave the
bright intermediate flare on 2001 April 18. SGR 1900+14 was active
during the LIGO S5 science run, but its activity was concentrated to
a short period in 2006 March encompassing the storm event (see
Figure\,\ref{fig:burstsamplehist}). At the time of this writing, the
last GCN report of burst activity from SGR 1900+14 was published on
2006 June 10, though this does not necessarily mean activity has
cased.

SGR 1627--41 was discovered in 1998, emitting about 100 bursts in
six weeks\,\cite{woods99b} and then fell silent.  It has recently
shown a renewal of its activity after 9.8 years of
quiescence\,\cite{gcn7777, atel1549}.  The new activity includes a
burst storm of dozens of events.  The BAT light curve of this storm
is shown in Figure\,\ref{fig:1627storm}.  No giant flare has been
observed from SGR 1627--41.

A few days before this writing, on 2008 August 22, SGR 0501+4516
was discovered by the BAT detector aboard the swift satellite
through three ``discovery bursts''\,\cite{gcn8112, gcn8113,
gcn8115}. The spin period has already been measured to be $5.769 \pm
0.004$\,s based on a 600\,s observation with the RXTE
satellite\,\cite{gcn8118} and $5.7620697 \pm 0.0000015$\,s with the
Swift XRT detector\,\cite{gcn8146}.  An association with SNR
G160.9+2.6, located 1.5\,kpc distant, is plausible\,\cite{gcn8149,
leahy95}. A 1.4 GHz radio image of the SNR taken from the Canadian
Galactic Plane Survey is shown in
Figure\,\ref{fig:0501snr}\,\cite{gcn8149}.  This SGR gave 56 bursts
in the first 87 hours since discovery, though activity declined in
the subsequent 43 hours\,\cite{atel1683}.  A few of these bursts
have been remarkably bright\,\cite{gcn8139}.  It will be interesting
to see how this new SGR behaves in the future.

Finally, we mention a candidate SGR, 1801-23\,\cite{cline00}.  This
candidate has only given two bursts (on one day in 1997) and so the
IPN localization is poor and not much else can be said.  The
characteristics of the two bursts were otherwise consistent with SGR
bursts, and this is probably an SGR which has entered an inactive
state\,\cite{hurley00}.

\begin{figure}[!t]
\begin{center}
\includegraphics[angle=0,width=140mm, clip=false]{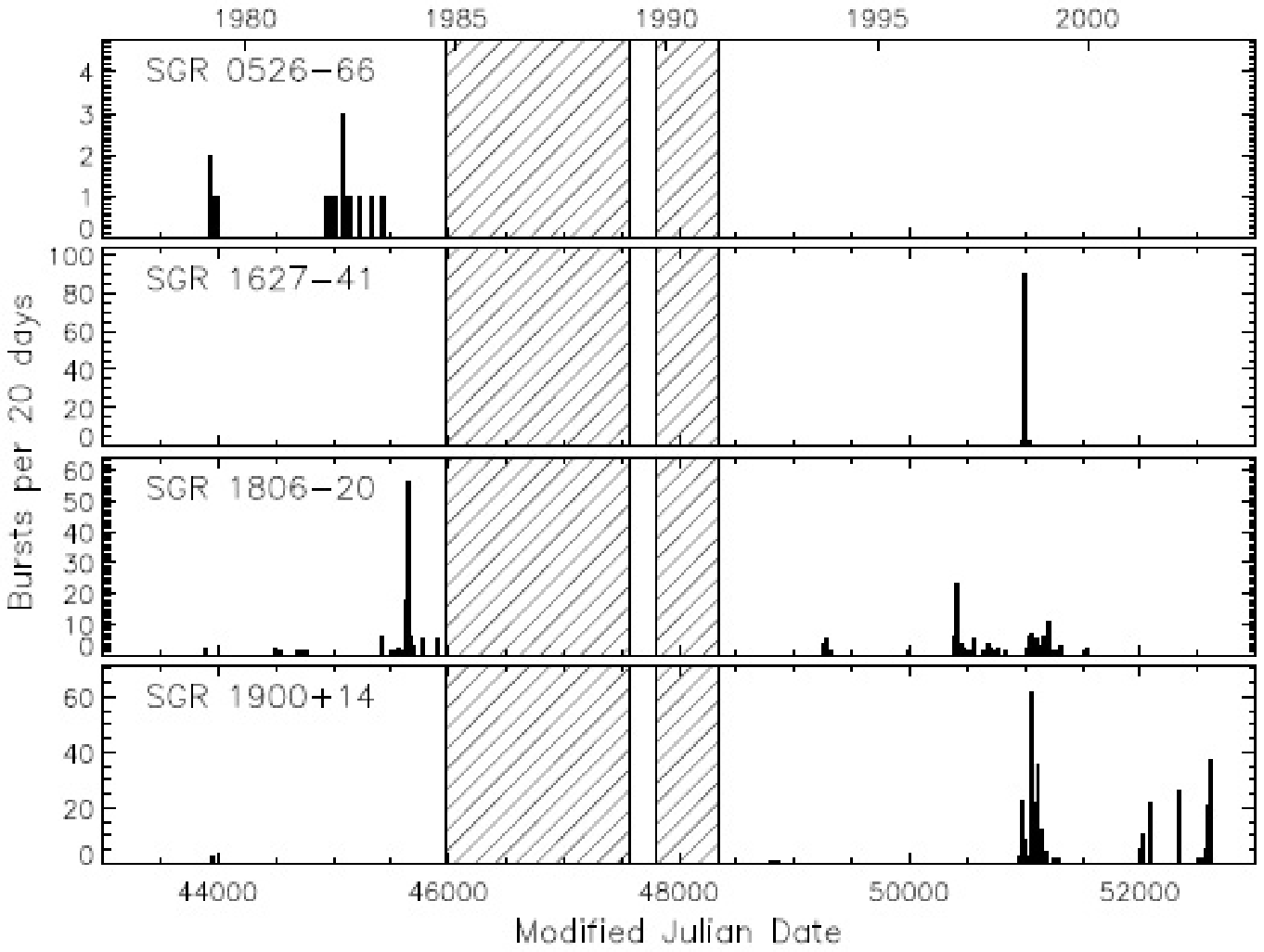}
\caption[Burst histories of the four SGRs to 2005] { Burst histories
of the four confirmed SGRs to 2005, with Julian date on the lower
axis and calendar years on the upper axis. Bursts were identified
with various detectors in the IPN network, which have different
sensitivities. No detectors were observing during the shaded epochs.
Figure taken from\,\cite{woods04b}. } \label{fig:burstHistory}
\end{center}
\end{figure}

\begin{figure}[!t]
\begin{center}
\includegraphics[angle=0,width=140mm, clip=false]{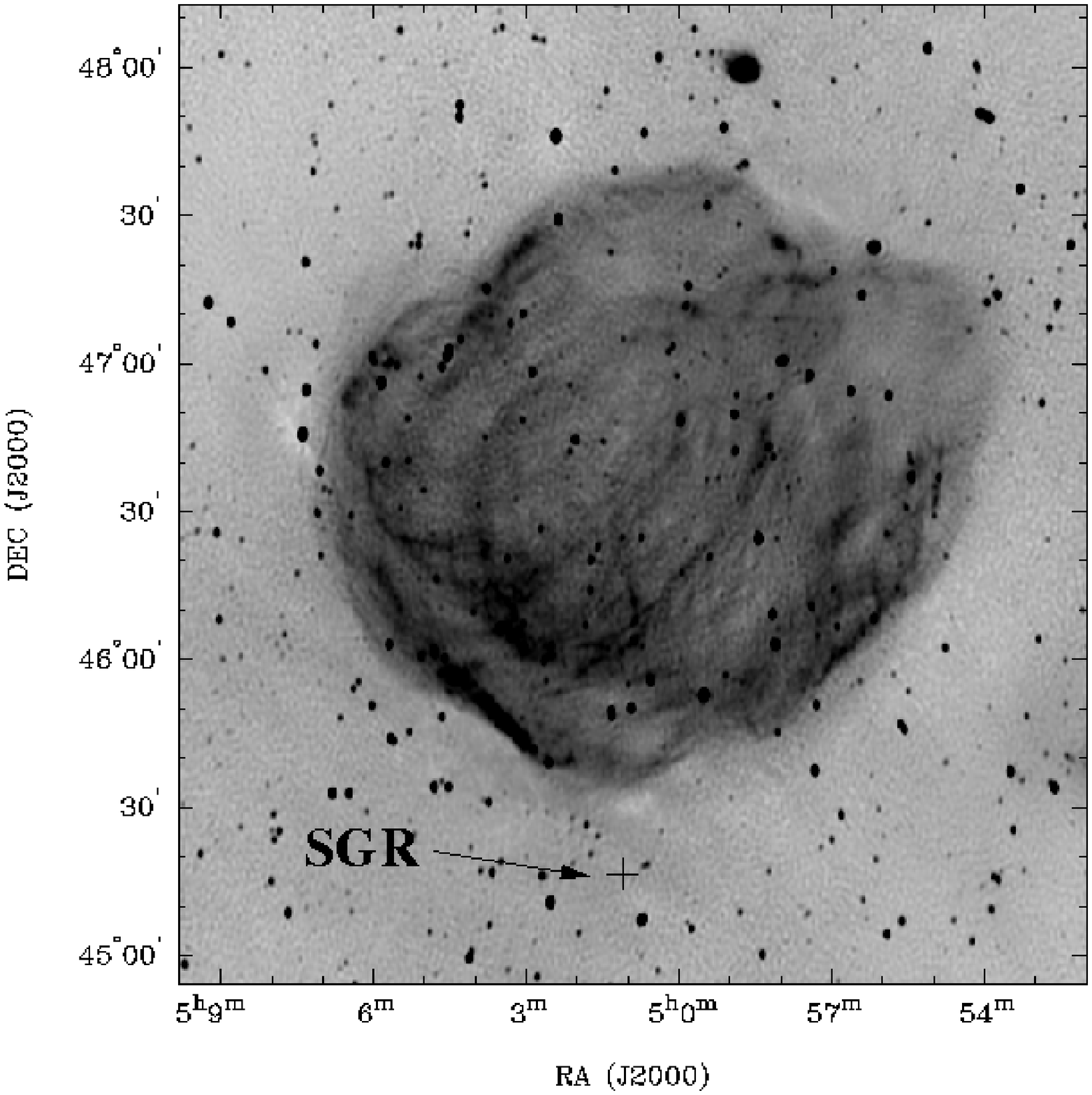}
\caption[SNR near SGR 0501+4516] { 1.4 GHz radio image of the SNR
G160.9+2.6 taken from the Canadian Galactic Plane Survey, showing
the location of SGR 0501+4516\,\cite{gcn8149}.  The SNR is
$\sim$1.5\,kpc distant\,\cite{leahy95}. } \label{fig:0501snr}
\end{center}
\end{figure}

\section{Other observed properties of SGRs}

We have so far focused on properties of SGRs which are directly
relevant to our gravitational wave search.  Here we briefly present
some of their other properties.

\subsection{Quiescent X-ray emission}

The persistent X-ray emission mentioned above, $10^{34}$ to
$10^{36}$\,erg/s in the 1--10\,keV band, exhibits a pulse shape
recurring at the SGR's rotation frequency and the spectra can
generally be fit by a blackbody plus power law
model\,\cite{woods04a}. The two components in the model, blackbody
and power law, can vary independently.  When an SGR is not bursting,
the blackbody temperature is relatively constant in time and between
sources\,\cite{marsden01, oosterbroek98} but the power law component
shows relatively large variations\,\cite{marsden01, woods04b}. Pulse
profiles tend to be roughly sinusoidal\,\cite{woods04a}.

An interesting property of the emission is that it can apparently be
affected by the transient bursts\,\cite{woods01, woods04b}.  After
its period of activity in 1998, SGR 1900+14 displayed an increase in
persistent X-ray emission, accompanied by changes in the spectrum.
The SGR 1900+14 persistent emission pulse profile in particular
became significantly more sinusoidal after the giant flare, and this
change appears to be permanent\,\cite{woods04a}.  This change is
evidence for a reconfiguration in the magnetic field at the time of
the giant flare\,\cite{woods01}.  Subtle changes have also been
observed in SGR pulse profiles during periods of common burst
activity. Changes in behavior observed between epochs before and
after burst active periods could potentially shed light on the
nature of the burst mechanism, but greater continuity of observation
would be necessary to understand the relationship between bursting
and X-ray variability.

\subsection{Timing} \label{section:sgrTiming}

SGR spin rates can be measured via pulsations in the X-Ray emission
or in modulations of tails in intermediate bursts or giant flares.
SGRs tend to spin significantly more slowly than radio pulsars, with
periods in the range 5--10\,s (Table\,\ref{table:sgrsummary}).

SGR timing noise is relatively large\,\cite{woods04b}.  The
spin-down rates of SGRs have shown substantial variability over
short timescales. For example, SGR 1806--20 shows at least a factor
of $\sim$6 in spin-down torque\,\cite{woods02, woods04b}.  In
general, changes in $\dot{P}$ do not correlate to periods of burst
activity, with the exception of the 1998 August 27 giant flare from
SGR 1900+14, in which a significant ``anti-glitch'' (i.e. a decrease
in spin frequency) was observed with $\Delta P/P =
10^{-4}$\,\cite{woods99}.  During an $\sim$3 month period containing
the giant flare, $\dot{P}$ increased by a factor of more than
2\,\cite{woods99}.  Due to observational sampling limitations,
however, it cannot be ruled out that this change in rotation of SGR
1900+14 occurred in the months leading up to the giant flare.   No
glitch or anti-glitch was observed in the 2004 December 27 giant
flare from SGR 1806--20, which was more than two orders of magnitude
more energetic, although significant changes were observed in the
months leading up to this event\,\cite{woods07}.

Lack of correlation between changes in $\dot{P}$ and burst active
periods\,\cite{woods02} has implications for models of SGR
activity\,\cite{woods04b}. Models for both the SGR burst mechanism
and torque variability invoke seismic activity or motion in the
crust.  In the context of such models, lack of correlation between
these observables implies either that the underlying seismic or
crustal activities are uncorrelated, or that one or both of the
observables are not caused by underlying seismic or crustal
activity.

Finally, glitches are observed in SGRs which are typically larger
and occur more frequently than in radio pulsars of comparable spin
periods\,\cite{mereghetti08}.

\subsection{Quasiperiodic oscillations}

QPOs were observed in the tail of the SGR 1806--20 giant flare by
detectors on two satellites, RXTE and RHESSI\,\cite{israel05,
watts06}. After this discovery the tail of the SGR 1900+14 giant
flare was re-analyzed, and QPOs were discovered there as well, at
28, 54, 84 and 155\,Hz\,\cite{strohmayer05}. Frequencies and
durations of QPOs from the SGR 1806--20 giant flare are given in
Table\,\ref{table:qpos}. It is possible that a feature at 43\,Hz in
the tail of the SGR 0526--66 was also a QPO\,\cite{barat83}.

QPOs are also observed in the Earth after earthquakes.  QPOs in SGR
giant flare tails are probably due to seismic oscillations in the
star, either in the crust or involving the entire star, probably in
the toroidal modes. These motions could couple to the magnetic field
thereby affecting the X-Ray emission.  It may be possible to extract
information about star parameters and possibly the star's EOS from
these QPOs.  A LIGO search for gravitational waves associated with
the SGR 1806--20 QPOs has been performed\,\cite{matone07}; no
gravitational wave detection was claimed.

\begin{table*}
\begin{center}
\caption[QPOs observed in giant flares from SGR 1806--20]{Summary of
the most significant QPOs observed in the pulsating tail of SGR
$1806-20$ during the 27 December 2004 hyperflare. The period of
observation for the QPO transient is measured with respect to the
flare peak, the frequencies are given from the Lorenzian fits of the
data and the width corresponds to the Full-Width-at-Half-Maximum
(FWHM) of the given QPO band.  Table from\,\cite{matone07}. }
\begin{tabular}{ccccc}

  {\bf Frequency} & {\bf FWHM} & {\bf Period} & {\bf Satellite} & {\bf References}\\

 {\bf [Hz]}  & {\bf [Hz]} & {\bf [s]} &  & \\

\hline \hline \\

 17.9 $\pm$ 0.1 & 1.9 $\pm$ 0.2 & 60-230 & RHESSI & \cite{watts06}\\ \\

  25.7 $\pm$ 0.1 & 3.0 $\pm$ 0.2 & 60-230 & RHESSI & \cite{watts06}\\ \\

  29.0 $\pm$ 0.4 & 4.1 $\pm$ 0.5 & 190-260 & RXTE  & \cite{strohmayer06}\\ \\

  92.5 $\pm$ 0.2 & $1.7^{+ 0.7}_{-0.4}$ & 170-220 & RXTE & \cite{israel05}\\

 "               & "                    & 150-260 & "    & \cite{strohmayer06}\footnote{Ref. \cite{strohmayer06} makes an adjustment to the observation period of Ref. \cite{israel05}} \\

 92.7 $\pm$ 0.1 & 2.3 $\pm$ 0.2 & 150-260 & RHESSI & \cite{watts06}\\

  92.9 $\pm$ 0.2 & 2.4 $\pm$ 0.3 & 190-260 & RXTE & \cite{strohmayer06}\\ \\

  150.3 $\pm$ 1.6 & 17 $\pm$ 5 & 10-350 & RXTE & \cite{strohmayer06}\\ \\

  626.46 $\pm$ 0.02 & 0.8 $\pm$ 0.1 & 50-200 & RHESSI & \cite{watts06} \\

  625.5 $\pm$ 0.2 & 1.8 $\pm$ 0.4 & 190-260 & RXTE & \cite{strohmayer06}\\ \\

 1837 $\pm$ 0.8 & 4.7 $\pm$ 1.2 & 230-245 & RXTE & \cite{strohmayer06}\\ \\

\hline \hline
\end{tabular}
\end{center} \label{table:qpos}
\end{table*}

\subsection{Association with supernova remnants}

All five of the confirmed SGRs might be associated with supernova
remnants, evidence for a neutron star progenitor for SGRs.  SGR
0501+4516 is thought to be associated with SNR HB9\,\cite{gcn8149}.
SGR 0526-66 is thought to be associated within the N49
SNR\,\cite{evans80}. SGR 1627-40 lies near to SNR G337.0-0.1.  SGR
1806--20 is thought to be associated with SNR
G10.0-0.3\,\cite{hurley00}. SGR 1900+14 is thought to be associated
with SNR G42.8+0.6\,\cite{hurley99b}.

\section{SGR distances and locations} \label{section:sgrDistances}

In the search for gravitational waves described in
Chapter\,\ref{chapter:search} we are interested in SGR 1806--20 and
SGR 1900+14.  Knowing distances and sky locations for these sources
is important for setting upper limits on isotropic gravitational
wave burst emission energy.

Understanding SGRs through their electromagnetic radiation has been
an endeavor involving multiple wavelengths.  Quiescent X-ray
counterparts can be localized on the sky to sub-arcsecond precision,
allowing for deep follow-up observations at other wavelengths, which
have borne fruit in the form of discoveries of low-probability
associations to other astrophysical objects such as supernova
remnants and clusters of massive stars.  Follow-up observations also
allow distances to be estimated using a variety of methods.

\subsection{SGR 1806--20} \label{section:distanceTo1806}

The X-Ray source associated with SGR 1806--20 coincides with a
supernova remnant (SNR)\,\cite{kulkarni94}, a luminous blue
variable\,\cite{kulkarni95, vankerkwijk95} (LBV 1806--20, which is
one of the most luminous stars in the local
group\,\cite{eikenberry04} and possibly the most luminous star in
our Galaxy\,\cite{eikenberry01}), and a massive star
cluster\,\cite{fuchs99} within a 10'' circle\,\cite{kaplan02a}.
These items are all rare. For example, only about a dozen luminous
blue variables are known in the local group of
galaxies\,\cite{fuchs99}.  Based on the association between SGR
1806--20 and a massive star cluster, and similar associations
observed for SGR 1900+14 and SGR 0526--66, it is plausible that
these SGR progenitors are massive stars belonging to the
clusters\,\cite{mereghetti08}.  Similarly, associations with an SNR
(also observed for  It is still unclear what connection, if any,
exists between the SGR and the LBV, but the chance line-of-sight
coincidence of these two objects is exceedingly
small\,\cite{fuchs99}.

Triangulation of eight bursts from this source occurring between
1996 and 1999 using Ulysses, BATSE and Konus-Wind led to a position
of right ascension $18^{\mathrm{h}} 08^{\mathrm{m}}
39.4^{\mathrm{s}}$ and declination $-20^\circ 24' 38.6''$ with a
3$\sigma$ ellipse of 230 arcsec$^2$\,\cite{hurley99}.

\cite{kaplan02a} gives an improved position estimate for SGR
1806--20 based on the X-ray counterpart of right ascension
$18^{\mathrm{h}} 08^{\mathrm{m}} 39.32^{\mathrm{s}}$ and declination
$-20^\circ 24' 39.5''$ with rms uncertainties of 0.3 arc second in
each coordinate, based on Chandra observations.

Several distance estimates for SGR 1806--20 have been made by
different methods, ranging from 6.4--9.8\,kpc\,\cite{cameron05} to
15.1\,kpc\,\cite{eikenberry04, corbel04, corbel97}.

\cite{cameron05} gives a distance range from 6.4--9.8\,kpc obtained
using the fading radio counterpart from the SGR 1806--20 2004 giant
flare.  Their estimate used a high resolution 21\,cm radio spectrum
tracing intervening interstellar neutral hydrogen clouds.

\cite{fuchs99} gives a distance of 14.5\,kpc by assuming association
with a cluster of giant massive stars enshrouded in a dense cloud of
dust. The distance to the dust cloud was estimated from mid-infrared
observations made with the Infrared Space Observatory.

\cite{corbel04} gives a distance of 15.1$^{+1.8-1.3}$\,kpc to the
radio nebula G10.0-0.3 which is powered by the wind of LBV 1806--20
and associated with the SGR.  This distance was estimated using
millimeter and infrared spectroscopic observations of CO emission
lines and NH$_3$ absorption features from molecular clouds along the
line of sight, as well as optical extinction of LBV 1806--20.
\cite{corbel97} had previously given a distance of $14.5 \pm
1.4$\,kpc based on a distance to the SNR G10.0-0.3 and other
molecular clouds including one of the brightest H II regions in the
Galaxy, W31 using CO observations.

For the remainder of this work, we use the position
from\,\cite{kaplan02a}, right ascension $18^{\mathrm{h}}
08^{\mathrm{m}} 39.32^{\mathrm{s}}$ and declination $-20^\circ 24'
39.5''$, and a nominal distance of 10\,kpc.  The sky position is
used to calculate LIGO interferometric detector antenna factors used
in gravitational wave searches, and the uncertainties in the sky
position lead to uncertainties in antenna factors which are
insignificant to the search results.  Energy upper limit estimates
produced by the search can easily be rescaled to other distances
$d_{\mathrm{alt}}$ by using the relation
 \be
 E_{\mathrm{alt}} = E_{\mathrm{nom}} \left(\frac{d_{\mathrm{alt}}}{10 \: \mathrm{ kpc}}\right)^2.
 \ee

\subsection{SGR 1900+14} \label{section:distanceTo1900}

As with SGR 1806--20, the X-Ray source associated with SGR 1900+14
also coincides with a high-mass star cluster\,\cite{vrba00}.

VLA observations of an associated fading radio source performed
after the SGR 1900+14 1998 giant flare led to the first
sub-arcsecond precision localization of the SGR, of right ascension
$19^{\mathrm{h}} 07^{\mathrm{m}} 14.33^{\mathrm{s}}$ and declination
$9^\circ 19' 20.1''$ with an uncertainty of 0.15''\,\cite{frail99}.

Triangulation of six bursts from this source led to a position of
right ascension $19^{\mathrm{h}} 07^{\mathrm{m}} 14.3^{\mathrm{s}}$
and declination $9^\circ 19' 19''$ with a 3$\sigma$ ellipse of 600
arcsec$^2$\,\cite{hurley99}.  This position is less precise than the
VLA position given in\,\cite{frail99}, and was used primarily as a
check on the statistical method for triangulating the position of
SGR 1806--20 using IPN satellite observations of bursts.

\cite{vasisht94} gives a distance of $\sim$5\,kpc to SGR 1900+14
assuming association with SNR G42.8+0.6.  The distance SNR G42.8+0.6
is estimated using the $\Sigma$--D relationship relating radio
surface brightness to diameter\,\cite{milne79}.

\cite{hurley99b} gives an independent distance estimate to SGR
1900+14 of 5.7\,kpc using measurement of 21\,cm neutral hydrogen
column density\,\cite{kalberla05}.  \cite{kaplan02b} supports this
distance estimate.

\cite{vrba00} and \cite{vrba96} give a distance estimate to SGR
1900+14 of 12--15\,kpc based on association with a cluster of
supergiant stars.  Distances to these stars were obtained using
spectral typing and optical extinction arguments, and astrometric
determination of proper motions over a 1.3\,year period to help
exclude the possibility that the cluster is composed of giant stars
and not supergiant stars.  The association is of the cluster with
SGR 1900+14 is bolstered by the similarity to the cluster observed
in coincidence with SGR 1806--20.

For the remainder of this work, we use the SGR 1900+14 position
from\,\cite{frail99}, right ascension $19^{\mathrm{h}}
07^{\mathrm{m}} 14.33^{\mathrm{s}}$ and declination $9^\circ 19'
20.1''$, and a nominal distance of 10\,kpc.  As with SGR 1806--20,
uncertainties in the sky position lead to uncertainties in antenna
factors which are insignificant to the search results. Energy upper
limit estimates produced by the search can be rescaled as discussed
at the end of Section\,\ref{section:distanceTo1806}.

\section{Magnetar model} \label{section:magnetarModel}

We now turn to a discussion of the most popular SGR model.  Under
the \emph{magnetar} model\,\cite{duncan92, thompson95} SGRs are
neutron stars with exceptionally strong internal toroidal magnetic
fields $\sim\nolinebreak10^{15}$\,G\,\cite{duncan92} or possibly
$\gtrsim 10^{16}$\,G\,\cite{ioka01}. The magnetar model attempts to
explain the observed properties of two classes of astrophysical
objects, the SGRs and the AXPs (Anomalous X-ray Pulsars).  The model
posits that the energy expenditure observed in burst and quiescent
emission is provided by the decay of the strong magnetic field.

Evidence that SGR progenitors were likely neutron stars came early.
As mentioned in Section\,\ref{section:giantFlares}, the 1979 SGR
0526-66 giant flare was too energetic to be anything but a compact
object; and evidence of 8.1\,s periodicity ruled out a black hole.
 Furthermore, two of the four known SGRs, SGR 1806--20 and SGR
0526-66, are associated with young supernova
remnants~\cite{kouveliotou98, kulkarni93, kulkarni94}, further
strengthening the case for neutron stars as SGR progenitors.

Magnetar fields are some $10^3$ times stronger than fields in
typical radio pulsars, and are much stronger than the quantum
critical value at which energy between Landau levels equals the
electron rest mass given by fundamental constants,
  \be
  B_{\mathrm{QED}}\equiv \frac{m^2 c^3}{\hbar e} = \sci{4.4}{13}
  \: \mathrm{Gauss}.
  \ee
In fields this strong, electrons are propelled at nearly the speed
of light around magnetic field lines and the vacuum itself becomes
birefringent, like a calcite crystal\,\cite{duncan00}. The strong
field is thought to originate in ``dynamo action'' which operates
until about 10\,s after the star's birth.  Dynamo action requires
convection within the star and a high initial spin rate greater than
about 200\,Hz; if the initial spin rate is too low, the dynamo will
fail. The magnetar model posits that garden variety radio pulsars,
with magnetic fields on the order of $10^{12}$\,G, were born with
spin rates too low for the dynamo to operate effectively. SGRs and
AXPs on the other hand were spinning sufficiently rapidly for
strong fields to be generated.

Evidence for the atypically strong fields comes from several
sources.  First, the SGRs and AXPs spin rate range of 2--12\,s is
slow compared to the spin rates of typical millisecond pulsars.
Though the magnetar model predicts rapid initial spins, the strong
field generated by the dynamo will spin the star down much more
rapidly than the ordinary neutron star magnetic fields.  The
magnetic dipole braking relationship, relating the star's external
magnetic field to period and spin down rate for isolated radio
pulsars is\,\cite{michel91}
 \be
 B \sim \sci{3}{19} \sqrt{P \dot{P}} \: \mathrm{Gauss}.
 \ee
The periods of SGR 1806--20 and SGR 1900+14 are 7.5\,s and 5.2\,s
respectively and their measured spin down rates are between
$10^{-11}$ and $10^{-10}$\,s\,s$^{-1}$, implying the strong magnetar
fields (e.g. $B\sim \sci{8}{14}$G for SGR 1900+14) if the above
relation holds for SGRs\,\cite{kouveliotou98, kouveliotou99}. This
implies an additional observational prediction, that spin down
powered beams are narrow or nonexistent in magnetars.

Second, the persistent X-ray emission from these objects is more
than can be supplied by a neutron star's rotational energy.   The
magnetic field can transfer energy to heating of the neutron star,
potentially through more than one mechanism.  First, ongoing seismic
activity caused by the strong field which can churn the star's
interior causes heating\,\cite{thompson96}. Second, changing fields
in the star's magnetosphere lead to currents of charged particles
which can transfer energy to X-rays.  Finally, these currents
terminate at the star's surface, resulting in heating.  A field in
excess of about $10^{15}$\,G is needed to power X-ray emission at
$10^{35}$\,erg/s for the typical magnetar age of $10^4$
years\,\cite{woods04a}.

Third, the extreme luminosities of SGR bursts can be explained by
magnetic field suppression of the electron scattering cross section
in the neutron star magnetosphere.

Fourth, the SGR bursts themselves, especially giant flares, require
an energy source greater than the rotational energy of the star.
Because magnetars spin slowly, rotational energy is not sufficient
to power the observed SGR activity.  Ordinary pulsars are powered by
rotational energy losses coupled through the $\sim10^{12}$\,G
magnetic field.  However, the much larger magnetar magnetic field
itself could act as the energy source. In addition it can supply a
mechanism for the sporadic bursting through its interaction with the
neutron star crust.    Burst emission occurs when the crust fails
suddenly due to increasing stresses from the internal toroidal
magnetic field (crustquake) as suggested in
Figure\,\ref{fig:crustquakeCartoon}, releasing in a burst of plasma
into the magnetosphere\,\cite{thompson95}. Giant flares may occur
when sudden large scale magnetic field reconfigurations cause
catastrophic failures of the crust\,\cite{schwartz05, rea06}.  The field
must be at least about $10^{14}$\,G before being capable of causing
crust failure.

\begin{figure}[!h]
\begin{center}
\includegraphics[angle=0,width=110mm, clip=false]{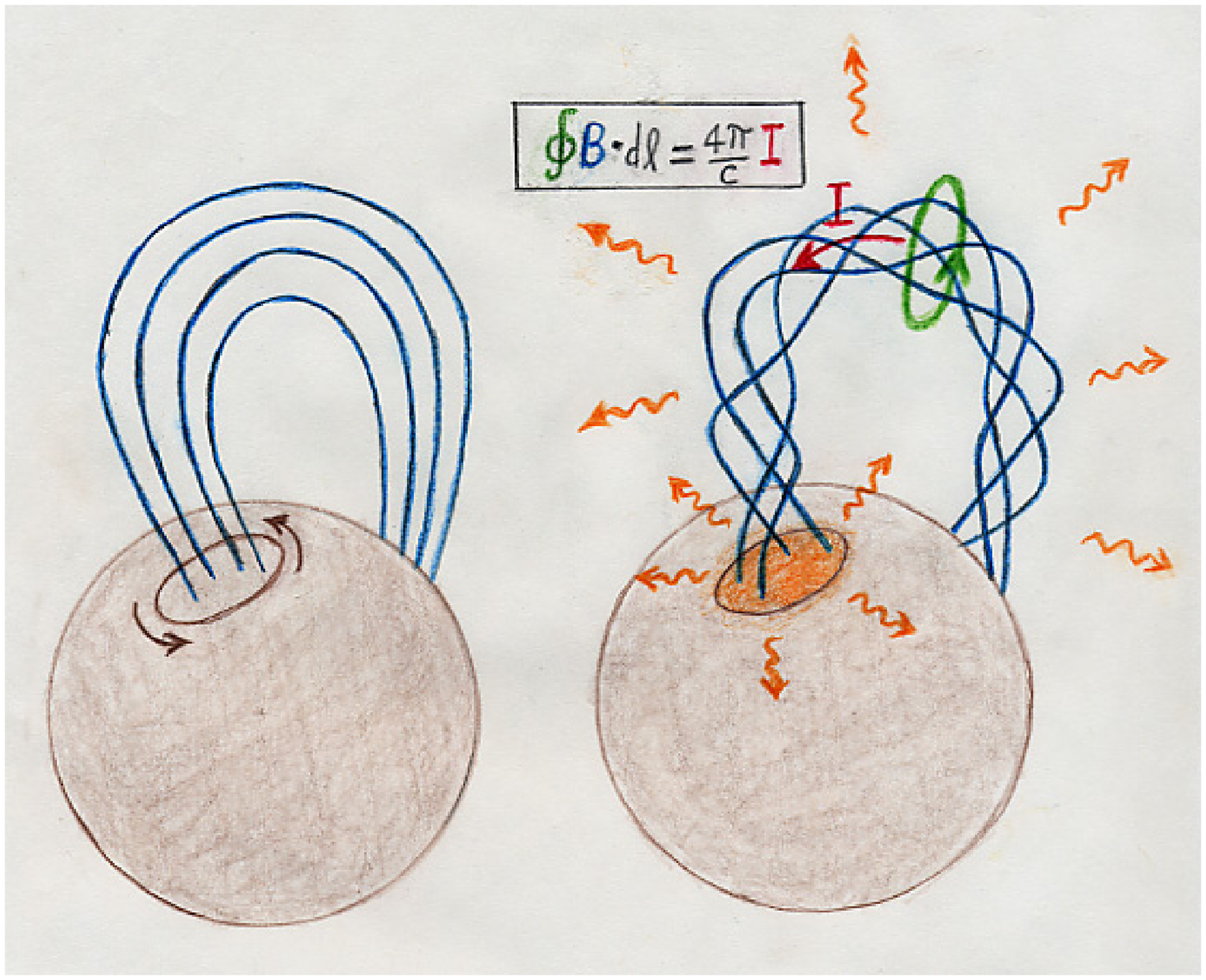}
\caption[Untwisting internal field stresses a magnetar's crust] {An
untwisting internal magnetic field may lead to twisted field lines
in the star's magnetosphere, which might contribute to the observed
persistent X-ray emission.  The mechanism may also stress a
magnetar's solid crust until it fails irreversibly. Drawing: R.
Duncan\,\cite{duncanWebsite}.} \label{fig:crustquakeCartoon}
\end{center}
\end{figure}

Finally, the fading tails observed after giant flares could be
explained by magnetically trapped fireballs in the star's magnetosphere. A field strength of about
$\sci{4}{14}$\,G is required for magnetically trapped fireballs.

We note that there is evidence that SGRs can be modeled as
relaxation systems in which a continuous input of energy from the
magnetic field causes sporadic and unpredictable releases of energy
when the crust fails\,\cite{palmer99}.  This supports the magnetar
crustquake model.  Other examples of relaxation systems include
avalanches and earthquakes.

\section{Emission of gravitational waves}
\label{section:sgrEmissionOfGW}

The question of whether and how SGRs emit gravitational waves is
unfortunately still murky.  We hope that gravitational wave
observational results, such as those discussed in this dissertation,
will stimulate further theoretical work in this area.

The observation that SGR burst events are apparently the output of
relaxation systems\,\cite{palmer99} suggests that burst events
plausibly begin with violent activity in the neutron star crust.
This is further reinforced by the prediction that the star's
interior can store much larger magnetic fields than the star's
exterior, which may indicate a burst mechanism beginning with a
crust event rather than reconnection of an external magnetic
field\,\cite{woods04a}.

Crustquakes could  excite the star's nonradial $f$-modes damped by
gravitational waves\,\cite{andersson97, pacheco98, ioka01}, making
SGRs interesting candidates for gravitational wave
emission\,\cite{horvath05, pacheco98}.  Evidence of QPOs in the
tails of SGR giant flares, which may be caused by seismic
oscillations in the star which are excited after the large
fracture\,\cite{israel05, strohmayer05} suggests that excitation of
$f$-modes may also occur.

There are few papers directly addressing the question of emission of
gravitational waves coincident with SGR bursts\,\cite{ioka01,
pacheco98, horvath05}. The most detailed model, which also allows
for the most gravitational wave energy to be emitted, is
Ioka's\,\cite{ioka01}. We are indebted to B. Owen for discussions
informing what follows.

Ioka's model is motivated by the observed increase in period of SGR
1900+14 associated with the 1998 August 27 giant flare
(Section\,\ref{section:sgrTiming}).  Thus, one problem with the
model, published in 2001, is that no similar anti-glitch was
observed in the SGR 1806--20 2004 giant flare
(Section\,\ref{section:sgrTiming}).  However, it is possible that
timing noise could hide some glitches, and it may be possible to
decouple the model from the anti-glitch, at least in part.

The angular velocity of a neutron star can be written
 \be
 \Omega = \mathcal{J} / \mathcal{I},
 \ee
where $\mathcal{I}$ is the moment of inertia and $\mathcal{J}$ is
the angular momentum.  The Ioka model explains spin down as an
increase in $\mathcal{I}$, as opposed to a change in $\mathcal{J}$
as in other models\,\cite{thompson00}.  $\mathcal{I}$ is affected by
changes in the star's strong field; Ioka assumes the
deformation is elliptical like the rotational deformation, causing
elongation along the rotational axis. At magnetar field strengths
the magnetic deformation should dominate the rotational deformation.
Ioka estimates that the fractional change in $\mathcal{I}$ due to
the magnetic deformation is of order $\delta$, the ratio of magnetic
energy to gravitational energy (about $10^{-4}$), and that the
fractional change in gravitational energy is roughly of order
$\delta^2$, making the change in energy of order $10^{45}$\,erg.
Therefore global rearrangements of the magnetic field would cause
fractional changes in $\Omega$ of $\sim 10^{-4}$ and would release
energies of $\sim 10^{45}$\,erg. Ioka was inspired to work out his
model by this correspondence with observations from the SGR 1900+14
giant flare.

\begin{figure}[!t]
\begin{center}
\includegraphics[angle=0,width=110mm, clip=false]{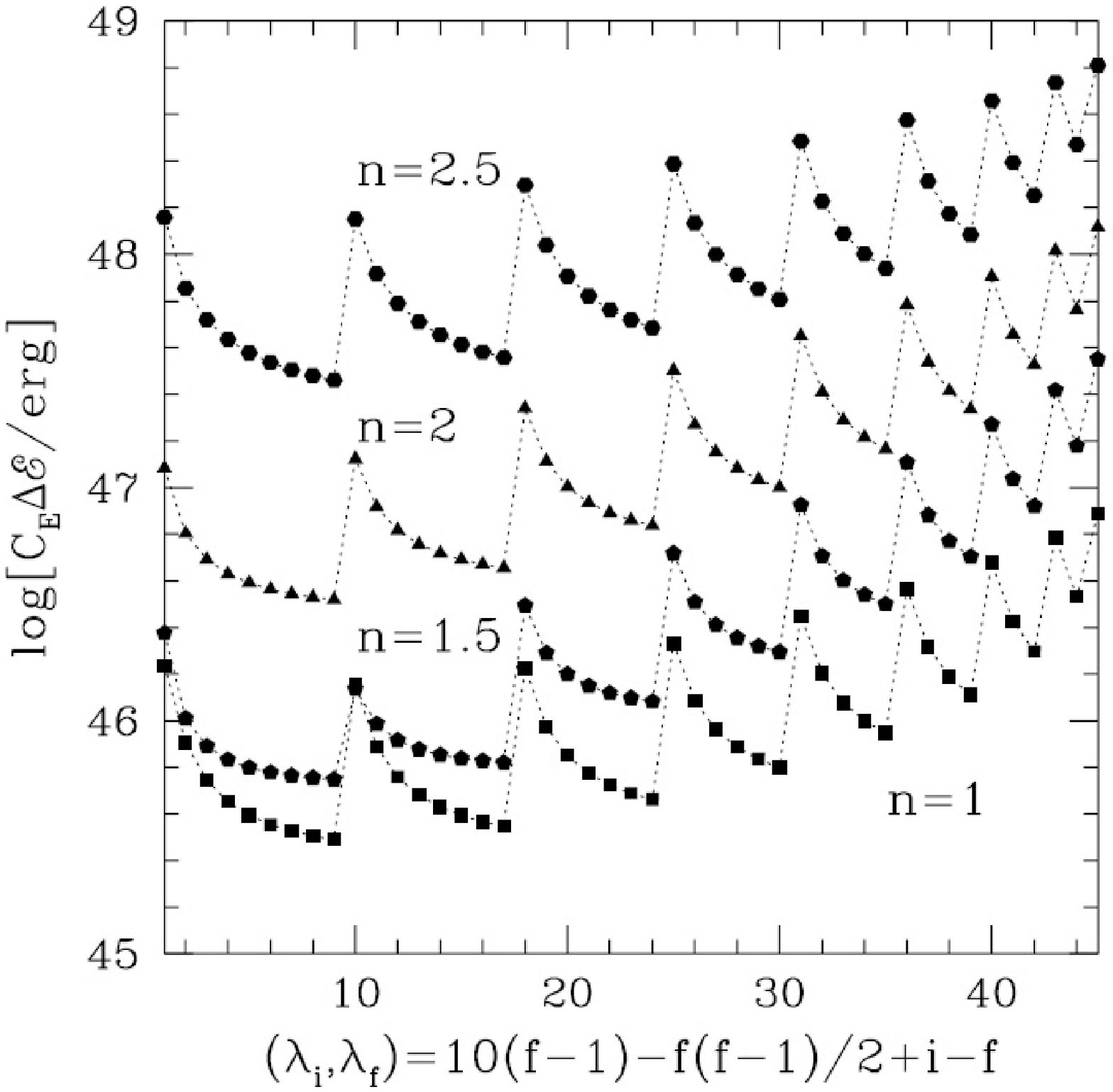}
\caption[Gravitational wave energies emitted in Ioka's model] { The
released gravitational wave energy as a function of the various sets
of the initial and final state in Ioka's model. Figure
from\,\cite{ioka01}. } \label{fig:iokaFigure3}
\end{center}
\end{figure}

Ioka calculates equilibria for neutron stars with different masses,
radii, and equations of state, finding that these equilibria are
characterized by discrete energy states which are related to the
number of loops of the magnetic field. Transitions between energy
states correspond to relatively small changes in the magnetic field
energy but large changes in the gravitational potential energy, and
would be observed as SGR bursts. A plot of the the gravitational
potential energy difference between equilibrium states is shown in
Figure\,\ref{fig:iokaFigure3}.  Ioka assumes most of this energy
goes into gravitational waves; it is possible in this model for
gravitational wave energy release to greatly exceed the gamma ray
energy release.  We see that in Ioka's model a transition between
even adjacent states could release $10^{46}$\,erg in gravitational
waves for an n=1 polytrope.  More esoteric equations of state and
transition between non-adjacent states could result in up to
$10^{49}$\,erg in gravitational wave emission under this model.

We can define the ratio
 \be
    \gamma = \egwn/E_{\mathrm{EM}},
 \ee
which is a measure of the coupling between the EM emission mechanism
and the gravitational wave emission mechanism.  This figure of merit
will be useful in our search described in
Chapter\,\ref{chapter:search}. Ioka does not explicitly address such
a ratio.  Indeed, for part of the paper he sets $E_{\mathrm{EM}}=0$
which he justifies since $E_{\mathrm{EM}}$ turns out to be a small
fraction of $\egw$. Unfortunately, Ioka's range of $E_{\mathrm{EM}}$
is never clear. However, Ioka keeps $E_{\mathrm{EM}} < 10^{45}$\,erg
(the isotropic electromagnetic energy release measured in the SGR
1900+14 giant flare), and with this $E_{\mathrm{EM}}$ and assuming a
largest $\egw$ from the model of $10^{49}$\,erg, $\gamma$ as large
as $10^4$ might fall in the range of model predictions.
\clearpage

\chapter[Search for GWs from Individual SGR Bursts]{Search for Gravitational Waves from Individual SGR Bursts}
\label{chapter:search}

In this chapter we describe a search for gravitational waves
associated with individual SGR bursts using the Flare pipeline, and
give results for the SGR 1806--20 giant flare and the first year of
S5 (S5y1)\,\cite{s5y1sgr}.  This work was published in \,\cite{s5y1sgr}.

We have searched for neutron star ringdowns and also for unmodeled
short-duration gravitational wave emission.  This decision was
motivated by predictions from some models of neutron star $f$-modes
damped by gravitational waves\,\cite{benhar04, thorne83,
andersson97, pacheco98, ioka01, andersson02}. The detectors' most
sensitive region, 100--1000\,Hz, was searched for unmodeled
short-duration gravitational wave emission.

\section{The sample of SGR bursts}

The SGR burst sample was provided by the gamma ray satellites of the
third interplanetary network (IPN)\,\cite{ipn}, and includes the
2004 December 27 SGR 1806--20 giant flare and 214 confirmed
IPN-listed SGR events occurring during the first year of LIGO's
fifth science run from 14 November 2005 to 14 November 2006.
The sample includes 152 SGR 1806--20 bursts (74 with three LIGO
detectors observing at the event time, 41 with two detectors, 18
with a single detector, and 19 with no detector) and 62 SGR 1900+14
bursts (43 with three detectors observing at the time, 12 with two
detectors, 2 with a single detector, and 5 with no detector).  One
of the SGR 1900+14 events was a storm lasting
$\sim$32\,s\,\cite{israel08} and consisting of multiple bursts, and
one of the SGR 1800--20 events was a burst series consisting of two
fairly bright bursts and four weaker bursts. Significant activity
during S5 from the other galactic SGRs has not been observed.
Including the SGR 1806--20 giant flare, analysis was possible for a
total of 191 listed SGR events.  

The S5 SGR burst events are not distributed uniformly in time.
Figure\,\ref{fig:burstsamplehist} shows histograms of the listed
burst events which occurred during S5.

\begin{figure}[!t]
\begin{center}
\subfigure{
\includegraphics[angle=0,width=100mm,clip=false]{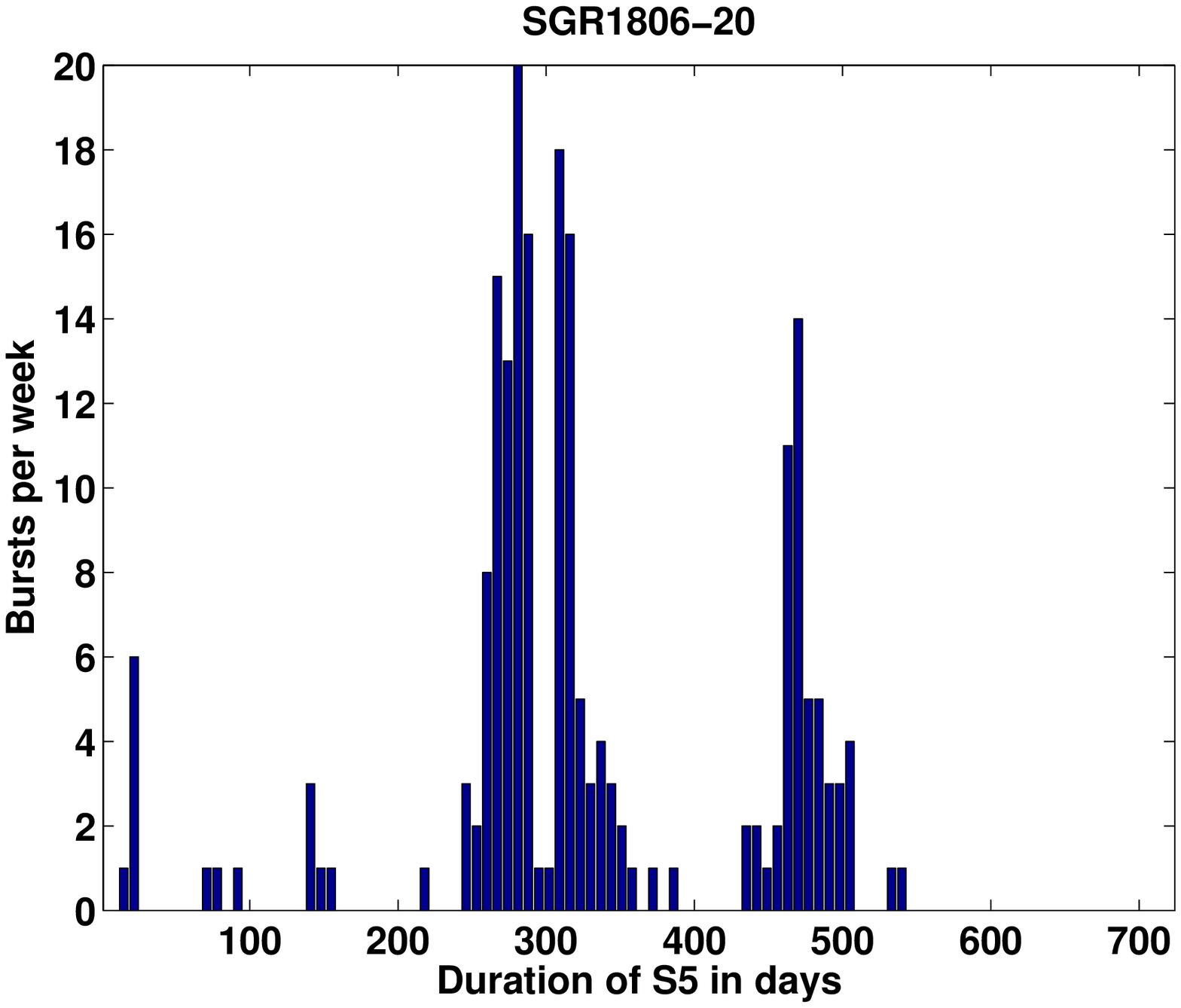}}
\subfigure{
\includegraphics[angle=0,width=100mm,clip=false]{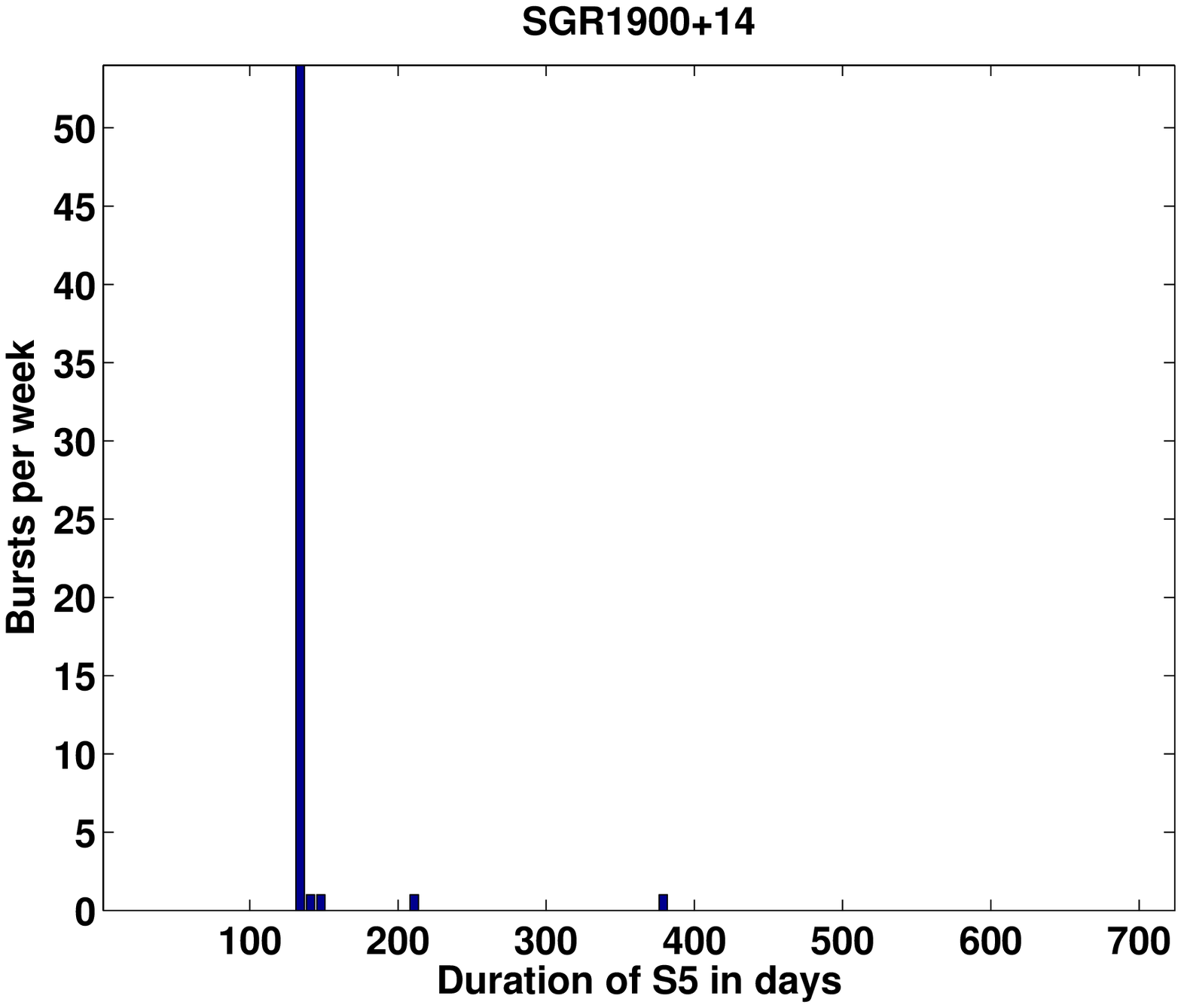}}
\caption[S5 SGR burst sample histograms.]{Histograms of the listed
burst events which occurred during S5.  Top: SGR 1806--20.  Bottom:
SGR 1900+14. Note the very different time distributions of activity
from these two SGRs.} \label{fig:burstsamplehist}
\end{center}
\end{figure}

\begin{table}[h]
\begin{center}
\caption[LIGO detectors available for analyzing 214 S5y1
bursts]{Number of the 214 S5y1 burst events occurring with triple,
double, and single LIGO detector availability after application of
data quality cuts.}
\begin{tabular}{c|rrrr}
\hline \textbf{source} & \textbf{triple} &
   \textbf{double} & \textbf{single} & \textbf{none}\\
   \hline
SGR 1806--20 & 74 & 41 & 18 & 19 \\
SGR 1900+14 &  43 &  12 & 2 & 5 \\

\end{tabular}
\label{table:burstsample}
\end{center}
\end{table}

Of the 214 bursts which occurred during S5y1, trigger times were
provided by IPN satellites as given in Table\,\ref{table:ipn}.  For
a gamma ray burst event to be considered a confirmed SGR 1806--20 or
SGR 1900+14 event, its localization must be consistent with the SGR
position. The localization can come from an IPN annulus, or from a
single detector.  HETE, INTEGRAL, and Swift sometimes image bursts
to arc minute accuracy.  The event spectrum and burst duration must
also be consistent with SGR events (that is, soft and short).  In
addition to the 214 confirmed events in S5y1, there are many
unconfirmed events listed.  A burst is considered unconfirmed if no
localization was obtained.  We do not include unconfirmed events in
the search. Finally, given the distribution of electromagnetic
energies of observed SGR events as discussed in
Section\,\ref{section:sgrbursts}, we note that there are likely many
SGR events which occurred below the detection threshold of the
satellite network. In addition, SGR bursts may fail to be detected
by any satellite due to occultation, field of view limitations,
or detector downtime.

The events list for the S5 first year SGR search was made from a set
of SGR electromagnetic burst trigger times provided by the IPN.   It should be reasonably complete for
bursts with fluences above $10^{-6}$\,erg
cm$^{-2}$\,\cite{khPrivateCompleteness}, but as mentioned above,
there can be bursts which occurred above this threshold and were not
observed by any satellite.  The
UTC times in this list are given to the nearest second, and are
triggering times of the detector at the satellite. (Recall that
the rise to peak flux of SGR events is rapid.)   These UTC
times were converted to GPS times.   Burst event times at the
KONUS/Wind satellite were propagated to the geocenter (see Section\,\ref{section:eventTriggerTimes}).

SGR bursts listed in the GCN\,\cite{gcnWeb} comprise a small subset
of the more electromagnetically spectacular bursts in the sample.    Table\,\ref{table:brightestbursts} lists all bursts in
the S5y1 sample for which fluences are given in a GCN report.
Durations are also given when available.

\begin{table}[h]
\begin{center}
\caption[IPN satellites providing trigger times for the 214 S5y1 SGR
bursts]{Breakdown of IPN satellites providing trigger times for the
214 S5y1 SGR bursts.  (The SGR 1806--20 giant flare time was provided
by RHESSI\,\cite{rhessi}, and cross-checked with the INTEGRAL
arrival time.) Altitude refers to distance above the surface of the
earth.  Max delay is the max light travel time between the
satellite and the geocenter. }
\begin{tabular}{lllrcr}
\hline
 satellite & detector &  & S5y1 triggers & max altitude [km] & max delay [ms] \\
   \hline
Swift & BAT &\cite{swift} & 172 & $\sci{6.0}{2}$  & 23 \\
Wind & Konus & \cite{wind} & 21 & $\sci{1.7}{6}$ & 5700 \\
INTEGRAL &  & \cite{integral} & 20 & $\sci{1.6}{5}$ & 520 \\
Suzaku &  & \cite{suzaku} & 1 & $\sci{5.7}{2}$  &  23\\

\end{tabular}
\label{table:ipn}
\end{center}
\end{table}

\subsection{Multi-episodic storm from SGR 1900+14}

One of the most interesting events in the S5y1 sample was a
multi-episodic storm from SGR 1900+14\,\cite{israel08}.  This event
occurred on 26 Mar. 2006 after a few days of activity.  As is
evident in Figure\,\ref{fig:burstsamplehist}, this storm and its
vicinity accounted for most of the S5 activity from
SGR 1900+14.  The storm itself lasted only $\sim30$\,s; the
Swift/BAT light curve is shown in Figure\,\ref{fig:stormLightcurve}.

The total isotropic electromagnetic energy of the storm was about $\eem=\sci{6}{41}$\,erg at a nominal distance of 10\,kpc.  This estimate is based on a conservative estimate of the storm fluence by the Konus-Wind team, in the 20--200\,keV range, of $\sci{(1-2)}{-4}$\, erg cm$^{-2}$\,\cite{gcn4946}.

\begin{figure}[!t]
\includegraphics[angle=0,width=130mm, clip=false]{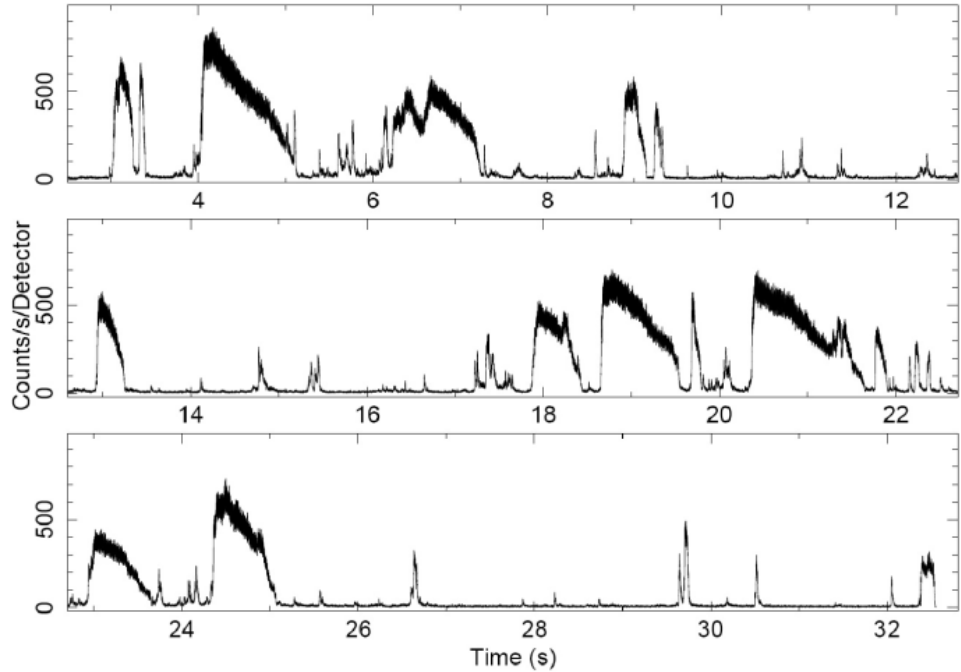}
\caption[Swift/BAT light curve from the SGR 1900+14 storm]{15\,keV
to 100\,keV BAT light curves with a time resolution of 1\,ms
obtained during the burst ``storm'' of 2006 March 29.  Times on the
x-axis are relative to 02:53:09 UT.  Figure from\,\cite{israel08}.}
\label{fig:stormLightcurve}
\end{figure}

\subsection{060806 burst series from SGR 1806--20}
\label{section:060806}

The electromagnetically brightest S5y1 SGR burst in the sample
occurred in a burst series emitted by SGR 1806--20 on 2006 August 06
(hereafter ``060806 event'')\,\cite{gcn5416, gcn5419}.  The
Konus-Wind light curve\,\cite{grb060806personal} is shown in
Figure\,\ref{fig:grb060806}.  The event was a series of two large
bursts and four small bursts occurring within about two minutes. The
fluence of the largest burst measured by Konus-Wind, in the energy
range $>18$\,keV, was
$\sci{2.4}{-4}$\,erg\,cm$^{-2}$\,\cite{gcn5426}. This gives a lower
limit on the isotropic electromagnetic energy at a distance of
10\,kpc of $\eem=\sci{2.9}{42}$\,erg.  Aside from the SGR 1806--20
giant flare (Section\,\ref{section:giantflare}), the largest burst
in this series was the burst in the sample with the smallest values
of $\gamma = \egwn/E_{\mathrm{EM}}$.  The fluence of the second
largest burst at the beginning of the series was
$\sci{6.0}{-5}$\,erg\,cm$^{-2}$\,\cite{grb060806personal}.

At the time of the 060806 event, the light travel time for a
wavefront arriving from SGR 1806--20 from the geocenter to the
Konus-Wind satellite was 5.051\,s.  The time axis in
Figure\,\ref{fig:grb060806} gives the light crossing time at the
geocenter; i.e. the 5.051\,s has already been applied.  The time for
the start of this series listed by the IPN gives the light crossing
time at the Konus-Wind satellite.

\begin{figure}[!t]
\includegraphics[angle=0,width=120mm, clip=false]{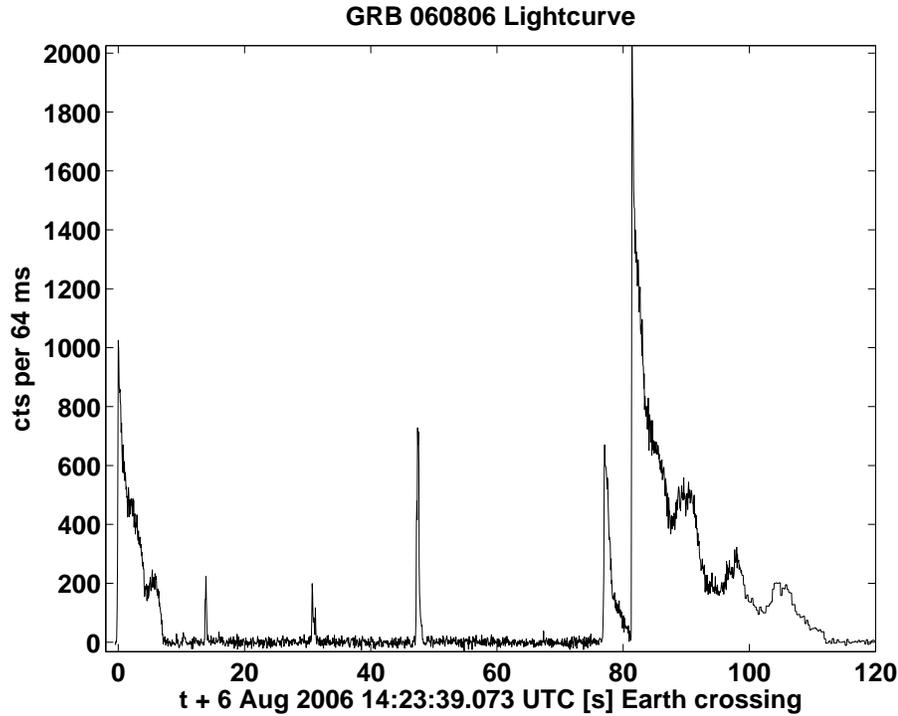}
\caption[060806 event light curve]{Konus-Wind light curve for the
060806 event from SGR 1806--20.  } \label{fig:grb060806}
\end{figure}

\subsection{The SGR 1806--20 giant flare} \label{section:giantflare}

The giant flare light curve was resolved by several instruments, and its
timescales are well known.  We use the light curve timescales as
a means to limit the parameter space of the search. The timescale of
the initial rise was $\lesssim$1\,ms, possibly associated with
propagation and reconnection in the magnetosphere~\cite{palmer05,
schwartz05}. There was also an intermediate rise with an e-folding
time of $\sim$5\,ms (see Figure\,\ref{fig:gftimescales}), which
could be explained in the magnetar model if the rise rate is limited
by propagation of a large ($\sim$5\,km) crustal fracture in the
neutron star\,\cite{schwartz05, thompson01}. The prompt flare
duration was $\sim$100\,ms, possibly the Alfven crossing time in the
star's interior~\cite{schwartz05}.  Repeated injections of energy on
this $\sim$100\,ms timescale are also observed in the GEOTAIL
data\,\cite{terasawa05}.  Finally, the tail observed after the flare
had a duration of minutes.  This longer timescale is not considered
relevant to our search for transient burst signals, though it is
important for a search for gravitational waves associated with QPO
oscillations in the tail~\cite{matone07}.

The inferred isotropic electromagnetic energy for the event,
assuming a distance of 10\,kpc, was
$\sci{1.6}{46}$\,erg\,\cite{hurley05}.

\begin{sidewaystable}[h]
\begin{center}
\caption[Electromagnetically bright bursts in the S5y1 sample]{The
bursts in the S5y1 sample, with largest electromagnetic fluences
reported, ordered by fluence. The information was taken from GCN
reports.  No other fluences were given for these sources in GCN
reports though other bursts were reported with no fluence given.
Other events may have occurred with larger fluences that were not
reported in the GCN.  Konus-Wind fluences are for photons in the
range 20--200\,keV. INTEGRAL fluences are for photons in the range
15--100\,keV. Long durations are caused by ``tails.'' The LIGO
gravitational wave detector network collecting analyzable data at
the time is also given.  If no LIGO detector was collecting data at
the time a dash is shown.  Event times given are times listed in the
GCN report and in the event list obtained by K. Hurley. On-source
times given are the times at the center of the on-source region in
the gravitational wave analysis.}
\begin{tabular}{ll|lllrrr}
 \hline
 \textbf{GCN} & \textbf{satellite} & \textbf{source} & \textbf{event time} & \textbf{on-source} & \textbf{fluence [erg cm$^{-2}$}] & \textbf{duration [s]} & \textbf{network} \\
 \hline

 5426 & Konus-Wind & SGR1806-20 & 2006 Aug 06 14:23:44 & 14:25:00.5 & $\sci{2.4}{-4}$ max & series    & H1H2   \\ 
 4312 & Konus-Wind & SGR1806-20 & 2005 Dec 03 11:43:24 & 11:43:28.3 & $\sci{1.5}{-4}$ & 22.0    & L1H1H2 \\ 
 4310 & Konus-Wind & SGR1806-20 & 2005 Dec 01 09:59:25 & 09:59:30.2 & $\sci{2.0}{-5}$  & 3.6     & L1H1H2 \\ 
 4946 & Konus-Wind & SGR1900+14 & 2006 Mar 29 02:53:08 & 02:53:24.0 & $\sci{1-2}{-4}$ sum  & series  & L1H1H2 \\ 
 4946 & Konus-Wind & SGR1900+14 & 2006 Mar 29 02:45:28 & ---        & $\sci{1.5}{-6}$ max & series  & ---    \\ 
 4936 & Konus-Wind & SGR1900+14 & 2006 Mar 28 09:03:00 & ---        & $\sci{1.1}{-6}$  & 0.13    & ---    \\ 
 5490 & INTEGRAL   & SGR1806-20 & 2006 Aug 29 22:10:28 & 22:10:28   & $\sci{1}{-6}$    & 1.2     & L1H2   \\ 
 4946 & Konus-Wind & SGR1900+14 & 2006 Mar 29 01:28:03 & 01:28:04   & $\sci{7.2}{-7}$  &  0.07   & L1H1H2 \\ 
 5490 & INTEGRAL   & SGR1806-20 & 2006 Aug 29 21:57:43 & 21:57:43.2 & $\sci{7}{-7}$    &  2.0    & L1     \\ 
 4965 & Swift-BAT  & SGR1900+14 & 2006 Apr 14 04:35:28 & ---        & $\sci{1.3}{-8}$    &      & ---    \\

\end{tabular}
\label{table:brightestbursts}
\end{center}
\end{sidewaystable}

\subsection{Event trigger times} \label{section:eventTriggerTimes}

The critical attribute of an external trigger in our search is the
trigger time.  Listed event times refer to wavefront arrival at a
satellite.  The arrival times are satellite detector trigger times,
and do not in general correspond to either the beginning or the peak
of the electromagnetic event. For short SGR bursts, lasting
typically 100~ms, trigger times are adequately close to the event
peaks for the purposes of this gravitational wave burst search,
which uses 4\,s on-source regions which account for this uncertainty.

The times in the list of SGR bursts are satellite trigger times at
the detector rounded to the nearest second.  Using these times as
listed in the externally triggered gravitational wave search
introduces two significant sources of error: $\pm 0.5$\,s error from
rounding and error from not considering the gravitational wave
travel time from the satellite to the detector.  Satellite timing
uncertainties are insignificant compared to these two sources, as
each spacecraft has an associated clock uncertainty, typically in
the several ms range and rarely exceeding ~100 ms.

As shown in Table\,\ref{table:ipn}, all satellites except Konus-Wind
reporting times in the SGR sample have maximum light travel times to
the geocenter of no more than about 0.5\,s.  Konus-Wind can have
light travel times greater than 5\,s, and thus we propagate light
crossing times for events with Konus-Wind trigger times, as
described in Appendix\,\ref{appendix:propagation}.

\section{On-source region} \label{sec:onsource}

To analyze a given SGR burst we divided gravitational wave data into
an on-source time region (in which gravitational waves associated
with the burst could be expected) and a background time region (in
which we do not expect an associated gravitational waves, but in
which the noise is statistically similar to the on-source region).
We require an on-source region large enough to account for satellite
timing uncertainties and wavefront propagation times to Earth, and
most SGR flare model predictions. Gravitational wave emission is
expected to occur almost simultaneously with the electromagnetic
burst\,\cite{ioka01}.

As stated above, Konus-Wind light crossing times are propagated the
to the geocenter, and other satellites providing burst times can
have no more than a 0.5\,s light travel time to the geocenter. Thus,
the two primary systematic errors in the trigger times --- rounding
of the times in the SGR burst events list and light travel time from
the satellite --- add up to at most about a second. For isolated
bursts we choose a 4\,s long on-source region $\pm2$\,s centered on
the SGR burst. This accounts for the systematic errors, and
conservatively accounts for uncertainty in coincidence of
electromagnetic and gravitational wave emission.

There are three special cases: 1) for two SGR 1900+14 bursts which
occurred within 4\,s of each other a combined 7\,s on-source region
was chosen; 2) for the SGR 1900+14 storm a 40\,s on-source region
was used; 3) for the 060806 event from SGR 1806$-20$
(Section\,\ref{section:060806}), two 4\,s on-source regions were
used, centered on the two distinct bright bursts comprising the
event.

Identical data quality cuts (Section\,\ref{section:dq}) were applied to both on-source and
background regions. On-source regions subject to a cut were
excluded.

We note here that using a smaller on-source region duration would
not significantly improve upper limits on average.  As discussed in
Section\,\ref{section:onsourceDependence} lowering the on-source
region duration from 180\,s to 4\,s only improved upper limits by
20\% on average.  As discussed in
Section\,\ref{section:stackOnSource} lowering the on-source
region duration from 4\,s to 2\,s in the multiple SGR burst search only improved upper limits by
2\% on average.

\section{Parameters for the ringdown search}
\label{sec:ringdownparams}

The fundamental problem in setting model-dependent upper limits in a
gravitational wave burst search is to effectively explore the signal
parameter space given computational limitations.  In general the
signal parameters we typically wish to explore are central
frequency, bandwidth, duration, and polarization state (circular,
linear, or elliptical). For every upper limit, the polarization
angle $\psi$ was chosen randomly for each simulation.   All parameters for
all searches were chosen using simulated signals added into
background data, before searching on-source regions.

Ringdowns are characterized by a single frequency. The parameter
space for ringdowns consists of frequency $f_0$, exponential decay
time constant $\tau$, and polarization state. Model predictions from reference\,\cite{benhar04} for ten realistic
neutron star equations of state give $f$-mode RD frequencies in the
range 1.5--3\,kHz and damping times in the range 100--400\,ms. We
used a search band 1--3\,kHz for RD searches (to include stiffer
equations of state), and found a 250\,ms time window (Fourier
transform length) for the Flare pipeline to be optimal for these
ringdowns.

Within this frequency range we chose specific ad hoc frequencies of
1090\,Hz, 1590\,Hz, 2090\,Hz and 2590\,Hz.  We note that in this
frequency range shot noise amplitude scales linearly with frequency,
which makes Flare pipeline strain upper limits also directly
proportional to frequency for the most part, so the choice of
particular frequencies within the range is not critical.
(Narrow-band noise sources in the detector spectra will cause
deviation from this proportionality at some frequencies.) As
discussed in Section\,\ref{section:ringdownDuration}, strain upper
limits from other values of $\tau$ in the range 100-300\,ms are
within $15$\% of the 200\,ms value.  Separate upper limits were
obtained from ringdown simulations with linear and circular
polarization states.

\section{Parameters for the unmodeled search}
\label{sec:unmodeledparams}

For the unmodeled search we choose to use band- and time-limited WNB simulations as the
most general unmodeled signals.  WNBs are primarily characterized by
their central frequency, their frequency band, and their duration.
Other decisions on the structure of WNB simulations are discussed in
Section\,\ref{section:makingWNBs}.

In choosing duration, we note the physical timescales of the giant
flare light curve: $\sim$1\,ms, $\sim$5\,ms, and $\sim$100\,ms. The
timescale of the typical duration of common SGR flares may depend on
the sensitivity of the detector used to construct the SGR flare
sample, as there is some correlation between fluence and flare
duration~\cite{gogus01}. In the sample presented in\,\cite{gogus01},
the mean durations for SGR~1806--20, SGR~1900+14, and the AXP
1E~2259+586 were 162, 94, and 99~ms respectively. We use these SGR
electromagnetic timescales to limit the space of plausible target
gravitational wave signals, assuming that gravitational wave signals
will have durations in the range $\sim$1--200\,ms.

For upper limits estimated via WNBs, two durations (11 and 100\,ms)
were used for simulations; other durations in the range 5--200\,ms
are detected using the flare pipeline with at most 20\% strain
sensitivity degradation (see Section\,\ref{section:flareDurations}).

We use the detectors' sensitive region to set the WNB central
frequencies and bandwidths.  We choose two distinct spectral regimes
for the WNBs: 100--200\,Hz and 100--1000\,Hz.  For each of these
regimes we match the band of the search to the band of the WNB
simulation.  This allows us to both set limits using the most
sensitive region 100--200\,Hz only, and also to cover the entire
frequency range from 100--3000\,Hz in conjunction with the ringdown
search.

\section{ Gravitational-wave data} \label{section:s5}

The SGR 1806--20 giant flare occurred during a period of detector
commissioning between science runs, and was observed under LIGO's
``Astrowatch'' program\,\cite{astrowatch}, which attempts to collect
as much high quality data as is practical during commissioning
breaks.

Neither strain-calibrated data nor data quality flags were produced
for the stretch of Astrowatch data containing the SGR 1806--20 giant
flare trigger.  Thus, Flare pipeline analysis of this event requires
calibration of the raw gravitational wave signal.  We decided to limit the background data to the H1
detector ``lock stretch'' containing the event, avoiding the edges
of this data stretch.

The S5 events in this search were analyzed using strain-calibrated
h(t) data (Section\,\ref{section:hOfT}).

\section{Uncertainties and errors in upper limits}

A general treatment of uncertainties in the Flare pipeline was
given in Section\,\ref{section:uncertaintiesMethod}.  Here we
discuss uncertainties and errors specific to the S5y1 plus SGR
1806--20 giant flare individual SGR search.

\subsection{Detector calibration for giant flare}

The H1 detector calibration for the time of the giant flare was
produced specifically for that
event\,\cite{landryPrivateCommunication}. The response function is
stored in version control\,\cite{flareCVSastrowatchCal}. The H1 response
function at a fiducial time was propagated to the time of the giant
flare as described in Section\,\ref{section:calibration}. This
response function has estimated 20\% statistical uncertainty (at 1
sigma) and 6\% systematic error (towards worse sensitivity).

\subsection{Detector calibration for S5}
\label{section:s5calibration}

The calibration committee cites 8.1\%, 7.2\%, and 6.0\% 1-sigma
statistical uncertainties in amplitude and 3, 2, and 2\,degree
1-sigma statistical uncertainties in phase for the H1, H2, and L1
detectors, respectively for the V3 S5
calibration\,\cite{V3uncertainty} used in the search, for
frequencies below 2\,kHz. For simulated waveforms below 2\,kHz we
conservatively treat the amplitude uncertainty as follows: 8.1\% is
added to the strain upper limits for any search which includes H1;
7.2\% is added to the strain upper limit for L1 or L1H2 searches;
and 6.0\% is added to the strain upper limit for H2-only searches.

For frequencies between 2\,kHz and 3\,kHz the 1-sigma statistical
amplitude uncertainties are 10\%, 6\%, and 6\%, and the 1-sigma
phase uncertainties are 2, 1, and 1\,degree for H1, H2 and L1
respectively.  For searches involving pairs of detectors, the larger
amplitude uncertainty of the pair is conservatively used in the
search as in the low frequency case.

The effect of phase uncertainty was determined via a Monte Carlo
simulation as described in
Section\,\ref{section:calibrationUncertainty}. We conservatively
used a phase calibration uncertainty standard deviation of
4\,degrees for all interferometers and all frequencies, rounding up
$1.28\times 3$\,degrees. We performed the experiment twice for each
simulation type, once with simulations of $\hrss=1$\,$\rthz $and
once with simulations of $\hrss=10$\,$\rthz $. Each experiment used
250 trials, giving errors on the means of $\sim$1\%. At this level
no significant effect was found for any simulation type, and we do
not include uncertainty from this source in the quadrature sum with
amplitude statistical uncertainty and statistical uncertainty from a
finite number of simulations, both of which are of order 10\%.  In
Figure\,\ref{fig:phaseUncert2590} we show histograms for the error
and control distributions in the Monte Carlo for 2590\,Hz linearly
polarized ringdowns with $\hrss=10\rthz$ and $\hrss=1$\,$\rthz$.  We
would expect an effect from phase uncertainty to be most pronounced
in the simulations with the highest frequency.  In
Figure\,\ref{fig:phaseUncert2590deg20} we show the same experiment
repeated with a phase uncertainty of 20\,degrees for comparison.

\begin{figure}[!t]
\begin{center}
\subfigure{
\includegraphics[angle=0,width=90mm, clip=false]{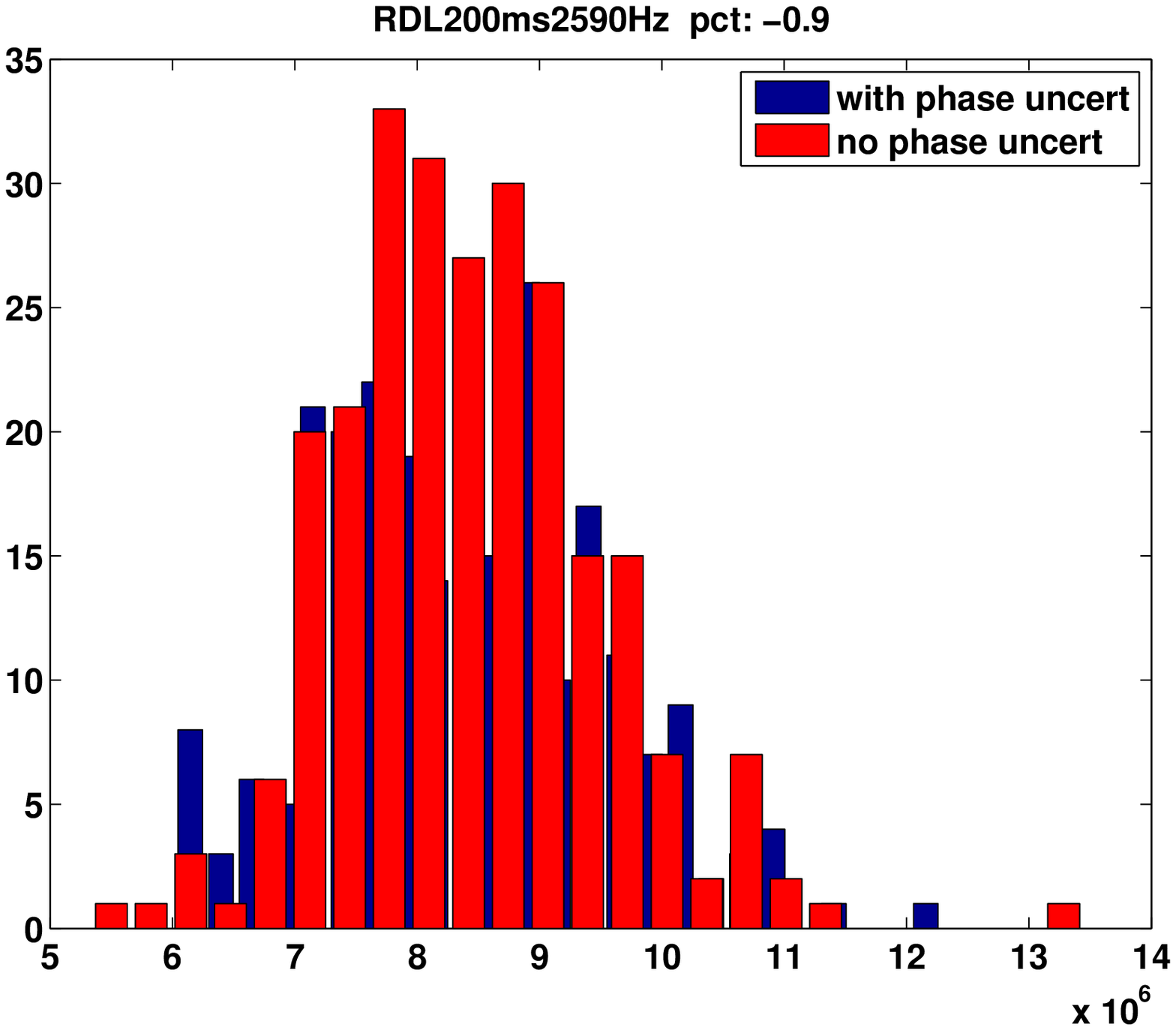}}
\subfigure{
\includegraphics[angle=0,width=90mm, clip=false]{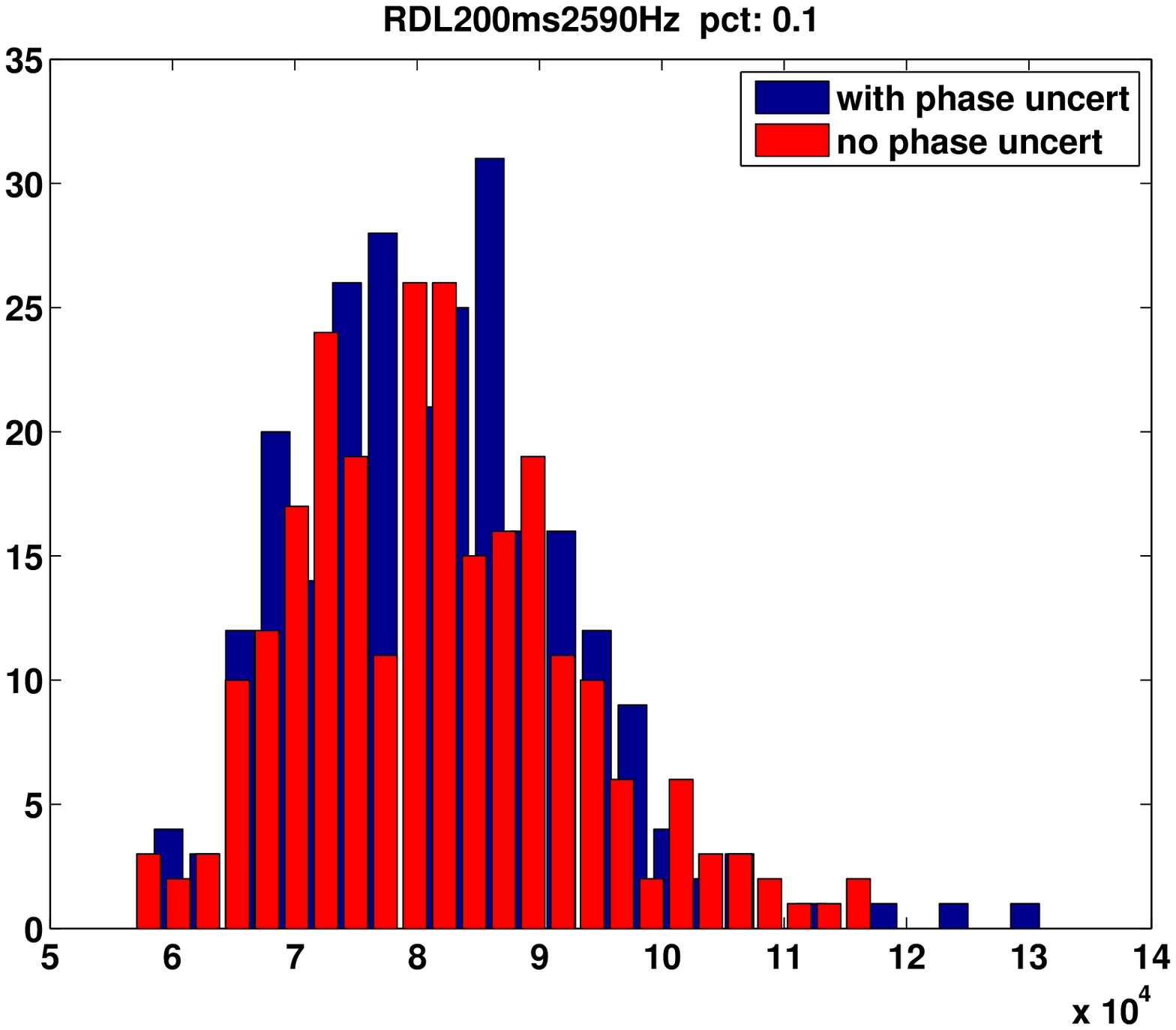}}
\caption[Monte Carlo results for 4 degrees of phase
uncertainty] { Monte Carlo results for 4 degrees of phase
uncertainty for linearly polarized 2590\,Hz ringdowns with
$\hrss=10$\,$\rthz$ (top) and $\hrss=1$\,$\rthz$ (bottom) injected
into white noise with standard deviation of 1.  The x-axis gives the
loudness $Z$ for the recovered simulation analysis events.  25
histogram bins and 250 trials were used. The percent change in the
means of the distributions are 0.9\% and 0.1\% (the ``wrong'' way)
respectively, consistent with no effect. Similar results were
obtained for the other 11 simulation types in the search.}
\label{fig:phaseUncert2590}
\end{center}
\end{figure}

\begin{figure}[!t]
\begin{center}
\includegraphics[angle=0,width=100mm, clip=false]{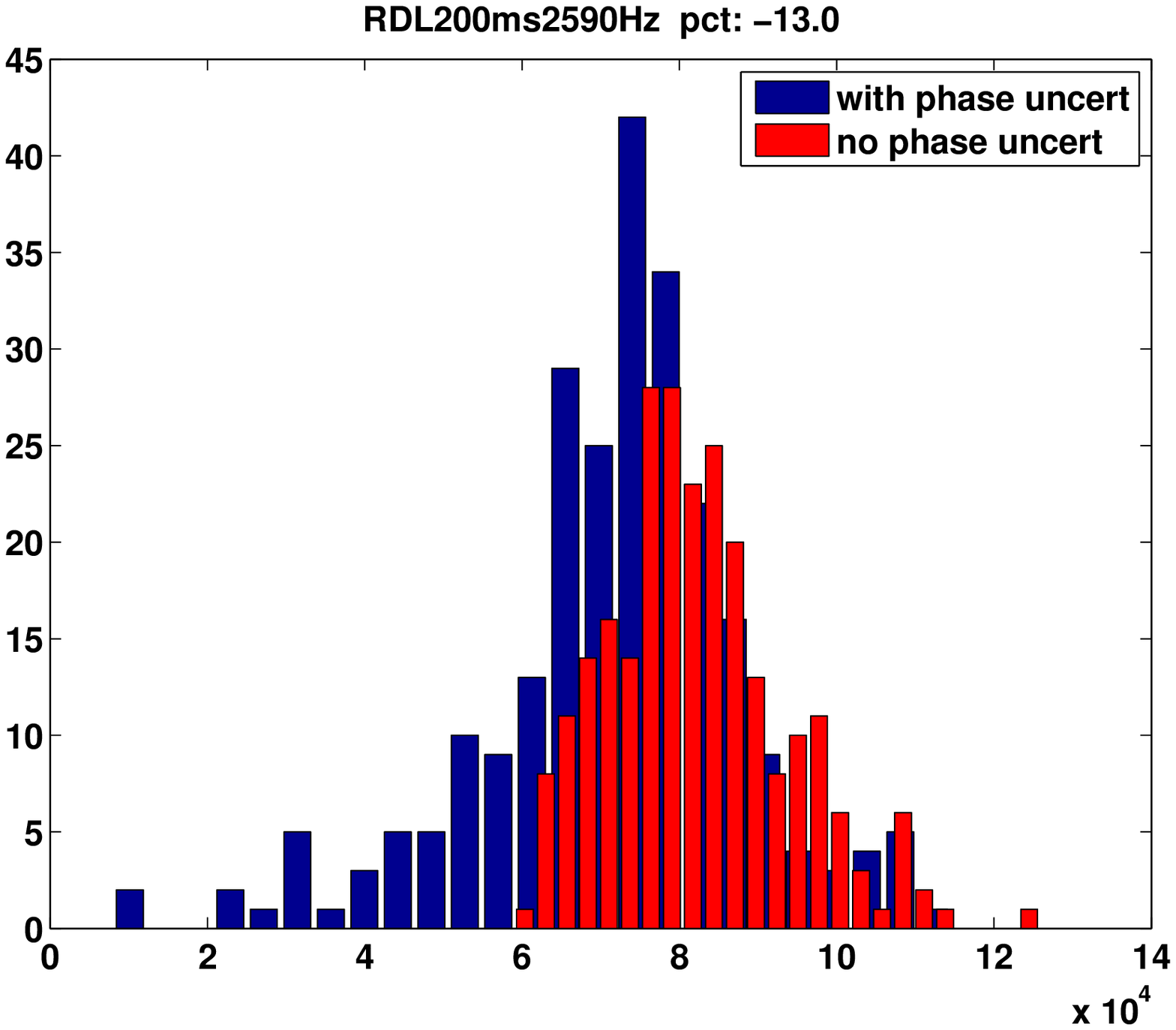}
\caption[Monte Carlo results for 20 degrees of phase
uncertainty] { For comparison we show Monte Carlo results
for 20 degrees of phase uncertainty and $\hrss=1$\,$\rthz$ with all
other variables the same as in Figure\,\ref{fig:phaseUncert2590}.
The x-axis gives the loudness $Z$ for the recovered simulation
analysis events.} \label{fig:phaseUncert2590deg20}
\end{center}
\end{figure}

The calibration committee also declared up to a 20\,$\mu$s timing
error between detectors in the S5 V3
calibration\,\cite{oreillyPersonalCommunication}, which was
subsequently revised up to 30\,$\mu$s\,\cite{v3errors}. We performed
a study to determine the effect of this error on upper limits for
each of the twelve simulation types in the search. The study was
similar to the one performed for the effect of phase uncertainty
described in Section\,\ref{section:calibrationUncertainty} but
simpler, as there was no need to explore a statistical distribution
with a Monte Carlo.

We found that the maximum effect was for linear and circular
2590\,Hz ringdowns, which suffered a 13\% degradation in loudness of
associated simulation analysis event. Table\,\ref{table:timingError}
gives the propagated systematic error, a degradation in loudness of
associated simulation analysis event, for each of the twelve
simulation types, for a 30\,$\mu$s relative timing error.
Figure\,\ref{fig:timingErrorRDL} and
Figure\,\ref{fig:timingErrorWNB} show the propagated systematic
error for 2590\,Hz linear RDs and 100\,ms duration 100--200\,Hz
WNBs, respectively. This systematic error only applies to searches
with two detectors.

\begin{table}[h]
\begin{center}
\caption[Effect of calibration systematic error on simulation
recovery]{Effect of a 30\,$\mu$s relative timing error given as a
percentage degradation in loudness of associated simulation analysis
event, for each of the twelve simulation types.  This systematic
error applies to two detector searches only.  Results were obtained
from Monte Carlo simulations with 200 trials.}
\begin{tabular}{lc}
 \hline
 \textbf{Simulation type} & \textbf{Degradation} \\
 \hline\
 WNB11ms100-200Hz & 0\% \\
 WNB100ms100-200Hz & 0\% \\
 WNB11ms100-1000Hz & 0\% \\
 WNB100ms100-1000Hz & 1\% \\
 \hline
 RDC 200 ms 1090 Hz & 2\% \\
 RDC 200 ms 1590 Hz & 4\% \\
 RDC 200 ms 2090 Hz & 8\% \\
 RDC 200 ms 2590 Hz & 13\% \\
 RDL 200 ms 1090 Hz & 3\% \\
 RDL 200 ms 1590 Hz & 5\% \\
 RDL 200 ms 2090 Hz & 11\% \\
 RDL 200 ms 2590 Hz & 13\% \\
\end{tabular}
\label{table:timingError}
\end{center}
\end{table}

\begin{figure}[!t]
\begin{center}
\includegraphics[angle=0,width=120mm, clip=false]{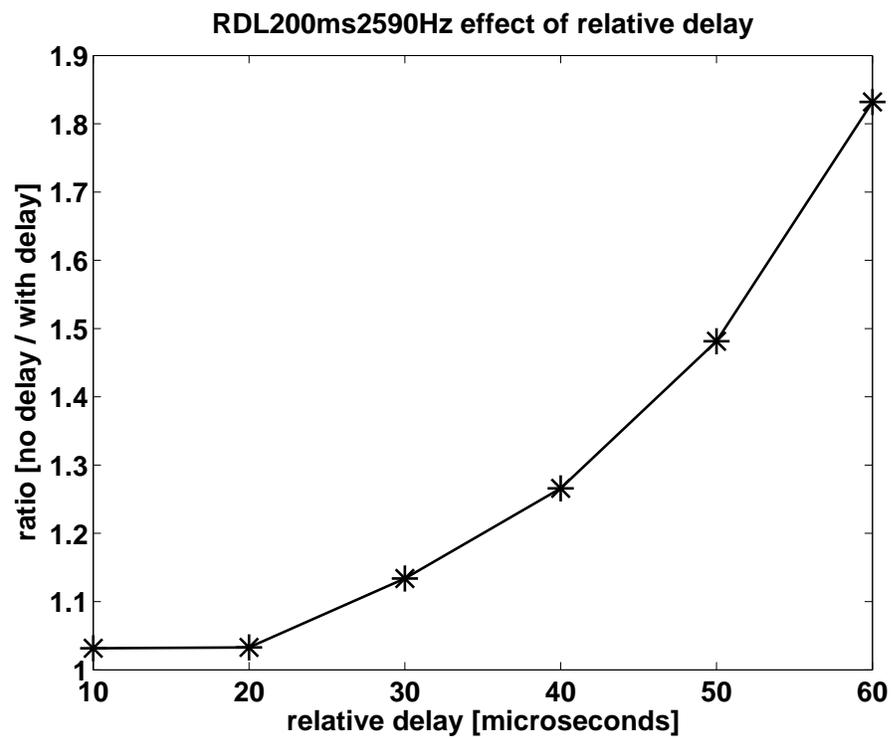}
\caption[Effect of 30\,$\mu$s detector relative timing systematic
error on 2590\,Hz linear RDs] { Effect of 30\,$\mu$s detector
relative timing calibration systematic error on 2590\,Hz linear RDs
in a two-detector search, as a function of error in relative
detector timing.  Results were obtained from Monte Carlo simulations
with 200 trials.  } \label{fig:timingErrorRDL}
\end{center}
\end{figure}

\begin{figure}[!t]
\begin{center}
\includegraphics[angle=0,width=120mm, clip=false]{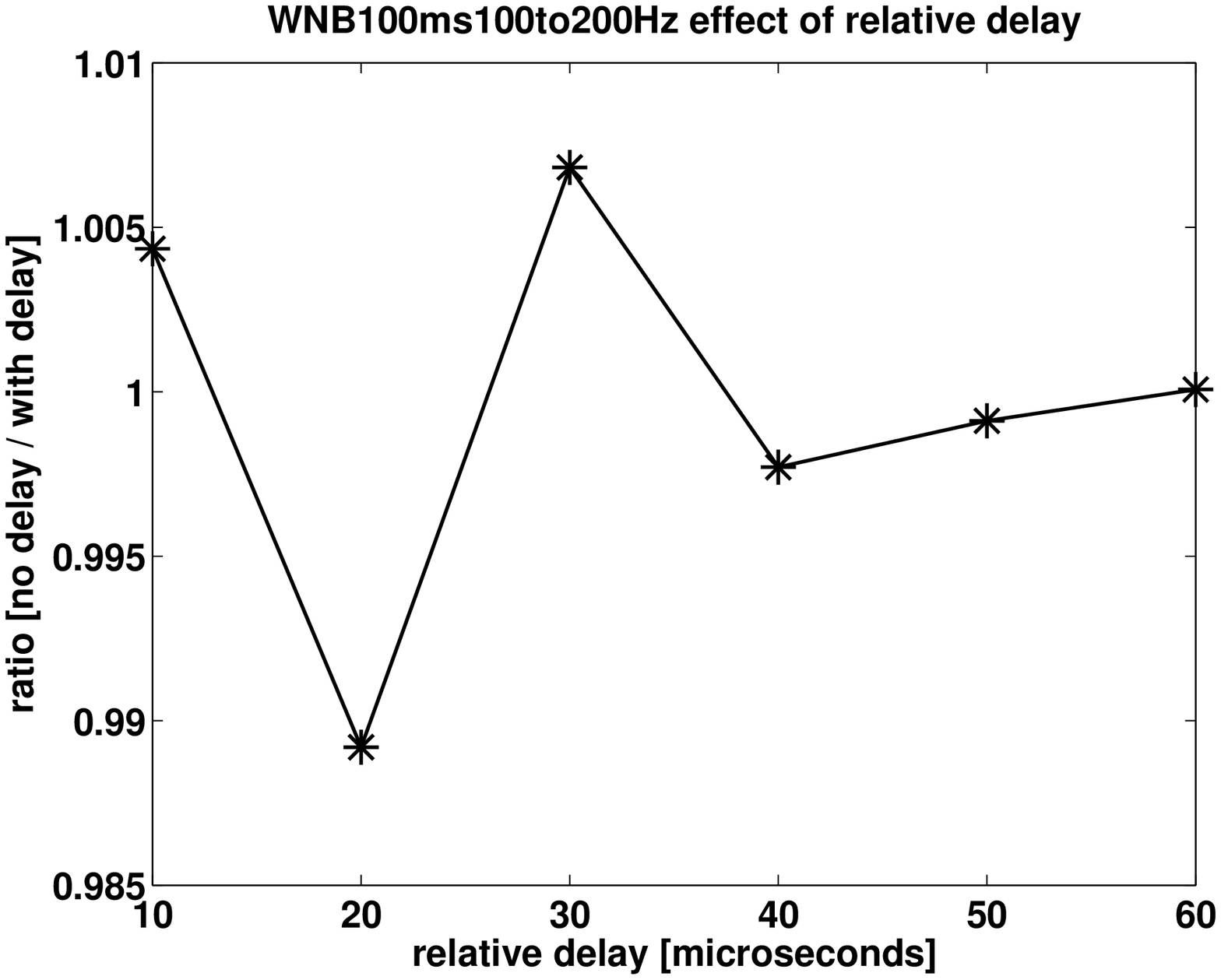}
\caption[Effect of 30\,$\mu$s timing systematic error on 100\,ms
duration 100--200\,Hz WNBs] {Effect of 30\,$\mu$s detector relative
timing calibration systematic error on 100\,ms duration 100--200\,Hz
WNBs in a two-detector search, as a function of error in relative
detector timing.  Results were obtained from Monte Carlo simulations
with 200 trials.  This plot suggests that there is no discernible
effect for this simulation type. } \label{fig:timingErrorWNB}
\end{center}
\end{figure}

Finally, the calibration committee gave correction factors for DC
systematic errors in the amplitude of V3 strain-calibrated h(t) data
for S5 of 1.074, 1.062, and $-1.040$ for the H1, H2, and L1
detectors, respectively\,\cite{V3uncertainty}.   These amplitude
scaling factors were applied to all h(t) data in the search
immediately after it was retrieved.  No additional adjustment to
upper limits was necessary.

\subsection{Statistical uncertainty from a finite number of
simulations}

Statistical uncertainty from a finite number of simulations is
estimated for both strain and energy upper limits via the bootstrap
method discussed in Section\,\ref{section:bootstrap} with $N=200$
ensembles.

\section{Closed box results for individual bursts}

We performed extensive ``closed box'' searches with both Flare
pipeline and cWB pipeline.  The
results were used to perfect the pipelines, tune pipeline
parameters, and review the pipelines before opening the box.  The
Flare pipeline was shown to be significantly more sensitive, on
average, than cWB to the waveforms studied in the search.

In order to keep the box closed, we ran searches on times near
to, but offset from, the actual event trigger times.  In other respects the searches were identical to real searches.  For a randomly selected set of events which include single, double,
and triple coincidence events, we performed two closed-box searches
for every actual event, for each of the two pipelines.  One search was run using a time 580~s before
the actual trigger, and the other was run using a time 580~s after
the trigger.  These searches used one-sided background (off-source)
data regions, providing a consistency check between search separations of only 1160~s (about 20 minutes).  Since these were
loudest event searches, there was variation caused by fluctuations in loudest events
used as efficiency curve thresholds.  Also, differences
in rms antenna factors can be significant over 20 minutes.  In
the group of six randomly chosen events the largest antenna factor
change over 20 minutes was about 13\%.

\section{Open box results for individual bursts}

For each of the three search bands 100--200\,Hz, 100--1000\,Hz and
1--3\,kHz, we searched a total of 191 on-source regions, for a total
of 803\,s of on-source data in the sample.  Twelve simulation types
were used to set upper limit estimates within these three bands, for
$12 \times 191 = 2292$ unique upper limit estimates.

No on-source analysis event was found to have a false alarm rate
(estimated from the background) less than $\sci{1.35}{-3}$\,Hz (1
per 741\,s), which is consistent with the expectation for the 799\,s
of on-source data in the sample. We thus find no evidence for
gravitational waves associated with any of the SGR burst events in
the sample.  A significancegram for the on-source region with the
loudest on-source analysis event is given in
Figure\,\ref{fig:mostSignificant}.  A rate versus loudness plot for
this on-source region with its corresponding background region is
given in Figure\,\ref{fig:mostSignificantRate}.  We have estimated
strain and energy upper limits $\hrssn$ and $\egwn$ using the
loudest on-source analysis event for each SGR burst. Upper limits
depend on detector sensitivity and antenna factors at the time of
the burst, the loudest on-source analysis event, and the simulation
waveform type used.

\begin{figure}[!t]
\begin{center}
\includegraphics[angle=0,width=130mm, clip=false]{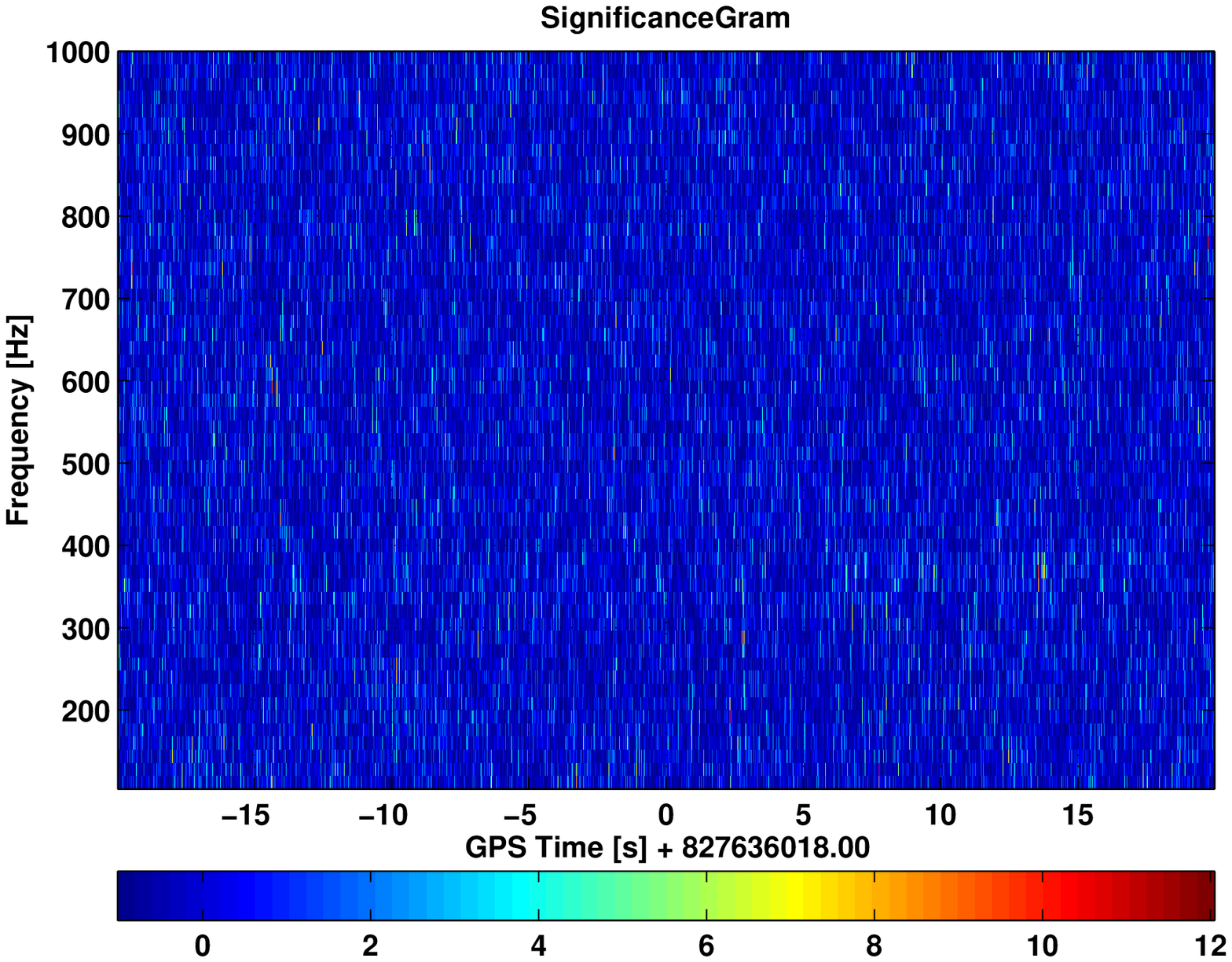}
\caption[Significancegram for the loudest on-source analysis event ]
{ Significancegram for the SGR 1900+14 storm event, the on-source
region with the loudest on-source analysis event.  The loudest
on-source event occurs at GPS 827636031.70 or 2006 March 29
02:53:37.70 UTC (13.7\,s after the center of the on-source region),
and has a duration of 290\,ms, a central frequency of 369\,Hz, and a
bandwidth of 32\,Hz.  It is comprised of 33 tiles and is faintly
visible in this significancegram.} \label{fig:mostSignificant}
\end{center}
\end{figure}

\begin{figure}[!t]
\begin{center}
\includegraphics[angle=0,width=130mm, clip=false]{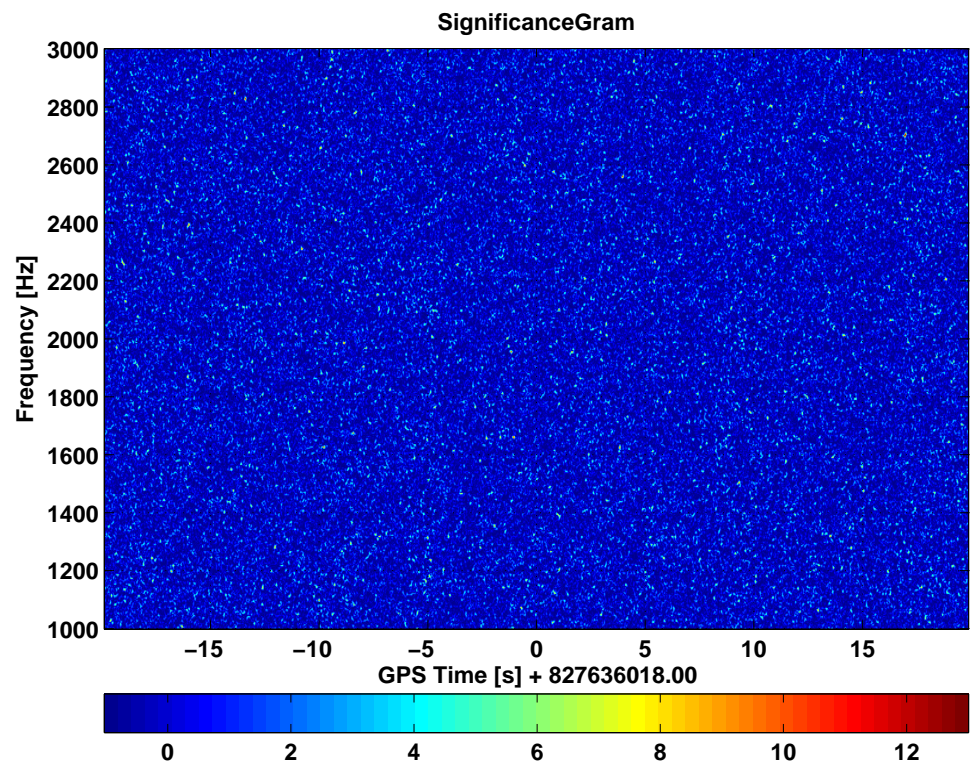}
\caption[Significancegram for the SGR 1900+14 storm event 1--3\,kHz]
{ Significancegram for the SGR 1900+14 storm event, the on-source
region with the loudest on-source analysis event, in the 1--3\,kHz
band. The loudest on-source event in this band occurs at GPS
827636034.95 or 2006 March 29 02:53:40.95 UTC (16.9\,s after the
center of the on-source region), and has a duration of 100\,ms, a
central frequency of 2702\,Hz, and a bandwidth of 12\,Hz. It is
comprised of 10 tiles and is faintly visible in this significance
gram.} \label{fig:mostSignificantHighBand}
\end{center}
\end{figure}

\begin{figure}[!t]
\begin{center}
\includegraphics[angle=0,width=110mm, clip=false]{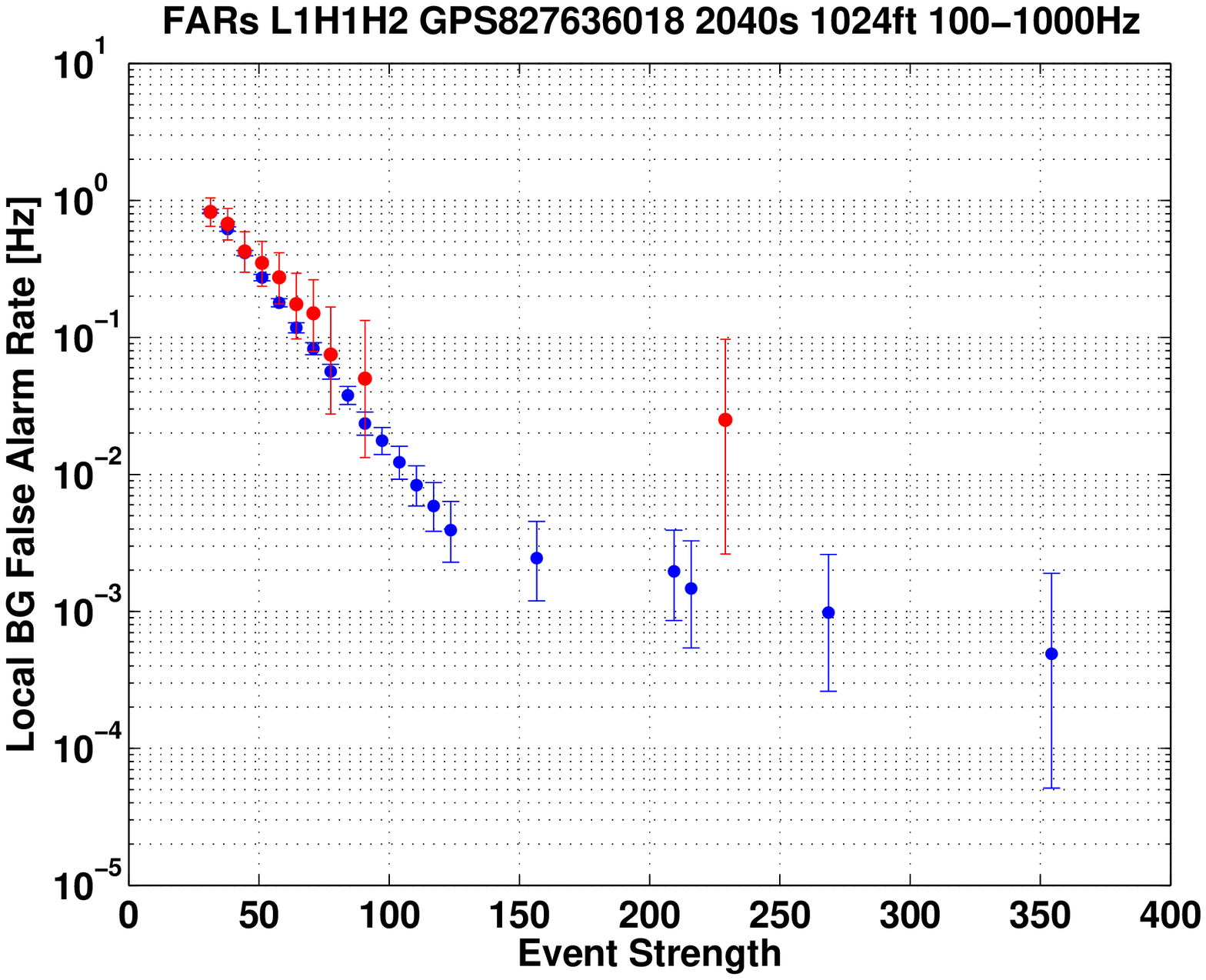}
\caption[Rate versus loudness plot for the loudest on-source
analysis event] {Rate versus loudness plot for the SGR 1900+14 storm
event, the on-source region with the loudest on-source analysis
event. Black points give the cumulative histogram for the
background, while red points give the cumulative histogram for the
on-source region.} \label{fig:mostSignificantRate}
\end{center}
\end{figure}

Complete upper limit results are listed in Table\,\ref{table:results}.) Table\,\ref{table:bestresults} lists upper limits for the SGR
1806$-20$ giant flare and the 060806 event (SGR 1806$-20$ burst
series)\,\cite{gcn5426}. These selected bursts have small values of
$\gamma = \egwn/E_{\mathrm{EM}}$
(Section\,\ref{section:sgrEmissionOfGW}).  At the time of the giant
flare maximum (2004 December 27 21:30:26.643 UTC) the LIGO Hanford
4\,km detector was taking data during a commissioning period (LIGO
Astrowatch) and had noise amplitude higher than during S5 by a
factor of $\sim$3; rms antenna factor for SGR 1806$-20$ was
$(F_{+}^2 + F_{\times}^2)^{1/2}=0.3$. The S5 event began at
2006 August 6 14:23:39 UTC; the two Hanford detectors were observing, with
rms antenna factor for SGR 1806$-20$ of 0.5. Times are for wavefront
arrival at the Hanford detectors. Isotropic electromagnetic energies
for the events, assuming a distance of 10\,kpc, were
$\sci{1.6}{46}$\,erg\,\cite{hurley05} and at least
$\sci{2.9}{42}$\,erg\,\cite{gcn5426}, respectively.

Superscripts in Table\,\ref{table:bestresults} and
Table\,\ref{table:results} give a systematic error and
uncertainties at 90\% confidence. The first and second superscripts
account for systematic error and statistical uncertainty in
amplitude and phase of the detector calibrations, estimated via
Monte Carlo simulations, respectively. The third is a statistical
uncertainty arising from using a finite number of injected
simulations, estimated with the bootstrap method using 200
ensembles\,\cite{efron79}.  The systematic error and the quadrature
sum of the statistical uncertainties are added to the upper limit
estimates. Figure\,\ref{fig:egw90All} shows $\egwn$ and $\hrssn$
upper limits for the waveforms considered, for the entire SGR burst
sample. The lowest upper limit in the sample, $\egwn=\sci{2.9}{45}$
erg, was obtained for an SGR 1806$-20$ burst at 2006 July 21
17:10:56.6 UTC.  Efficiency curves for this upper limit are given in
Figure\,\ref{fig:bestEfficiency}.

\begin{longtable}{@{\extracolsep{\fill}}l|r@{}llr@{}l|r|c}
\caption[$\hrssn$ and $\egwn$ for SGR 1806$-20$ giant flare and the
060806 event]{Gravitational wave strain and energy upper limit
estimates at 90\% detection efficiency ($\hrssn$ and $\egwn$) for
the SGR 1806$-20$ giant flare and the S5 SGR burst with the smallest
limits on the ratio $\gamma = \egwn/E_{\mathrm{EM}}$ for various
circularly/linearly polarized RD (RDC/RDL) and white noise burst
(WNB) simulations, and 4\,s on-source regions. Uncertainties (given
in superscripts for strain upper limits and explained in the text)
are folded into the final limit estimates.  The fluences used to
calculate $\gamma$ values for the brightest and second brightest
peaks of the 060806 event were $\sci{2.4}{-4}$\,erg\,cm$^{-2}$ and
$\sci{6.0}{-5}$\,erg\,cm$^{-2}$ respectively (Section\,\ref{section:060806}). } \\
    \hline \hline
 & \multicolumn{7}{c}{SGR 1806$-20$ Giant Flare} \\
 Simulation type & \multicolumn{5}{c}{$ \hrssn [ 10^{-22}~ \rthz $] }  & $\egwn$ [erg] & $\gamma$  \\
 \hline
 \endfirsthead

 \multicolumn{7}{c}{{\bfseries \tablename\ \thetable{} -- continued from previous page}} \\
 \hline
 Simulation type & \multicolumn{5}{c}{$ \hrssn [ 10^{-22}~ \rthz $] }  & $\egwn$ [erg] & $\gamma$  \\
 \hline
 \endhead
 \hline \multicolumn{7}{|r|}{{Continued on next page}} \\
 \hline
 \endfoot
 \hline \hline
 \endlastfoot

 WNB 11ms  100-200 Hz   & 22& & $^{ +1.3 ~ +5.6 ~ +1.2}$ &  $= 29$ &  & $\sci{7.3}{47}$ & $\sci{5}{1}$  \\
 WNB 100ms  100-200 Hz     & 18& & $^{ +1.1 ~ +4.6 ~ +0.5}$ &  $= 24$ & & $\sci{4.9}{47}$ & $\sci{3}{1}$ \\
 WNB 11ms  0.1-1 kHz   & 50& & $^{ +3.0 ~ +13 ~ +1.3}$ &  $= 66$ & & $\sci{5.4}{49}$ & $\sci{3}{3}$  \\
 WNB 100ms  0.1-1 kHz   & 45& & $^{ +2.7 ~ +12 ~ +1.1}$ &  $= 59$ & & $\sci{3.7}{49}$ & $\sci{2}{3}$ \\
 RDC 200ms  1090 Hz   & 59& & $^{ +3.6 ~ +15 ~ +1.7}$ &  $= 78$ &  & $\sci{2.6}{50}$ & $\sci{2}{4}$ \\
 RDC 200ms  1590 Hz   & 93& & $^{ +5.6 ~ +24 ~ +2.8}$ &  $= 120$ & & $\sci{1.4}{51}$ & $\sci{9}{4}$ \\
 RDC 200ms  2090 Hz   & 120& & $^{ +7.4 ~ +32 ~ +3.5}$ &  $= 160$ & & $\sci{4.2}{51}$ & $\sci{3}{5}$ \\
 RDC 200ms  2590 Hz   & 150& & $^{ +9.1 ~ +39 ~ +4.1}$ &  $= 200$ &  & $\sci{9.8}{51}$ & $\sci{6}{5}$ \\
 RDL 200ms  1090 Hz   & 170& & $^{ +10 ~ +44 ~ +36}$ &  $= 240$ &  & $\sci{2.6}{51}$ & $\sci{2}{5}$ \\
 RDL 200ms  1590 Hz   & 260& & $^{ +16 ~ +68 ~ +32}$ &  $= 360$ &  & $\sci{1.2}{52}$ & $\sci{7}{5}$ \\
 RDL 200ms  2090 Hz   & 390& & $^{ +23 ~ +99 ~ +46}$ &  $= 520$ &  & $\sci{4.4}{52}$ & $\sci{3}{6}$ \\
 RDL 200ms  2590 Hz   & 440& & $^{ +26 ~ +110 ~ +63}$ &  $= 600$ &  & $\sci{8.9}{52}$ & $\sci{6}{6}$ \\
\hline \hline
& \multicolumn{7}{c}{SGR 1806$-20$ 060806 Event Main Peak} \\
\hline

 WNB 11ms 100-200Hz & $3.4$ && $^{ +0.0 ~ +0.4 ~ +0.2}$ &  $=3.8$ &  & $\sci{1.3}{46}$ & $\sci{4}{3}$ \\
 WNB 100ms 100-200Hz  & $2.9$ && $^{ +0.0 ~ +0.3 ~ +0.1}$ &  $=3.3$ &  & $\sci{9.1}{45}$ & $\sci{3}{3}$ \\
 WNB 11ms 100-1000Hz  & $7.5$ && $^{ +0.0 ~ +0.8 ~ +0.3}$ &  $=8.3$ &  & $\sci{8.3}{47}$ & $\sci{3}{5}$ \\
 WNB 100ms 100-1000Hz & $7.0$ && $^{ +0.1 ~ +0.7 ~ +0.2}$ &  $=7.9$ &  & $\sci{6.8}{47}$ & $\sci{2}{5}$ \\
 RDC 200ms 1090Hz &  $10$ && $^{ +0.2 ~ +1.1 ~ +0.4}$ &  $=12$ &  & $\sci{5.8}{48}$ & $\sci{2}{6}$ \\
 RDC 200ms 1590Hz  & $15$ && $^{ +0.6 ~ +1.5 ~ +0.5}$ &  $=17$ &  & $\sci{2.5}{49}$ & $\sci{8}{6}$ \\
 RDC 200ms 2090Hz  & $20$ && $^{ +1.6 ~ +2.5 ~ +0.6}$ &  $=24$ &  & $\sci{8.9}{49}$ & $\sci{3}{7}$ \\
 RDC 200ms 2590Hz  & $24$ && $^{ +3.1 ~ +3.0 ~ +0.9}$ &  $=30$ &  & $\sci{2.2}{50}$ & $\sci{7}{7}$ \\
 RDL 200ms 1090Hz  & $33$ && $^{ +1.0 ~ +3.4 ~ +3.5}$ &  $=39$ &  & $\sci{6.7}{49}$ & $\sci{2}{7}$ \\
 RDL 200ms 1590Hz  & $44$ && $^{ +2.2 ~ +4.6 ~ +6.3}$ &  $=54$ &  & $\sci{2.8}{50}$ & $\sci{9}{7}$ \\
 RDL 200ms 2090Hz  & $64$ && $^{ +7.0 ~ +8.1 ~ +9.1}$ &  $=83$ &  & $\sci{1.1}{51}$ & $\sci{4}{8}$ \\
 RDL 200ms 2590Hz  & $79$ && $^{ +10 ~ +10 ~ +9.7}$ &  $=100$ &  & $\sci{2.6}{51}$ & $\sci{9}{8}$ \\

\hline \hline
& \multicolumn{7}{c}{SGR 1806$-20$ 060806 Event Initial Peak} \\
\hline

WNB 11ms 100-200Hz  & $3.4$ && $^{ +0.0 ~ +0.3 ~ +0.2}$ &  $=3.8$ &  & $\sci{1.3}{46}$ & $\sci{2}{4}$ \\
WNB 100ms 100-200Hz  & $2.8$ && $^{ +0.0 ~ +0.3 ~ +0.1}$ &  $=3.2$ &  & $\sci{8.7}{45}$ & $\sci{1}{4}$ \\
WNB 11ms 100-1000Hz  & $7.1$ && $^{ +0.0 ~ +0.7 ~ +0.2}$ &  $=7.9$ &  & $\sci{7.5}{47}$ & $\sci{1}{6}$ \\
WNB 100ms 100-1000Hz  & $7.1$ && $^{ +0.1 ~ +0.7 ~ +0.2}$ &  $=8.0$ &  & $\sci{7.0}{47}$ & $\sci{8}{5}$ \\
RDC 200ms 1090Hz  & $9.7$ && $^{ +0.2 ~ +1.0 ~ +0.4}$ &  $=11$ &  & $\sci{5.3}{48}$ & $\sci{7}{6}$ \\
RDC 200ms 1590Hz  & $14$ && $^{ +0.5 ~ +1.4 ~ +0.5}$ &  $=16$ &  & $\sci{2.3}{49}$ & $\sci{3}{7}$ \\
RDC 200ms 2090Hz  & $20$ && $^{ +1.6 ~ +2.5 ~ +0.8}$ &  $=24$ &  & $\sci{9.2}{49}$ & $\sci{1}{8}$ \\
RDC 200ms 2590Hz  & $23$ && $^{ +3.0 ~ +3.0 ~ +0.8}$ &  $=29$ &  & $\sci{2.1}{50}$ & $\sci{3}{8}$ \\
RDL 200ms 1090Hz  & $27$ && $^{ +0.8 ~ +2.8 ~ +3.4}$ &  $=32$ &  & $\sci{4.6}{49}$ & $\sci{6}{7}$ \\
RDL 200ms 1590Hz  & $40$ && $^{ +2.0 ~ +4.1 ~ +6.3}$ &  $=50$ &  & $\sci{2.3}{50}$ & $\sci{3}{8}$ \\
RDL 200ms 2090Hz  & $54$ && $^{ +6.0 ~ +7.0 ~ +7.5}$ &  $=71$ &  & $\sci{8.0}{50}$ & $\sci{1}{9}$ \\
RDL 200ms 2590Hz  & $58$ && $^{ +7.5 ~ +7.4 ~ +9.2}$ &  $=77$ &  & $\sci{1.5}{51}$ & $\sci{2}{9}$ \\

 \hline \label{table:bestresults}
 \end{longtable}

\begin{figure}[!t]
\begin{center}
\includegraphics[angle=0,width=120mm,clip=false]{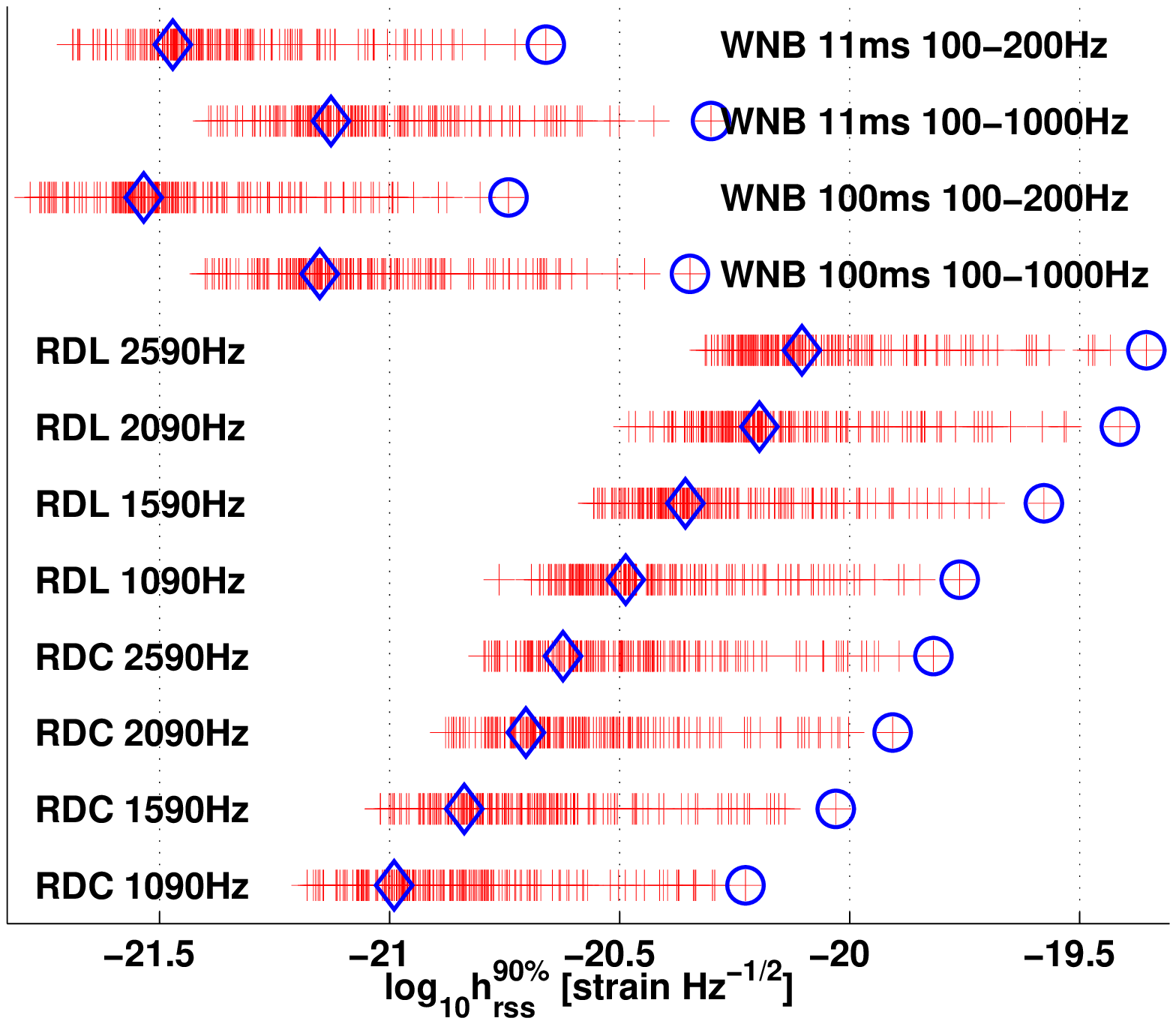}
\caption[$hrssn$ upper limits for the entire SGR burst sample]{
$\hrssn$ upper limits for the entire SGR burst sample (giant flare
and S5y1) for various circularly/linearly polarized RD (RDC/RDL) and
white noise burst (WNB) simulations.  The limits shown in
Table\,\ref{table:bestresults}, for the giant flare and the 060806
event, are indicated in the figure by circles and diamonds,
respectively.} \label{fig:egw90All}
\end{center}
\end{figure}

\begin{figure}[!t]
\begin{center}
\includegraphics[angle=0,width=120mm,clip=false]{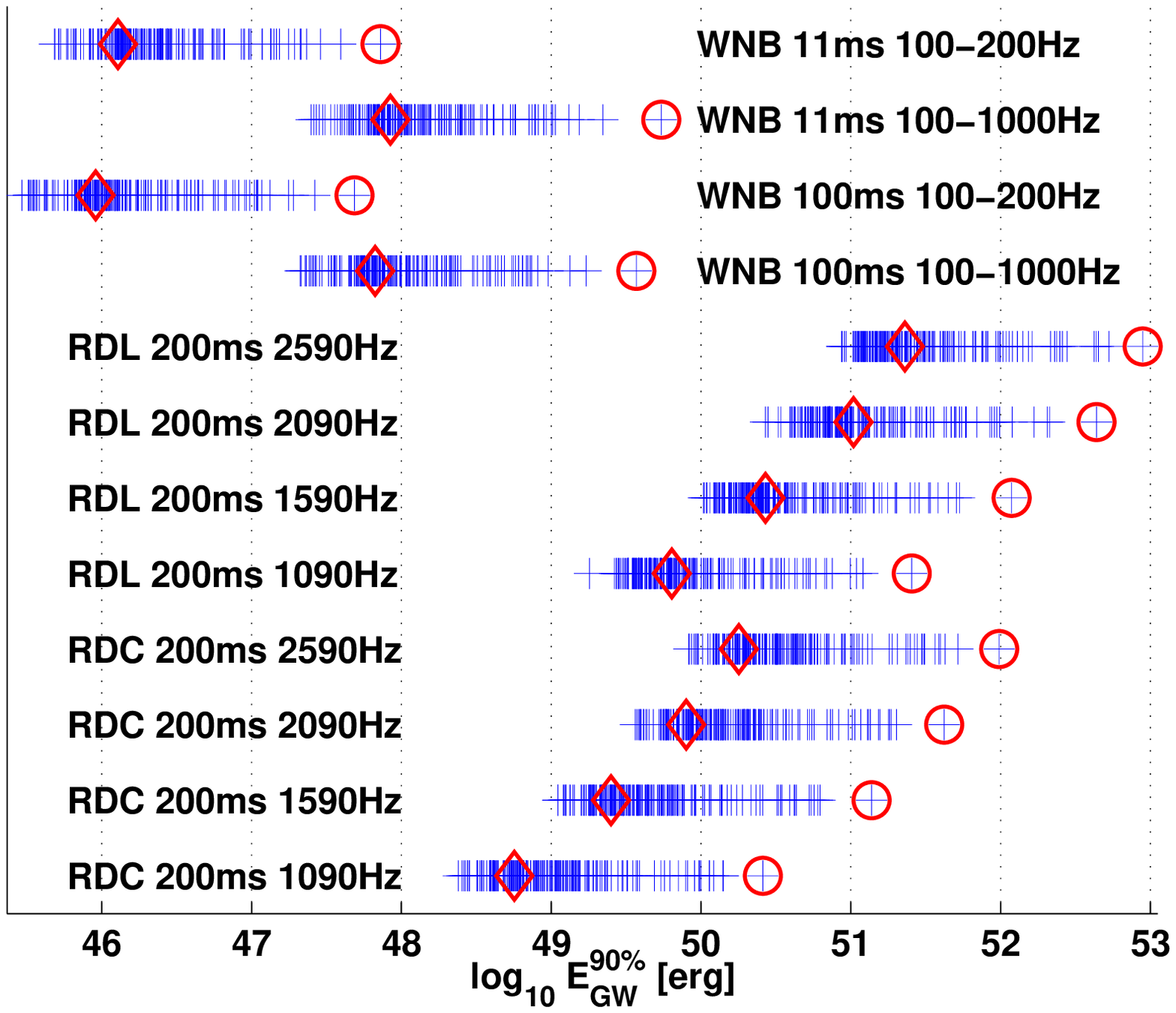}
\caption[$\egwn$ upper limits for the entire SGR burst sample]{
$\egwn$ upper limits for the entire SGR burst sample (giant flare
and S5y1) for various circularly/linearly polarized RD (RDC/RDL) and
white noise burst (WNB) simulations.  The limits shown in
Table\,\ref{table:bestresults}, for the giant flare and the 060806
event, are indicated in the figure by circles and diamonds,
respectively.} \label{fig:egw90All}
\end{center}
\end{figure}

\begin{figure}[!t]
\begin{center}
\subfigure{
\includegraphics[angle=0,width=100mm,clip=false]{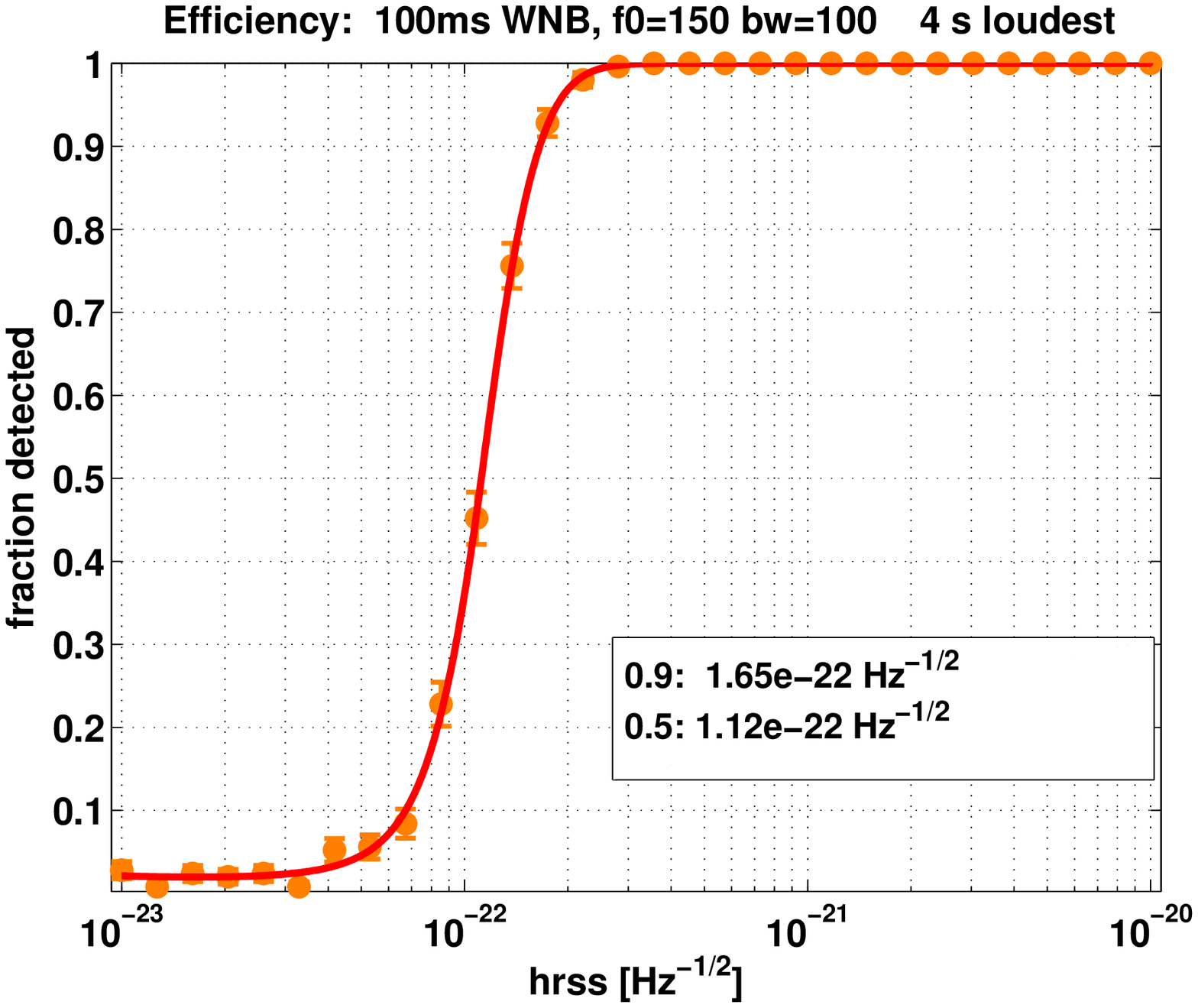}}
\subfigure{
\includegraphics[angle=0,width=100mm,clip=false]{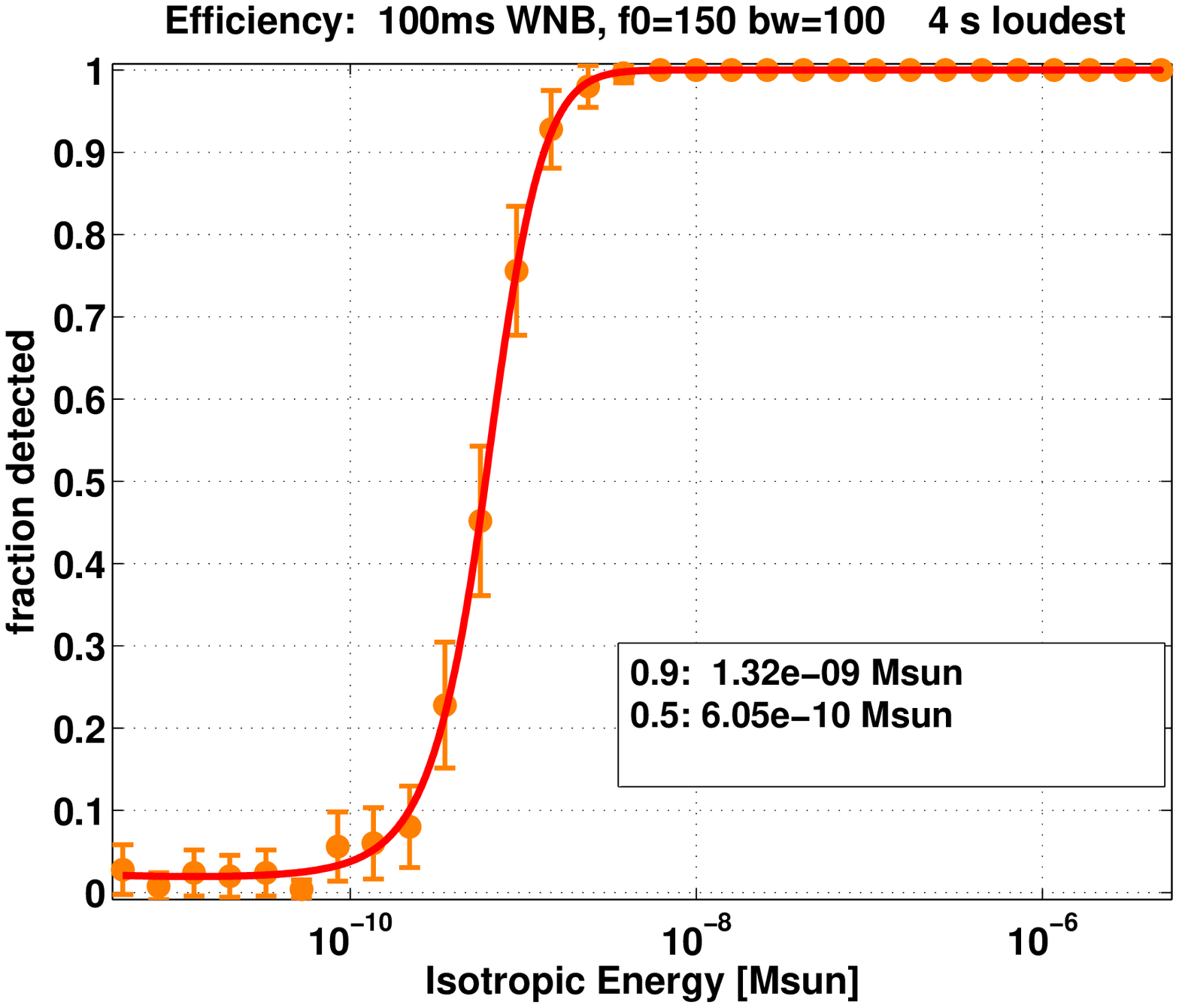}}
\caption[$\egw$ and $\hrss$ efficiency curves for the lowest upper
limit]{ \textbf{(a)} $\hrss$ and \textbf{(b)}$\egw$ efficiency
curves for the lowest upper limit in the entire sample, for an SGR
1806$-20$ burst at 2006 July 21 17:10:56.6 UTC.  The lowest upper
limits were for the WNB 100ms 100-200Hz simulation type.  Note that these number have not yet had uncertainties folded into them.}
\label{fig:bestEfficiency}
\end{center}
\end{figure}

\begin{table}[h]
\begin{center}
\caption[The most significant loudest on-source analysis events]{The
twenty most significant loudest on-source analysis events from
amongst the on-source regions searched.  For every SGR burst trigger
there was an on-source region, which was analyzed three times with
different search bands and values of Fourier transform length. The
search bands were 1--3\,kHz for RDs, 100-1000\,Hz for large band
WNBs and 100-200\,Hz for small band WNBs.  The Fourier transform
lengths were 1/4\,s for RDs and 1/16\,s for WNBs. Event times given
to a tenth of a second correspond to listed events timed by
Konus-Wind, whose light crossing times were propagated from the
satellite to the geocenter. If there is a GCN report describing the
trigger event it is given.  About 50 listed burst events were also
mentioned in the GCN reports.  Table\,\ref{table:brightestbursts}
lists the 10 bursts out of these $\sim$50 for which fluences were
explicitly given.  Two of those bright bursts appear here. }
\begin{small}
\begin{tabular}{rcllccc}

 \hline
 & \textbf{GCN} & \textbf{Trigger UTC time} & \textbf{GPS} & \textbf{Network} & \textbf{Band [Hz]} & \textbf{FAR [Hz]} \\
 \hline

 1 & GCN 4946 & Mar 29 2006 02:53:24  & 827636018 & L1H1H2 & 100--1000 & 1.35e-03 \\
 2 & GCN 5490 & Aug 30 2006 03:16:47  & 840943021 & L1 & 100--200 &  1.51e-03 \\
 3 & GCN 5490 & Aug 30 2006 03:16:47  & 840943021 & L1 & 1000--3000 & 1.55e-03 \\
 4 & GCN 5490 & Aug 30 2006 03:16:47  & 840943021 & L1 & 100--1000 &  2.03e-03 \\
 5 &  & Aug 10 2006 13:42:39  & 839252573 & L1H1H2 & 1000--3000 &  2.48e-03 \\
 6 &  & Mar 29 2006 02:49:42  & 827635796 & L1H1H2 & 100--1000 & 2.80e-03 \\
 7 &  & Aug 03 2006 11:13:51  & 838638845 & H1H2 & 100--200 &  3.27e-03 \\
 8 &  & Oct 27 2006 15:20:53  & 845997667 & H1H2 & 1000--3000 &  3.71e-03 \\
 9 &  & Aug 12 2006 21:51:12  & 839454686 & L1H1H2 & 100--200 & 3.82e-03 \\
 10 &  & Aug 07 2006 01:03:21.0  & 838947815.0 & H1H2 & 1000--3000 &  4.21e-03 \\
 11 &  & Aug 23 2006 15:03:16  & 840380610 & L1H1H2 & 100--1000 & 5.47e-03 \\
 12 & GCN 4946 & Mar 29 2006 02:53:24  & 827636018 & L1H1H2 & 1000--3000 &  5.67e-03 \\
 13 & GCN 5490 & Aug 29 2006 21:57:43.2  & 840923877.2 & L1 & 100--200 &  5.83e-03 \\
 14 & GCN 5490 & Aug 26 2006 04:25:25  & 840601539 & H1H2 & 100--200 &  7.20e-03 \\
 15 &  & Sep 30 2006 06:03:33  & 843631427 & L1H1H2 & 1000--3000 &  7.47e-03 \\
 16 &  & Aug 21 2006 16:17:27  & 840212261 & H1H2 & 1000--3000 &  7.51e-03 \\
 17 &  & Aug 23 2006 15:03:16  & 840380610 & L1H1H2 & 100--200 &  8.20e-03 \\
 18 &  & Mar 29 2006 04:39:46  & 827642400 & L1H1H2 & 100--1000 &  8.92e-03 \\
 19 & GCN 5426 & Jul 28 2006 08:21:36.7  & 838110110.7 & L1H1H2 & 1000--3000 &  9.70e-03 \\
 20 & GCN 4312 & Dec 03 2005 11:43:28.3  & 817645421.3 & L1H1H2 & 100--200 &  9.92e-03 \\

\end{tabular}
\end{small}
\label{table:timingError}
\end{center}
\end{table}

\clearpage
\chapter[Search for GWs from Multiple SGR Bursts]{Search for Gravitational Waves from Multiple SGR Bursts}
\label{chapter:stack}

As we have seen, SGRs have unique properties that make them intriguing gravitational wave targets.  They are nearby, their burst emission  mechanism may involve neutron star crust fractures and excitation of non-radial modes, and they burst repeatedly and sometimes spectacularly.

In Chapter\,\ref{chapter:search} we described a search for transient gravitational waves associated with almost 200 individual electromagnetic SGR triggers.  That search did not detect gravitational waves, but it did place the most stringent upper limits on transient gravitational wave amplitudes at the time it was published, and set isotropic emission energy upper limits that fell 
within the theoretically predicted range of some SGR models.  

In this chapter we extend that work and describe a search method for gravitational waves from multiple SGR bursts.  
The method builds upon the Flare pipeline described in Chapter\,\ref{chapter:flare} by attempting to ``stack" potential 
gravitational wave signals from multiple SGR bursts.  We assume that variation in the time difference between the peak in electromagnetic emission and the peak in potential gravitational 
wave emission in SGR bursts is small relative to the  gravitational wave signal duration, and we time-align gravitational wave excess power time-frequency tilings containing individual 
burst triggers to their corresponding electromagnetic peaks.    We plan to use this method in the near future to perform out gravitational wave searches which we believe will offer a 
significantly higher chance for a first detection than the individual burst method used in Chapter\,\ref{chapter:search} and \,\cite{s5y1sgr}.

This chapter is organized as follows.  In Section\,\ref{section:stackStrategy} we discuss aspects of the multiple SGR burst search strategy.  In Section\,\ref{section:stackMethod} we  describe two complementary incarnations of the new analysis pipeline (``Stack-a-flare''), both of which are built upon
the Flare pipeline.   We then characterize the two methods using simulations in
white noise, demonstrating the strengths and weaknesses of each, and showing that relatively weak signals which could not be detected in the individual burst search can
easily be detected in the multiple burst search.  We show that gains in gravitational wave energy sensitivity of $N^{1/2}$ are feasible, where $N$ is the number of stacked SGR bursts.  Finally, in Section\,\ref{section:stacksim}
we present estimated search sensitivities for a simulated search for gravitational waves from the SGR 1900+14 storm of 2006 March 29, for two stacking scenarios: the ``fluence-
weighted" scenario and the ``flat" (unweighted) scenario.

\section{Strategy} \label{section:stackStrategy}

The major goals of the multiple SGR search are the same as those of
the individual search\,\cite{s5y1sgr} upon which it is based: make a detection statement, set
upper limits, and place these results into an astrophysical context.  However, we hope to improve the search sensitivity by combining potential
gravitational wave signals from separate bursts in an attempt to
increase the signal-to-noise ratio, increasing the probability of detection and placing more stringent constraints on theoretical models via upper limits. In this section we outline the strategy of the multiple SGR search
and discuss choices needed to achieve these goals.  Because this search is an extension of the individual SGR burst search with the same goals, many key 
decisions will be the same.

\subsection{Search signal parameter space}

We do not make any new assumptions about the nature of individual bursts of gravitational waves from SGRs.  Therefore, as in the individual SGR search, the multiple SGR search will target neutron star fundamental mode
ringdowns (RDs) predicted in\,\cite{thorne83,
andersson97, pacheco98, ioka01, andersson02} as well as unmodeled
short-duration gravitational wave signals.   As in Chapter\,\ref{chapter:search} and \,\cite{s5y1sgr}, we correspondingly focus on two 
distinct regions in the target signal time-frequency parameter space:  $\sim$100-400\,ms duration signals in the 1--3\,kHz band, which includes $f$-mode ringdown (RD) signals  predicted in\,\cite{benhar04}
for ten realistic neutron star equations of state; and $
\sim$5--200\,ms duration signals in the 100--1000\,Hz band.   We again choose a search band of 1--3\,kHz for RD searches, with a 250\,ms time window which we found to give optimal search sensitivity (see Section\,\ref{sec:ringdownparams}).  The search for unmodeled signals uses time windows set by prompt SGR burst timescales (5--200\,ms) and frequency bands
set by the detector's sensitivity.  We again search in two
bands: 100--200\,Hz (probing the region in which the detectors are
most sensitive) and 100--1000\,Hz (for full spectral coverage below the ringdown search band) using a 125\,ms time window (see Section\,\ref{sec:unmodeledparams}).

We also use the same twelve simulated waveform types used for setting upper limits 
in the individual SGR burst search, described in Chapter\,\ref{chapter:search} and\,\cite{s5y1sgr}:  linearly and circularly polarized RDs with $\tau=200$\,ms and frequencies in the range 1--3\,kHz; and band- and time-limited white noise bursts (WNBs) with durations of 11\,ms and 100\,ms and frequency bands matched to the two low frequency search bands.  Polarization angle is chosen randomly for every injection.

It seems plausible to assume that, for a given neutron star, $f$-mode
frequencies and damping timescales would be similar from event to
event.    However, the major motivation for the low frequency unmodeled portion of the search is
stochastic gravitational wave emission arising from violent events
in the neutron star crust.  Therefore, we will not
assume similar waveforms from event to event in the unmodeled
search, although we will  assume similar central
frequencies and durations.

\subsection{On-source region} \label{section:stackOnSource}

As in\,\cite{s5y1sgr}, we divide the gravitational wave data into an on-source
time region, in which gravitational waves associated with a given burst could be
expected, and a background time region.   On-source and background segments are
analyzed identically, including data quality cuts, resulting in
lists of ``analysis events.''  

For the individual SGR search, there was no need for
millisecond timing precision for the event trigger times.
Precision on the order of a second led us to 4\,s on-source
regions which did not degrade upper limit results significantly. For
a multiple SGR search, significantly higher precision in \emph{relative} trigger times between burst events in the stack will be required.  A common bias in trigger times shared by all bursts in the stacking set can be handled with an adequately large (e.g. 4\,s) on-source region, as before.
In the individual SGR search, imprecision in trigger times came primarily from two sources: satellite
to geocenter light crossing delay and arbitrariness of the satellite
trigger point in the light curve.  If necessary, light crossing times at the
satellites can be propagated to the geocenter (and subsequently to
any given interferometer) using the appropriate ephemeris.  If satellite data is public, we can also obtain light curves and
produce trigger times standardized to a specific point in the light
curve (e.g. start of the steep rise or the peak itself).  This latter procedure would probably dominate the timing uncertainty budget.

Increased timing precision could allow us to use smaller on-source regions
with durations set by theoretical predictions of time delay between
electromagnetic and gravitational wave emission from SGR bursts.
For the individual SGR search, the limit we placed for such a delay
was on the order of 100\,ms, which was insignificant compared to the
padding built into the on-source region duration due to timing
uncertainties.  If timing
uncertainties can be reduced to the millisecond level, then on-source regions could potentially be reduced to this scale. However, this could exclude some models with larger timing delays, and it turns out there is little benefit to be gained.  We have performed Monte Carlo simulations comparing $\pm2$\,s and $\pm1$\,s on-source regions.  Reducing the on-source region from 4\,s to 2\,s resulted in a meager 2\% reduction in amplitude upper limits, on average over 24 trials with various waveform types.

\subsection{Background region}

As with the individual burst search, the background region
serves three purposes:
 \ben
 \i it is used to estimate statistics of the power tiling as a function of
 frequency for use in the Flare pipeline (see\,\ref{chapter:flare});
 \i it provides FAR estimates from which the significance of the loudest
 on-source analysis event can be determined;
  \i it provides a substrate into which simulated waveforms can be
 injected for estimating upper limits.
 \een

 In the course of validating the individual SGR search we showed that 1000\,s of data on either side of an on-source region produce sufficient estimates of the power tiling statistics {(Section\,\ref{sec:offsourceSize}).   This requirement and the estimation procedure are unchanged in the multiple SGR search, so $\pm1000$\,s of data will again suffice for this purpose.   The background region required for injecting simulations to estimate upper limits may depend on the system being modeled and the desired statistical precision;  for the hypothetical SGR 1900+14 ``storm'' search we describe below, $\pm1000$\,s of background is sufficient.  The background region required for FAR estimates depends primarily on the range of FAR estimates desired.  Estimating the FAR of a very large on-source analysis event  requires a larger background than estimating the FAR of a small on-source analysis event, for a given level of precision.

\subsection{Stacking scenarios}
\label{section:groupingAndWeighting}

Two new decisions unique to the multiple burst search are 1) which
bursts to include in the set and 2) how to weight them.  As with the individual burst search, we assume that the SGR burst
sample is comprised of bursts occurring within some specified time
range defined by the observatory's science run schedule.  We will refer to a set of SGR bursts to be included in the multiple burst search, along with a weighting strategy, a ``stacking scenario.''

We could use Occam's razor to select  stacking scenarios, such as the
following:
 \begin{enumerate}[s1.]
 \i use every detected and confirmed burst from a given SGR source within the time range, with equal
 weighting (``flat scenario'');
 \i use every detected and confirmed burst within the time range, from any SGR source, with equal
 weighting (``generic scenario'');
 \i use every detected and confirmed burst from a given SGR source within the time range, weighted proportional to fluence (``proportional scenario'');
 \i use a subset of component bursts from a multi-episodic burst event such as
 the SGR 1900+14 storm, with some weighting scheme.
 \end{enumerate}
 
We note that the so-called generic scenario could benefit from a search method that was insensitive to variations
between SGR sources.  For example, we would expect two different sources to emit from $f$-modes at different frequencies, which may not brighten corresponding pixels in a time-frequency tiling.  The method we describe in this paper does not attempt to solve this problem.

Stacking scenarios based on arguments from theoretical
considerations could also be compelling.  One such scenario could use every detected and
confirmed burst from a given SGR source for which fluence has been
measured, weighted by a model-dependent predicted function of
fluence.  However, theoretical understanding of gravitational wave
emission from SGR bursts will probably need to be significantly
advanced before such a scenario could be implemented.  Furthermore, such
a specific model-dependent choice, while being well-suited to probing its progenitor model, would lead to reduced sensitivity if it
happens to be incorrect.

A theory may predict that there is no
correlation between $\eem$ and $\egw$.  Such a prediction could be implemented with the flat scenario in our search.

A theory may predict that the time delay between
electromagnetic and gravitational emission varies from burst to
burst.  If the predicted variation was greater than the target signal durations of tens or hundreds of milliseconds, it would bely the fundamental
assumption in this search that bursts from a given SGR source emit
gravitational waves similarly from burst to burst.  Although such a prediction 
could potentially be treated by sweeping over some range of time
delays for each burst, we will consider this possibility no further
unless well-founded theoretical predictions are made that indicate it.

We will neglect other considerations which would complicate the multiple burst search, such as:  multiple injections of energy into
a single burst, with possible correlation in gravitational wave emission
energy;  qualitatively different gravitational wave emission in the
case of intermediate flares and common bursts (see
Figure\,\ref{fig:stormMarkedZoom}); beaming issues; and so forth.

\section{Analysis method} \label{section:stackMethod}

Both incarnations of the Stack-a-flare pipeline, ``T-Stack'' and ``P-Stack,''
consist of thin extension layers built around the Flare pipeline (Chapter \,\ref{chapter:flare}).

\subsection{T-Stack incarnation}

The T-Stack pipeline combines burst events in the time domain.  Except for the addition of the
time-domain stacking layer the T-Stack pipeline is the same as the Flare pipeline.

For each of $N$ burst events a trigger time is determined.  For a given gravitational wave
detector,  $N$ time series containing those trigger times are then
aligned to the trigger times, weighted according to antenna factor,
and added together.  The resulting time series (either one or two,
depending on how many detectors are included in the search) are then
fed to the Flare pipeline.

As will be described below, the T-Stack pipeline has the advantage
of achieving  optimal sensitivity in white noise, but the
disadvantage of being sensitive to timing inaccuracies.  This makes
it a potentially viable choice for analyzing multi-episodic events ---
in which a single contiguous 100\,$\mu$s-binned light curve might
provide adequate timing precision --- but a poor choice for
analyzing isolated burst events or incoherent signals such as
band-limited WNBs. 

\begin{figure*}[!t]
\begin{center}
\includegraphics[angle=0,width=140mm, clip=false]{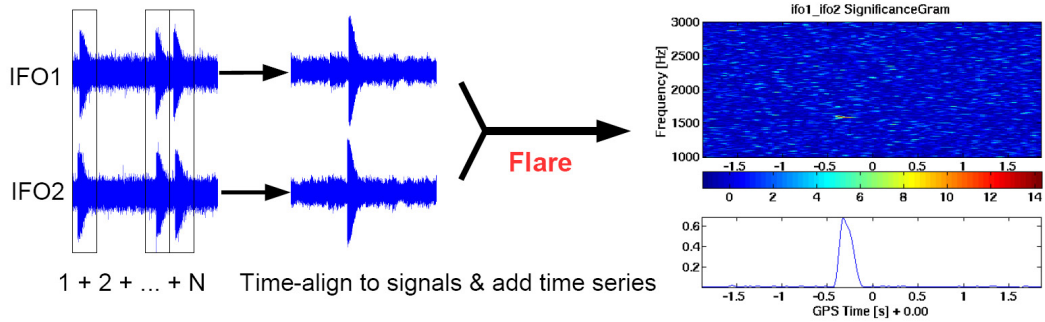}
\caption[Diagram of T-Stack pipeline] { Diagram of the T-Stack
version of the Stack-a-flare pipeline.  The T-Stack pipeline has a
thin layer added before the  Flare pipeline in which
gravitational wave data time series containing SGR burst event
triggers are aligned on the trigger times and added together. These stacked time series are made for each detector and then run through the Flare pipeline as normal.}
\label{fig:tstackFlow}
\end{center}
\end{figure*}

\subsection{P-Stack incarnation}

The P-Stack pipeline combines burst events in the frequency domain.   Except for the addition of the
frequency-domain stacking layer the P-Stack pipeline is the same as the  Flare pipeline.

For each of $N$ burst events a trigger time is determined.  Each of
$N$ timeseries containing those triggers is processed with the Flare pipeline, up to the
clustering algorithm, exactly as in an individual SGR burst search.
Antenna factors are applied at this time. The result is $N$ time-frequency significance tilings. The $N$
significance tilings are then aligned to the trigger time and added
together.  The combined significance tiling is then fed through the
Flare pipeline clustering algorithm with a fixed fraction of tiles to
include in the clustering (e.g. 0.1\%).  A fixed fraction of tiles is used
instead of a fixed loudness threshold value because  the variance of the tile loudness
distribution at a given frequency increases with $N$.  (In fact, clustering on a fixed fraction of tiles may be the better choice for the individual burst search as
well:  computer memory use is predictable even when large instrumental glitches are encountered.)

As will be described below, the P-Stack pipeline has the advantage
of being relatively insensitive to timing inaccuracies or
differences in waveform from burst to burst, but it has less sensitivity than the T-Stack pipeline for the (possibly unrealistic) precisely-known
timing case, with deterministic waveforms. 

\begin{figure*}[!t]
\begin{center}
\includegraphics[angle=0,width=140mm, clip=false]{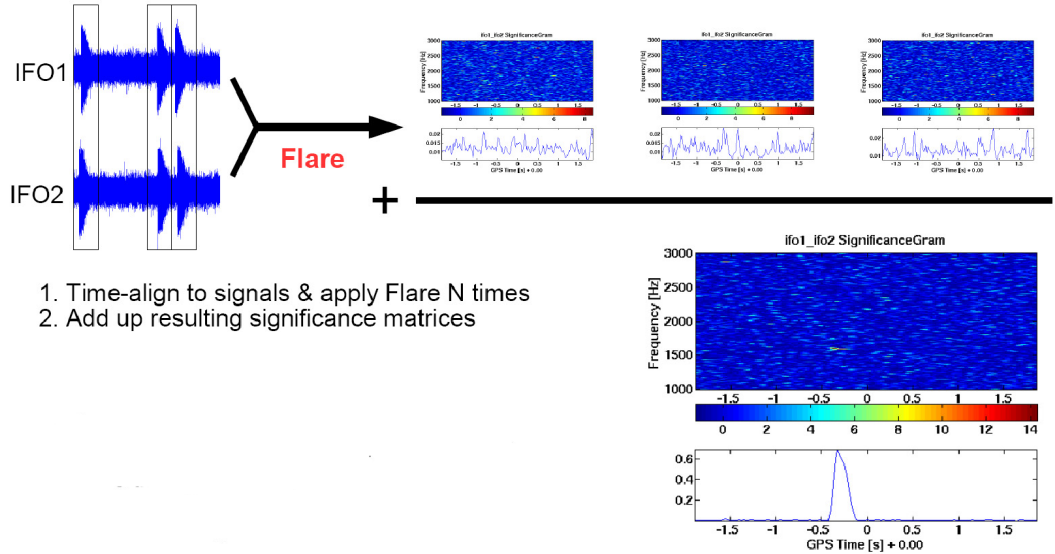}
\caption[Diagram of P-Stack pipeline] { Diagram of the P-Stack
version of the Stack-a-flare pipeline.  The P-Stack pipeline has a
thin layer added \emph{after} the  Flare pipeline in which
gravitational wave data significance tilings containing SGR burst
event triggers are aligned on the trigger times and added together.  Stacked significance tilings can then be run through the Flare pipeline clustering algorithm.
} \label{fig:fstackFlow}
\end{center}
\end{figure*}

\subsection{Loudest event upper limits}

As in the individual SGR search, in the absence of a detection we still estimate
loudest event upper limits\,\cite{brady04} on gravitational wave root-sum-squared strain $\hrss$ incident
at the detector, and gravitational wave energy emitted isotropically from the source assuming a nominal source distance.  

The procedure for estimating loudest event upper limits in the individual burst search is detailed in Section\,\ref{section:upperLimits}.   In brief, the upper limit is computed in a frequentist framework following the commonly used procedure of
injecting simulated signals in the background data and recovering them using
the search pipeline\,(see for example \cite{S2inspiral,S2S3S4GRB}).  An analysis event is associated with each injected simulation, and compared to the loudest on-source analysis event.  The gravitational wave strain or isotropic energy at e.g. 90\% detection efficiency is the strain or isotropic energy at which 90\% of injected simulations have associated events louder than the loudest on-source event.      

We can follow the same procedure for the multiple burst search.  The only difference is the need to measure the $\hrss$ or $\egw$ of a compound injection, instead of a simple (single) injection.

\subsection{Sensitivity dependence on $N$}
\label{section:sensitivityDependenceOnN}

The matched filter amplitude signal-to-noise ratio (SNR) is defined
in the frequency domain as\,\cite{S1inspiral}
 \be
    \rho = \left[ 4 \int_0^\infty \  \frac{\tilde{h}(f)^2}{S_n (f)}
    df \right]^{1/2},
 \ee
where $\tilde{h}(f)$ is the Fourier transform of the signal time
series and $S_n (f)$ is the noise power spectral density.  Here, the
numerator is the square root of the power in the signal.  In white
noise with zero mean, $S_n (f) = \sigma^2$, a constant. Since the
standard deviation $\sigma$ of white noise goes as the square root
of $N$ and the amplitude of identical stacked signals goes as $N$,
we expect the SNR of the optimal T-Stack algorithm for the recovery
of identical signals from noise to go as $N^{1/2}$.

While the T-Stack pipeline stacks amplitude, the P-Stack pipeline
stacks power.  The background tiles in the power tiling at each
individual frequency bin can be modeled as Gamma-distributed noise,
for which the variance also goes as $N$, so we expect the power
signal-to-noise ratio to increase as $N^{1/2}$. Since the amplitude
goes as the square root of the power, we expect the P-Stack
amplitude sensitivity to increase as $N^{1/4}$.

We tested these predictions by  injecting $N$ stacked 1590\,Hz 200\,ms $\tau$ ringdowns into white noise with
$\sigma=1$. We then constructed efficiency curves in the usual
manner, determining the injection $\hrss$ at 50\% and 90\% detection efficiency.  Each
efficiency curve was constructed using 20 amplitude scaling factors
and 20 trials at each $\hrss$ amplitude (see Section \,\ref{section:makingEfficiencyCurves}).  These are bare-bone statistics, but they turned out to be effective at characterizing the Stack-a-flare pipelines.  An example
efficiency curve is shown in Figure\,\ref{fig:stackOfNSample}.

We then fit the 50\% and 90\% detection efficiency level results as
functions of $N$ to a two-parameter power law of the form $y=A N^B$.
The results for both the T-Stack and P-Stack pipelines are shown in
Figure\,\ref{fig:ndepend}.  The fit for the T-Stack pipeline gives a sensitivity dependence
in amplitude at both detection efficiency levels of  nearly
$N^{1/2}$, confirming our prediction.  This corresponds to an improvement in \emph{energy} of a
factor of $N$.  The fit for the P-Stack pipeline gives a sensitivity dependence in
amplitude at both detection efficiency levels of nearly
$N^{1/4}$, confirming our prediction.  This corresponds to an improvement in \emph{energy} of a
factor of $N^{1/2}$.

\begin{figure}[!t]
\begin{center}
\includegraphics[angle=0,width=110mm, clip=false]{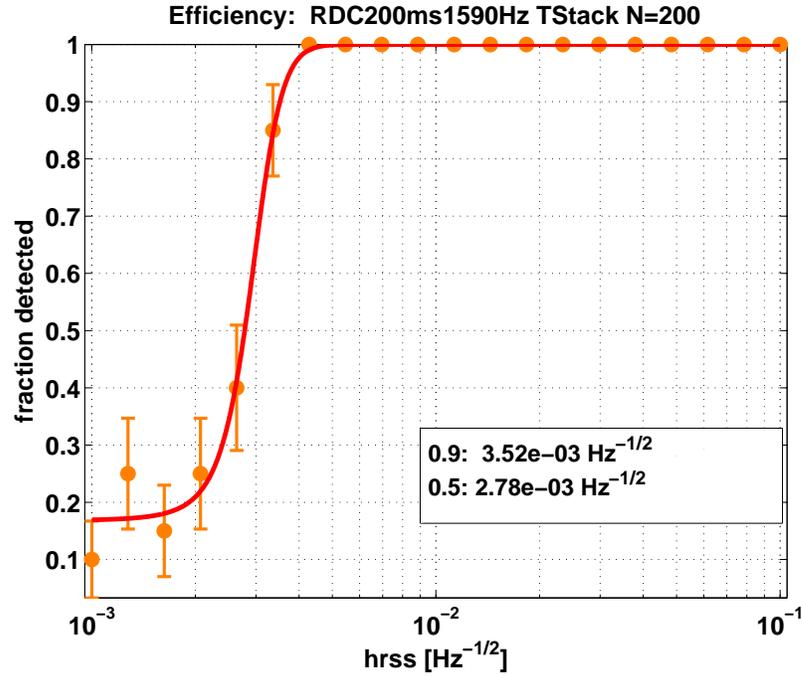}
\caption[Example efficiency curve  for Stack-a-flare sensitivity vs.
$N$] { Example efficiency curve generated for the Monte Carlo experiment investigating Stack-a-flare sensitivity vs.
$N$. This example curve is for $N=200$ T-Stack pipeline
using 1590\,Hz circularly polarized ringdowns.   Each efficiency
curve was constructed using 20 amplitude scaling factors and 20
trials at each $\hrss$ amplitude. } \label{fig:stackOfNSample}
\end{center}
\end{figure}

\begin{figure}[!t]
\begin{center}
\subfigure[]{
\includegraphics[angle=0,width=90mm,clip=false]{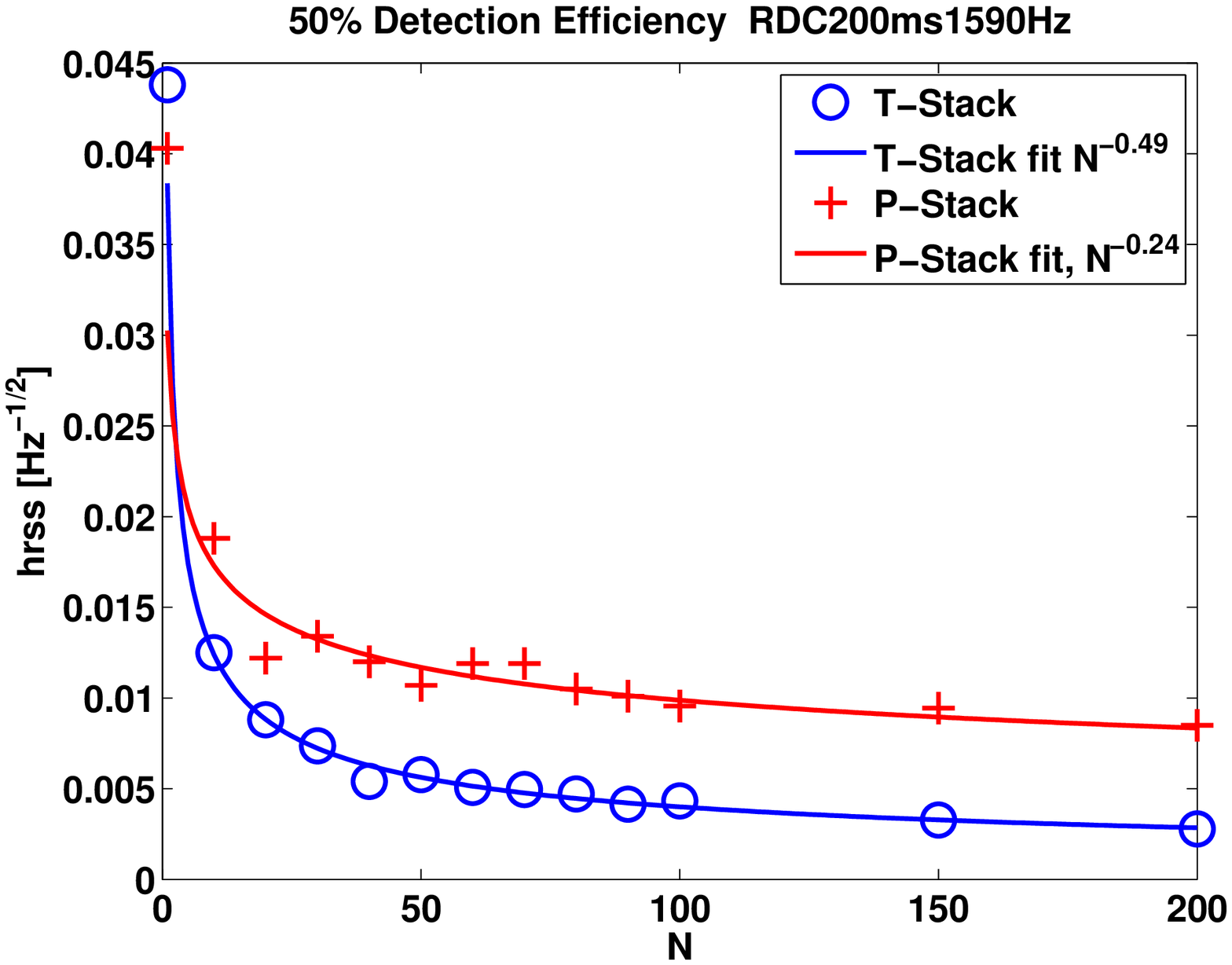}}
\subfigure[]{
\includegraphics[angle=0,width=90mm,clip=false]{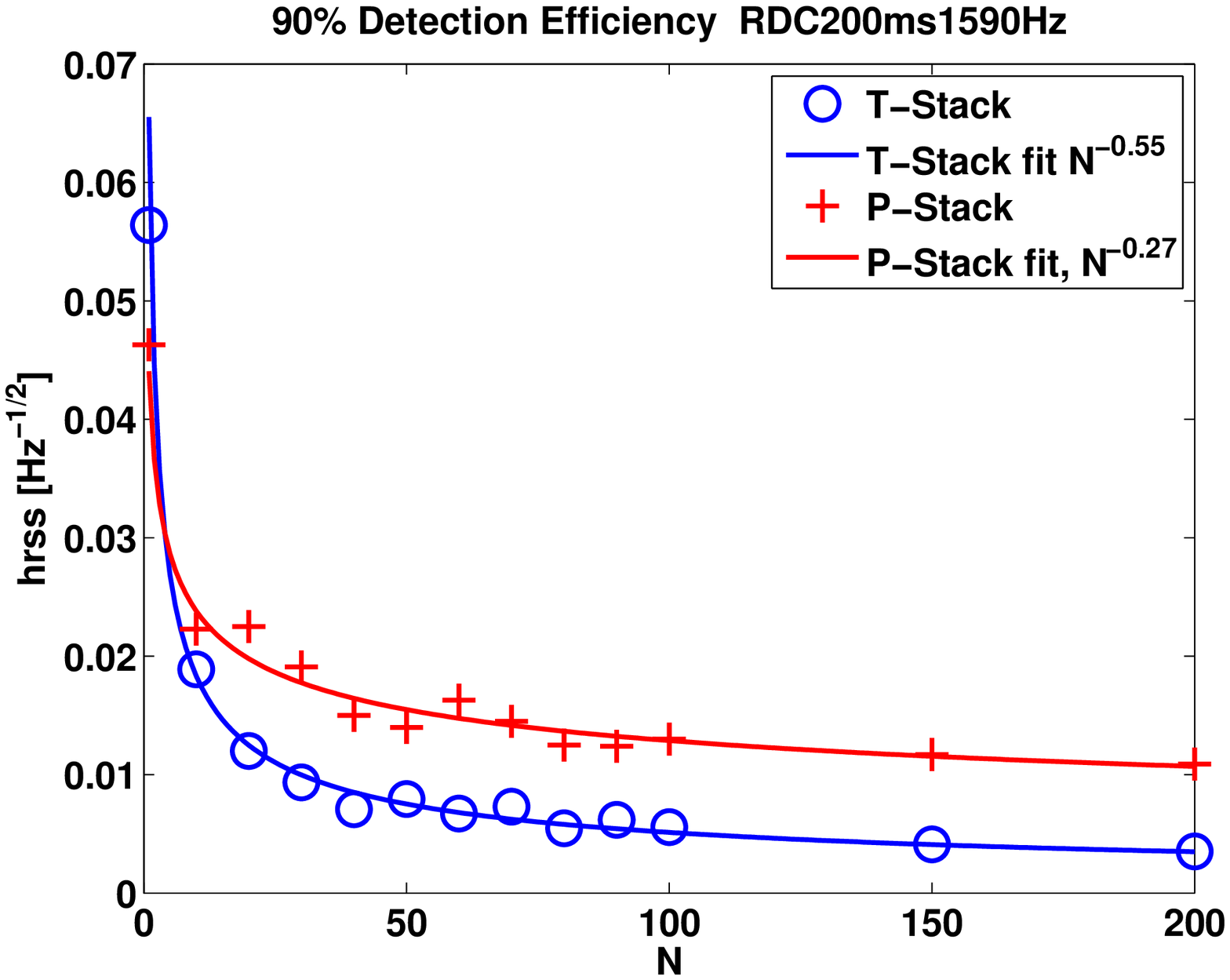}}
\caption[T-Stack and P-Stack ringdown sensitivity dependence on $N$]
{T-Stack and P-Stack sensitivity dependence on $N$, 50\% (top) and
90\% (bottom) detection efficiency, for 1590\,Hz $\tau$=200\,ms
ringdowns in white noise with $\sigma=1$.   The results for the
T-Stack pipeline show a sensitivity dependence at both detection
efficiency levels of  nearly $N^{1/2}$ ($N^{0.49}$ and
$N^{0.55}$ for 50\% and 90\% detection efficiencies respectively),
and the results for the P-Stack pipeline show a sensitivity
dependence at both detection efficiency levels of nearly
$N^{1/4}$ ($N^{0.24}$ and $N^{0.27}$ for 50\% and 90\% detection
efficiencies respectively). All fits excluded the point $N=1$.}
\label{fig:ndepend}
\end{center}
\end{figure}

We repeated the experiment for 100\,ms duration 100--1000\,Hz
band-limited WNBs.  In this case, we expected the coherent T-Stack
pipeline to underperform the the P-Stack pipeline on these
independently-generated stochastic incoherent signals.
As expected, we found that the T-Stack pipeline shows no improvement
as $N$ increases, while the P-Stack pipeline show the same $N^{1/4}$
sensitivity dependence seen in the coherent ringdown case.  The
results are shown in Figure\,\ref{fig:ndependWNB}; they illustrate
the relative model-independence of the P-Stack pipeline.

\begin{figure}[!t]
\begin{center}
\subfigure[]{
\includegraphics[angle=0,width=90mm,clip=false]{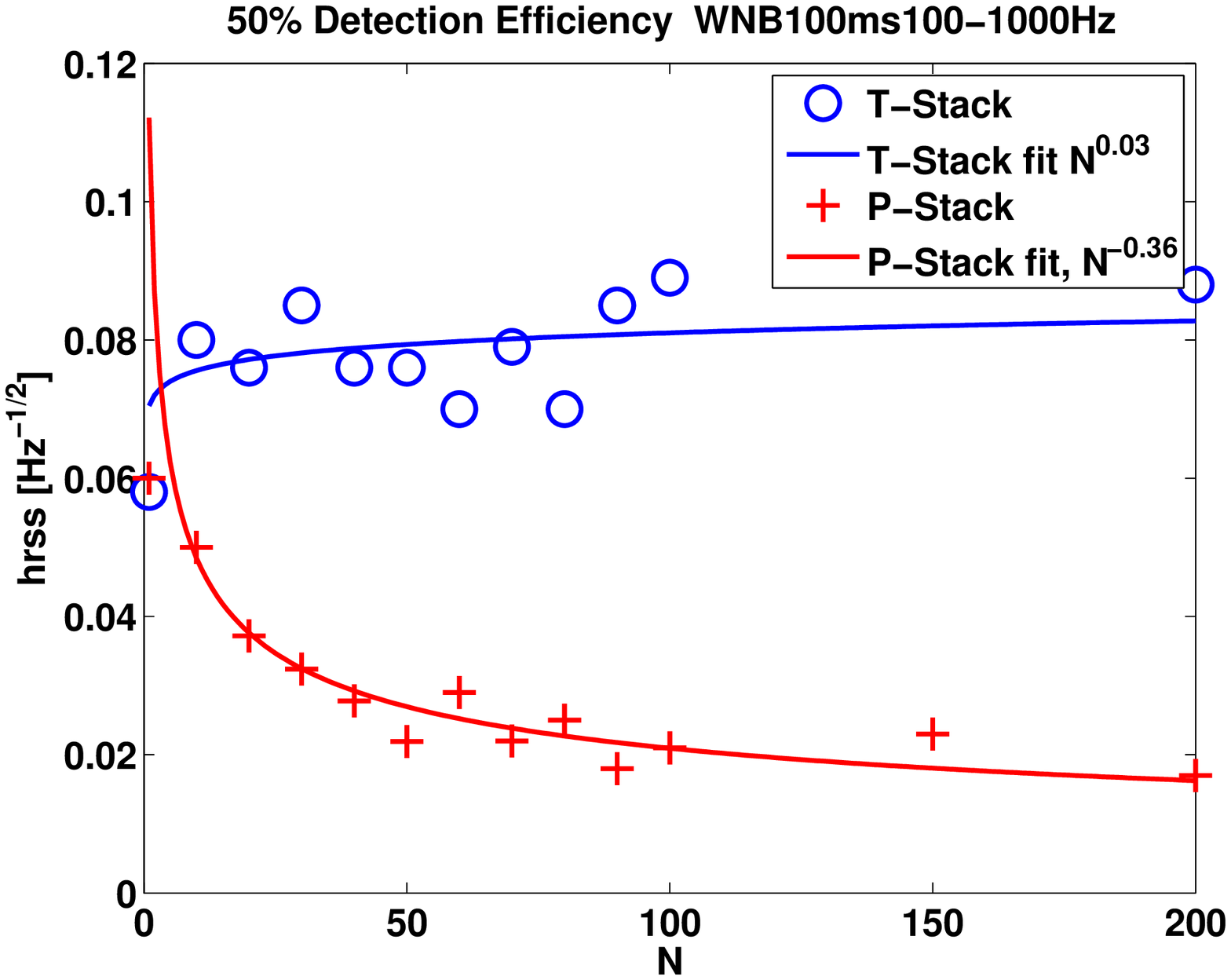}}
\subfigure[]{
\includegraphics[angle=0,width=90mm,clip=false]{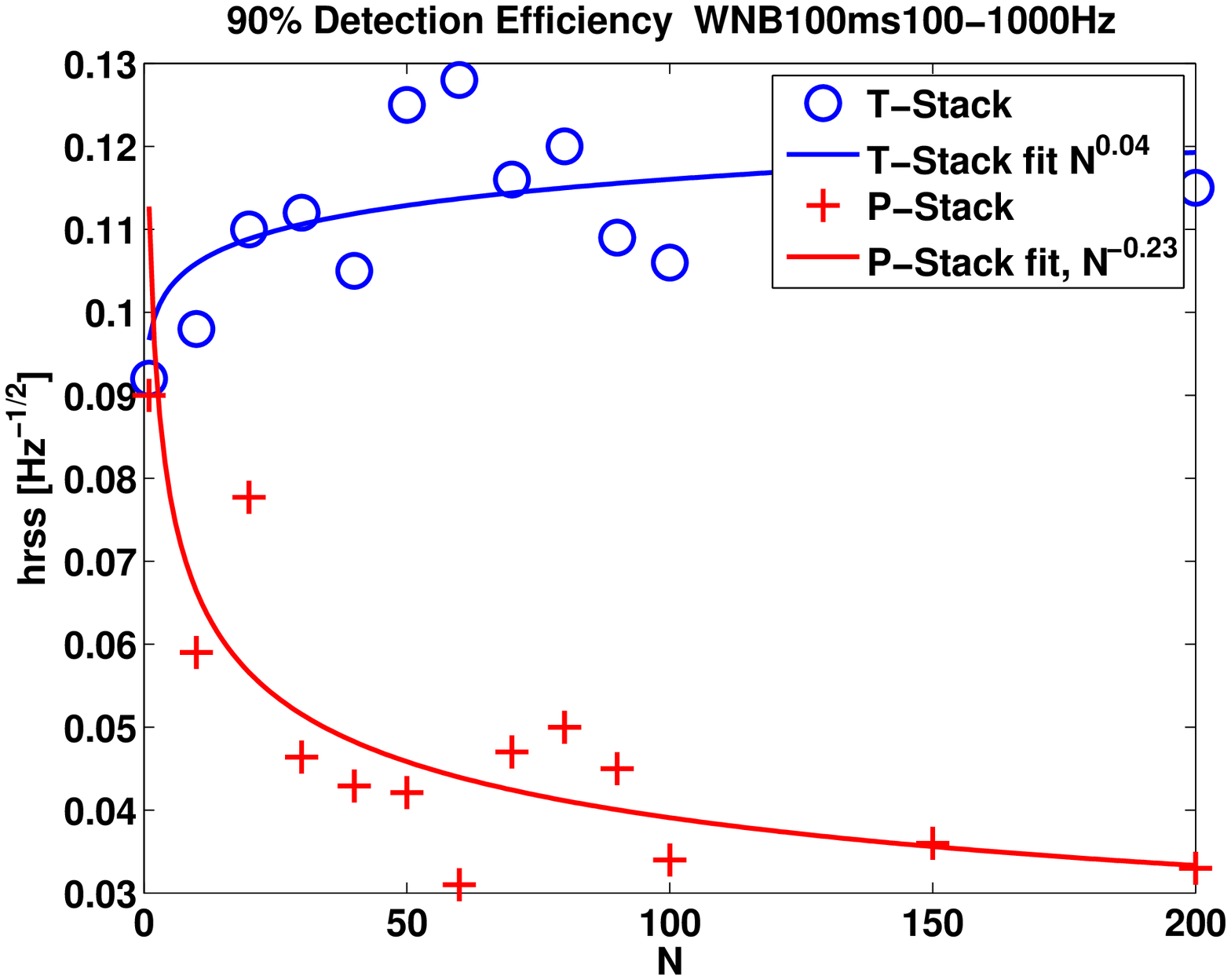}}
\caption[T-Stack and P-Stack WNB sensitivity dependence on $N$]
{T-Stack and P-Stack sensitivity dependence on $N$, 50\% (top) and
90\% (bottom) detection efficiency, for 100--1000\,Hz 100\,ms
duration white noise bursts in white noise with $\sigma=1$. The
results for the T-Stack pipeline show a sensitivity dependence at
both detection efficiency levels of nearly $N^{0}$ (flat
dependence), and the results for the P-Stack pipeline show a
sensitivity dependence at both detection efficiency levels of nearly
$N^{1/4}$ ($N^{0.36}$ and $N^{0.23}$ for 50\% and 90\% detection
efficiencies respectively), as in the coherent ringdown case. All
fits excluded the point $N=1$.} \label{fig:ndependWNB}
\end{center}
\end{figure}

\subsection{Sensitivity dependence on timing errors}

The T-Stack pipeline attains optimal sensitivity gains with
increasing $N$ because it performs a phase coherent addition of signals.  We have shown that the
P-Stack pipeline attains its
$N^{1/2}$ energy sensitivity performance even in the case of stacked signals that are
not coherent such as independently-generated white noise bursts.

In the case of identical signals such as ringdowns, an error in the
relative times between stacked signals will cause breakdown of phase
coherence.  For a constant timing error which is small relative to
the duration of one ringdown cycle the coherence breakdown will
increase with frequency.  Therefore we expect the T-Stack pipeline
to be more sensitive to timing errors than the P-Stack pipeline.

To begin quantifying this effect we performed a Monte Carlo
with a simulated burst series roughly modeled after the
SGR 1900+14 storm of 2006 March 29, with simulated 1590\,Hz $\tau=200$\,ms ringdown
signals of equal amplitude corresponding to each of the 18 largest
bursts in the storm. Timing errors were randomly chosen for each
ringdown from a normal distribution with $\sigma=100$\,$\mu$s, and
were applied as a timing shift (``wiggle'') to the given ringdown.
Results from a search with wiggles are then compared to the
identical search with no wiggles. A $\sigma=100$\,$\mu$s
distribution was chosen for the first tests because the BAT light
curve time bin size for the SGR 1900+14 storm is 100\,$\mu$s, which
might approximate the relative timing error between bursts in the
storm.

In the T-Stack case the timing degradation is approximately a factor
of 1.7.  In the P-Stack case no degradation was observed in this
preliminary low-statistics Monte Carlo.  This preliminary test
implies that if timing error cannot be reduced below 100\,$\mu$s the
P-Stack pipeline may perform better than the T-Stack pipeline for a
multiple burst SGR 1900+14 storm search.

\begin{figure}[!t]
\begin{center}
\subfigure[]{
\includegraphics[angle=0,width=90mm,clip=false]{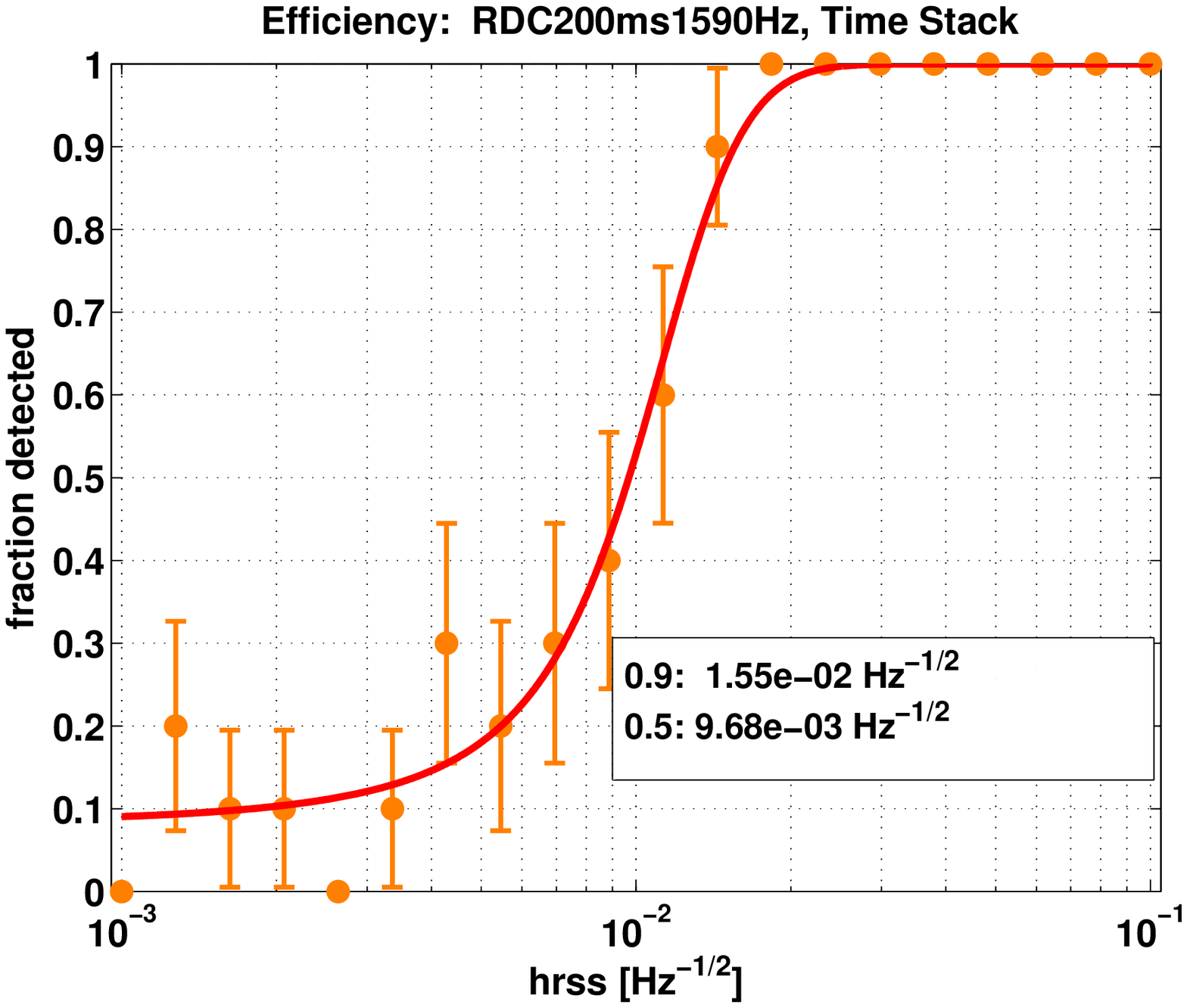}}
\subfigure[]{
\includegraphics[angle=0,width=90mm,clip=false]{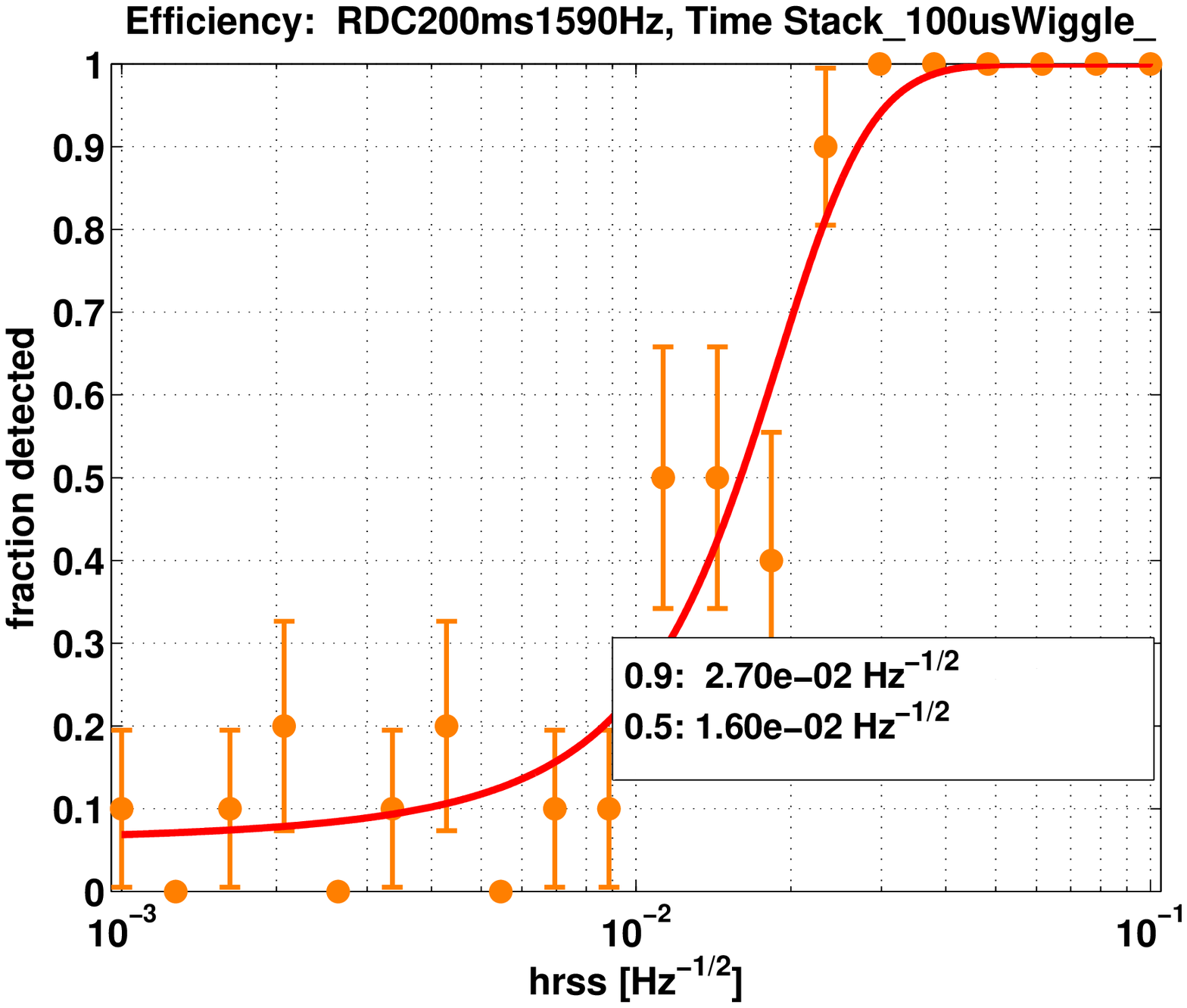}}
\caption[T-Stack dependence on 100\,$\mu$s timing error] {T-Stack
efficiency curves for 1590\,Hz 200\,ms $\tau$ ringdowns, with
perfect timing (top) and timing errors randomly chosen from a normal
distribution with $\sigma=100$\,$\mu$s (bottom).  The timing
degradation in this case is approximately a factor of 1.7.  Timing
degradation in the T-Stack pipeline increases with simulation
frequency. } \label{fig:tstackTimeDepend}
\end{center}
\end{figure}

We further quantified this effect with additional Monte Carlo
simulations, using a simulated burst series with $N=20$
equal-amplitude ringdowns, and allowing the timing error to range.
We performed the Monte Carlo with two ringdown types, 1090\,Hz
$\tau=200$\,ms and 2590\,Hz $\tau=200$\,ms circularly polarized
ringdowns, corresponding to the low and high frequency ranges in the
signal parameter space. Timing errors were randomly chosen for each
ringdown from a normal distribution with $\sigma$ ranging from
10\,$\mu$s to 100\,ms, and were applied as a timing shift to the
given ringdown. We also included the perfect timing case (no
wiggle). The tests were performed with both the T-Stack and P-Stack
pipelines.

As before, each efficiency curve was constructed using 20 amplitude
scaling factors and 20 trials at each $\hrss$ amplitude.  These low statistics efficiency curves turned out to
be adequate for characterizing the Stack-a-flare
pipelines. An example efficiency curve for the timing precision
Monte Carlos is shown in Figure\,\ref{fig:stackOfESample}.

\begin{figure}[!t]
\begin{center}
\includegraphics[angle=0,width=110mm, clip=false]{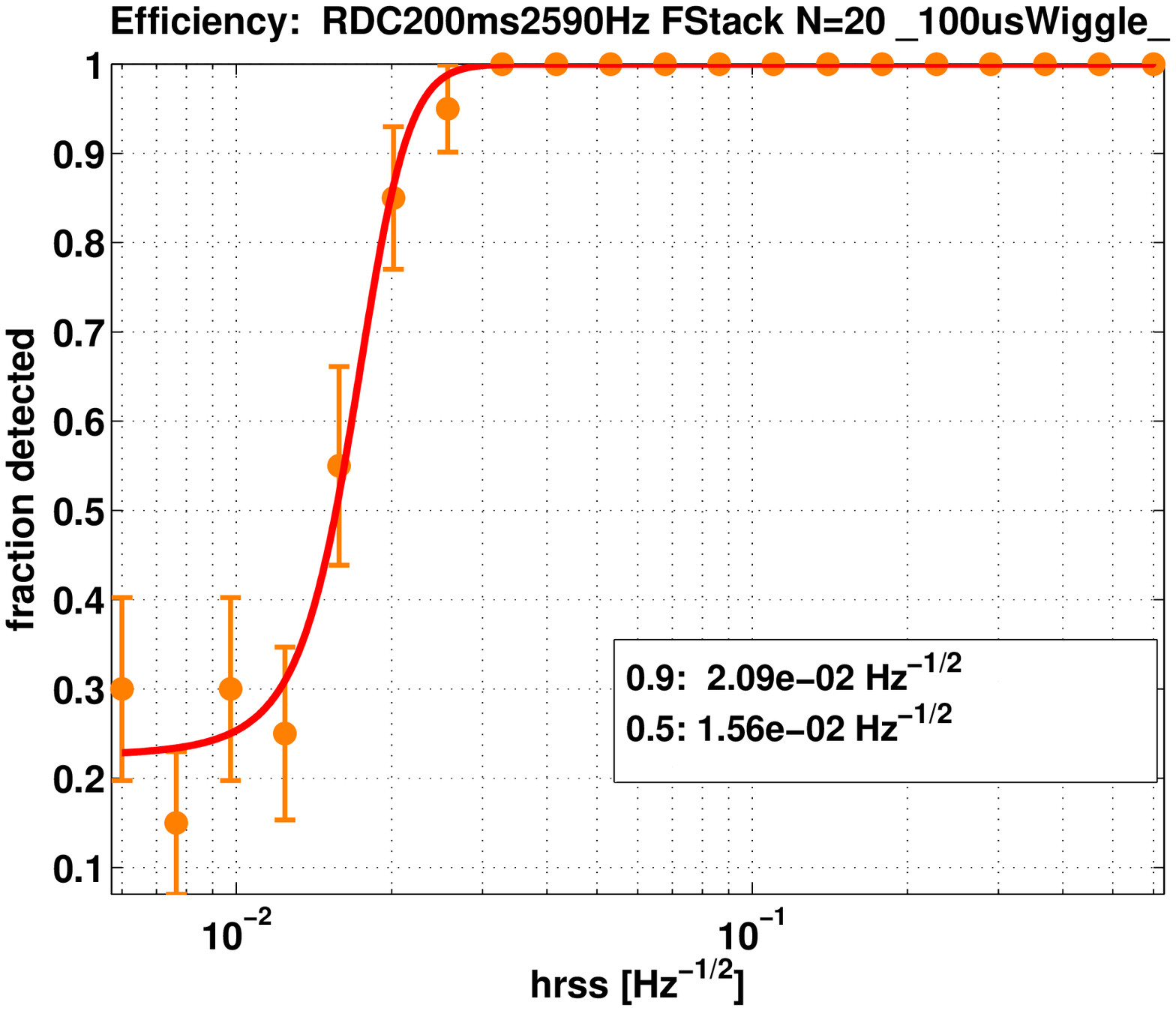}
\caption[Efficiency curve example for Stack-a-flare sensitivity vs.
timing error] { Efficiency curve example for Stack-a-flare
sensitivity vs. timing error Monte Carlos. This example curve is for
$N=20$ P-Stack pipeline with 100\,$\mu$s timing error (at
1-$\sigma$) using 2590\,Hz circularly polarized ringdowns. Each
efficiency curve was constructed using 20 amplitude scaling factors
and 20 trials at each $\hrss$ amplitude. }
\label{fig:stackOfESample}
\end{center}
\end{figure}

The results are displayed in Figure\,\ref{fig:stackOfE1090}
(1090\,Hz ringdowns) and Figure\,\ref{fig:stackOfE2590} (2590\,Hz
ringdowns) at both the 50\% and 90\% detection efficiencies.  As
expected, the P-Stack method is independent of timing error, up
until large timing errors on the order of the signal duration.
The T-Stack pipeline, on the other hand, shows a pronounced
dependence on timing error, which is more pronounced in the case of
high frequency simulations.  Each plot shows data for both T-Stack
and P-Stack pipelines, and finds the equal-sensitivity timing error
(P-Stack and T-Stack curve intersection point) using polynomial
fits.

For the T-Stack pipeline to be effective at 1090\,Hz, apparently,
timing error must be $\lesssim100$\,$\mu$s at 1-$\sigma$.  For the
T-Stack pipeline to be effective at 2590\,Hz, timing error must be
$\lesssim50$\,$\mu$s at 1-$\sigma$.  For the $N=20$ case shown, the
T-Stack pipeline is a factor of about 1.5 more sensitivity than the
P-Stack pipeline, with no or small timing errors.  These
precision requirements are close to the actual precision
available from a continuous BAT light curve, as in the case of a
storm event (Section\,\ref{section:stormTimingPrecision}).

\begin{figure}[!t]
\begin{center}
\subfigure[]{
\includegraphics[angle=0,width=90mm,clip=false]{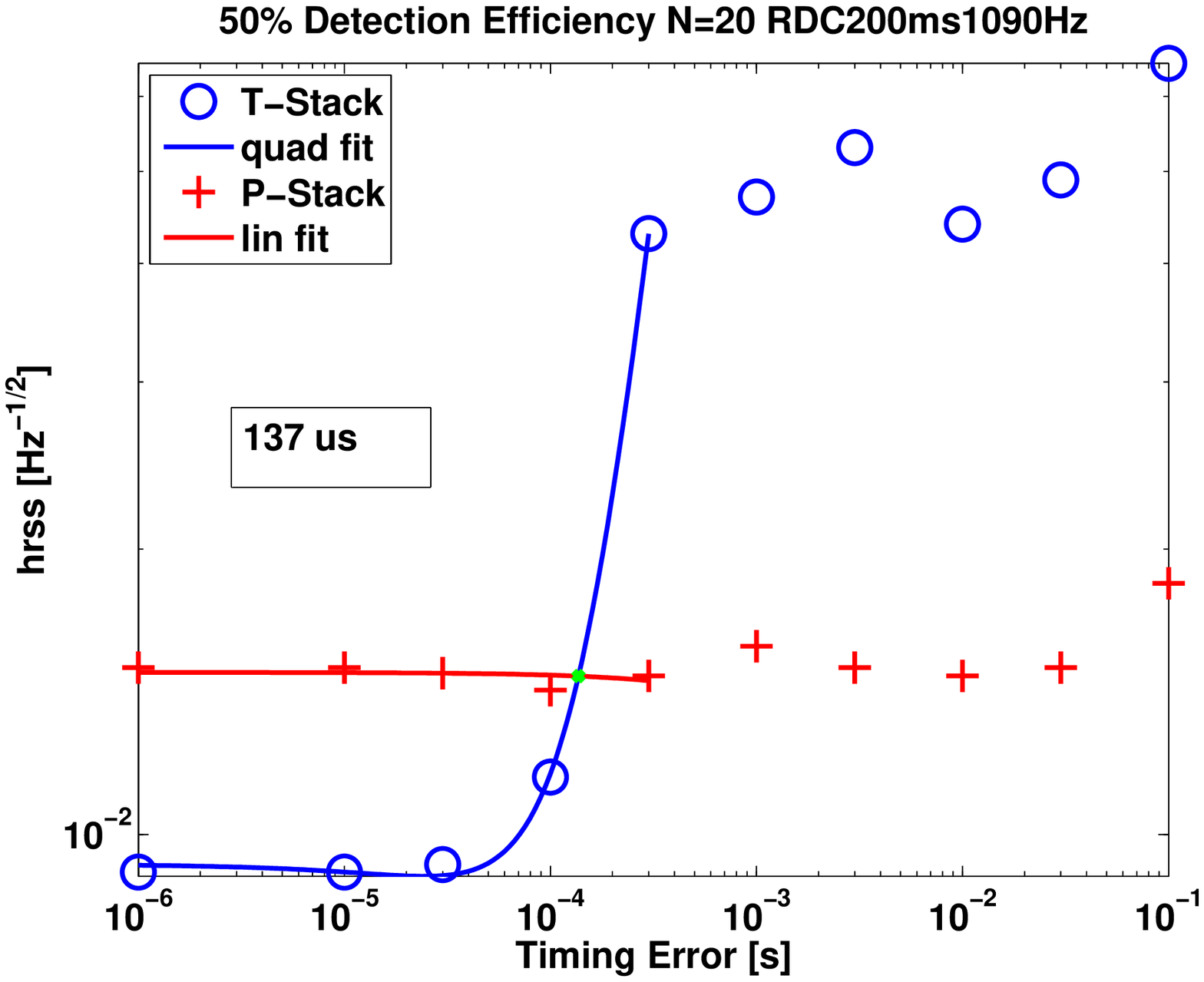}}
\subfigure[]{
\includegraphics[angle=0,width=90mm,clip=false]{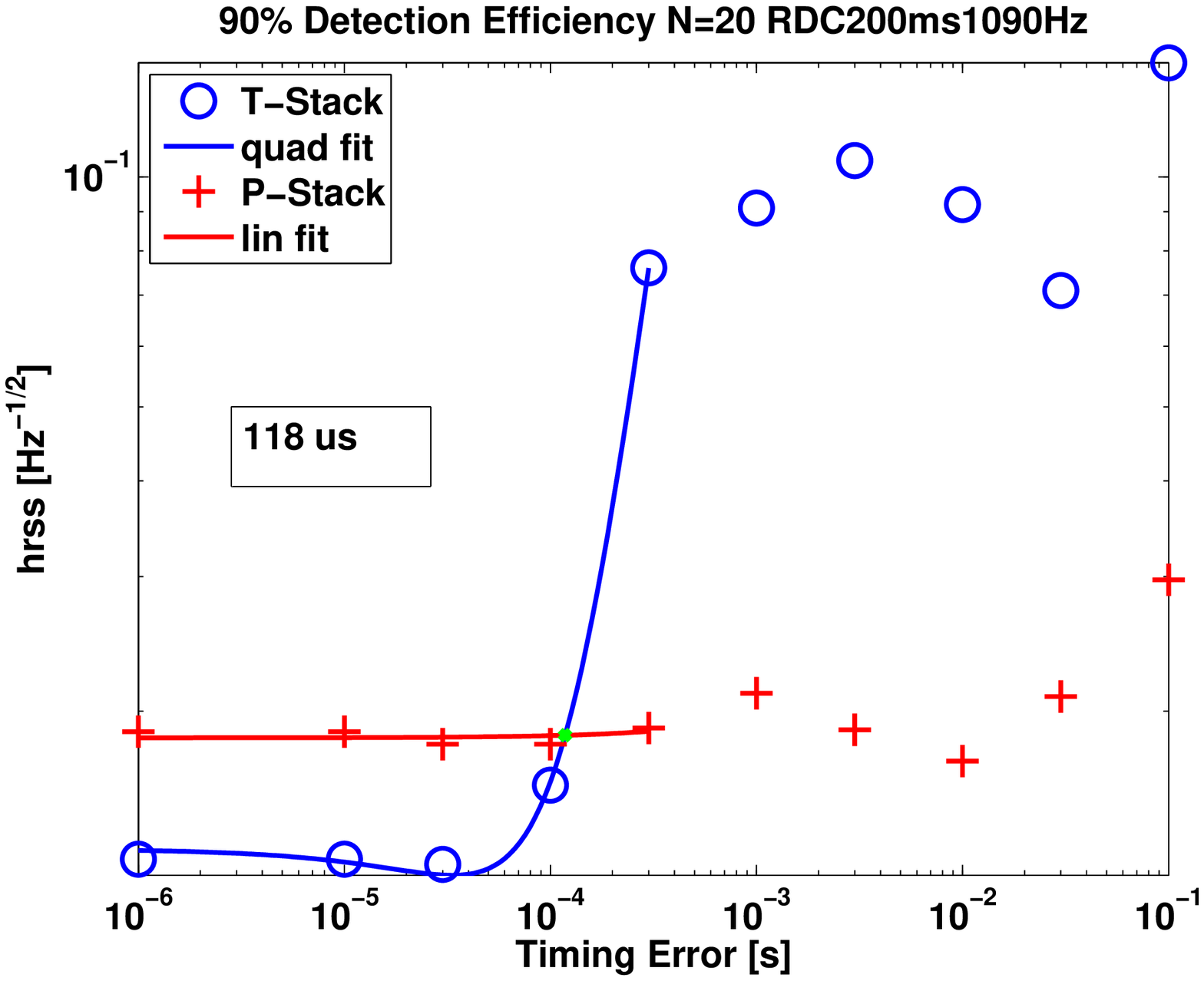}}
\caption[T-Stack and P-Stack sensitivity vs. timing error, 1090\,Hz
RDs] { T-Stack and P-Stack sensitivity versus timing error, for 1090\,Hz $\tau=200$\,ms circularly polarized
RD,  $N=20$.  Top
plot shows results for $\hrssf$ and bottom plot
shows results for $\hrssn$.  T-Stack
is more sensitive for small timing
errors, but degrades.  The crossover
point is noted;  for T-Stack to be
effective at 1090\,Hz, apparently timing error must be
$\lesssim100$\,$\mu$s at 1-$\sigma$.  The leftmost point on the
plots (at a timing error value of $\sci{1}{-6}$\,s) was actually
made with no timing error.  T-Stack results level off at high timing errors (greater
than $\sim \sci{2}{-4}$, or $\sim$90 degrees of phase) because the Monte Carlo effectively randomizes the phases
of the stacked signals.} \label{fig:stackOfE1090}
\end{center}
\end{figure}

\begin{figure}[!t]
\begin{center}
\subfigure[]{
\includegraphics[angle=0,width=90mm,clip=false]{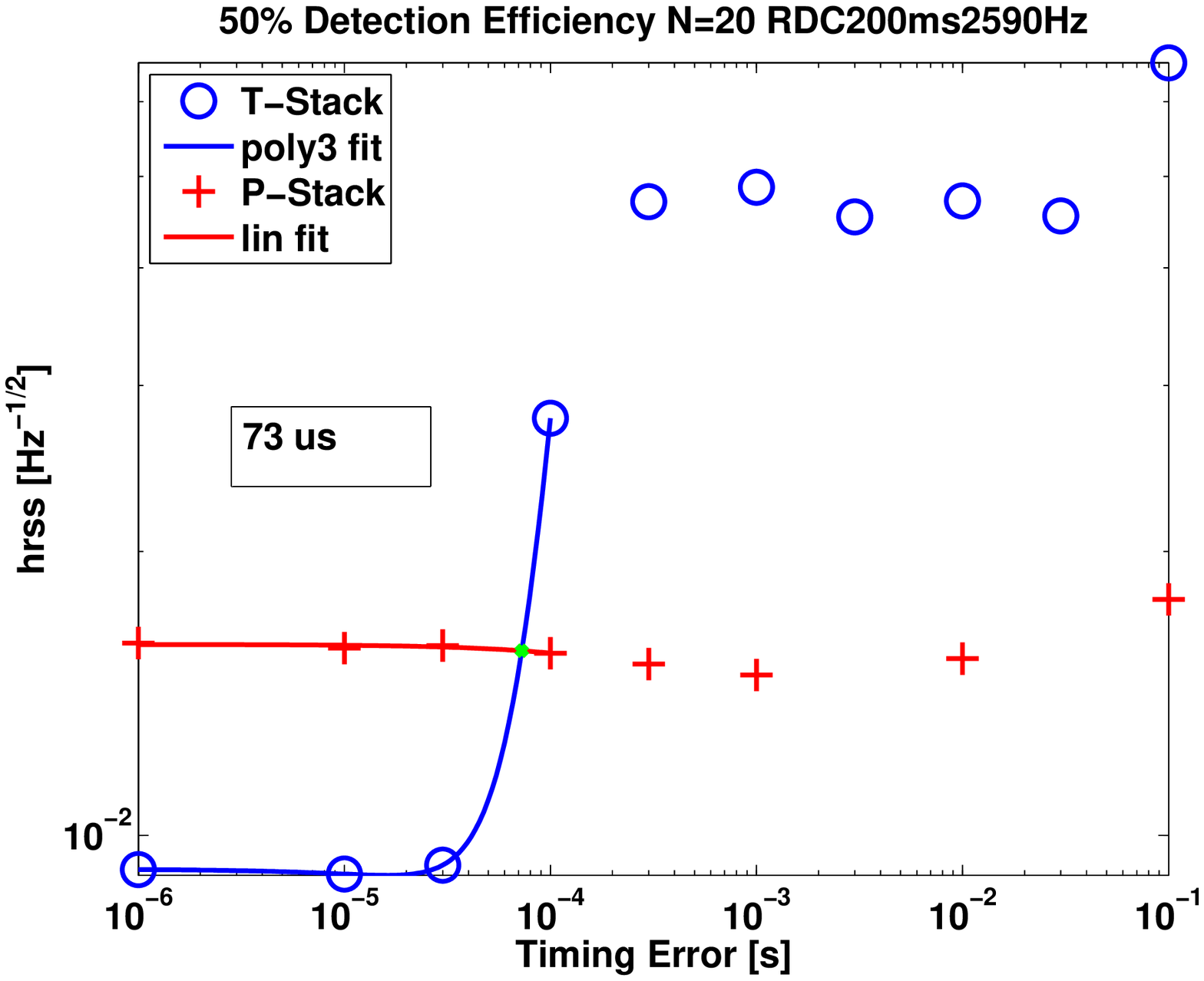}}
\subfigure[]{
\includegraphics[angle=0,width=90mm,clip=false]{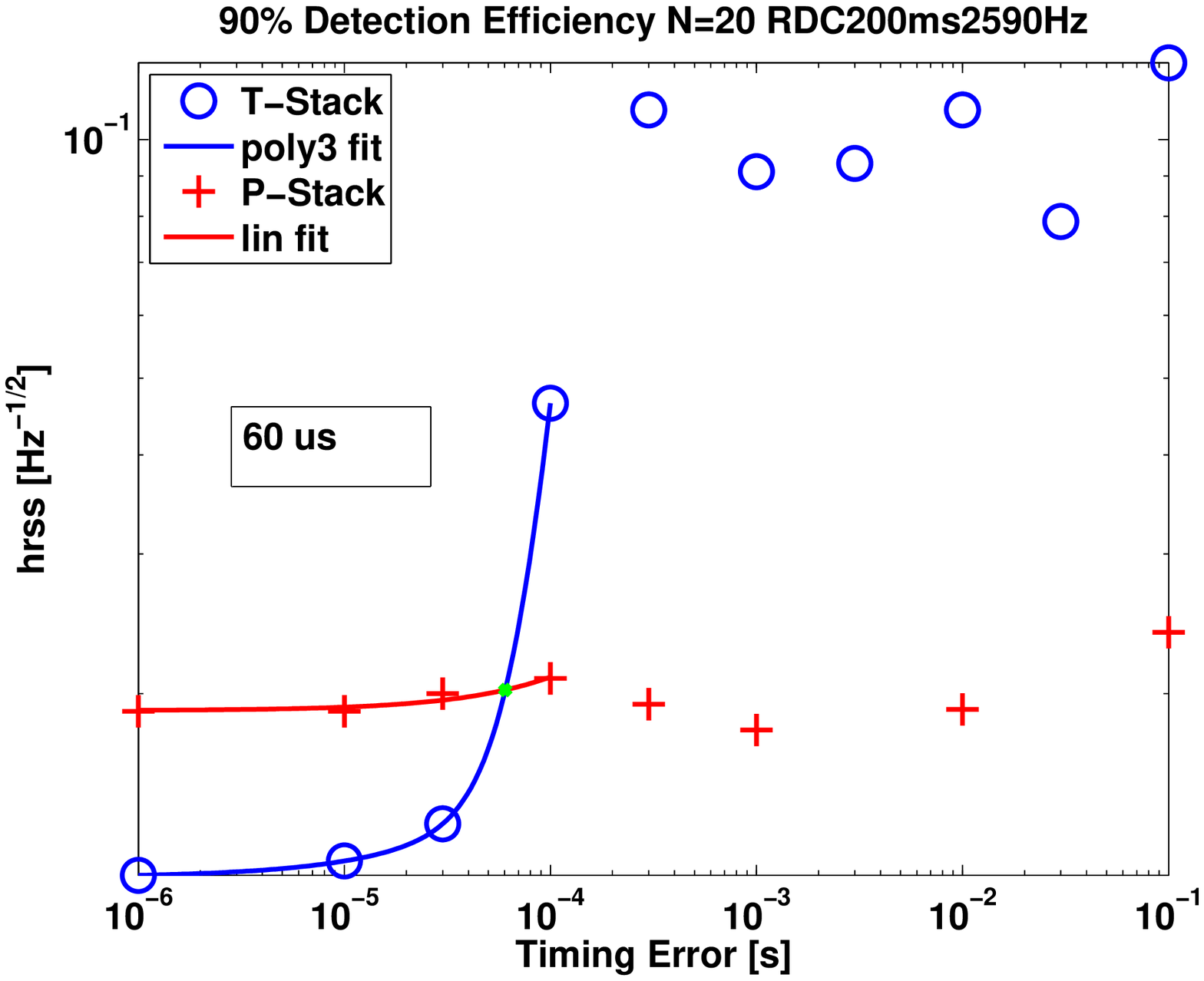}}
\caption[T-Stack and P-Stack sensitivity vs. timing error, 2590\,Hz
RDs] { T-Stack and P-Stack sensitivity versus timing error, for 2590\,Hz $\tau=200$\,ms circularly polarized
RD,  $N=20$.  Top
plot shows results for $\hrssf$ and bottom plot
shows results for $\hrssn$. T-Stack
is more sensitive for small timing
errors, but degrades.  The crossover
point is noted;  for T-Stack to be
effective at 2590\,Hz, apparently, timing error must be
$\lesssim50$\,$\mu$s at 1-$\sigma$.  The leftmost point on the plots
(at a timing error value of $\sci{1}{-6}$\,s) was actually made with
no timing error.
T-Stack results level off at high timing errors (greater than $\sim
\sci{1}{-4}$, or $\sim$90 degrees of phase) because the
Monte Carlo effectively randomizes the phases of the stacked
signals.} \label{fig:stackOfE2590}
\end{center}
\end{figure}

\subsection{Optimal use of the pipelines}

We summarize the implications from characterizing the two Stack-a-flare incarnations, T-Stack and P-Stack.  We envision four possible types of stacked SGR searches:
 \ben
 \i High frequency (1000--3000\,Hz) searches for ringdown burst
 emission,  for single SGR storm events (ringdown upper limits);
 \i Low frequency (100--1000\,Hz) searches for stochastic burst
 emission,  for single SGR storm events (band- and time-limited WNB upper limits);
 \i High frequency (1000--3000\,Hz) searches for ringdown burst
 emission, for isolated, time-separated SGR bursts (ringdown upper limits);
 \i Low frequency (100--1000\,Hz) searches for stochastic burst
 emission,  for isolated, time-separated SGR bursts (band- and time-limited WNB upper limits).
\een

We have found that the P-Stack pipeline can be used effectively in
any of these cases, with an energy sensitivity gain over the
individual burst search of approximately $N^{1/2}$. The T-Stack pipeline shows an energy sensitivity improvement of approximately $N$, but
only if the target signal is deterministic, and only if the relative
timing between SGR gravitational wave burst events can be known to high precision,
$\lesssim$100\,$\mu$s at 1090\,Hz and $\lesssim$50\,$\mu$s at
2590\,Hz or if gravitational wave data streams are time shifted
relative to each other at vastly increased computational expense and additional code complexity. The T-Stack pipeline might be practical only in the first case.

\section{ SGR 1900+14 storm mock search} \label{section:stacksim}

\subsection{BAT light curve for the SGR 1900+14 storm}
\label{section:stormTimingPrecision}

Data from the BAT detector on the Swift satellite are publicly
available.  In Figure\,\ref{fig:stormMarked}, we show the storm
light curve with 100\,$\mu$s bins.  The red crosses mark burst peak
heights and time locations.  Times of intermediate flares were assigned at the center
of the steep rising edge.  Times of common bursts were assigned to
the brightest bin.  Figure\,\ref{fig:stormMarkedZoom} shows a detail
of the light curve.  The two major types of bursts are clearly
visible: longer duration intermediate flares, and shorter duration
common bursts. 

It may be possible to fit the rising edges of the peaks in the light
curve and perhaps obtain relative timing precision of better than a
bin resolution. Even so, as we have discussed the
model-dependent nature of the T-Stack method brings other
disadvantages. For example, if the rise timescales determined by the
fit are not the same, we would face a decision about which part of
the rising edge to use when lining up stacked time series.

\subsection{Results}

The Stack-a-flare pipeline was implemented to run on
either real LIGO data or simulated data.  In this section we present the results of runs with the P-Stack pipeline simulating a search for gravitational waves associated with the 2006 March 29 SGR 
1900+14 storm, using simulated data.  At the time of the storm, all three LIGO detectors were taking science-quality data.  Simulated data modeled from real data from the two LIGO 4\,km detectors were created from white noise by matching power spectra with LIGO data in the frequency domain.  Therefore, the 
sensitivity estimates should be close to upper limit estimates from real data.  Although no attempt has been made to model correlations in time and frequency (e.g. glitches) which do occur in the 
real data, we expect agreement with results from real data to be better than 10\,\% on average.  In fact, comparison between the mock search amplitude sensitivity estimates and closed box amplitude upper limits (presented in Table\,\ref{table:stackclosed} in Section\,\ref{section:stackClosed} before adding uncertainties shows average agreement to better than 1\%, validating the simulated data. 

For this mock search, we consider two stacking scenarios. The first scenario was an unweighted stack of the 11 bursts in the storm with the largest fluences.  A histogram of integrated counts under each burst in the light curve, a measure approximately proportional to fluence,  shows a clear separation between these 11 bright bursts and the rest of the set (see Figure\,\ref{fig:stormFluenceHist}).  The second stacking scenario was the fluence-weighted scenario, which included 77 bursts in the storm light curve (all but the weakest, which would not contribute significantly).  In the fluence-weighted scenario, we assume that gravitational wave emission energy is proportional to fluence, so the compound simulations are weighted to the square root of the fluence measure, and then normalized so that the total $\hrss$ is the same as the total $\hrss$ of the unweighted case.  Then the significance tilings are weighted according to the fluence measure before being stacked.

\begin{figure}[!t]
\begin{center}
\includegraphics[angle=0,width=110mm, clip=false]{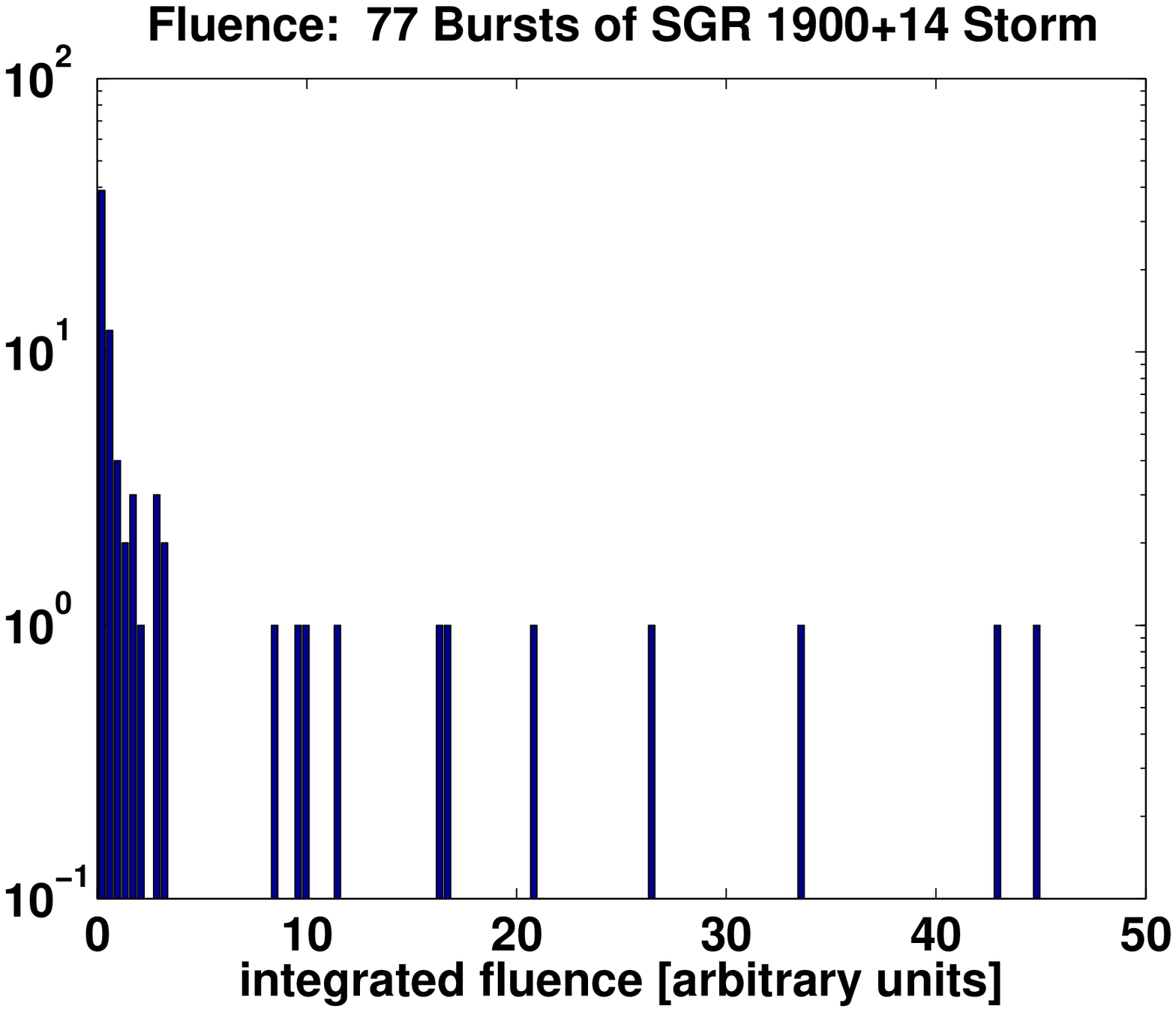}
\caption[Histogram of relative fluence of bursts in the SGR 1900+14 storm ] {  A histogram of integrated counts under each burst in the SGR 1900+14 storm light curve from the BAT detector on the Swift satellite, a measure approximately proportional to fluence,  shows a clear separation between these 11 bright bursts and the rest of the set.}
\label{fig:stormFluenceHist}
\end{center}
\end{figure}

In Figure\,\ref{fig:exampleStackFAR} we show example cumulative histograms showing false alarm rates versus analysis event loudness for the background and the stacked on-source region.  There are three such plots for each scenario, one per search band.  Since the stacked livetime is 4\,s here, the loudest on-source event occurs once per 4\,s, and is plotted at a y-value of 0.25\,Hz.    We can estimate the FAR of this loudest on-source analysis from the background.  

\begin{figure}[!t]
\begin{center}
\includegraphics[angle=0,width=110mm, clip=false]{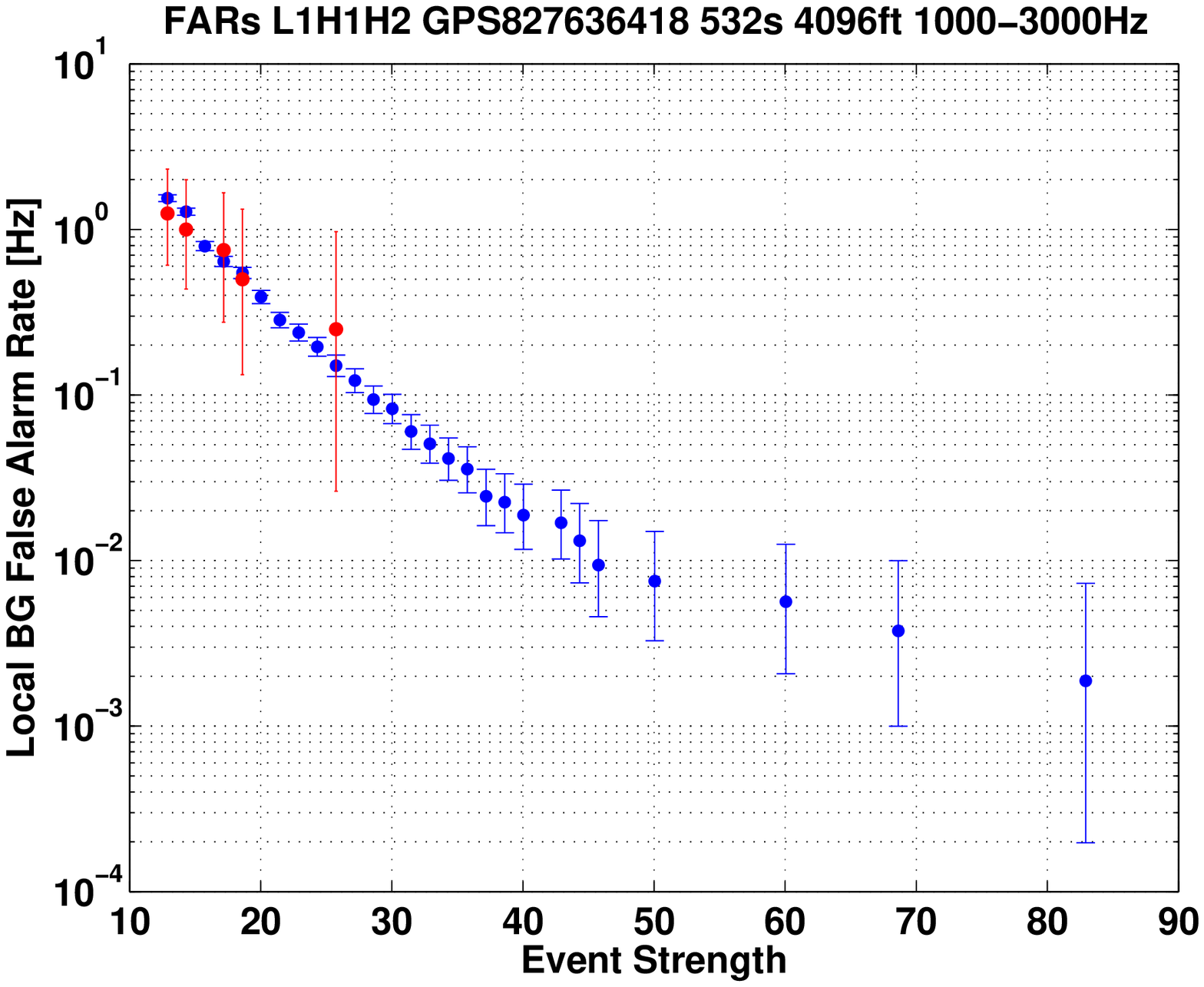}
\caption[Rate versus analysis event loudness] { Example cumulative histograms showing false alarm rates versus analysis event loudness for the background (blue) and the stacked on-source region (red).  There are three such plots for each scenario, one per search band.  Since the stacked livetime is 4\,s here, the loudest on-source event occurs once per 4\,s, and is plotted at a y-value of 0.25\,Hz.    We can estimate the FAR of this loudest on-source analysis from the background.  }
\label{fig:exampleStackFAR}
\end{center}
\end{figure}

Table\,\ref{table:stacksim} shows sensitivity estimates, at 90\% detection efficiency, for the $N=11$ flat scenario and the fluence-weighted scenario.  Sensitivity estimates with $N=1$ are also shown for 
comparison; when $N=1$ the Stack-a-flare pipeline reduces to the individual burst search pipeline (Flare pipeline).   Superscripts in Table\,\ref{table:stacksim} give a
systematic error and uncertainties at 90\% confidence. The first and
second superscripts account for systematic error and statistical
uncertainty in amplitude and phase of the detector calibrations,
estimated via Monte Carlo simulations, respectively. The third is a
statistical uncertainty arising from using a finite number of
injected simulations, estimated with the bootstrap method using 200
ensembles\,\cite{efron79}.  The systematic error and the quadrature
sum of the statistical uncertainties are added to the final upper
limit estimates.  We also present the sensitivity estimates (with uncertainties folded in) in Figures\,\ref{fig:stacksimegw} and\,\ref{fig:stacksimamp}.

\begin{figure}[!t]
\begin{center}
\includegraphics[angle=0,width=110mm, clip=false]{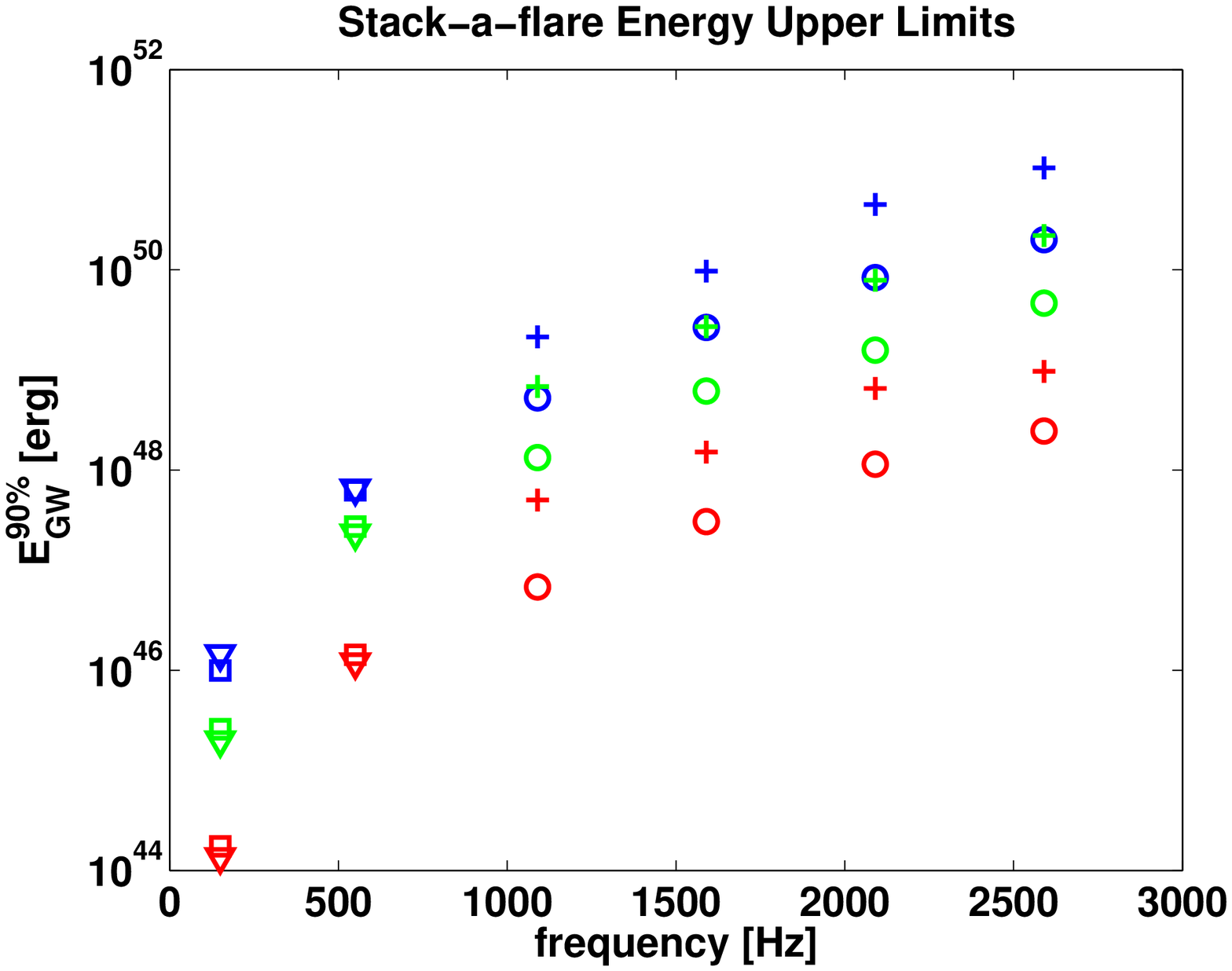}
\caption[Stack-a-flare simulated data energy sensitivity estimates] { Stack-a-flare simulated data energy sensitivity estimates for the 29 March 2006 storm from SGR 1900+14, for 
the $N=11$ flat and the fluence-weighted scenarios.  The $N=1$ scenario is shown for reference.  Uncertainty estimates have been folded in, as tabulated in Table\,
\ref{table:stacksim}.  Crosses and circles indicate linearly and circularly polarized RDs, respectively.  Triangles and squares represent 11\,ms and 100\,ms band- and time-limited WNBs, respectively, and are placed at the WNB central frequency.  Color indicates the stacking scenario:  Blue indicates $N=1$,  green indicates $N=11$ flat, and red indicates fluence-weighted.}
\label{fig:stacksimegw}
\end{center}
\end{figure}

\begin{figure}[!t]
\begin{center}
\includegraphics[angle=0,width=110mm, clip=false]{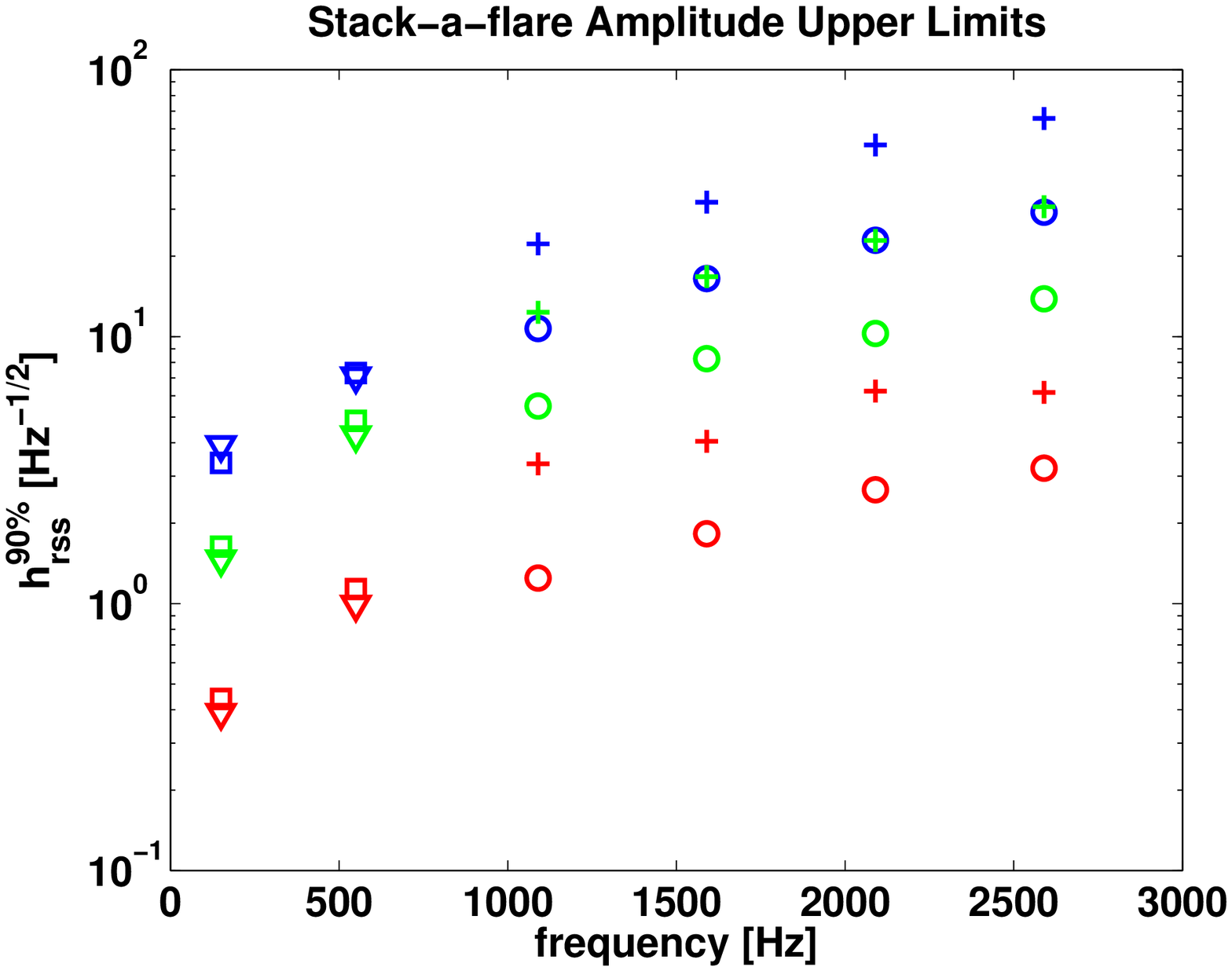}
\caption[Stack-a-flare simulated data amplitude sensitivity estimates] { Stack-a-flare simulated data amplitude sensitivity estimates for the 29 March 2006 storm from SGR 
1900+14, for the $N=11$ flat and the fluence-weighted scenarios.  The $N=1$ scenario is shown for reference.  Uncertainty estimates have been folded in, as tabulated in Table\,
\ref{table:stacksim}.  Crosses and circles indicate linearly and circularly polarized RDs, respectively.  Triangles and squares represent 11\,ms and 100\,ms band- and time-limited WNBs, respectively, and are placed at the WNB central frequency.  Color indicates the stacking scenario:  Blue indicates $N=1$,  green indicates $N=11$ flat, and red indicates fluence-weighted.}
\label{fig:stacksimamp}
\end{center}
\end{figure}

\section{SGR 1900+14 storm closed box search} \label{section:stackClosed}

We have repeated the search described in the Section\,\ref{section:stacksim}, but with real LIGO data instead of simulated LIGO-like data.  This ``closed box'' search, done in preparation for ``opening the box,''  avoids analysis of the actual on-source region by adding 400\,s to the electromagnetic trigger time.  In other respects is identical to an actual search.   The  results are presented in Table\,\ref{table:stackclosed}.   The 1\% average agreement of amplitude results between the closed box and simulated data runs mentioned in the previous section indicates that the search does not strongly depend on glitchiness in the real LIGO data.     Because of the close agreement, we do not show closed box figures corresponding to Figures\,\ref{fig:stacksimegw} and\,\ref{fig:stacksimamp}.

The results in Table\,\ref{table:stackclosed}, while using real LIGO data, had not been reviewed by the LIGO Scientific Collaboration at the time of writing.  Therefore, that table does not reflect the scientific opinion 
of the LSC.

\section{Conclusion and future plans}

We have presented a method for searching for gravitational waves associated with multiple SGR bursts that extends the individual SGR burst search presented in Chapter\,\ref{chapter:search} and \,\cite{s5y1sgr}.   We have characterized both the T-Stack and the P-Stack incarnations of the pipeline, demonstrating sensitivity dependence on stacking number $N$ and uncertainty in relative timing between bursts.  The P-Stack pipeline is robust to timing errors, and we have used it to estimate search sensitivities for a mock SGR 1900+14 storm multiple SGR search, using simulated data modeled after real LIGO data.    

In the near future, we plan to perform the actual multiple burst search for gravitational waves associated with the SGR 1900+14 storm using real LIGO data.  The real search will be very similar to the mock search in simulated data presented here, although we may choose to explore different or additional stacking scenarios.    We are also considering  a multiple SGR burst search on isolated bursts spanning months or years.   The Advanced LIGO detectors promise an improvement in $\hrss$ by more
than a factor of 10 over S5, corresponding to an improvement in
energy sensitivity by more than a factor of 100.  We hope to continue improving this search method, and to use it to perform a searches using Advanced LIGO data. 

\begin{sidewaysfigure}[!t]
\includegraphics[angle=0,width=240mm,clip=false]{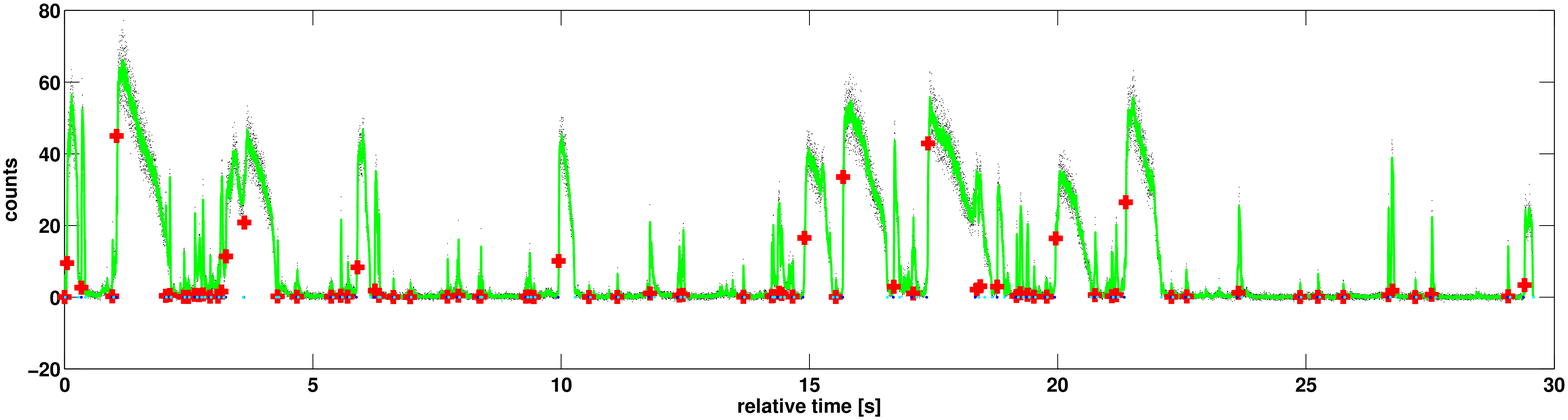}
\caption[BAT light curve of the SGR 1900+14 storm event] { BAT light
curve of the SGR 1900+14 storm event, 100\,$\mu$s bins.  The light
curve shows the BAT event data, from sequence 00203127000, from
approximately 20\,s after the start of the sequence to its end. The
red crosses mark burst integrated counts, a measure which is approximately proportional to fluence which were used in the fluence-weighted scenario.  Times of intermediate flares were
assigned at the center of the steep rising edge.  Times of common
bursts were assigned to the brightest bin.   }
\label{fig:stormMarked}
\end{sidewaysfigure}

\begin{sidewaysfigure}
 \centering
\includegraphics[angle=0,width=240mm,clip=false]{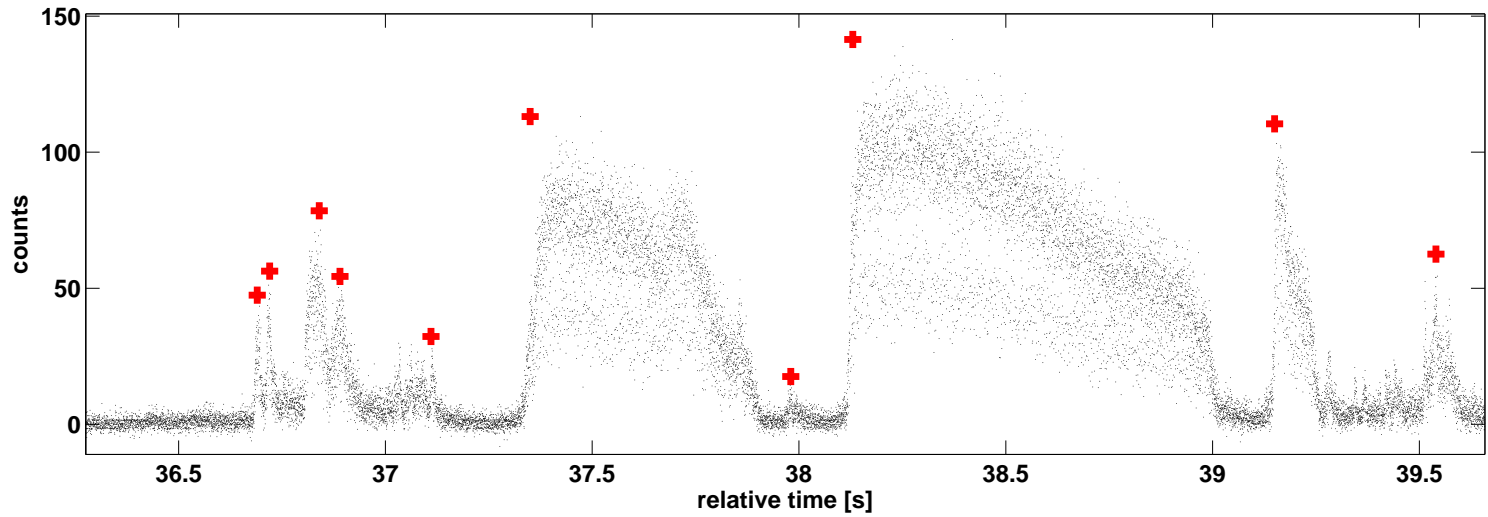}
\caption[BAT light curve of the SGR 1900+14 storm event, detail] {
Detail of BAT light curve of the SGR 1900+14 storm event shown in
Figure\,\ref{fig:stormMarked}.  The two major types of bursts are
clearly visible:  longer duration intermediate flares, and shorter
duration common bursts. However, there were some bursts that fall
somewhere in the middle, such as the burst at 39.2\,s. Red
crosses mark approximate burst peak heights and time locations.  } \label{fig:stormMarkedZoom}
\end{sidewaysfigure}

\begin{sidewaystable*}
\caption[Sensitivity estimates for a mock SGR 1900+14 storm stacking search ]{Sensitivity estimates for a mock SGR 1900+14 storm search in simulated LIGO noise, for two stacking scenarios ($N=11$ flat scenario and fluence weighted scenario).  The $N=1$ case is shown for comparison.  Results are shown for various ringdown and band- and time-limited white noise burst target signal classes.  }
\begin{scriptsize}
\begin{tabular}{@{\extracolsep{\fill}}lrlr|r||rlr|r||rlr|r}

 \hline \hline
 & \multicolumn{4}{c}{$N=1$} & \multicolumn{4}{c}{$N=11$ Flat} & \multicolumn{4}{c}{Fluence-weighted} \\ 
 Simulation type & \multicolumn{3}{c}{$ \hrssn [ 10^{-22}~ \rthz $] }  & $\egwn$ [erg]   & \multicolumn{3}{c}{$ \hrssn [ 10^{-22} ~ \rthz $] }  & $\egwn$ [erg]  & \multicolumn{3}{c}{$ 
\hrssn [ 10^{-22} ~ \rthz $] }  & $\egwn$ [erg] \\ 
 \hline 
 WNB 11ms 100-200 Hz   & 3.2 & $^{ +0.0 ~ +0.34 ~ +0.61}$ &  $= 3.9$ & $\sci{1.5}{46}$  & 1.3 & $^{ +0.0 ~ +0.13 ~ +0.11}$ &  $= 1.5$ & $\sci{2.0}{45}$  & 0.34 & $^{ +0.0 ~ 
+0.036 ~ +0.027}$ &  $= 0.39$ & $\sci{1.4}{44}$  \\
 WNB 100ms 100-200 Hz   & 3.0 & $^{ +0.0 ~ +0.31 ~ +0.24}$ &  $= 3.4$ & $\sci{1.0}{46}$  & 1.4 & $^{ +0.0 ~ +0.14 ~ +0.23}$ &  $= 1.6$ & $\sci{2.6}{45}$  & 0.39 & $^{ +0.0 ~ 
+0.040 ~ +0.028}$ &  $= 0.44$ & $\sci{1.7}{44}$  \\
 WNB 11ms 100-1000 Hz   & 6.3 & $^{ +0.0 ~ +0.65 ~ +0.50}$ &  $= 7.1$ & $\sci{6.5}{47}$  & 3.8 & $^{ +0.0 ~ +0.40 ~ +0.23}$ &  $= 4.3$ & $\sci{2.3}{47}$  & 0.89 & $^{ +0.0 ~ 
+0.092 ~ +0.054}$ &  $= 0.99$ & $\sci{1.2}{46}$  \\
 WNB 100ms 100-1000 Hz   & 6.2 & $^{ +0.062 ~ +0.64 ~ +0.83}$ &  $= 7.3$ & $\sci{6.3}{47}$  & 4.3 & $^{ +0.043 ~ +0.44 ~ +0.31}$ &  $= 4.8$ & $\sci{2.7}{47}$  & 0.98 & 
$^{ +0.0098 ~ +0.10 ~ +0.099}$ &  $= 1.1$ & $\sci{1.4}{46}$  \\
 RDC 200ms 1090 Hz   & 9.1 & $^{ +0.18 ~ +0.94 ~ +1.1}$ &  $= 11$ & $\sci{5.3}{48}$  & 4.8 & $^{ +0.096 ~ +0.50 ~ +0.35}$ &  $= 5.5$ & $\sci{1.3}{48}$  & 1.1 & $^{ +0.022 ~ 
+0.11 ~ +0.082}$ &  $= 1.2$ & $\sci{6.8}{46}$  \\
 RDC 200ms 1590 Hz   & 14 & $^{ +0.54 ~ +1.4 ~ +1.9}$ &  $= 17$ & $\sci{2.7}{49}$  & 7.1 & $^{ +0.28 ~ +0.73 ~ +0.56}$ &  $= 8.3$ & $\sci{6.2}{48}$  & 1.5 & $^{ +0.058 ~ +0.15 ~ 
+0.27}$ &  $= 1.8$ & $\sci{3.1}{47}$  \\
 RDC 200ms 2090 Hz   & 19 & $^{ +1.5 ~ +2.4 ~ +1.3}$ &  $= 23$ & $\sci{8.4}{49}$  & 8.2 & $^{ +0.66 ~ +1.1 ~ +0.87}$ &  $= 10$ & $\sci{1.6}{49}$  & 2.2 & $^{ +0.17 ~ +0.28 ~ 
+0.20}$ &  $= 2.7$ & $\sci{1.1}{48}$  \\
 RDC 200ms 2590 Hz   & 22 & $^{ +2.9 ~ +2.8 ~ +3.2}$ &  $= 29$ & $\sci{2.0}{50}$  & 11 & $^{ +1.4 ~ +1.4 ~ +0.61}$ &  $= 14$ & $\sci{4.7}{49}$  & 2.4 & $^{ +0.32 ~ +0.31 ~ 
+0.35}$ &  $= 3.2$ & $\sci{2.5}{48}$  \\
 RDL 200ms 1090 Hz   & 18 & $^{ +0.54 ~ +1.9 ~ +3.2}$ &  $= 22$ & $\sci{2.1}{49}$  & 10 & $^{ +0.30 ~ +1.1 ~ +1.6}$ &  $= 12$ & $\sci{6.8}{48}$  & 2.7 & $^{ +0.082 ~ +0.28 ~ 
+0.46}$ &  $= 3.3$ & $\sci{5.0}{47}$  \\
 RDL 200ms 1590 Hz   & 25 & $^{ +1.3 ~ +2.6 ~ +4.4}$ &  $= 32$ & $\sci{9.7}{49}$  & 14 & $^{ +0.70 ~ +1.5 ~ +1.3}$ &  $= 17$ & $\sci{2.7}{49}$  & 3.3 & $^{ +0.16 ~ +0.34 ~ 
+0.51}$ &  $= 4.1$ & $\sci{1.5}{48}$  \\
 RDL 200ms 2090 Hz   & 37 & $^{ +4.1 ~ +4.8 ~ +9.7}$ &  $= 52$ & $\sci{4.5}{50}$  & 17 & $^{ +1.9 ~ +2.2 ~ +3.3}$ &  $= 23$ & $\sci{7.9}{49}$  & 4.8 & $^{ +0.53 ~ +0.62 ~ 
+0.67}$ &  $= 6.2$ & $\sci{6.5}{48}$  \\
 RDL 200ms 2590 Hz   & 47 & $^{ +6.2 ~ +6.1 ~ +11}$ &  $= 66$ & $\sci{1.0}{51}$  & 23 & $^{ +3.0 ~ +2.9 ~ +3.7}$ &  $= 31$ & $\sci{2.2}{50}$  & 4.7 & $^{ +0.62 ~ +0.61 ~ +0.55}$ 
&  $= 6.2$ & $\sci{9.7}{48}$  \\
 \hline \label{table:stacksim} 

 \end{tabular} 
  \end{scriptsize}
 \end{sidewaystable*}

\begin{sidewaystable*}
\caption[Closed box upper limit estimates for a closed box SGR 1900+14 storm stacking search ]{Closed box upper limit estimates for a closed box SGR 1900+14 storm search, for two stacking scenarios ($N=11$ flat scenario and fluence weighted scenario).  The $N=1$ case is shown for comparison.  Electromagnetic trigger time was offset by 400\,s from the actual time.  Results are shown for various ringdown and band- and time-limited white noise burst target signal classes.  Note that the results in this table, while using real LIGO data, had not been reviewed by the LIGO Scientific Collaboration at the time of writing.  Therefore, this table does not reflect the scientific opinion of the LSC.  }
\begin{scriptsize}
\begin{tabular}{@{\extracolsep{\fill}}l||rlr|r||rlr|r||rlr|r}
 \hline \hline
 & \multicolumn{4}{c}{N=1} & \multicolumn{4}{c}{N=11 Flat} & \multicolumn{4}{c}{Fluence-weighted} \\ 
 Simulation type & \multicolumn{3}{c}{$ \hrssn [ 10^{-22}~ \rthz $] }  & $\egwn$ [erg]   & \multicolumn{3}{c}{$ \hrssn [ 10^{-22} ~ \rthz $] }  & $\egwn$ [erg]  & \multicolumn{3}{c}{$ \hrssn [ 10^{-22} ~ \rthz $] }  & $\egwn$ [erg] \\ 
 \hline 
 WNB 11ms 100-200 Hz   & 2.7 & $^{ +0.0 ~ +0.28 ~ +0.33}$ &  $= 3.1$ & $\sci{9.4}{45}$  & 1.4 & $^{ +0.0 ~ +0.15 ~ +0.13}$ &  $= 1.6$ & $\sci{2.4}{45}$  & 0.32 & $^{ +0.0 ~ +0.033 ~ +0.022}$ &  $= 0.35$ & $\sci{1.2}{44}$  \\
 WNB 100ms 100-200 Hz   & 2.8 & $^{ +0.0 ~ +0.29 ~ +0.20}$ &  $= 3.2$ & $\sci{8.6}{45}$  & 1.5 & $^{ +0.0 ~ +0.15 ~ +0.11}$ &  $= 1.6$ & $\sci{2.3}{45}$  & 0.32 & $^{ +0.0 ~ +0.034 ~ +0.015}$ &  $= 0.36$ & $\sci{1.2}{44}$  \\
 WNB 11ms 100-1000 Hz   & 5.4 & $^{ +0.0 ~ +0.56 ~ +0.55}$ &  $= 6.1$ & $\sci{4.8}{47}$  & 3.2 & $^{ +0.0 ~ +0.33 ~ +0.31}$ &  $= 3.7$ & $\sci{1.7}{47}$  & 0.80 & $^{ +0.0 ~ +0.083 ~ +0.061}$ &  $= 0.91$ & $\sci{1.0}{46}$  \\
 WNB 100ms 100-1000 Hz   & 5.1 & $^{ +0.051 ~ +0.53 ~ +0.73}$ &  $= 6.1$ & $\sci{4.3}{47}$  & 4.0 & $^{ +0.040 ~ +0.42 ~ +0.38}$ &  $= 4.6$ & $\sci{2.3}{47}$  & 0.92 & $^{ +0.0092 ~ +0.096 ~ +0.073}$ &  $= 1.1$ & $\sci{1.2}{46}$  \\
 RDC 200ms 1090 Hz   & 11 & $^{ +0.21 ~ +1.1 ~ +0.94}$ &  $= 12$ & $\sci{6.3}{48}$  & 4.7 & $^{ +0.094 ~ +0.49 ~ +0.31}$ &  $= 5.4$ & $\sci{1.3}{48}$  & 1.2 & $^{ +0.023 ~ +0.12 ~ +0.12}$ &  $= 1.4$ & $\sci{8.3}{46}$  \\
 RDC 200ms 1590 Hz   & 14 & $^{ +0.57 ~ +1.5 ~ +2.1}$ &  $= 17$ & $\sci{2.8}{49}$  & 7.5 & $^{ +0.30 ~ +0.77 ~ +0.63}$ &  $= 8.8$ & $\sci{6.9}{48}$  & 1.7 & $^{ +0.067 ~ +0.17 ~ +0.12}$ &  $= 1.9$ & $\sci{3.5}{47}$  \\
 RDC 200ms 2090 Hz   & 20 & $^{ +1.6 ~ +2.5 ~ +2.0}$ &  $= 24$ & $\sci{9.5}{49}$  & 9.0 & $^{ +0.72 ~ +1.2 ~ +0.89}$ &  $= 11$ & $\sci{2.0}{49}$  & 2.4 & $^{ +0.19 ~ +0.31 ~ +0.20}$ &  $= 3.0$ & $\sci{1.4}{48}$  \\
 RDC 200ms 2590 Hz   & 20 & $^{ +2.6 ~ +2.6 ~ +2.8}$ &  $= 27$ & $\sci{1.7}{50}$  & 11 & $^{ +1.5 ~ +1.5 ~ +0.94}$ &  $= 15$ & $\sci{5.2}{49}$  & 2.8 & $^{ +0.37 ~ +0.36 ~ +0.19}$ &  $= 3.6$ & $\sci{3.1}{48}$  \\
 RDL 200ms 1090 Hz   & 19 & $^{ +0.56 ~ +1.9 ~ +3.7}$ &  $= 23$ & $\sci{2.5}{49}$  & 8.9 & $^{ +0.27 ~ +0.93 ~ +1.8}$ &  $= 11$ & $\sci{5.5}{48}$  & 2.0 & $^{ +0.061 ~ +0.21 ~ +0.37}$ &  $= 2.5$ & $\sci{3.0}{47}$  \\
 RDL 200ms 1590 Hz   & 29 & $^{ +1.4 ~ +3.0 ~ +6.6}$ &  $= 37$ & $\sci{1.2}{50}$  & 12 & $^{ +0.58 ~ +1.2 ~ +2.5}$ &  $= 15$ & $\sci{2.1}{49}$  & 4.1 & $^{ +0.21 ~ +0.43 ~ +0.52}$ &  $= 5.0$ & $\sci{2.4}{48}$  \\
 RDL 200ms 2090 Hz   & 45 & $^{ +5.0 ~ +5.8 ~ +8.0}$ &  $= 60$ & $\sci{6.0}{50}$  & 18 & $^{ +2.0 ~ +2.4 ~ +2.5}$ &  $= 24$ & $\sci{9.1}{49}$  & 5.1 & $^{ +0.56 ~ +0.65 ~ +1.3}$ &  $= 7.1$ & $\sci{7.7}{48}$  \\
 RDL 200ms 2590 Hz   & 50 & $^{ +6.5 ~ +6.4 ~ +14}$ &  $= 71$ & $\sci{1.2}{51}$  & 18 & $^{ +2.3 ~ +2.2 ~ +5.8}$ &  $= 26$ & $\sci{1.4}{50}$  & 5.6 & $^{ +0.73 ~ +0.72 ~ +0.75}$ &  $= 7.4$ & $\sci{1.4}{49}$  \\
 \hline \label{table:stackclosed} 
 \end{tabular} 
 \end{scriptsize} 
 \end{sidewaystable*}

\clearpage

\chapter{Conclusion} \label{chapter:conclusion}

\section{Summary}

We have described work done preparing photon calibrators for use in
LIGO detectors.  We have described the discovery of a significant
discrepancy between calibration via photon calibrators and the coil
calibration, which led to an improved understanding of LIGO detector
calibration.  We have described how the photon calibrators were used
to discover a significant error in the detector timing calibration.

We have described the Flare pipeline, a simple but
powerful tool for performing externally triggered gravitational wave
searches.  Though we have focused on SGR bursts in this work, the Flare pipeline is a general tool which could be used effectively in searches with other astrophysical targets.  We have described work done characterizing and validating
the pipeline, including analysis of GRB 070201 and  comparisons to
two other major LSC burst pipelines, and the LSC CBC matched filter
pipeline.  These are the first comparisons between coherent burst
pipelines and matched filter CBC pipelines of which we are aware.   

We have described a search for gravitational waves associated with the
SGR 1806--20 giant flare and 214 SGR bursts in the first year of
LIGO's fifth science run\,\cite{s5y1sgr}. The loudest events from the
on-source regions analyzed are consistent with no detection. Twelve strain and twelve
$\egw$ upper limits were set for each on-source region at 90\%
detection efficiency, one for each of twelve simulated signal types.

We have described a method for a powerful follow-up search which
stacks individual SGR bursts in order to increase the chances for a first gravitational wave detection and significantly improve upper
limits in a reasonably model-independent way.

\section{Discussion of individual SGR burst search}

Two searches for gravitational waves associated with SGR events had
been published before we published our work; neither claimed detection. The AURIGA collaboration
searched for gravitational wave bursts associated with the SGR
1806$-20$ giant flare in the band 850--950~Hz with damping time
100~ms, setting upper limits on the gravitational wave energy of
$\sim10^{49}$~erg\,\cite{auriga05}. The LIGO collaboration also
published on the same giant flare, targeting times and frequencies
of the quasi-periodic oscillations in the flare's x-ray tail as well
as other frequencies in the detector's band, setting upper limits on
gravitational wave energy as low as $8\times10^{46}$~erg for
quasi-periodic signals lasting tens of seconds\,\cite{matone07}.

In addition to the giant flare from SGR 1806--20, the search
described in Chapter\,\ref{chapter:search} covers 214 smaller flares which occurred during the
LIGO S5 data run, when the LIGO amplitude noise was typically
$\sim1/3$ the value at the time of the giant flare. This was the
first search sensitive to the $f$-modes, which are usually
considered the most efficient gravitational wave
emitters\,\cite{andersson97}; we also searched the entire frequency
band of best detector sensitivity.  We have done this with
unprecedentedly sensitive gravitational wave detectors, and an
unprecedentedly sensitive triggered burst analysis pipeline.  Not
surprisingly our upper limits on gravitational wave strain amplitude
were the best  published to date for a short-duration gravitational wave
burst search. Our upper limits on gravitational wave emission energy
(Figure\,\ref{fig:egw90All}) overlap the range of electromagnetic
energies $\sim10^{44}$--$10^{46}$~erg seen in SGR giant
flares\,\cite{hurley05, palmer05} and more than one third are below
the $\sim10^{49}$~erg maximum gravitational wave energy predicted in
some theoretical models\,\cite{ioka01}. Our best upper limits on
$\gamma$ are within the theoretically predicted range implied
in\,\cite{ioka01}.

\section{Future work}

Three new analyses are planned for the near future.   First, we plan to analyze individual SGR bursts in the remainder of the S5 LIGO
science run and the Virgo VSR1 science run using the Flare pipeline.  Second, we plan to analyze individual SGR bursts from the newly-discovered SGR 0501+4516, which may be an order of magnitude closer to Earth than SGR 1900+14 and SGR 1806--20, using Astrowatch data from the LIGO H2 detector which was taken after the end of S5.  These projects should both be straightforward, as the
method is already completely implemented and reviewed.  Finally, we have already begun a Stack-a-flare search for gravitational waves associated with the S5 SGR 1900+14 storm, and we expect to finish the analysis and publish the results in 2009.

As existing detectors become increasingly
sensitive and new detectors in the global network come online,
the prospects for detecting gravitational waves from SGRs improve. We plan to make scientific statements about gravitational waves from SGRs using data from two upcoming science runs.  The enhanced LIGO detectors will double the amplitude sensitivity of S5, giving an improvement in energy sensitivity and therefore $\gamma$ of a factor of four or more.  A sixth LIGO science run (S6) with the enhanced detectors is scheduled to begin in mid-2009.  Further in the future, the Advanced LIGO detectors promise an improvement in $\hrss$ by more than a factor of
10 over S5, corresponding to an improvement in energy sensitivity
(and therefore $\gamma$) by more than a factor of 100.  

The methods described here could usefully analyze SGR activity in future science
runs.  However, we will continue to search for better methods.  For example,  the Flare pipeline could be redesigned with the capacity to use more than two detectors.  Also, the T-Stack prototype exhibits energy sensitivity gains that scale with the first power of $N$ in sandbox conditions; we would like to find a way to actualize those gains in real Stack-a-flare searches, which currently gain sensitivity as $N^{1/2}$.

In conclusion, SGRs are promising sources for a first gravitational wave detection.  Our analysis tools have proven to be valuable additions to the collaboration's data analysis battery.  Our work has ensured that externally triggered searches on SGR bursts will continue to be a significant part of the LSC's science output as the detectors continue to improve. 
\clearpage

\bibliographystyle{unsrt}

\begin{thebibliography}{100}

\bibitem{einstein1916}
A.~{Einstein}.
\newblock {Die Grundlage der allgemeinen Relativitätstheorie}.
\newblock {\em Annalen der Physik}, 49:769--822, 1916.

\bibitem{hulse1975}
R.~A. {Hulse} and J.~H. {Taylor}.
\newblock {Discovery of a pulsar in a binary system}.
\newblock {\em Astrophysical Journal Letters}, 195:L51--L53, January 1975.

\bibitem{taylor79}
J.~H. {Taylor}, Jr., L.~A. {Fowler}, and P.~M. {McCulloch}.
\newblock {Measurements of general relativistic effects in the binary pulsar
  PSR 1913+16}.
\newblock {\em Nature}, 277:437, 1979.

\bibitem{taylor82}
J.~H. {Taylor} and J.~M. {Weisberg}.
\newblock {A new test of general relativity - Gravitational radiation and the
  binary pulsar PSR 1913+16}.
\newblock {\em \apj}, 253:908--920, February 1982.

\bibitem{taylor94}
J.~H. {Taylor}, Jr.
\newblock {Binary pulsars and relativistic gravity}.
\newblock {\em Reviews of Modern Physics}, 66:711--719, July 1994.

\bibitem{will06}
Clifford~M. Will.
\newblock The confrontation between general relativity and experiment.
\newblock {\em Living Reviews in Relativity}, 9(3), 2006.

\bibitem{weisberg04}
J.~M. {Weisberg} and J.~H. {Taylor}.
\newblock {The Relativistic Binary Pulsar B1913+16: Thirty Years of
  Observations and Analysis}.
\newblock In F.~A. {Rasio} and I.~H. {Stairs}, editors, {\em Binary Radio
  Pulsars}, volume 328 of {\em Astronomical Society of the Pacific Conference
  Series}, pages 25--+, July 2005.

\bibitem{S5}
{B.~Abbott} et~al.
\newblock {LIGO: The Laser Interferometer Gravitational-Wave Observatory}.
\newblock {\em ArXiv e-prints}, November 2007.

\bibitem{aligosite}
{http://www.ligo.caltech.edu/advLIGO/}.

\bibitem{ligoWeb}
{http://www.ligo.caltech.edu/}.

\bibitem{geo}
{http://www.ligo.caltech.edu/}.

\bibitem{virgo}
{http://www.virgo.infn.it/}.

\bibitem{tama}
{http://tamago.mtk.nao.ac.jp/}.

\bibitem{lcgt}
{http://www.icrr.u-tokyo.ac.jp/gr/gre.html}.

\bibitem{lisaNasa}
{http://lisa.nasa.gov/}.

\bibitem{lisaAei}
{http://www.lisa.aei-hannover.de/}.

\bibitem{thompson95}
C.~{Thompson} and R.~C. {Duncan}.
\newblock {The soft gamma repeaters as very strongly magnetized neutron stars -
  I. Radiative mechanism for outbursts}.
\newblock {\em \mnras}, 275:255--300, July 1995.

\bibitem{schwartz05}
S.~J. {Schwartz} et~al.
\newblock {The Gamma-Ray Giant Flare from SGR 1806--20: Evidence of Crustal
  Cracking via Initial Timescales}.
\newblock {\em \apjl}, 627:L129--L132, July 2005.

\bibitem{andersson97}
N.~{Andersson} and K.~D. {Kokkotas}.
\newblock {Towards gravitational wave asteroseismology}.
\newblock {\em \mnras}, 299:1059--1068, October 1998.

\bibitem{pacheco98}
J.~A. {de Freitas Pacheco}.
\newblock {Do soft gamma repeaters emit gravitational waves?}
\newblock {\em Astronomy and Astrophysics}, 336:397--401, August 1998.

\bibitem{ioka01}
K.~{Ioka}.
\newblock {Magnetic deformation of magnetars for the giant flares of the soft
  gamma-ray repeaters}.
\newblock {\em \mnras}, 327:639--662, October 2001.

\bibitem{andersson02}
N.~{Andersson}.
\newblock {TOPICAL REVIEW: Gravitational waves from instabilities in
  relativistic stars}.
\newblock {\em \cqg}, 20:105--+, April 2003.

\bibitem{horvath05}
J.~E. {Horvath}.
\newblock {Energetics of the Superflare from SGR1806--20 and a Possible
  Associated Gravitational Wave Burst}.
\newblock {\em Modern Physics Lett. A}, 20:2799--2804, 2005.

\bibitem{ju00}
L.~{Ju}, D.~G. {Blair}, and C.~{Zhao}.
\newblock {Detection of gravitational waves.}
\newblock {\em Reports of Progress in Physics}, 63:1317--1427, 2000.

\bibitem{coccia98}
E.~Coccia, V.~Fafone, G.~Frossati, J.~A. Lobo, and J.~A. Ortega.
\newblock A hollow sphere as a detector of gravitational radiation.
\newblock {\em Physical Review D}, 57:2051, 1998.

\bibitem{schutz}
B.~F. {Schutz}.
\newblock {\em {A First Course in General Relativity}}.
\newblock Cambridge, UK: Cambridge University Press, February 1985.

\bibitem{carroll}
S.~M. {Carroll}.
\newblock {\em {Spacetime and geometry. An introduction to general
  relativity}}.
\newblock San Francisco, CA, USA: Addison Wesley, 2004.

\bibitem{misner}
C.~W. {Misner}, K.~S. {Thorne}, and J.~A. {Wheeler}.
\newblock {\em {Gravitation}}.
\newblock San Francisco: W.H.~Freeman and Co., 1973.

\bibitem{sigg98}
D.~{Sigg}.
\newblock {Gravitational waves}.
\newblock In {\em Proceedings of TASI, 98, Boulder, CO}, 1998.
\newblock LIGO-P980007-00-D.

\bibitem{rotor}
L.~{Matone}, P.~{Raffai}, S.~{Marka}, R.~{Grossman}, P.~{Kalmus}, Z.~{Marka},
  J.~{Rollins}, and V.~{Sannibale}.
\newblock {Benefits of Artificially Generated Gravity Gradients for
  Interferometric Gravitational-Wave Detectors}.
\newblock {\em ArXiv General Relativity and Quantum Cosmology e-prints},
  January 2007.

\bibitem{300years}
K.~S. {Thorne}.
\newblock Gravitational radiation.
\newblock In S.~W. {Hawking} and W~{Israel}, editors, {\em 300 Years of
  Gravitation}, page 417. Cambridge University Press, Cambridge, 1987.

\bibitem{S5GRB070201}
{LIGO Scientific Collaboration} and K.~{Hurley}.
\newblock {Implications for the Origin of GRB 070201 from LIGO Observations}.
\newblock {\em \apj}, 681:1419--1430, July 2008.

\bibitem{ott04}
C.~D. {Ott}, A.~{Burrows}, E.~{Livne}, and R.~{Walder}.
\newblock {Gravitational Waves from Axisymmetric, Rotating Stellar Core
  Collapse}.
\newblock {\em \apj}, 600:834--864, January 2004.

\bibitem{dimmelmeier02}
H.~{Dimmelmeier}, J.~A. {Font}, and E.~{M{\"u}ller}.
\newblock {Relativistic simulations of rotational core collapse II. Collapse
  dynamics and gravitational radiation}.
\newblock {\em \aap}, 393:523--542, October 2002.

\bibitem{zwerger97}
T.~{Zwerger} and E.~{Mueller}.
\newblock {Dynamics and gravitational wave signature of axisymmetric rotational
  core collapse.}
\newblock {\em \aap}, 320:209--227, April 1997.

\bibitem{S1inspiral}
B.~Abbott et~al.
\newblock Analysis of ligo data for gravitational waves from binary neutron
  stars.
\newblock {\em \prd}, 69(12):122001, Jun 2004.

\bibitem{S2inspiral}
B.~Abbott et~al.
\newblock Search for gravitational waves from galactic and extra-galactic
  binary neutron stars.
\newblock {\em \prd}, 72(8):082001, 2005.

\bibitem{S2macho}
B.~Abbott et~al.
\newblock Search for gravitational waves from primordial black hole binary
  coalescences in the galactic halo.
\newblock {\em \prd}, 72(8):082002, 2005.

\bibitem{S2inspiralLigoTama}
B.~Abbott et~al.
\newblock Joint ligo and tama300 search for gravitational waves from
  inspiralling neutron star binaries.
\newblock {\em \prd}, 73(10):102002, 2006.

\bibitem{S2inspiralBBH}
B.~Abbott et~al.
\newblock Search for gravitational waves from binary black hole inspirals in
  ligo data.
\newblock {\em \prd}, 73(6):062001, 2006.

\bibitem{S3S4inspiral}
B.~Abbott et~al.
\newblock Search for gravitational waves from binary inspirals in s3 and s4
  ligo data.
\newblock {\em \prd}, 77(6):062002, 2008.

\bibitem{S3spinningInspiral}
B.~Abbott et~al.
\newblock Search of s3 ligo data for gravitational wave signals from spinning
  black hole and neutron star binary inspirals.
\newblock 2007.

\bibitem{cutler93}
C.~{Cutler}, T.~A. {Apostolatos}, L.~{Bildsten}, L.~S. {Finn}, E.~E.
  {Flanagan}, D.~{Kennefick}, D.~M. {Markovic}, A.~{Ori}, E.~{Poisson}, and
  G.~J. {Sussman}.
\newblock {The last three minutes - Issues in gravitational-wave measurements
  of coalescing compact binaries}.
\newblock {\em \prl}, 70:2984--2987, May 1993.

\bibitem{apostolatos95}
T.~A. {Apostolatos}.
\newblock {Search templates for gravitational waves from precessing,
  inspiraling binaries}.
\newblock {\em \prd}, 52:605--620, July 1995.

\bibitem{bildsten92}
L.~{Bildsten} and C.~{Cutler}.
\newblock {Tidal interactions of inspiraling compact binaries}.
\newblock {\em \apj}, 400:175--180, November 1992.

\bibitem{blanchet96}
L.~{Blanchet}, B.~R. {Iyer}, C.~M. {Will}, and A.~G. {Wiseman}.
\newblock {Gravitational waveforms from inspiralling compact binaries to second
  post-Newtonian order.}
\newblock {\em \cqg}, 13:575--584, April 1996.

\bibitem{droz99}
S.~{Droz}, D.~J. {Knapp}, E.~{Poisson}, and B.~J. {Owen}.
\newblock {Gravitational waves from inspiraling compact binaries: Validity of
  the stationary-phase approximation to the Fourier transform}.
\newblock {\em \prd}, 59(12):124016--+, June 1999.

\bibitem{blanchet06}
L.~{Blanchet}.
\newblock {Gravitational Radiation from Post-Newtonian Sources and Inspiralling
  Compact Binaries}.
\newblock {\em Living Reviews in Relativity}, 9:4--+, June 2006.

\bibitem{buonanno04}
Alessandra Buonanno, Yanbei Chen, Yi~Pan, and Michele Vallisneri.
\newblock {Quasiphysical family of gravity-wave templates for precessing
  binaries of spinning compact objects: Application to double-spin precessing
  binaries}.
\newblock {\em \prd}, 70(10):104003, Nov 2004.

\bibitem{kalogera04}
V.~{Kalogera}, C.~{Kim}, D.~R. {Lorimer}, M.~{Burgay}, N.~{D'Amico},
  A.~{Possenti}, R.~N. {Manchester}, A.~G. {Lyne}, B.~C. {Joshi}, M.~A.
  {McLaughlin}, M.~{Kramer}, J.~M. {Sarkissian}, and F.~{Camilo}.
\newblock {The Cosmic Coalescence Rates for Double Neutron Star Binaries}.
\newblock {\em \apjl}, 601:L179--L182, February 2004.

\bibitem{postnov06}
Lev R.~Yungelson Konstantin A.~Postnov.
\newblock The evolution of compact binary star systems.
\newblock {\em Living Reviews in Relativity}, 9(6), 2006.

\bibitem{S5strain}
{LIGO Scientific Collaboration}.
\newblock {Best Strain Sensitivities for the LIGO interferometers}.
\newblock {\em LIGO Internal Note G060009-02}, June 2006.

\bibitem{S5crab}
B.~Abbott et~al.
\newblock {Beating the spin-down limit on gravitational wave emission from the
  Crab pulsar}.
\newblock {\em ArXiv e-prints}, 805, May 2008.

\bibitem{S1pulsar}
B.~{Abbott} et~al.
\newblock {Setting upper limits on the strength of periodic gravitational waves
  from PSR J1939+2134 using the first science data from the GEO 600 and LIGO
  detectors}.
\newblock {\em \prd}, 69(8):082004--+, April 2004.

\bibitem{lisaMDC}
S.~{Babak}, J.~G. {Baker}, M.~J. {Benacquista}, N.~J. {Cornish}, J.~{Crowder},
  S.~L. {Larson}, E.~{Plagnol}, E.~K. {Porter}, M.~{Vallisneri}, A.~{Vecchio},
  K.~{Arnaud}, L.~{Barack}, A.~{B{\l}aut}, C.~{Cutler}, S.~{Fairhurst},
  J.~{Gair}, X.~{Gong}, I.~{Harry}, D.~{Khurana}, A.~{Kr{\'o}lak}, I.~{Mandel},
  R.~{Prix}, B.~S. {Sathyaprakash}, P.~{Savov}, Y.~{Shang}, M.~{Trias},
  J.~{Veitch}, Y.~{Wang}, L.~{Wen}, and J.~T. {Whelan}.
\newblock {The Mock LISA Data Challenges: from Challenge 1B to Challenge 3}.
\newblock {\em ArXiv e-prints}, 806, June 2008.

\bibitem{confusionNoise}
Peter~L Bender and Dieter Hils.
\newblock Confusion noise level due to galactic and extragalactic binaries.
\newblock {\em \cqg}, 14(6):1439--1444, 1997.

\bibitem{S3stochastic}
B.~Abbott et~al.
\newblock Upper limits on a stochastic background of gravitational waves.
\newblock {\em \prl}, 95(22):221101, 2005.

\bibitem{battye96}
R.~A. {Battye} and E.~P.~S. {Shellard}.
\newblock {Primordial gravitational waves : a probe of the early universe}.
\newblock {\em ArXiv Astrophysics e-prints}, April 1996.

\bibitem{fritschel01}
Peter Fritschel, Rolf Bork, Gabriela Gonz\'{a}lez, Nergis Mavalvala, Dale
  Ouimette, Haisheng Rong, Daniel Sigg, and Michael Zucker.
\newblock Readout and control of a power-recycled interferometric
  gravitational-wave antenna.
\newblock {\em Appl. Opt.}, 40(28):4988--4998, 2001.

\bibitem{abramovici92}
D.~{Abramovici} et~al.
\newblock {\em Science}, 256:325, 1992.

\bibitem{ligoNoiseRuns}
{http://www.ligo.caltech.edu/docs/G/G060009-03/}.

\bibitem{savage98}
R.L. {Savage}, P.J. {King}, and S.U. {Seel}.
\newblock {A highly stabilized 10-watt Nd:YAG laser for the laser
  interferometer gravitational-wave observatory (LIGO)}.
\newblock {\em Laser Phys.}, 8:679, 1998.

\bibitem{drever83}
R.W.P. {Drever} et~al.
\newblock {Laser phase and frequency stabilization using an optical resonator}.
\newblock {\em Appl. Phys. B: Photophys. Laser Chem.}, 31:97--105, 1983.

\bibitem{ruldiger81}
A.~{Ruldiger} et~al.
\newblock {A mode selector to suppress fluctuations in laser beam geometry}.
\newblock {\em Optica Acta}, 28:641, 1981.

\bibitem{meers88}
Brian~J. Meers.
\newblock Recycling in laser-interferometric gravitational-wave detectors.
\newblock {\em \prd}, 38(8):2317--2326, Oct 1988.

\bibitem{callen51}
H.~B. {Callen} and T.~A. {Welton}.
\newblock {Irreversibility and generalized noise}.

\bibitem{stan03}
{LIGO Scientific Collaboration: B.~Abbott et al.}
\newblock {Detector Description and Performance for the First Coincidence
  Observations between LIGO and GEO}.
\newblock {\em Nucl. Instrum. Methods A}, 517, 2004.

\bibitem{barish99}
B.~{Barish} and R.~{Weiss}.
\newblock {LIGO and the Detection of Gravitational Waves}.
\newblock {\em Phys. Today}, 52, October 1999.

\bibitem{sigg03}
D.~{Sigg}.
\newblock {Strain Calibration in LIGO}.
\newblock {\em LIGO Internal Note T970101-B-D}, 2003.

\bibitem{S4}
A.~{Dietz} et~al.
\newblock {Calibration of the LIGO detectors for S4}.
\newblock {\em LIGO Internal Note T050262-00-D}, 2006.

\bibitem{S1}
R.~{Adhikari}, G.~{Gonz{\'a}lez}, M.~{Landry}, B.~{O'Reilly}, and {the LIGO
  Scientific Collaboration}.
\newblock {Calibration of the LIGO detectors for the First LIGO Science Run}.
\newblock {\em \cqg}, 20:903--+, September 2003.

\bibitem{S2}
G.~{Gonz{\'a}lez}, M.~{Landry}, B.~{O'Reilly}, and H.~{Radkins}.
\newblock {Calibration of the LIGO detectors for S2}.
\newblock {\em LIGO Internal Note T040060-01-D}, 2004.

\bibitem{S3}
G.~{Gonz{\'a}lez}, M.~{Landry}, B.~{O'Reilly}, and X.~{Siemens}.
\newblock {Calibration of the LIGO detectors for S3}.
\newblock {\em LIGO Internal Note T050059-01-D}, 2005.

\bibitem{fringe}
{R. Adhikari et al.}
\newblock {Input test mass (ITM) absolute calibrations: fringe counting, fringe
  fitting, and sign toggling methods}.
\newblock {\em LIGO Internal Note T020141-01-D}, 2002.

\bibitem{siemens04}
X.~{Siemens} et~al.
\newblock {Making h(t) for LIGO}.
\newblock {\em \cqg}, 21:S1723, May 2004.

\bibitem{v3S5dq}
{http://lancelot.mit.edu/$\sim$cadonati/S5/DQkleinFullS5-v3/Summary.html}.

\bibitem{saulson}
P.~R. {Saulson}.
\newblock {\em {Fundamentals of interferometric gravitational wave detectors}}.
\newblock Singapore: World Scientific Publishers, 1994.

\bibitem{anderson01}
W.~G. {Anderson}, P.~R. {Brady}, J.~D. {Creighton}, and {\'E}.~{\'E}.
  {Flanagan}.
\newblock {Excess power statistic for detection of burst sources of
  gravitational radiation}.
\newblock {\em Phys.~Rev.~D}, 63(4):042003--+, February 2001.

\bibitem{eligo}
{{Landry}~M.}
\newblock {Enhanced LIGO Status - PAC 24 Meeting @ LIGO Hanford Observatory,
  June 24-25, 2008}.
\newblock {\em LIGO Internal Note G080387-00-D}, 2008.

\bibitem{aligo}
{{Fritschel}~P. and {Coyne}~D.}
\newblock {Technical Status of Advanced LIGO}.
\newblock {\em LIGO Internal Note G080391-00}, 2008.

\bibitem{yoichiThesis}
Y.~{Aso}.
\newblock {\em Active Vibration Isolation for a Laser Interferometric
  Gravitational Wave Detector using a Suspension Point Interferometer}.
\newblock {PhD thesis}, University of Tokyo, 2008.

\bibitem{justice03}
J.~{Bruursema}.
\newblock {Calibration of the LIGO Interferometer Using the Recoil of Photons}.
\newblock {\em LIGO Internal Note G030513-00-D}, 2003.

\bibitem{goetz04}
E.~{Goetz}.
\newblock {Commissioning of the Photon Calibrators}.
\newblock {\em LIGO Internal Note T040196-00-D}, 2004.

\bibitem{goetz07}
E.~{Goetz}, P.~{Kalmus}, and R.~{Savage}.
\newblock {Commissioning of the Photon Calibrators}.
\newblock {\em LIGO Internal Note T070026-00-W}, 2007.

\bibitem{clubley01}
D.~A. {Clubley}, G.~P. {Newton}, K.~D. {Skeldon}, and J.~{Hough}.
\newblock {Calibration of the Glasgow 10 m prototype laser interferometric
  gravitational wave detector using photon pressure}.
\newblock {\em Physics Letters A}, 283:85--88, May 2001.

\bibitem{geo06}
K.~{Mossavi}, M.~{Hewitson}, S.~{Hild}, F.~{Seifert}, U.~{Weiland}, J.~R.
  {Smith}, H.~{L{\"u}ck}, H.~{Grote}, B.~{Willke}, and K.~{Danzmann}.
\newblock {A photon pressure calibrator for the GEO 600 gravitational wave
  detector}.
\newblock {\em Physics Letters A}, 353:1--3, April 2006.

\bibitem{kalmus05}
P.~{Kalmus}.
\newblock {State of the Photon Calibrators}.
\newblock {\em {LIGO Internal Note G05433-00-I}}, 2005.
\newblock {Talk presented at 2005 August LSC meeting at Hanford}.

\bibitem{tdsresp}
{code written by M. Evans}.

\bibitem{scientechEmail}
{Personal communication with D. Froman, Scientech Inc.}

\bibitem{dmt}
{http://www.ligo-wa.caltech.edu/gds/}.

\bibitem{kalmus06commissioning}
P.~{Kalmus}.
\newblock {Discrepancy Between Photon Calibration and Official Calibration}.
\newblock {\em {LIGO Internal Note G060687-00-I}}, 2006.
\newblock {Talk presented at 2006 May 15 Commissioning Meeting}.

\bibitem{winkler91}
W.~{Winkler}, K.~{Danzmann}, A.~{R{\"u}diger}, and R.~{Schilling}.
\newblock {Heating by optical absorption and the performance of interferometric
  gravitational-wave detectors}.
\newblock {\em Phys.~Rev.~A}, 44:7022--7036, December 1991.

\bibitem{raoThesis}
S.~{Rao}.
\newblock {\em Mirror Thermal Noise in Interferometric GravitationalWave
  Detectors}.
\newblock {PhD thesis}, California Institute of Technology, 2003.

\bibitem{kalmus06}
E.~{Goetz} and P.~{Kalmus}.
\newblock {State of the Photon Calibrators}.
\newblock {\em {LIGO Internal Note G060686-00-I}}, 2006.
\newblock {Talk presented at August 2006 LSC meeting at LSU}.

\bibitem{goetz06}
E.~{Goetz}, P.~{Kalmus}, and R.~{Savage}.
\newblock {Status of the Photon Calibrators}.
\newblock {\em {LIGO Internal Note G060647-00-I}}, 2006.
\newblock {Talk presented at 2006 December 11 Commissioning Telecon}.

\bibitem{hild07}
S.~{Hild}, M.~{Brinkmann}, K.~{Danzmann}, H.~{Grote}, M.~{Hewitson},
  J.~{Hough}, H.~{L{\"u}ck}, I.~{Martin}, K.~{Mossavi}, N.~{Rainer}, S.~{Reid},
  J.~R. {Smith}, K.~{Strain}, M.~{Weinert}, P.~{Willems}, B.~{Willke}, and
  W.~{Winkler}.
\newblock {Photon-pressure-induced test mass deformation in gravitational-wave
  detectors}.
\newblock {\em \cqg}, 24:5681--5688, November 2007.

\bibitem{goldStandardNIST}
{National Institute of Standards and Technology}.
\newblock {Report of calibration (42110CA/42111CA)}.
\newblock {\em LIGO Internal Note T070245-00-W}, 2008.

\bibitem{goetz08}
E.~{Goetz} and R.~{Savage}.
\newblock {Investigation of discrepancies between Photon calibrator, VCO and
  Official (coil) calibration techniques}.
\newblock {\em {LIGO Internal Note G080216-00-Z}}, 2008.
\newblock {Talk presented at March 2008 LSC-VIRGO meeting}.

\bibitem{aso08}
Y.~{Aso} et~al.
\newblock {Accurate measurement of the time delay in the response of the LIGO
  gravitational wave detectors}.
\newblock 2008.
\newblock {submitted to CQG}.

\bibitem{smith08}
N.~{Smith} and D.~{Sigg}.
\newblock {\em LIGO Internal Note T080039-00-D}, 2008.

\bibitem{aligoPcal}
M.~{Smith} and P.~{Willems}.
\newblock {Advanced LIGO Photon Calibrator Conceptual Design}.
\newblock {\em {LIGO Internal Note T070167-01-D}}, 2007.

\bibitem{kalmus07}
P.~{Kalmus} et~al.
\newblock {Search method for unmodeled transient gravitational waves associated
  with SGR Flares}.
\newblock {\em \cqg}, 24:S659--S669, October 2007.

\bibitem{flanagan98a}
{\'E}.~{\'E}. {Flanagan} and S.~A. {Hughes}.
\newblock {Measuring gravitational waves from binary black hole coalescences.
  I. Signal to noise for inspiral, merger, and ringdown}.
\newblock {\em Phys.~Rev.~D}, 57:4535--4565, April 1998.

\bibitem{flanagan98b}
{\'E}.~{\'E}. {Flanagan} and S.~A. {Hughes}.
\newblock {Measuring gravitational waves from binary black hole coalescences.
  II. The waves' information and its extraction, with and without templates}.
\newblock {\em Phys.~Rev.~D}, 57:4566--4587, April 1998.

\bibitem{brady04}
P.~R. {Brady}, J.~D.~E. {Creighton}, and A.~G. {Wiseman}.
\newblock {Upper limits on gravitational-wave signals based on loudest events}.
\newblock {\em \cqg}, 21:S1775, May 2004.

\bibitem{S4burstAllSky}
B.~Abbott et~al.
\newblock Search for gravitational-wave bursts in ligo data from the fourth
  science run.
\newblock {\em \cqg}, 24(22):5343--5369, 2007.

\bibitem{hamming98}
R.~Hamming.
\newblock {\em Digital Filters}.
\newblock Dover Publications, New York, 1998.

\bibitem{khan07}
R.~{Khan} and S.~{Chatterji}.
\newblock {Enhancing the capabilities of LIGO time-frequency plane searches
  through clustering}.
\newblock {\em in progress}, 2007.

\bibitem{burstmdc}
K~{Thorne}.
\newblock https://gravity.psu.edu/$\sim$psurg/sims/burstmdc/index.html.

\bibitem{shapiro83}
S.~{Shapiro} and S.~{Teukolsky}.
\newblock {\em Black Holes, White Dwarfs, and Neutron Stars}.
\newblock Wiley, New York, 1983.

\bibitem{xphome}
P~{Sutton}.
\newblock http://www.ligo.caltech.edu/\~{}psutton/protected/ \\
  xpipeline/xpipeline.html.

\bibitem{efron79}
B.~{Efron}.
\newblock {Bootstrap methods: another look at the jackknife}.
\newblock {\em Ann. Statist.}, 7:1--26, 1979.

\bibitem{S4HardwareInjections}
{http://lhocds.ligo-wa.caltech.edu/scirun/S4/HardwareInjection/}.

\bibitem{S5HardwareInjections}
{http://lhocds.ligo-wa.caltech.edu/scirun/S5/HardwareInjection/}.

\bibitem{noiseCurveEarlyS5}
{LIGO Scientific Collaboration}.
\newblock {Best Strain Sensitivities for the LIGO Interferometers, Early S5
  Performance}.
\newblock {\em LIGO Internal Note G060010}, January 2006.

\bibitem{s4allsky}
B.~Abbott et~al.
\newblock {Search for gravitational-wave bursts in LIGO data from the fourth
  science run}.
\newblock {\em ArXiv e-prints}, 704, April 2007.

\bibitem{hurley99WhereIs}
K.~{Hurley} et~al.
\newblock {Where is SGR 1806--20?}
\newblock {\em \apjl}, 523:L37--L40, September 1999.

\bibitem{gcn6088}
S.~{Golentskii} et~al.
\newblock {IPN localization of very intense short GRB 070201}.
\newblock {\em GRB Coordinates Network}, 6088, 2008.

\bibitem{cWBnote}
S.~{Klimenko}.
\newblock {Coherent WaveBurst}.
\newblock {\em LIGO note T060282-00-Z}, 2006.

\bibitem{klimenko08}
S.~{Klimenko}, I.~{Yakushin}, A.~{Mercer}, and G.~{Mitselmakher}.
\newblock {A coherent method for detection of gravitational wave bursts}.
\newblock {\em \cqg}, 25(11):114029--+, June 2008.

\bibitem{klimenko05}
S.~Klimenko, S.~Mohanty, Malik Rakhmanov, and G.~Mitselmakher.
\newblock Constraint likelihood analysis for a network of gravitational wave
  detectors.
\newblock {\em \prd}, 72:122002, 2005.

\bibitem{cvswat}
S.~{Klimenko}.
\newblock http://www.lsc-group.phys.uwm.edu/cgi-bin/Analysis/Waveburst/S4/wat.

\bibitem{WBnoteS5}
S.~{Klimenko}, I.~{Yakushin}, and G.~{Mitselmakher}.
\newblock {WaveBurst: S5 version}.
\newblock {\em LIGO note T060112-00-Z}, 2006.

\bibitem{chatterji05}
Shourov Chatterji, Albert Lazzarini, Leo Stein, Patrick~J. Sutton, Antony
  Searle, and Massimo Tinto.
\newblock Coherent network analysis technique for discriminating
  gravitational-wave bursts from instrumental noise.
\newblock {\em \prd}, 74(8):082005, 2006.

\bibitem{beauville07}
F.~Beauville et~al.
\newblock {A comparison of methods for gravitational wave burst searches from
  LIGO and Virgo}.
\newblock {\em \cqg}, 25:045002, 2008.

\bibitem{hurley05}
K.~{Hurley} et~al.
\newblock {An exceptionally bright flare from SGR 1806--20 and the origins of
  short-duration {$\gamma$}-ray bursts}.
\newblock {\em Nature}, 434:1098--1103, April 2005.

\bibitem{palmer05}
D.~M. {Palmer} et~al.
\newblock {A giant {$\gamma$}-ray flare from the magnetar SGR 1806 - 20}.
\newblock {\em Nature}, 434:1107--1109, April 2005.

\bibitem{terasawa05}
T.~{Terasawa}, Y.~T. {Tanaka}, Y.~{Takei}, N.~{Kawai}, A.~{Yoshida},
  K.~{Nomoto}, I.~{Yoshikawa}, Y.~{Saito}, Y.~{Kasaba}, T.~{Takashima},
  T.~{Mukai}, H.~{Noda}, T.~{Murakami}, K.~{Watanabe}, Y.~{Muraki},
  T.~{Yokoyama}, and M.~{Hoshino}.
\newblock {Repeated injections of energy in the first 600ms of the giant flare
  of SGR1806 - 20}.
\newblock {\em Nature}, 434:1110--1111, April 2005.

\bibitem{RXTEGCN}
E.~{Smith}, J.~{Swank}, C.~{Markwardt}, Y.~{Rephaeli}, D.~{Gruber},
  M.~{Persic}, and R.~{Rothschild}.
\newblock {SGR1806--20: RXTE-PCA observation of the 041227 super-flare.}
\newblock {\em GRB Coordinates Network}, 2927, 2005.

\bibitem{KonusWindGCN}
E.~{Mazets}, S.~{Golenetskii}, R.~{Aptekar}, D.~{Frederiks}, V.~{Pal'Shin}, and
  T.~{Cline}.
\newblock {The giant outburst from SGR 1806--20.}
\newblock {\em GRB Coordinates Network}, 2922, 2004.

\bibitem{IntegralGCN}
J.~{Borkowski}, D.~{Gotz}, S.~{Mereghetti}, N.~{Mowlavi}, S.~{Shaw}, and
  M.~{Turler}.
\newblock {Giant flare from SGR 1806--20 detected by INTEGRAL.}
\newblock {\em GRB Coordinates Network}, 2920, 2004.

\bibitem{RHESSIGCN}
S.~{Boggs}, K.~{Hurley}, D.~M. {Smith}, R.~P. {Lin}, G.~{Hurford}, W.~{Hajdas},
  and C.~{Wigger}.
\newblock {SGR 1806--20, RHESSI observations of the 041227 giant flare.}
\newblock {\em GRB Coordinates Network}, 2936, 2005.

\bibitem{SwiftGCN}
D.~{Palmer} et~al.
\newblock {SGR1806--20: Swift-BAT observation of the 041227 super-flare.}
\newblock {\em GRB Coordinates Network}, 2925, 2004.

\bibitem{inan07}
U.~S. {Inan}, N.~G. {Lehtinen}, R.~C. {Moore}, K.~{Hurley}, S.~{Boggs}, D.~M.
  {Smith}, and G.~J. {Fishman}.
\newblock {Massive disturbance of the daytime lower ionosphere by the giant
  {$\gamma$}-ray flare from magnetar SGR 1806-20}.
\newblock {\em \grl}, 34:8103--+, April 2007.

\bibitem{MoonGCN}
S.~{Golenetskii}, R.~{Aptekar}, E.~{Mazets}, V.~{Pal'Shin}, D.~{Frederiks}, and
  T.~{Cline}.
\newblock {Detection of the SGR 1806--20 giant outburst back-scattered by the.}
\newblock {\em GRB Coordinates Network}, 2923, 2004.

\bibitem{mereghetti05}
S.~{Mereghetti}, D.~{G{\"o}tz}, I.~F. {Mirabel}, and K.~{Hurley}.
\newblock {INTEGRAL discovery of persistent hard X-ray emission from the Soft
  Gamma-ray Repeater SGR 1806--20}.
\newblock {\em Astronomy and Astrophysics}, 433:L9--L12, April 2005.

\bibitem{IPNGCN}
K.~{Hurley}, T.~{Cline}, I.~{Mitrofanov}, S.~{Charyshnikov}, V.~{Grinkov},
  A.~{Kozyrev}, M.~{Litvak}, A.~{Sanin}, W.~{Boynton}, C.~{Fellows},
  K.~{Harshman}, C.~{Shinohara}, and R.~{Starr}.
\newblock {IPN localization of giant flare from SGR1806--20.}
\newblock {\em GRB Coordinates Network}, 2921, 2004.

\bibitem{mereghetti08}
S.~{Mereghetti}.
\newblock {The strongest cosmic magnets: Soft Gamma-ray Repeaters and Anomalous
  X-ray Pulsars}.
\newblock {\em \aapr}, 15:225--287, 2008.

\bibitem{woods04a}
P.~M. {Woods} and C.~{Thompson}.
\newblock Soft gamma repeaters and anomalous x-ray pulsars: Magnetar
  candidates.
\newblock In W.~G.~H. {Lewin} and M.~{van der Klis}, editors, {\em Compact
  Stellar X-Ray Sources}. Cambridge Univ. Press, Cambridge, 2004.

\bibitem{gcn8112}
S.~T. {Holland } et~al.
\newblock {GRB 080822: Swift detection of a short burst}.
\newblock {\em GRB Coordinates Network}, 8112, 2008.

\bibitem{kouveliotou98}
C.~{Kouveliotou}, S.~{Dieters}, T.~{Strohmayer}, J.~{van Paradijs}, G.~J.
  {Fishman}, C.~A. {Meegan}, K.~{Hurley}, J.~{Kommers}, I.~{Smith}, D.~{Frail},
  and T.~{Murakami}.
\newblock {An X-ray pulsar with a superstrong magnetic field in the soft
  gamma-ray repeater SGR 1806--20.}
\newblock {\em Nature}, 393:235--237, 1998.

\bibitem{gcn8113}
S.~{Barthelmy } et~al.
\newblock {New Soft Gamma Repeater 0501+4516 was GRB 080822}.
\newblock {\em GRB Coordinates Network}, 8113, 2008.

\bibitem{gcn8115}
D.~{Palmer} and S.~{Barthelmy}.
\newblock {Third event from SGR 0501+4516}.
\newblock {\em GRB Coordinates Network}, 8115, 2008.

\bibitem{gcn8149}
B.~M. {Gaensler } and S.~{Chatterjee}.
\newblock {SGR 0501+4516: Proximity to supernova remnant HB9}.
\newblock {\em GRB Coordinates Network}, 8149, 2008.

\bibitem{leahy95}
D.~A. {Leahy} and B.~{Aschenbach}.
\newblock {ROSAT X-ray observations of the supernova remnant HB 9.}
\newblock {\em \aap}, 293:853--858, January 1995.

\bibitem{mazets79}
E.~P. {Mazets} et~al.
\newblock {Observations of a flaring X-ray pulsar in Dorado}.
\newblock {\em Nature}, 282:587--589, December 1979.

\bibitem{hurley99}
K.~{Hurley} et~al.
\newblock {A giant periodic flare from the soft gamma-ray repeater SGR
  1900+14.}
\newblock {\em Nature}, 397:41--43, 1999.

\bibitem{gcn8148}
P.~A. {Evans } and J.~P. {Osborne}.
\newblock {Enhanced Swift-XRT position of SGR 0501+4516}.
\newblock {\em GRB Coordinates Network}, 8148, 2008.

\bibitem{gcn8166}
P.~M. {Woods }, E.~{Gogus}, and C.~{Kouveliotou}.
\newblock {SGR 0501+4516: Spin Down Rate and Inferred Dipole Magnetic Field}.
\newblock {\em GRB Coordinates Network}, 8166, 2008.

\bibitem{kaplan01}
D.~L. {Kaplan}, S.~R. {Kulkarni}, M.~H. {van Kerkwijk}, R.~E. {Rothschild},
  R.~L. {Lingenfelter}, D.~{Marsden}, R.~{Danner}, and T.~{Murakami}.
\newblock {Hubble Space Telescope Observations of SGR 0526-66: New Constraints
  on Accretion and Magnetar Models}.
\newblock {\em \apj}, 556:399--407, July 2001.

\bibitem{kulkarni03}
S.~R. {Kulkarni}, D.~L. {Kaplan}, H.~L. {Marshall}, D.~A. {Frail},
  T.~{Murakami}, and D.~{Yonetoku}.
\newblock {The Quiescent Counterpart of the Soft Gamma-Ray Repeater SGR
  0526-66}.
\newblock {\em \apj}, 585:948--954, March 2003.

\bibitem{klose04}
S.~{Klose}, A.~A. {Henden}, U.~{Geppert}, J.~{Greiner}, H.~H. {Guetter}, D.~H.
  {Hartmann}, C.~{Kouveliotou}, C.~B. {Luginbuhl}, B.~{Stecklum}, and F.~J.
  {Vrba}.
\newblock {A Near-Infrared Survey of the N49 Region around the Soft Gamma
  Repeater SGR 0526-66}.
\newblock {\em \apjl}, 609:L13--L16, July 2004.

\bibitem{wachter04}
S.~{Wachter}, S.~K. {Patel}, C.~{Kouveliotou}, P.~{Bouchet}, F.~{{\"O}zel},
  A.~F. {Tennant}, P.~M. {Woods}, K.~{Hurley}, W.~{Becker}, and P.~{Slane}.
\newblock {Precise Localization of the Soft Gamma Repeater SGR 1627-41 and the
  Anomalous X-Ray Pulsar AXP 1E1841-045 with Chandra}.
\newblock {\em \apj}, 615:887--896, November 2004.

\bibitem{kaplan02a}
D.~L. {Kaplan}, D.~W. {Fox}, S.~R. {Kulkarni}, E.~V. {Gotthelf}, G.~{Vasisht},
  and D.~A. {Frail}.
\newblock {Precise Chandra Localization of the Soft Gamma-Ray Repeater SGR
  1806--20}.
\newblock {\em \apj}, 564:935--940, January 2002.

\bibitem{corbel04}
S.~{Corbel} and S.~S. {Eikenberry}.
\newblock {The connection between W31, SGR 1806--20, and LBV 1806--20:
  Distance, extinction, and structure}.
\newblock {\em \aap}, 419:191--201, May 2004.

\bibitem{fuchs99}
Y.~{Fuchs}, F.~{Mirabel}, S.~{Chaty}, A.~{Claret}, C.~J. {Cesarsky}, and D.~A.
  {Cesarsky}.
\newblock {ISO observations of the environment of the soft gamma-ray repeater
  SGR 1806-20}.
\newblock {\em \aap}, 350:891--899, October 1999.

\bibitem{frail99}
D.~A. {Frail}, S.~R. {Kulkarni}, and J.~S. {Bloom}.
\newblock {An outburst of relativistic particles from the soft gamma-ray
  repeater SGR 1900+14}.
\newblock {\em Nature}, 398:127, 1999.

\bibitem{kaplan02b}
D.~L. {Kaplan}, S.~R. {Kulkarni}, D.~A. {Frail}, and M.~H. {van Kerkwijk}.
\newblock {Deep Radio, Optical, and Infrared Observations of SGR 1900+14}.
\newblock {\em \apj}, 566:378--386, February 2002.

\bibitem{vrba00}
F.~J. {Vrba}, C.~B. {Luginbuhl}, A.~A. {Henden}, H.~H. {Guetter}, and D.~H.
  {Hartmann}.
\newblock {Search for photometric variability in the vicinity of SGR 1900+14
  and discovery of a high-mass cluster}.
\newblock In R.~M. {Kippen}, R.~S. {Mallozzi}, and G.~J. {Fishman}, editors,
  {\em Gamma-ray Bursts, 5th Huntsville Symposium}, volume 526 of {\em American
  Institute of Physics Conference Series}, pages 809--813, September 2000.

\bibitem{aptekar01}
R.~L. {Aptekar}, D.~D. {Frederiks}, S.~V. {Golenetskii}, V.~N. {Il'inskii},
  E.~P. {Mazets}, V.~D. {Pal'shin}, P.~S. {Butterworth}, and T.~L. {Cline}.
\newblock {Konus Catalog of Soft Gamma Repeater Activity: 1978 to 2000}.
\newblock {\em \apj, Supplement}, 137:227--277, December 2001.

\bibitem{gogus01}
E.~{G{\"o}{\u g}{\"u}{\c s}}, C.~{Kouveliotou}, P.~M. {Woods}, C.~{Thompson},
  R.~C. {Duncan}, and M.~S. {Briggs}.
\newblock {Temporal and Spectral Characteristics of Short Bursts from the Soft
  Gamma Repeaters 1806-20 and 1900+14}.
\newblock {\em \apj}, 558:228--236, September 2001.

\bibitem{kouveliotou87}
C.~{Kouveliotou}, J.~P. {Norris}, T.~L. {Cline}, B.~R. {Dennis}, U.~D. {Desai},
  L.~E. {Orwig}, E.~E. {Fenimore}, R.~W. {Klebesadel}, J.~G. {Laros}, J.-L.
  {Atteia}, M.~{Boer}, K.~{Hurley}, M.~{Neil}, G.~{Vedrenne}, A.~V.
  {Kuznetsov}, R.~A. {Sunyaev}, and O.~V. {Terekhov}.
\newblock {SMM hard X-ray observations of the soft gamma-ray repeater 1806-20}.
\newblock {\em \apjl}, 322:L21--L25, November 1987.

\bibitem{cheng96}
B.~{Cheng}, R.~I. {Epstein}, R.~A. {Guyer}, and A.~C. {Young}.
\newblock {Earthquake-like behaviour of soft {$\gamma$}-ray repeaters}.
\newblock {\em Nature}, 382:518--520, August 1996.

\bibitem{gogus99}
E.~{G{\"o}{\u g}{\"u}{\c S} }, P.~M. {Woods}, C.~{Kouveliotou}, J.~{van
  Paradijs}, M.~S. {Briggs}, R.~C. {Duncan}, and C.~{Thompson}.
\newblock {Statistical Properties of SGR 1900+14 Bursts}.
\newblock {\em \apjl}, 526:L93--L96, December 1999.

\bibitem{woods04b}
P.~M. {Woods}.
\newblock {The dynamic behavior of soft gamma repeaters}.
\newblock {\em Advances in Space Research}, 33:630--637, 2004.

\bibitem{olive03}
J.-F. {Olive}, K.~{Hurley}, J.-P. {Dezalay}, J.-L. {Atteia}, C.~{Barraud},
  N.~{Butler}, G.~B. {Crew}, J.~{Doty}, G.~{Ricker}, R.~{Vanderspek}, D.~Q.
  {Lamb}, N.~{Kawai}, A.~{Yoshida}, Y.~{Shirasaki}, T.~{Sakamoto},
  T.~{Tamagawa}, K.~{Torii}, M.~{Matsuoka}, E.~E. {Fenimore}, M.~{Galassi},
  T.~{Tavenner}, T.~Q. {Donaghy}, and C.~{Graziani}.
\newblock {FREGATE observation of a strong burst from SGR1900+14}.
\newblock In G.~R. {Ricker} and R.~K. {Vanderspek}, editors, {\em Gamma-Ray
  Burst and Afterglow Astronomy 2001: A Workshop Celebrating the First Year of
  the HETE Mission}, volume 662 of {\em American Institute of Physics
  Conference Series}, pages 82--87, April 2003.

\bibitem{cline82b}
T.~L. {Cline}.
\newblock {A review of the 1979 March 5 transient}.
\newblock In R.~E. {Lingenfelter}, H.~S. {Hudson}, and D.~M. {Worall}, editors,
  {\em Gamma Ray Transients and Related Astrophysical Phenomena}, volume~77 of
  {\em American Institute of Physics Conference Series}, pages 17--33, 1982.

\bibitem{cline82}
T.~L. {Cline}, U.~D. {Desai}, B.~J. {Teegarden}, W.~D. {Evans}, R.~W.
  {Klebesadel}, J.~G. {Laros}, C.~{Barat}, K.~{Hurley}, M.~{Niel}, and M.~C.
  {Weisskopf}.
\newblock {Precise source location of the anomalous 1979 March 5 gamma-ray
  transient}.
\newblock {\em \apjl}, 255:L45--L48, April 1982.

\bibitem{golenetskii84}
S.~V. {Golenetskii}, V.~N. {Ilinskii}, and E.~P. {Mazets}.
\newblock {Recurrent bursts in GBS0526 - 66, the source of the 5 March 1979
  gamma-ray burst}.
\newblock {\em \nat}, 307:41--43, January 1984.

\bibitem{fenimore81}
E.~E. {Fenimore}, W.~D. {Evans}, R.~W. {Klebesadel}, J.~G. {Laros}, and
  J.~{Terrell}.
\newblock {Spectral evolution of the 5 March 1979 gamma burst}.
\newblock {\em \nat}, 289:42--+, January 1981.

\bibitem{inan99}
U.~S. {Inan}, N.~G. {Lehtinen}, S.~J. {Lev-Tov}, M.~P. {Johnson}, T.~F. {Bell},
  and K.~{Hurley}.
\newblock {Ionization of the lower ionosphere by {$\gamma$}-rays from a
  magnetar: Detection of a low energy (3-10 keV) component}.
\newblock {\em \grl}, 26:3357--3360, November 1999.

\bibitem{woods99}
P.~M. {Woods}, C.~{Kouveliotou}, J.~{van Paradijs}, M.~H. {Finger},
  C.~{Thompson}, R.~C. {Duncan}, K.~{Hurley}, T.~{Strohmayer}, J.~{Swank}, and
  T.~{Murakami}.
\newblock {Variable Spin-Down in the Soft Gamma Repeater SGR 1900+14 and
  Correlations with Burst Activity}.
\newblock {\em \apjl}, 524:L55--L58, October 1999.

\bibitem{rea06}
N.~{Rea}, G.~L. {Israel}, S.~{Mereghetti}, A.~{Tiengo}, S.~{Zane},
  R.~{Turolla}, and L.~{Stella}.
\newblock {Magnetars' Giant Flares: the Case of SGR 1806 20}.
\newblock {\em Chinese Journal of Astronomy and Astrophysics Supplement},
  6(1):010000--158, December 2006.

\bibitem{israel05}
G.~L. {Israel}, T.~{Belloni}, L.~{Stella}, Y.~{Rephaeli}, D.~E. {Gruber},
  P.~{Casella}, S.~{Dall'Osso}, N.~{Rea}, M.~{Persic}, and R.~E. {Rothschild}.
\newblock {The Discovery of Rapid X-Ray Oscillations in the Tail of the SGR
  1806--20 Hyperflare}.
\newblock {\em \apjl}, 628:L53--L56, July 2005.

\bibitem{gaensler05}
B.~M. {Gaensler}, C.~{Kouveliotou}, J.~D. {Gelfand}, G.~B. {Taylor},
  D.~{Eichler}, R.~A.~M.~J. {Wijers}, J.~{Granot}, E.~{Ramirez-Ruiz}, Y.~E.
  {Lyubarsky}, R.~W. {Hunstead}, D.~{Campbell-Wilson}, A.~J. {van der Horst},
  M.~A. {McLaughlin}, R.~P. {Fender}, M.~A. {Garrett}, K.~J. {Newton-McGee},
  D.~M. {Palmer}, N.~{Gehrels}, and P.~M. {Woods}.
\newblock {An expanding radio nebula produced by a giant flare from the
  magnetar SGR 1806--20}.
\newblock {\em Nature}, 434:1104--1106, April 2005.

\bibitem{boggs06}
S.~E. {Boggs}, A.~{Zoglauer}, E.~{Bellm}, K.~{Hurley}, R.~P. {Lin}, D.~M.
  {Smith}, C.~{Wigger}, and W.~{Hajdas}.
\newblock {The Giant Flare of December 27, 2004 from SGR 1806--20}.
\newblock {\em ArXiv e-prints}, November 2006.

\bibitem{watts06website}
A.~{Watts} and T.~{Strohmayer}.
\newblock {Neutron starquake shakes RHESSI}.
\newblock January 2006.

\bibitem{strohmayer05}
T.~E. {Strohmayer} and A.~L. {Watts}.
\newblock {Discovery of Fast X-Ray Oscillations during the 1998 Giant Flare
  from SGR 1900+14}.
\newblock {\em \apjl}, 632:L111--L114, October 2005.

\bibitem{guidorzi04}
C.~{Guidorzi}, F.~{Frontera}, E.~{Montanari}, M.~{Feroci}, L.~{Amati},
  E.~{Costa}, and M.~{Orlandini}.
\newblock {Comparative study of the two large flares from SGR1900+14 with the
  BeppoSAX Gamma-Ray Burst Monitor}.
\newblock {\em \aap}, 416:297--310, March 2004.

\bibitem{olive04}
J.-F. {Olive}, K.~{Hurley}, T.~{Sakamoto}, J.-L. {Atteia}, G.~{Crew},
  G.~{Ricker}, G.~{Pizzichini}, C.~{Barraud}, and N.~{Kawai}.
\newblock {Time-resolved X-Ray Spectral Modeling of an Intermediate Burst from
  SGR 1900+14 Observed by HETE-2 FREGATE and WXM}.
\newblock {\em \apj}, 616:1148--1158, December 2004.

\bibitem{hurley99c}
K.~{Hurley}, C.~{Kouveliotou}, P.~{Woods}, T.~{Cline}, P.~{Butterworth},
  E.~{Mazets}, S.~{Golenetskii}, and D.~{Frederics}.
\newblock {Reactivation and Precise Interplanetary Network Localization of the
  Soft Gamma Repeater SGR 1900+14}.
\newblock {\em \apjl}, 510:L107--L109, January 1999.

\bibitem{mazets99}
E.~P. {Mazets}, T.~L. {Cline}, R.~L. {Aptekar'}, P.~S. {Butterworth}, D.~D.
  {Frederiks}, S.~V. {Golenetskii}, V.~N. {Il'Inskii}, and V.~D. {Pal'Shin}.
\newblock {Activity of the soft gamma repeater SGR 1900 + 14 in 1998 from
  Konus-Wind observations: 1. Short recurrent bursts}.
\newblock {\em Astronomy Letters}, 25:628--634, October 1999.

\bibitem{israel08}
G.~L. {Israel}, P.~{Romano}, V.~{Mangano}, S.~{Dall'Osso}, G.~{Chincarini},
  L.~{Stella}, S.~{Campana}, T.~{Belloni}, G.~{Tagliaferri}, A.~J. {Blustin},
  T.~{Sakamoto}, K.~{Hurley}, S.~{Zane}, A.~{Moretti}, D.~{Palmer},
  C.~{Guidorzi}, D.~N. {Burrows}, N.~{Gehrels}, and H.~A. {Krimm}.
\newblock {A Swift Gaze into the 2006 March 29 Burst Forest of SGR 1900+14}.
\newblock {\em \apj}, 685:1114--1128, October 2008.

\bibitem{gcn7777}
D.~{Palmer} et~al.
\newblock {Reactivation of SGR 1627-41}.
\newblock {\em GRB Coordinates Network}, 7777, 2008.

\bibitem{atel1549}
P.M. {Woods}, C.~{Kouveliotou}, E.~{Gogos}, K.~{Hurley}, and J.~{Tomsick}.
\newblock {RXTE Observations of Renewed Burst Activity from SGR 1627-41}.
\newblock {\em Atel}, 1549, 2008.

\bibitem{golenetskii87}
S.~V. {Golenetskiy}, R.~L. {Aptekar}, Y.~A. {Guryan}, V.~N. {Ilinskiy}, and
  Y.~P. {Mazets}.
\newblock {Observations of gamma bursts from GBS 0526-66}.
\newblock {\em JPRS Report Science Technology USSR Space}, 5:54--+, November
  1987.

\bibitem{hurley00}
K.~{Hurley}.
\newblock {The 4.5+/-0.5 soft gamma repeaters in review}.
\newblock In R.~M. {Kippen}, R.~S. {Mallozzi}, and G.~J. {Fishman}, editors,
  {\em Gamma-ray Bursts, 5th Huntsville Symposium}, volume 526 of {\em American
  Institute of Physics Conference Series}, pages 763--770, September 2000.

\bibitem{woods99b}
P.~M. {Woods}, C.~{Kouveliotou}, J.~{van Paradijs}, K.~{Hurley}, R.~M.
  {Kippen}, M.~H. {Finger}, M.~S. {Briggs}, S.~{Dieters}, and G.~J. {Fishman}.
\newblock {Discovery of a New Soft Gamma Repeater, SGR 1627-41}.
\newblock {\em \apjl}, 519:L139--L142, July 1999.

\bibitem{gcn8118}
E.~{Gogus} et~al.
\newblock {Discovery of the Spin Period of the New Soft Gamma Repeater SGR
  0501+4516}.
\newblock {\em GRB Coordinates Network}, 8118, 2008.

\bibitem{gcn8146}
V.~{Mangano } et~al.
\newblock {SGR 0501+4516: Swift XRT measure of the spin period}.
\newblock {\em GRB Coordinates Network}, 8146, 2008.

\bibitem{atel1683}
D.~{Palmer}.
\newblock {SGR 0501+4516: Decline in Activity}.
\newblock {\em Atel}, 1683, 2008.

\bibitem{gcn8139}
G.~J. {Fishman } et~al.
\newblock {SGR 0501+4516: Fermi GBM (formerly GLAST Burst Monitor) observations
  of three exceptionally intense outbursts}.
\newblock {\em GRB Coordinates Network}, 8139, 2008.

\bibitem{cline00}
T.~{Cline}, D.~D. {Frederiks}, S.~{Golenetskii}, K.~{Hurley}, C.~{Kouveliotou},
  E.~{Mazets}, and J.~{van Paradijs}.
\newblock {Observations of a Possible New Soft Gamma Repeater, SGR 1801-23}.
\newblock {\em \apj}, 531:407--410, March 2000.

\bibitem{marsden01}
D.~{Marsden} and N.~E. {White}.
\newblock {Correlations between Spectral Properties and Spin-down Rate in Soft
  Gamma-Ray Repeaters and Anomalous X-Ray Pulsars}.
\newblock {\em \apjl}, 551:L155--L158, April 2001.

\bibitem{oosterbroek98}
T.~{Oosterbroek}, A.~N. {Parmar}, S.~{Mereghetti}, and G.~L. {Israel}.
\newblock {The two-component X-ray spectrum of the 6.4 S pulsar 1E
  1048.1-5937}.
\newblock {\em \aap}, 334:925--930, June 1998.

\bibitem{woods01}
P.~M. {Woods}, C.~{Kouveliotou}, E.~{G{\"o}{\u g}{\"u}{\c s}}, M.~H. {Finger},
  J.~{Swank}, D.~A. {Smith}, K.~{Hurley}, and C.~{Thompson}.
\newblock {Evidence for a Sudden Magnetic Field Reconfiguration in Soft Gamma
  Repeater 1900+14}.
\newblock {\em \apj}, 552:748--755, May 2001.

\bibitem{woods02}
P.~M. {Woods}, C.~{Kouveliotou}, E.~{G{\"o}{\u g}{\"u}{\c s}}, M.~H. {Finger},
  J.~{Swank}, C.~B. {Markwardt}, K.~{Hurley}, and M.~{van der Klis}.
\newblock {Large Torque Variations in Two Soft Gamma Repeaters}.
\newblock {\em \apj}, 576:381--390, September 2002.

\bibitem{woods07}
P.~M. {Woods}, C.~{Kouveliotou}, M.~H. {Finger}, E.~{G{\"o}{\u g}{\"u}{\c s}},
  C.~A. {Wilson}, S.~K. {Patel}, K.~{Hurley}, and J.~H. {Swank}.
\newblock {The Prelude to and Aftermath of the Giant Flare of 2004 December 27:
  Persistent and Pulsed X-Ray Properties of SGR 1806--20 from 1993 to 2005}.
\newblock {\em \apj}, 654:470--486, January 2007.

\bibitem{watts06}
A.~L. {Watts} and T.~E. {Strohmayer}.
\newblock {Detection with RHESSI of High-Frequency X-Ray Oscillations in the
  Tail of the 2004 Hyperflare from SGR 1806--20}.
\newblock {\em \apjl}, 637:L117--L120, February 2006.

\bibitem{barat83}
C.~{Barat}, R.~I. {Hayles}, K.~{Hurley}, M.~{Niel}, G.~{Vedrenne}, U.~{Desai},
  V.~G. {Kurt}, V.~M. {Zenchenko}, and I.~V. {Estulin}.
\newblock {Fine time structure in the 1979 March 5 gamma ray burst}.
\newblock {\em \aap}, 126:400--402, October 1983.

\bibitem{matone07}
B.~Abbott et~al.
\newblock {Search for gravitational wave radiation associated with the
  pulsating tail of the SGR 1806--20 hyperflare of 27 December 2004 using
  LIGO}.
\newblock {\em \prd}, 76:062003, 2007.

\bibitem{strohmayer06}
T.~E. {Strohmayer} and A.~L. {Watts}.
\newblock {The 2004 Hyperflare from SGR 1806--20: Further Evidence for Global
  Torsional Vibrations}.
\newblock {\em ArXiv Astrophysics e-prints}, August 2006.

\bibitem{evans80}
W.~D. {Evans}, R.~W. {Klebesadel}, J.~G. {Laros}, T.~L. {Cline}, U.~D. {Desai},
  B.~J. {Teegarden}, G.~{Pizzichini}, K.~{Hurley}, M.~{Niel}, and
  G.~{Vedrenne}.
\newblock {Location of the gamma-ray transient event of 1979 March 5}.
\newblock {\em \apjl}, 237:L7--L9, April 1980.

\bibitem{hurley99b}
K.~{Hurley}, P.~{Li}, C.~{Kouveliotou}, T.~{Murakami}, M.~{Ando},
  T.~{Strohmayer}, J.~{van Paradijs}, F.~{Vrba}, C.~{Luginbuhl}, A.~{Yoshida},
  and I.~{Smith}.
\newblock {ASCA Discovery of an X-Ray Pulsar in the Error Box of SGR 1900+14}.
\newblock {\em \apjl}, 510:L111--L114, January 1999.

\bibitem{kulkarni94}
S.~R. {Kulkarni}, D.~A. {Frail}, N.~E. {Kassim}, T.~{Murakami}, and
  G.~{Vasisht}.
\newblock {The Radio Nebula of the Soft Gamma-Ray Repeater 1806--20}.
\newblock {\em Nature}, 368:129--+, March 1994.

\bibitem{kulkarni95}
S.~R. {Kulkarni}, K.~{Matthews}, G.~{Neugebauer}, I.~N. {Reid}, M.~H. {van
  Kerkwijk}, and G.~{Vasisht}.
\newblock {Optical and infrared observations of SGR 1806-20}.
\newblock {\em \apjl}, 440:L61--L64, February 1995.

\bibitem{vankerkwijk95}
M.~H. {van Kerkwijk}, S.~R. {Kulkarni}, K.~{Matthews}, and G.~{Neugebauer}.
\newblock {A luminous companion to SGR 1806-20}.
\newblock {\em \apjl}, 444:L33--L35, May 1995.

\bibitem{eikenberry04}
S.~S. {Eikenberry}, K.~{Matthews}, J.~L. {LaVine}, M.~A. {Garske}, D.~{Hu},
  M.~A. {Jackson}, S.~G. {Patel}, D.~J. {Barry}, M.~R. {Colonno}, J.~R.
  {Houck}, J.~C. {Wilson}, S.~{Corbel}, and J.~D. {Smith}.
\newblock {Infrared Observations of the Candidate LBV 1806-20 and Nearby
  Cluster Stars1,}.
\newblock {\em \apj}, 616:506--518, November 2004.

\bibitem{eikenberry01}
S.~S. {Eikenberry}, M.~A. {Garske}, D.~{Hu}, M.~A. {Jackson}, S.~G. {Patel},
  D.~J. {Barry}, M.~R. {Colonno}, and J.~R. {Houck}.
\newblock {Possible Infrared Counterparts to the Soft Gamma-Ray Repeater SGR
  1806-20}.
\newblock {\em \apjl}, 563:L133--L137, December 2001.

\bibitem{cameron05}
P.~B. {Cameron}, P.~{Chandra}, A.~{Ray}, S.~R. {Kulkarni}, D.~A. {Frail}, M.~H.
  {Wieringa}, E.~{Nakar}, E.~S. {Phinney}, A.~{Miyazaki}, M.~{Tsuboi},
  S.~{Okumura}, N.~{Kawai}, K.~M. {Menten}, and F.~{Bertoldi}.
\newblock {Detection of a radio counterpart to the 27 December 2004 giant flare
  from SGR 1806--20}.
\newblock {\em Nature}, 434:1112--1115, April 2005.

\bibitem{corbel97}
S.~{Corbel}, P.~{Wallyn}, T.~M. {Dame}, P.~{Durouchoux}, W.~A. {Mahoney},
  O.~{Vilhu}, and J.~E. {Grindlay}.
\newblock {The Distance of the Soft Gamma Repeater SGR 1806-20}.
\newblock {\em \apj}, 478:624--+, March 1997.

\bibitem{vasisht94}
G.~{Vasisht}, S.~R. {Kulkarni}, D.~A. {Frail}, and J.~{Greiner}.
\newblock {Supernova remnant candidates for the soft gamma-ray repeater
  1900+14}.
\newblock {\em \apjl}, 431:L35--L38, August 1994.

\bibitem{milne79}
D.~K. {Milne}.
\newblock {A new catalogue of galactic SNRs corrected for distance from the
  galactic plane}.
\newblock {\em Australian Journal of Physics}, 32:83--92, March 1979.

\bibitem{kalberla05}
P.~M.~W. {Kalberla}, W.~B. {Burton}, D.~{Hartmann}, E.~M. {Arnal}, E.~{Bajaja},
  R.~{Morras}, and W.~G.~L. {P{\"o}ppel}.
\newblock {The Leiden/Argentine/Bonn (LAB) Survey of Galactic HI. Final data
  release of the combined LDS and IAR surveys with improved stray-radiation
  corrections}.
\newblock {\em \aap}, 440:775--782, September 2005.

\bibitem{vrba96}
F.~J. {Vrba}, C.~B. {Luginbuhl}, K.~C. {Hurley}, P.~{Li}, S.~R. {Kulkarni},
  M.~H. {van Kerkwijk}, D.~H. {Hartmann}, L.~E. {Campusano}, M.~J. {Graham},
  R.~G. {Clowes}, C.~{Kouveliotou}, R.~{Probst}, I.~{Gatley}, M.~{Merrill},
  R.~{Joyce}, R.~{Mendez}, I.~{Smith}, and A.~{Schultz}.
\newblock {The Double Infrared Source toward the Soft Gamma-Ray Repeater SGR
  1900+14}.
\newblock {\em \apj}, 468:225--+, September 1996.

\bibitem{duncan92}
R.~C. {Duncan} and C.~{Thompson}.
\newblock {Formation of very strongly magnetized neutron stars - Implications
  for gamma-ray bursts}.
\newblock {\em \apjl}, 392:L9--L13, June 1992.

\bibitem{kulkarni93}
S.~R. {Kulkarni} and D.~A. {Frail}.
\newblock {Identification of a supernova remnant coincident with the soft
  gamma-ray repeater SGR1806--20}.
\newblock {\em Nature}, 365:33--35, September 1993.

\bibitem{duncan00}
R.~C. {Duncan}.
\newblock {Physics in ultra-strong magnetic fields}.
\newblock In R.~M. {Kippen}, R.~S. {Mallozzi}, and G.~J. {Fishman}, editors,
  {\em Gamma-ray Bursts, 5th Huntsville Symposium}, volume 526 of {\em American
  Institute of Physics Conference Series}, pages 830--841, September 2000.

\bibitem{michel91}
F.~C. {Michel}.
\newblock {\em Theory of Neutron Star Magnetospheres}.
\newblock Univ. Chigaco Press, Chicago, 1991.

\bibitem{kouveliotou99}
C.~{Kouveliotou}, T.~{Strohmayer}, K.~{Hurley}, J.~{van Paradijs}, M.~H.
  {Finger}, S.~{Dieters}, P.~{Woods}, C.~{Thompson}, and R.~C. {Duncan}.
\newblock {Discovery of a Magnetar Associated with the Soft Gamma Repeater SGR
  1900+14}.
\newblock {\em \apjl}, 510:L115--L118, January 1999.

\bibitem{thompson96}
C.~{Thompson} and R.~C. {Duncan}.
\newblock {The Soft Gamma Repeaters as Very Strongly Magnetized Neutron Stars.
  II. Quiescent Neutrino, X-Ray, and Alfven Wave Emission}.
\newblock {\em \apj}, 473:322--+, December 1996.

\bibitem{duncanWebsite}
{http://solomon.as.utexas.edu/$\sim$duncan/magnetar.html}.

\bibitem{palmer99}
D.~M. {Palmer}.
\newblock {SGR 1806-20 Is a Set of Independent Relaxation Systems}.
\newblock {\em \apjl}, 512:L113--L116, February 1999.

\bibitem{thompson00}
C.~{Thompson}, R.~C. {Duncan}, P.~M. {Woods}, C.~{Kouveliotou}, M.~H. {Finger},
  and J.~{van Paradijs}.
\newblock {Physical Mechanisms for the Variable Spin-down and Light Curve of
  SGR 1900+14}.
\newblock {\em \apj}, 543:340--350, November 2000.

\bibitem{s5y1sgr}
B.~{Abbott} et~al.
\newblock Search for gravitational-wave bursts from soft gamma repeaters.
\newblock {\em \prl}, 101(21):211102, 2008.

\bibitem{benhar04}
O.~{Benhar}, V.~{Ferrari}, and L.~{Gualtieri}.
\newblock {Gravitational wave asteroseismology reexamined}.
\newblock {\em \prd}, 70(12):124015--+, December 2004.

\bibitem{thorne83}
B.~L. {Schumaker} and K.~S. {Thorne}.
\newblock {Torsional oscillations of neutron stars}.
\newblock {\em \mnras}, 203:457--489, May 1983.

\bibitem{ipn}
{http://ssl.berkeley.edu/ipn3}.

\bibitem{khPrivateCompleteness}
{Personal communication with K. Hurley}.

\bibitem{gcnWeb}
{http://gcn.gsfc.nasa.gov}.

\bibitem{rhessi}
{http://hessi.ssl.berkeley.edu/}.

\bibitem{swift}
{http://swift.gsfc.nasa.gov/docs/swift/swiftsc.html}.

\bibitem{wind}
{http://www-spof.gsfc.nasa.gov/istp/wind/}.

\bibitem{integral}
{http://integral.esac.esa.int/}.

\bibitem{suzaku}
{http://www.isas.jaxa.jp/e/enterp/missions/suzaku/index.shtml}.

\bibitem{gcn4946}
S.~{Golenetskii} et~al.
\newblock {Konus-Wind observation of the recent SGR 1900+14 activity}.
\newblock {\em GRB Coordinates Network}, 4946, 2006.

\bibitem{gcn5416}
K.~{Hurley} et~al.
\newblock {IPN triangulation of GRB060806 (long, exceptionally bright)}.
\newblock {\em GRB Coordinates Network}, 5416, 2006.

\bibitem{gcn5419}
K.~{Hurley} et~al.
\newblock {GRB060806 = SGR1806--20}.
\newblock {\em GRB Coordinates Network}, 5419, 2006.

\bibitem{grb060806personal}
{Personal communication with D. Frederiks}.

\bibitem{gcn5426}
S.~{Golenetskii} et~al.
\newblock {GRB060806 is similar to previous activity of SGR1806--20}.
\newblock {\em GRB Coordinates Network}, 5426, 2006.

\bibitem{thompson01}
C.~{Thompson} and R.~C. {Duncan}.
\newblock {The Giant Flare of 1998 August 27 from SGR 1900+14. II. Radiative
  Mechanism and Physical Constraints on the Source}.
\newblock {\em \apj}, 561:980--1005, November 2001.

\bibitem{astrowatch}
F.~{Raab}.
\newblock {"Astrowatch" at LIGO Hanford Observatory}.
\newblock {\em LIGO Internal Note G050122}, 2005.

\bibitem{landryPrivateCommunication}
{Personal communication with M. Landry}.

\bibitem{flareCVSastrowatchCal}
P.~{Kalmus}.
\newblock {\em mattapps Flare package, CVS tag ``open4-v3''}, 2008.
\newblock {input/calibration/H1response\_788218239.txt}.

\bibitem{V3uncertainty}
{2008 May 16 communication from B. O'Reilly to the LSC}.

\bibitem{oreillyPersonalCommunication}
{Personal communication with B. O'Reilly}.

\bibitem{v3errors}
{Calibration Team}.
\newblock {S5 V3 Error Budget}.
\newblock {\em LIGO Internal Note (document number not available)}, 2008.

\bibitem{S2S3S4GRB}
B.~Abbott et~al.
\newblock Search for gravitational waves associated with 39 gamma-ray bursts
  using data from the second, third, and fourth ligo runs.
\newblock {\em \prd}, 77(6):062004, 2008.

\bibitem{auriga05}
L.~Baggio, M.~Bignotto, M.~Bonaldi, M.~Cerdonio, L.~Conti, M.~De Rosa,
  P.~Falferi, P.~Fortini, M.~Inguscio, N.~Liguori, F.~Marin, R.~Mezzena,
  A.~Mion, A.~Ortolan, G.~A. Prodi, S.~Poggi, F.~Salemi, G.~Soranzo,
  L.~Taffarello, G.~Vedovato, A.~Vinante, S.~Vitale, and J.~P. Zendri~AURIGA
  Collaboration.
\newblock {Erratum: Upper Limits on Gravitational-Wave Emission in Association
  with the 27 Dec 2004 Giant Flare of SGR 1806--20}.
\newblock {\em \prl}, 95(13):139903, 2005.

\bibitem{review}
K.~{Cannon}, P.~{Kalmus}, and B.~{Owen}.
\newblock {SGR burst analysis review}.
\newblock {\em LIGO Internal Note T080256-00-Z}, 2008.

\bibitem{lichti00}
G.~G. {Lichti}, R.~{Georgii}, A.~{von Kienlin}, V.~{Sch{\"o}nfelder},
  C.~{Wunderer}, H.-J. {Jung}, and K.~{Hurley}.
\newblock {The {$\gamma$}-Ray Burst-Detection System of SPI}.
\newblock In M.~L. {McConnell} and J.~M. {Ryan}, editors, {\em American
  Institute of Physics Conference Series}, pages 722--+, 2000.

\bibitem{smithPrivateCommunication}
{Personal communication with D. Smith}.

\bibitem{lichtiPrivateCommunication}
{Personal communication with G. Lichti}.

\end{thebibliography}

\clearpage

\appendix
\chapter{Glossary} \label{appendix:abbreviations}

\begin{center}
\begin{longtable}{@{\extracolsep{\fill}}ll}
\caption[Abbreviations and terms relevant to the work]{Abbreviations
and terms relevant the work.} \\
    \hline \hline
Term & Definition \\
 \hline
 \endfirsthead
 \multicolumn{2}{c}
 {{\bfseries \tablename\ \thetable{} -- continued from previous page}} \\
 \hline
 Term & Definition \\
 \hline
 \endhead
 \hline \multicolumn{2}{|r|}{{Continued on next page}} \\
 \hline
 \endfoot
 \hline \hline
 \endlastfoot

ACIGA & Australian Consortium for Interferometric Gravitational
Astronomy \\
ADC & Analog to Digital Converter \\
AEG & Analysis Event Generator \\
AIGO & Australian Interferometric Gravitational Observatory \\
AIGRC & Australian International Gravitational Research Centre \\
AXP & Anomalous X-ray Pulsar \\
BAT & Burst Alert Telescope, detector on Swift satellite \\
BATSE & detector on NASA's Compton Gamma-Ray Observatory satellite \\
BBH & Binary Black Hole \\
BH & Black Hole \\
BNS & Binary Neutron Star \\
BP & Bandpass \\
BS & Beam Splitter \\
CBC & Compact Binary Coalescence \\
Chandra & Chandra X-ray observatory satellite  \\
cWB & Coherent WaveBurst analysis pipeline \\
D1 & Detector 1 \\
D2 & Detector 2 \\
DARM & Differential Arm \\
DAC & Digital to Analog Converter \\
DAQ & Data Acquisition \\
DCC & Document Control Center \\
DFT & Discrete Fourier Transform \\
DMT & Data Monitor Tool \\
DQ & Data Quality \\
EndX & X-arm ETM on an interferometer \\
EndY & Y-arm ETM on an interferometer \\
EOS & Equation of State \\
ETM & End Test Mass \\
GCN & GRB Coordinate Network \\
GR & General Theory of Relativity \\
GRB & Gamma Ray Burst \\
GW & Gravitational Wave \\
H1 & LIGO Hanford Observatory 4\,km detector\\
H1X & H1 detector X-arm end station \\
H1Y & H1 detector Y-arm end station \\
H2 & LIGO Hanford Observatory 2\,km detector\\
H2X & H2 detector X-arm end station \\
H2Y & H2 detector Y-arm end station \\
HR & Highly Reflective (optical coating) \\
HW & Hardware \\
IBAS & INTEGRAL Burst Alert System \\
IFO & interferometer \\
IPN & InterPlanetary Network \\
ITM & Input Test Mass \\
Konus & Gamma-ray detector on Wind satellite \\
L1 & LIGO Livingston Observatory 4\,km detector\\
L1X & L1 detector X-arm end station \\
L1Y & L1 detector Y-arm end station \\
LAL & LIGO Algorithm Library \\
LBV & Luminous Blue Variable \\
LCGT & Large Scale Cryogenic Gravitational Wave Telescope \\
LHO & LIGO Hanford Observatory \\
LIGO & Laser Interferometer Gravitational-Wave Observatory \\
LISA & Laser Interferometer Space Antenna \\
LLO & LIGO Livingston Observatory \\
LMC & Large Magellanic Cloud \\
LSC & LIGO Scientific Collaboration \\
LTF & Likelihood Time-Frequency map \\
MDC & Mock Data Challenge \\
NIST & National Institute of Standards and Technology \\
nfft & Fourier transform length in samples \\
NS & Neutron Star \\
OSEM & Optical Shadow Sensor and Electromagnetic Actuator \\
OTTB & Optically Thin Thermal Bremsstrahlung \\
overlap & Fourier transform overlap \\
pcal & photon calibrator \\
PD & Photodetector \\
PBH & Primordial Black Hole \\
PSL & Pre-Stabilized Laser \\
PSR & Pulsar \\
QPO & Quasiperiodic Oscillation \\
RD & Ringdown \\
RDL & Linear ringdown \\
RDC & Circular ringdown \\
RHESSI & Ramaty High Energy Solar Spectroscopic Imager (satellite) \\
RXTE & Rossi X-Ray Timing Explorer (satellite) \\
S5 & LIGO's fifth science run \\
S5y1 & First year of S5 14 Nov. 2005 to 14 Nov. 2006 \\
SG & Sine-Gaussian \\
SGR & Soft Gamma Repeater \\
SNR & Signal to Noise Ratio \\
SNR & Supernova Remnant \\
SPI & Suspension Point Interferometer \\
SR & Special Theory of Relativity \\
SRD & Science Requirements Document \\
Swift &  IPN satellite with BAT detector \\
TAMA & Tokyo Advanced Medium-Scale Antenna \\
TF & Time and Frequency \\
Ulysses &  solar observatory satellite \\
VLA & Very Large Array \\
WAT & Wavelet Analysis Toolbox \\
Wind & IPN satellite with Konus detector \\
WNB & band- and time-lmited White Noise Burst \\
XRT & X-Ray Telescope (on Swift satellite) \\
$Z$ & loudness statistic \\

\label{table:abbreviations}
\end{longtable}
\end{center}

\chapter{Technical Flare pipeline validations} \label{appendix:technicalValidations}

Here we present additional checks of the data
conditioning algorithm and time bookkeeping.  The Flare pipeline underwent a formal code review performed by an LSC review committee, and some of these checks were done as part of the formal review process.

\section{Data conditioning stage}
\label{section:conditioningValidation}

We first look at the effect of a 64--2048\,Hz bandpass filter on
white noise, in the frequency domain
(Figure~\ref{fig:bandpassfrequencycheck}).   This filter is representative of bandpass filters used in searches, which may have different pass bands.  In the middle of the
passband the filter gain is 1.

\begin{figure}[!h]
\includegraphics[angle=0,width=120mm, clip=false]{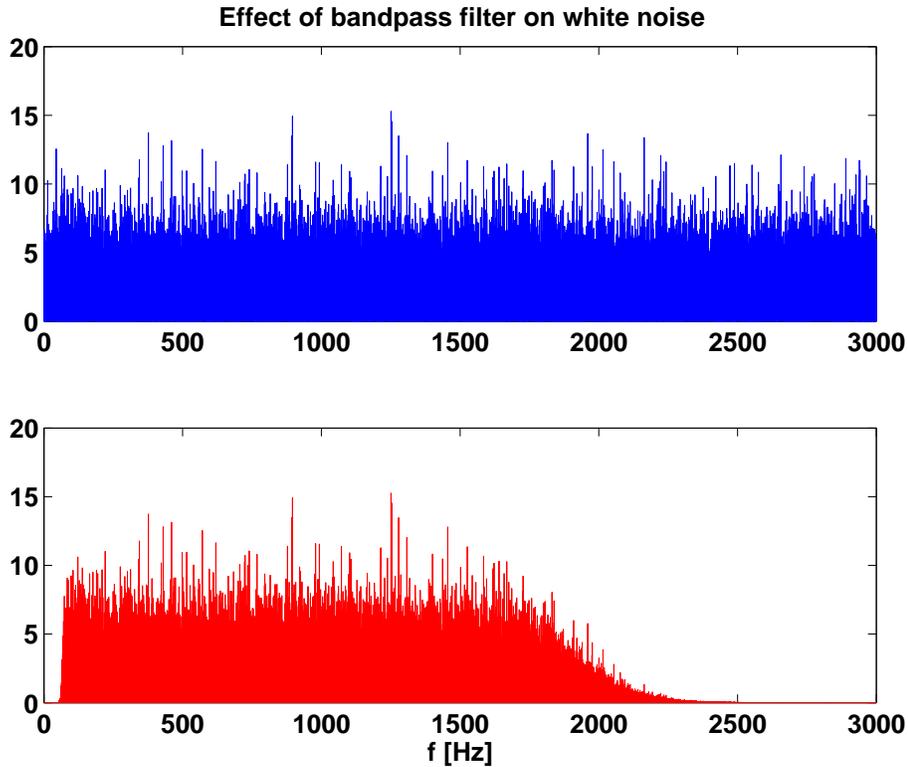}
\caption[Effect of 64--2048 bandpass filter on white noise]{Effect
of 64--2048 bandpass filter on white noise. In the middle of the
passband the filter gain is identically 1. This filter is representative of bandpass filters used in searches, which may have different pass bands.   The Flare pipeline uses any bandpass
filter specified in a configuration file.}
\label{fig:bandpassfrequencycheck}
\end{figure}

We next examine the effect of a representative notch filter.
Figure~\ref{fig:notchperiodogram} shows periodograms of unnotched
and notched simulated LIGO data (simulated data are described in
Section\,\ref{section:SimulatedData}.

Spectrograms showing the progression from bandpassed unnotched LIGO
noise, to bandpassed notched noise, to noise after whitening via the
Flare excess power algorithm described in
Section\,\ref{section:dataReductionDescription} are shown in
Figures~\ref{fig:unnotchedspectrogram}
through~\ref{fig:whitenedspectrogram}.  Several simulated large band
WNBs were injected into the noise in order to qualitatively
demonstrate the effect of filtering on signals.

\begin{figure}[!h]
\includegraphics[angle=0,width=120mm, clip=false]{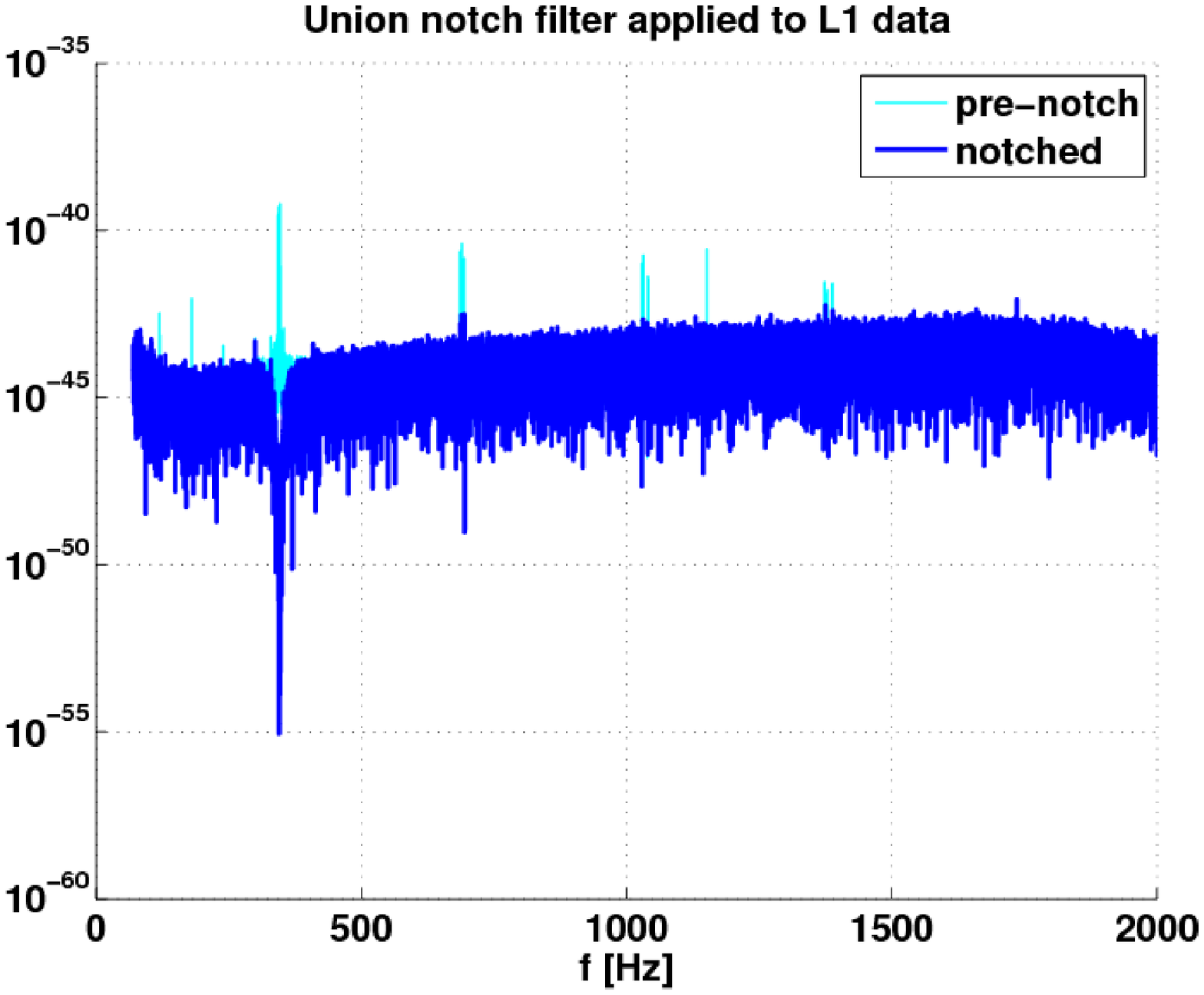}
\caption[Effect of notch filter]{Periodograms of unnotched and
notched L1 LIGO data. The notch filter was a union filter generated
automatically from LIGO H1 and L1 data. }
\label{fig:notchperiodogram}
\end{figure}

\begin{figure}[!h]
\includegraphics[angle=0,width=120mm, clip=false]{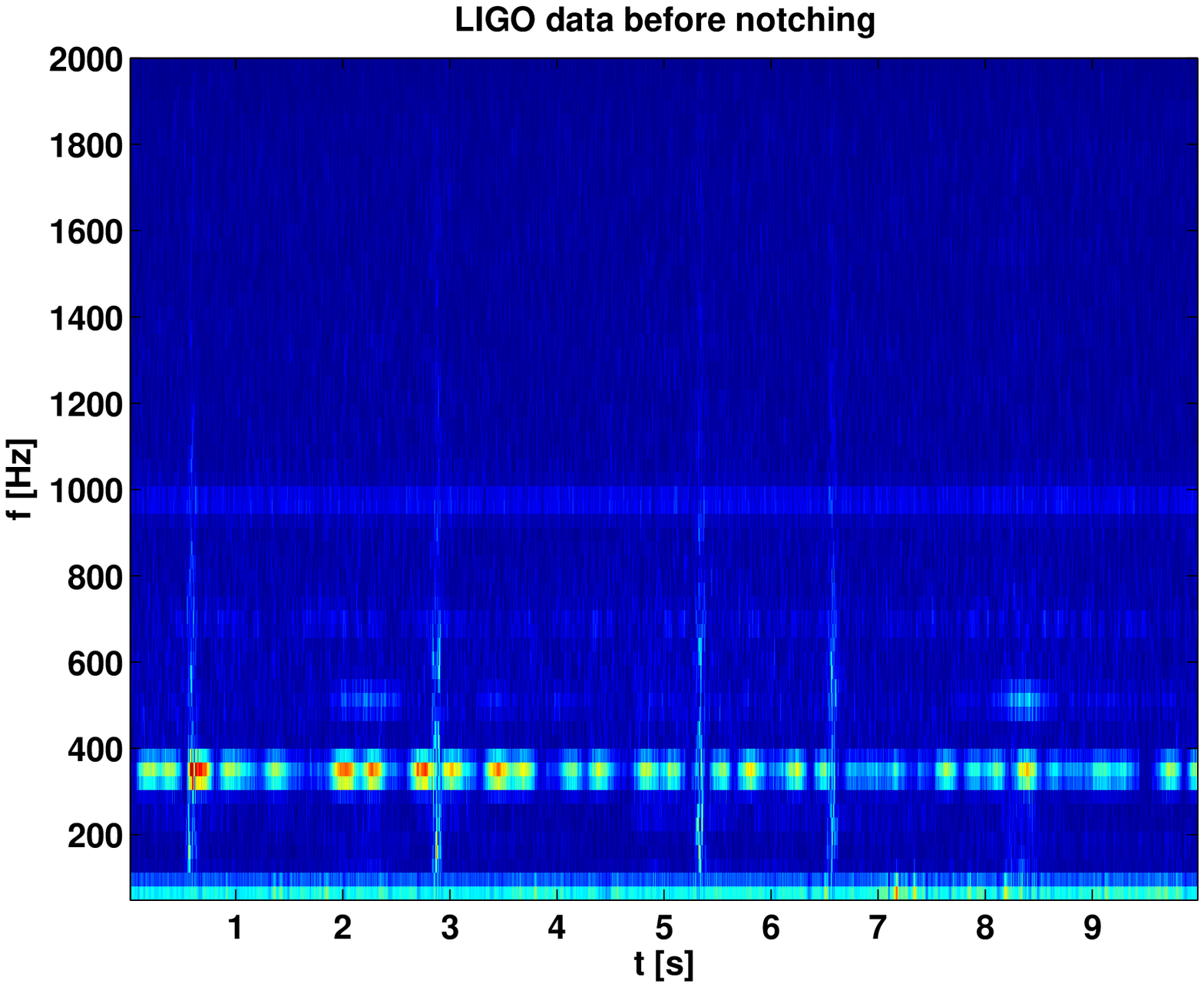}
\caption[Spectrogram of raw noise]{Spectrogram of some LIGO data
near the trigger before applying whitening or a notch filter.
Vertical stripes are white noise bursts intentionally added to the
data. } \label{fig:unnotchedspectrogram}
\end{figure}

\begin{figure}[!h]
\includegraphics[angle=0,width=120mm, clip=false]{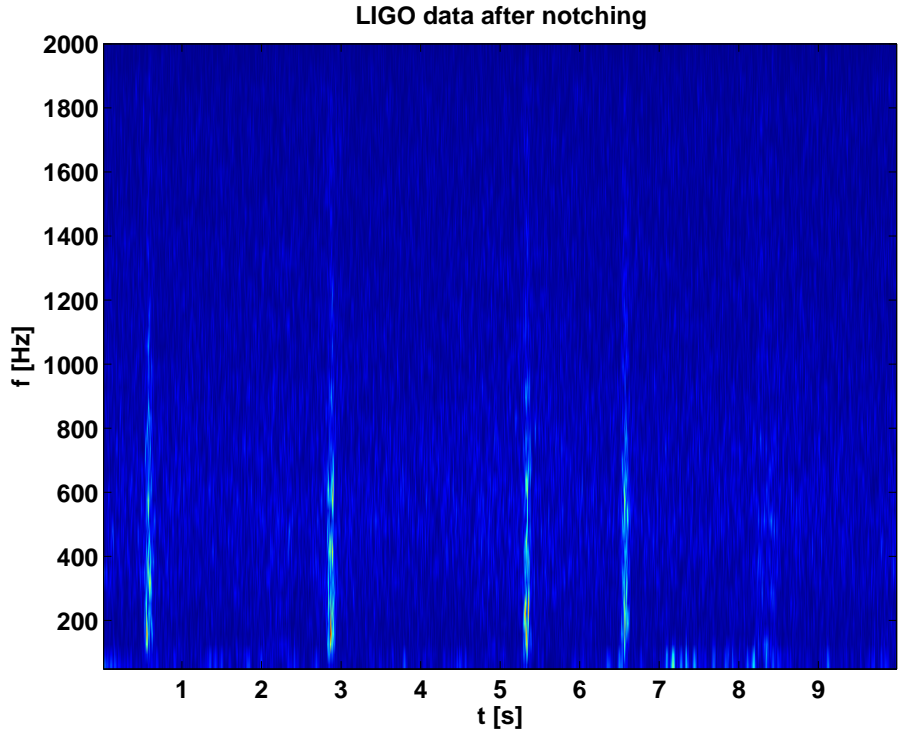}
\caption[Spectrogram after notching]{Spectrogram of the same LIGO
data near the trigger before applying whitening, and after applying
the notch filter. Vertical stripes are white noise bursts
intentionally added to the data. } \label{fig:notchedspectrogram}
\end{figure}

\begin{figure}[!h]
\includegraphics[angle=0,width=120mm, clip=false]{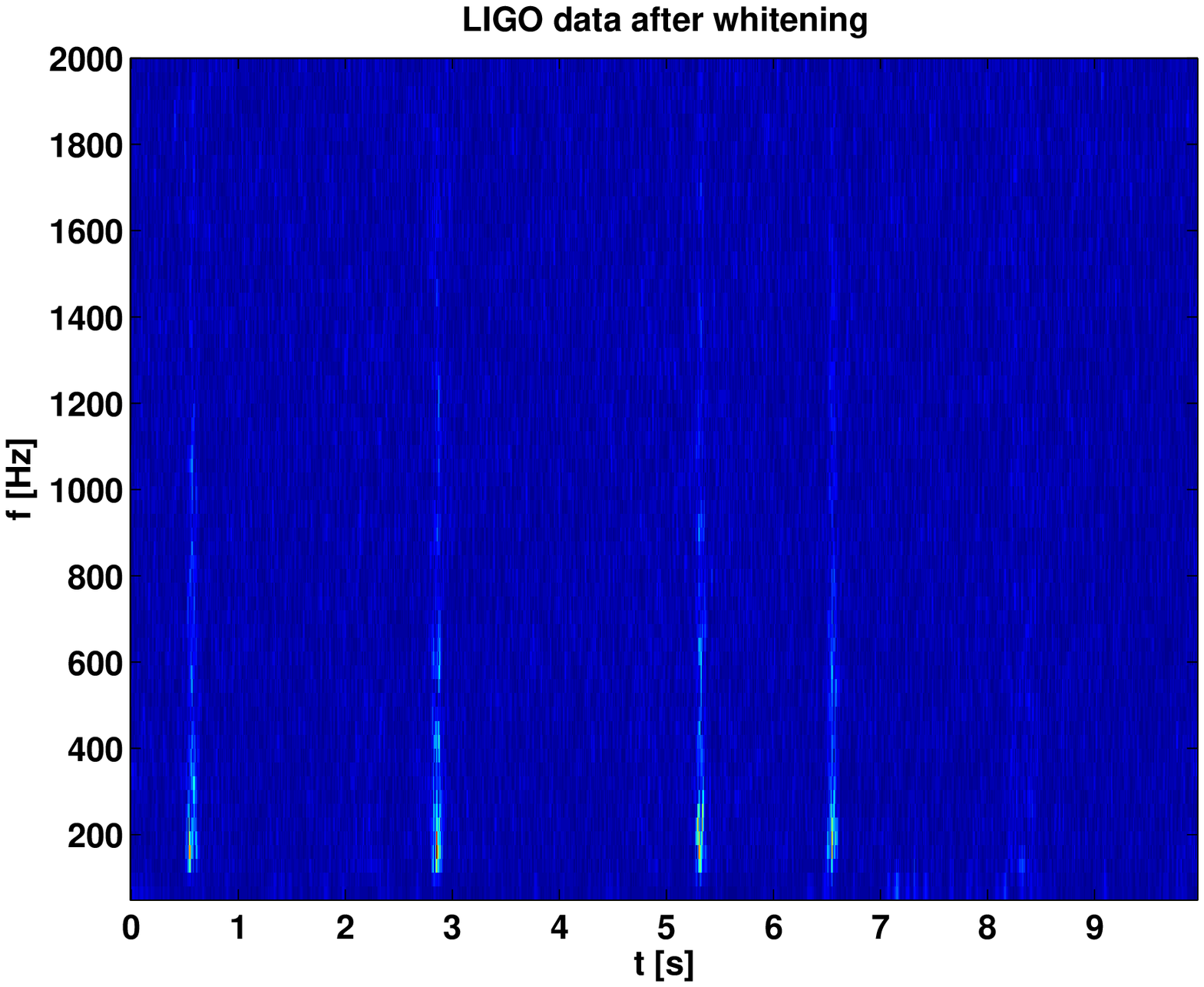}
\caption[Spectrogram after notching and whitening]{Spectrogram of
the same LIGO data near the trigger after applying whitening.
Vertical stripes are white noise bursts intentionally added to the
data. } \label{fig:whitenedspectrogram}
\end{figure}

\section{Validation of time bookkeeping} \label{subsection:timebookkepping}

We generate ASCII lists of analysis events for all simulation
recovery and on-source regions.  These lists are checked to insure
that time bookkeeping is performed correctly, that is, that analysis
event times are all contained in the expected region.  A script
produced by an LSC reviewer (K. Cannon) checks the analysis event time dumps from the
pipeline and verifies that no event lies outside of the segment in
which it was supposed to be found\,\cite{review}.  This test passes.

\section{Comparison of Flare and LAL simulations}
\label{section:flareLalSimComparison}

We produced samples of Flare pipeline simulations for
comparison to simulations produced by LAL simulation code.  A test
program written by an LSC reviewer (K. Cannon) loads two files produced by the Flare
pipeline:  a raw file containing the original injection plus- and
cross-polarization time series, and a post-injection dump file
containing the H1 and L1 time series\,\cite{review}. The test
program uses LAL injection code to convert the Flare plus- and
cross-polarization time series into a prediction of the H1 and
L1 strains.  The LAL and Flare versions of the instrument strains
are dumped together into two files, one for each instrument. Figure
\ref{fig:simLalVsFlare} shows the extent of the discrepancy.

\begin{figure}[!h]
\begin{center}
\subfigure{
\includegraphics[width=110mm,clip=false]{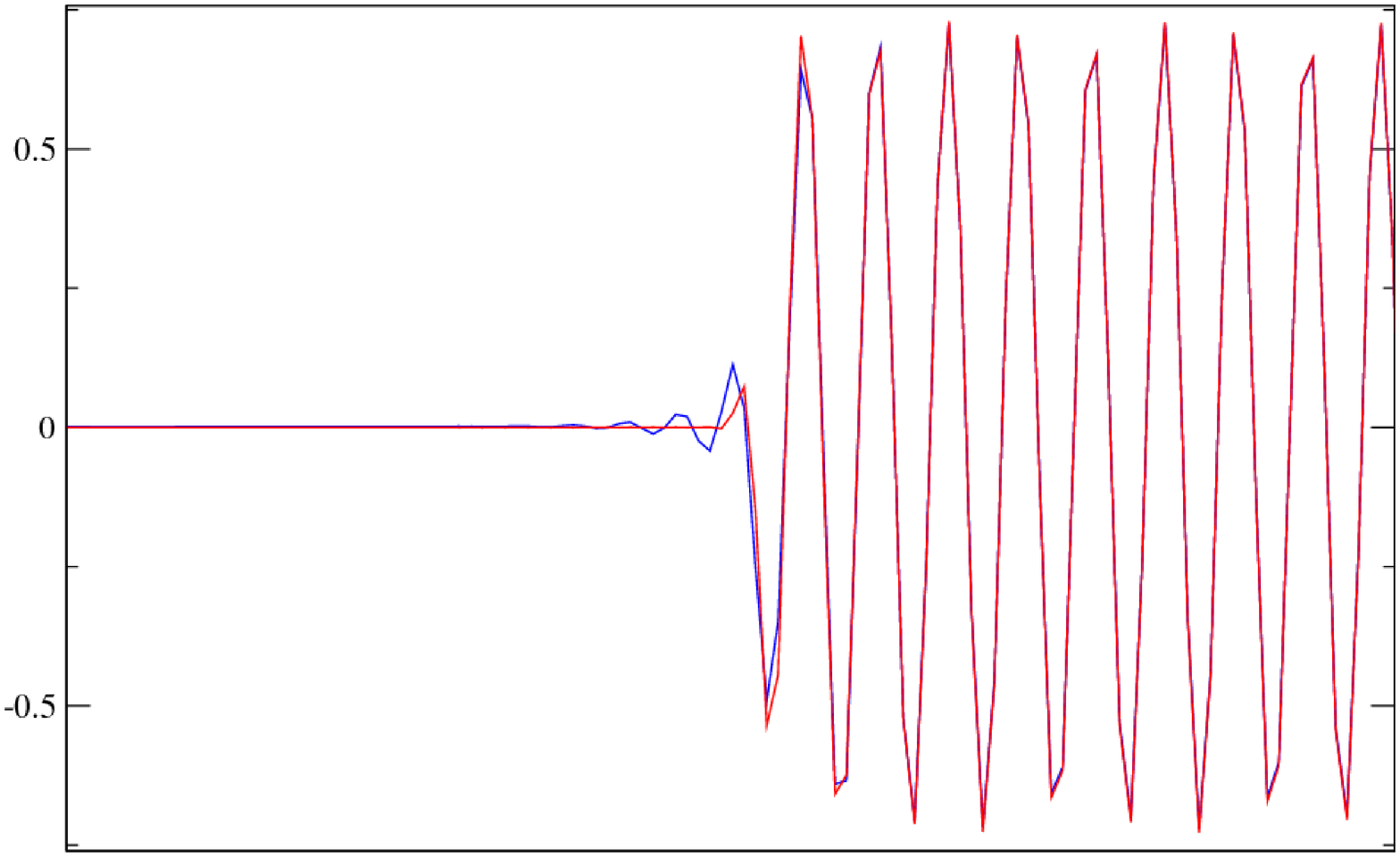}}
\subfigure{
\includegraphics[width=110mm,clip=false]{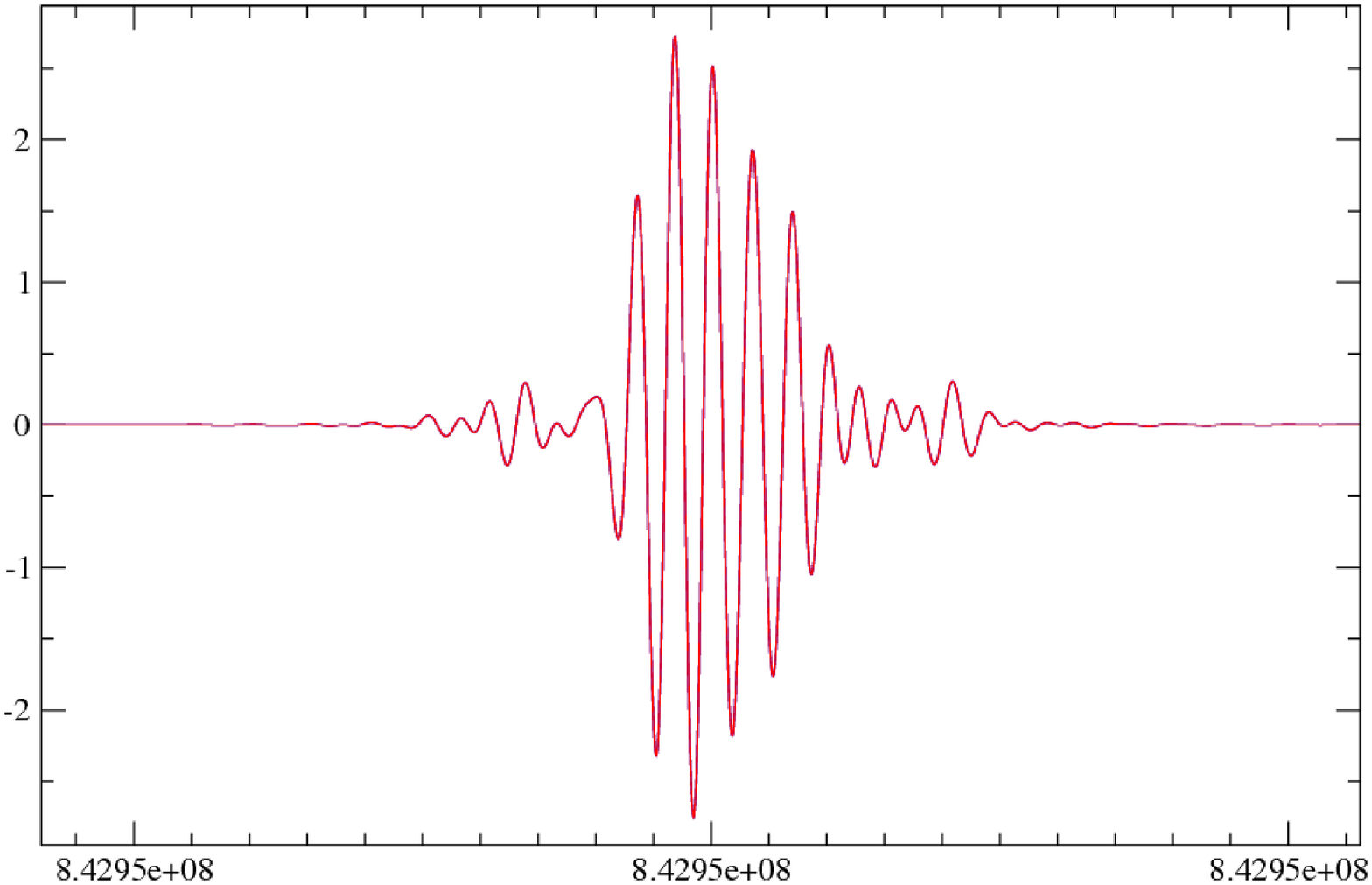}}
\caption[Comparison between simulation time series generated by
Flare pipeline and LAL]{ Comparison between simulation time series
generated by Flare pipeline and LAL.   The top plot is the
beginning of a 200\,ms 2590\,Hz ringdown.  The bottom plot is a
100\,ms 100--200\,Hz WNB.  Blue curves are the Flare pipeline time
series while red curves were produced by LAL.  These are the largest
discrepancies observed, and the plots are nearly indistinguishable.
Plots courtesy K. Cannon.} \label{fig:simLalVsFlare}
\end{center}
\end{figure}

\section{Validation of simulation $\hrss$ calculation}

Comparisons between $\hrss$ values for simulations computed by Flare
and LAL agree to within a few times double-precision epsilon.

\section{Validation of simulation $\egw/r^2$ calculation}

Flare pipeline calculated $\egw$ for the injection samples described in Section\,\ref{section:flareLalSimComparison}.  These calculations were compared to
reviewed LAL code.  The $\egw$ values computed for the waveforms
agree to within 0.1\% with the exception of the
lowest-frequency WNB waveforms, where Flare assigns an energy about
1\% higher than LAL does.  This is in the safe direction -- Flare is
either correct or computing a conservative upper limit if LAL is
correct -- and is small compared to other sources of uncertainty.
After investigation, the source of this 1\% discrepancy remains
unidentified.  

\section{Validation of antenna factor calculation}

The comparison in Section\,\ref{section:flareLalSimComparison} also provides stringent, if
qualitative, validation of Flare's antenna factor code.  In
addition, during the course of review we directly compared Flare and
LAL $\fp$ and $\fc$ antenna factors over a mesh of 167936
combinations of sky positions, polarizations, and times. The time
range spanned a UTC leap second where the correct GPS to GMST
conversion should make the Earth appear to rotate backwards
momentarily.

\section{Simulation time delays in Fourier space}

The comparison in Section\,\ref{section:flareLalSimComparison} also provides stringent, if
qualitative, validation of Flare's simulation time delay code.  In addition, the Flare pipeline code for performing time delays in
Fourier space on simulations has been incorporated into the BurstMDC
simulations engine\,\cite{burstmdc}, the primary external
simulations engine for the LSC Burst Working Group.  The Flare
pipeline code has given BurstMDC the capability of performing
subsample time delays. BurstMDC has since undergone an independent
review, further validating this function.

\section{Upper limit results from Flare vs. BurstMDC simulations}

In addition to comparisons with LAL, upper limits for 16 searches (2
GPS times and 8 simulations -- RDC and RDL) estimated using
simulations produced with Flare pipelines internal simulations
engine were compared to upper limits estimated using MDCs produced
by BurstMDC\,\cite{burstmdc}. Average agreement was better than 2\%.

\section{Validation of events list generation}
\label{section:eventsListValidation}

The Flare pipeline looks at on-source
regions which are fed to it in the events list.  Not only must Flare
do this correctly (as checked in Section\,\ref{subsection:timebookkepping}),
but the listed on-source regions must be correct as well. In addition, in order to estimate the on-source loudest event
significance and upper limits correctly, the background segment list
in the events list must be correct.

Creating the events list is described in
Section\,\ref{section:eventslist}.  There are several things to
check here.  UTC times from the IPN list must be correctly converted to
GPS times.   DQ segments must be collated correctly for the on-source and
background regions. For Konus-Wind events, timing must be propagated
correctly from satellite crossing time to Earth crossing time.

The events list was reproduced starting from the original upstream
files and using entirely independent LAL code (e.g. LAL light
travel-time functions, UTC-to-GPS conversions, segment algebra
functions etc.)\,\cite{review}. With the exception of 1\,s
discrepancies in background segment times consistent with rounding fluctuations, the
reproduction and original events lists are the same.

Additionally, in the course of procuring the SGR 1806--20 060806
event Konus-Wind light curve, D. Frederiks calculated the time
between light crossing at the geocenter and light crossing at the
satellite at 06 Aug 2006 14:23:44.12 UTC to be
5.051\,s\,\cite{grb060806personal}. We had previously calculated it
to be 5.0508\,s.  This provides external validation of the light
crossing time propagation technique. 

We further validated the
calculation using a second method based on 
trigonometry instead of rotations.
\clearpage
\chapter{Propagation of light crossing times} \label{appendix:propagation}

In this appendix we describe the method for propagation of satellite light crossing times.

In the individual SGR burst search described in Chapter\,\ref{chapter:search}, the giant flare event time was propagated from a satellite crossing
time to a crossing time at the Hanford LIGO site, while 21 of the
S5y1 events times were propagated from the Konus satellite to the
geocenter.

Propagation of the wavefront arrival time to the geocenter requires
knowledge of the SGR sky position and knowledge of the satellite
position relative to the geocenter. Propagation of the wavefront
arrival time to the gravitational wave detector requires knowledge of the position of the
detector on the Earth relative to the geocenter as well.  Positions of the
source are given in terms of right ascension and declination, and
positions of satellites an locations on earth are given in terms of
right-handed sky-fixed geocentered cartesian coordinates, with the
Z-axis pointing north and the X-axis pointing to the vernal equinox.
Converting detector latitudes and longitudes to these coordinates
requires knowledge of the sidereal time of the event.

Propagation was performed via rotation matrices.  The first rotation
around the Z-axis causes the Y-axis to be perpendicular to the
source-geocenter line, and the second rotation around the Y-axis
causes the X-axis to point to the source.  After these rotations are
applied,  we take the difference in satellite
and Earth X-coordinates and divide by the speed of light to get the
relative light crossing delay.  Validation of this technique is
described in Section\,\ref{section:eventsListValidation}.

As an example, we discuss the peak time of the giant flare event in
detail.  GCN reports 2920\,\cite{IntegralGCN} and
2936\,\cite{RHESSIGCN} assign times for the giant flare wavefront
arrival time at the INTEGRAL and RHESSI satellites respectively. The
RHESSI timing declares a 3\,ms systematic error, while the INTEGRAL
timing is considered to be accurate to within
$\sim$50\,ms~\cite{lichti00}. Therefore, we use the RHESSI time for
our trigger time, and the INTEGRAL time provides a cross check.

The time of the abrupt rise of the giant flare at RHESSI was
21:30:26.6376(30) UT on 2004 December
27\,\cite{smithPrivateCommunication}. This agrees with the time
cited in~\cite{boggs06}, 21:30:26.64, but disagrees with the time
cited in the GCN report which was
incorrect\,\cite{smithPrivateCommunication}. The light crossing time
at INTEGRAL was 21:30:26:55(5) UT\,\cite{lichtiPrivateCommunication}
which is consistent with the corrected RHESSI time given the
locations of the satellites and source.

We find that the wavefront arrived at Hanford 5.88 ms after it
arrived at the RHESSI satellite, and 81.0 ms after arrival at the
INTEGRAL satellite.

Therefore, we choose to center our search at 21:30:26.643(3) UT on
2004 December 27, corresponding to the arrival of the rising edge of
the giant flare at the detector.   This trigger time corresponds to
a GPS time of 788218239.643(3). We note that the individual burst
search does not require more than 1\,s precision in the external
trigger times; the above determination with millisecond precision
was performed before it was clear that the search was much more
tolerant of timing imprecision.
\clearpage
\oddsidemargin 0.20in
\chapter[Full Table of Upper Limits]{Full Table of Upper Limits for the Giant Flare and S5y1
flares} \label{appendix:results}

\begin{center}
\begin{scriptsize}
\begin{longtable}{@{\extracolsep{\fill}}cccclr|rcrr}
\caption[$\egwn$ and $\hrssn$ for the SGR 1806$-20$ giant flare and
the SGR bursts in S5y1]{ Strain and energy loudest event upper limit
estimates at 90\% detection efficiency ($\egwn$ and $\hrssn$) for
the SGR 1806$-20$ giant flare and the SGR bursts in the first year
of LIGO's fifth science run for various circularly/linearly
polarized RD (RDC/RDL) and WNB simulations.  Root mean square
antenna factor for a detector is defined as $F_{\mathrm{RMS}} \equiv
\sqrt{F_{+}^2 + F_{\times}^2}$. The times given in both UTC and GPS
are the geocenter crossing times of the on-source region centers.
 The ``o.s.'' column is the duration of the on-source region, centered around the respective geocenter crossing time.  Uncertainties (in
superscripts, the first and second due to detector calibration
statistical uncertainty and systematic error respectively, the third
due to monte carlo statistics) are folded into the final energy and
strain upper limit
estimates. }\\
    \hline \hline
    trigger & detector & $F_{\mathrm{rms}}^{\mathrm{H1}}$ & $F_{\mathrm{rms}}^{\mathrm{L1}}$ & type & o.s.  & \multicolumn{3}{c}{$ h_{\mathrm{rss}}^{90\%} \qquad [ 10^{-22} ~ ~ \mathrm{strain} \cdot \rthz $] }  & $E_{\mathrm{GW}}^{90\%}$ [erg]  \\
 \hline
 \endfirsthead
 \multicolumn{10}{c}
{{\bfseries \tablename\ \thetable{} -- continued from previous page}} \\ \hline    trigger & detector & $F_{\mathrm{rms}}^{\mathrm{H1}}$ & $F_{\mathrm{rms}}^{\mathrm{L1}}$ & type & o.s.  & \multicolumn{3}{c}{$ h_{\mathrm{rss}}^{90\%} \qquad [ 10^{-22} ~ ~ \mathrm{strain} \cdot \rthz $] }  & $E_{\mathrm{GW}}^{90\%}$ [erg]  \\
 \hline
 \endhead
 \hline \multicolumn{10}{|r|}{{Continued on next page}} \\
 \hline
 \endfoot
 \hline \hline
 \endlastfoot
\hline
SGR 1806$-20$ & H1 & 0.29 & -  & WNB 11ms 100-200Hz & 4 & $21.8$ & $^{ +1.3 ~ +5.6 ~ +1.2}$ &  $=28.9$ & $\sci{7.3}{47}$ \\
 & & &  & WNB 100ms 100-200Hz & 4 & $18.1$ & $^{ +1.1 ~ +4.6 ~ +0.5}$ &  $=23.9$ & $\sci{4.9}{47}$ \\
\multicolumn{3}{l}{GPS 788218239.643} &  & WNB 11ms 100-1000Hz & 4 & $50.0$ & $^{ +3.0 ~ +12.8 ~ +1.3}$ &  $=65.8$ & $\sci{5.4}{49}$ \\
\multicolumn{3}{l}{Dec 27 2004 21:30:26.643 UTC
} &  & WNB 100ms 100-1000Hz & 4 & $44.9$ & $^{ +2.7 ~ +11.5 ~ +1.1}$ &  $=59.2$ & $\sci{3.7}{49}$ \\
 & & &  & RDC 200ms 1090Hz & 4 & $59.4$ & $^{ +3.6 ~ +15.2 ~ +1.7}$ &  $=78.2$ & $\sci{2.6}{50}$ \\
 & & &  & RDC 200ms 1590Hz & 4 & $93.3$ & $^{ +5.6 ~ +23.9 ~ +2.8}$ &  $=122.9$ & $\sci{1.4}{51}$ \\
 & & &  & RDC 200ms 2090Hz & 4 & $123.9$ & $^{ +7.4 ~ +31.7 ~ +3.5}$ &  $=163.3$ & $\sci{4.2}{51}$ \\
 & & &  & RDC 200ms 2590Hz & 4 & $152.1$ & $^{ +9.1 ~ +38.9 ~ +4.1}$ &  $=200.4$ & $\sci{9.8}{51}$ \\
\end{longtable}
\end{scriptsize}
\end{center}
\setlength{\oddsidemargin}{0.60in}

\end{document}